\documentclass[prc,twocolumn,amssymb,amsmath,amsfonts,floatfix,aps]{revtex4-1}

\usepackage{graphicx}
\usepackage{multirow}
\DeclareMathAlphabet{\mathpzc}{OT1}{pzc}{m}{it}

\begin{document}

\title{QCD2019 Workshop Summary}

\newcounter{univ_counter}
\setcounter{univ_counter} {0}

\addtocounter{univ_counter} {1} 
\edef\SLAC{$^{\arabic{univ_counter}}$ } 

\addtocounter{univ_counter} {1} 
\edef\JLAB{$^{\arabic{univ_counter}}$ } 

\addtocounter{univ_counter} {1} 
\edef\NANJING{$^{\arabic{univ_counter}}$ } 

\addtocounter{univ_counter} {1} 
\edef\NANJINGINP{$^{\arabic{univ_counter}}$ } 

\addtocounter{univ_counter} {1} 
\edef\GWU{$^{\arabic{univ_counter}}$ } 

\addtocounter{univ_counter} {1} 
\edef\HEIDEL{$^{\arabic{univ_counter}}$ } 

\addtocounter{univ_counter} {1} 
\edef\LSU{$^{\arabic{univ_counter}}$ } 

\addtocounter{univ_counter} {1} 
\edef\GLASGOW{$^{\arabic{univ_counter}}$ } 

\addtocounter{univ_counter} {1} 
\edef\CUA{$^{\arabic{univ_counter}}$ } 

\addtocounter{univ_counter} {1} 
\edef\UCONN{$^{\arabic{univ_counter}}$ } 

\addtocounter{univ_counter} {1} 
\edef\INHA{$^{\arabic{univ_counter}}$ } 

\addtocounter{univ_counter} {1} 
\edef\ODU{$^{\arabic{univ_counter}}$ } 

\addtocounter{univ_counter} {1} 
\edef\SACLAY{$^{\arabic{univ_counter}}$ } 

\addtocounter{univ_counter} {1} 
\edef\INFN{$^{\arabic{univ_counter}}$ } 

\addtocounter{univ_counter} {1} 
\edef\HUELVA{$^{\arabic{univ_counter}}$ } 

\addtocounter{univ_counter} {1} 
\edef\SEVILLA{$^{\arabic{univ_counter}}$ }

\addtocounter{univ_counter} {1} 
\edef\IU{$^{\arabic{univ_counter}}$ } 

\addtocounter{univ_counter} {1} 
\edef\CRICA{$^{\arabic{univ_counter}}$ } 


\author{
S.J. Brodsky,\SLAC
V.D. Burkert,\JLAB
D.S. Carman,\JLAB
J.P. Chen,\JLAB
Z.-F. Cui,\NANJING$\!\!^,$\NANJINGINP
M. D\"oring,\JLAB$\!\!^,$\GWU \\
H.G. Dosch,\HEIDEL
J. Draayer,\LSU
L. Elouadrhiri,\JLAB
D.I. Glazier,\GLASGOW
A.N. Hiller Blin,\JLAB
T. Horn,\CUA
K. Joo,\UCONN \\
H.C. Kim,\INHA
V. Kubarovsky,\JLAB
S.E. Kuhn,\ODU
Y. Lu\NANJING$\!\!^,$\NANJINGINP
W. Melnitchouk,\JLAB \\
C. Mezrag,\SACLAY
V.I. Mokeev,\JLAB
J.W. Qiu,\JLAB
M. Radici,\INFN
D. Richards,\JLAB
C.D. Roberts,\NANJING$\!\!^,$\NANJINGINP \\
J. Rodr\'{\i}guez-Quintero,\HUELVA
J. Segovia,\NANJINGINP$\!\!^,$\SEVILLA
A.P. Szczepaniak,\JLAB$\!\!^,$\IU
G.F. de T\'eramond,\CRICA
D. Winney\IU \vspace{4mm}
} 

\affiliation{\SLAC SLAC National Accelerator Laboratory, Stanford University, Stanford, CA 94039}
\affiliation{\JLAB Thomas Jefferson National Accelerator Laboratory, Newport News, VA 23606}
\affiliation{\NANJING School of Physics, Nanjing University, Nanjing, Jiangsu 210093, China}
\affiliation{\NANJINGINP Institute for Nonperturbative Physics, Nanjing University, Nanjing, Jiangsu 210093, China}
\affiliation{\GWU Institute for Nuclear Studies and Department of Physics, The George Washington University, Washington, DC 20052}
\affiliation{\HEIDEL Institut f{\"u}r Theoretische Physik, Philosophenweg 16, 69120 Heidelberg, Germany}
\affiliation{\LSU Louisiana State University, Baton Rouge, LA 70803}
\affiliation{\GLASGOW University of Glasgow, Glasgow G12 8QQ, United Kingdom}
\affiliation{\CUA The Catholic University of America, Washington, DC 20064}
\affiliation{\UCONN The University of Connecticut, Storrs, CT 06269}
\affiliation{\INHA Department of Physics, Inha University, 22212 Incheon, The Republic of Korea}
\affiliation{\ODU Old Dominion University, Norfolk, VA 23529}
\affiliation{\SACLAY IRFU, CEA, Universit\'e Paris-Saclay, F-91191 Gif-sur-Yvette, France}
\affiliation{\INFN INFN Sezione di Pavia, via Bassi 6, I-27100 Pavia, Italy}
\affiliation{\HUELVA Department of Integrated Sciences and Center for Advanced Studies in Physics,
Mathematics and Computation; University of Huelva, E-21071 Huelva; Spain}
\affiliation{\SEVILLA Dpto. Sistemas F\'isicos, Qu\'imicos y Naturales, Univ.\ Pablo de Olavide, E-41013 Sevilla, Spain}
\affiliation{\IU Indiana University, Bloomington, IN 47405}
\affiliation{\CRICA Universidad de Costa Rica, 11501 San Pedro de Montes de Oca, Costa Rica}

\date{\today}

\thispagestyle{empty}

\begin{abstract}
  \vspace{5mm}
  The topical workshop {\it Strong QCD from Hadron Structure Experiments} took place at Jefferson Lab from Nov. 6-9, 2019. 
  Impressive progress in relating hadron structure observables to the strong QCD mechanisms has
  been achieved from the {\it ab initio} QCD description of hadron structure in a diverse array of methods in order to
  expose emergent phenomena via quasi-particle formation. The wealth of experimental data and the advances in
  hadron structure theory make it possible to gain insight into strong interaction dynamics in the regime of large
  quark-gluon coupling (the strong QCD regime), which will address the most challenging problems of the Standard Model
  on the nature of the dominant part of hadron mass, quark-gluon confinement, and the emergence of the ground and
  excited state hadrons, as well as atomic nuclei, from QCD. This workshop aimed to develop plans and to facilitate the
  future synergistic efforts between experimentalists, phenomenologists, and theorists working on studies of hadron
  spectroscopy and structure with the goal to connect the properties of hadrons and atomic nuclei available from data to
  the strong QCD dynamics underlying their emergence from QCD. These results pave the way for a future breakthrough
  extension in the studies of QCD with an Electron-Ion Collider in the U.S. 
\end{abstract}

\maketitle
~~
\vspace{8.0cm}
\thispagestyle{empty}

\clearpage

\vfil
\eject

\tableofcontents

\vfil
\eject

\newpage

\section{Preface}

Following successful meetings in France and South Korea, the third workshop on ``{\it Strong QCD from Hadron Structure Experiments}'', was held 
at Jefferson Laboratory (JLab) during the period of Nov. 6-9, 2019. It gathered a select group of world experts and early career researchers in 
hadron physics in order to forge an array of synergistic efforts. The workshop consisted of 42 invited talks and a long, concluding discussion 
session. This document is a summary of the workshop, based on contributions from the conveners of the topical sessions. It aims to provide the 
scientific background for future joint projects between the experts from these diverse areas in hadron physics.

Studies of the spectrum of hadrons and their structure in experiments with electromagnetic probes offer unique insight into many facets of the 
strong interaction in the regime of large running coupling of Quantum Chromodynamics (QCD). The experimental program that is part of the new 
12-GeV era in the four halls at Jefferson Lab in the U.S., as well as the ongoing and planned experiments at the European facilities ELSA, 
MAMI, and GSI, and the Asian facilities BES, SPring-8, and JPARC, have considerably extended the scope of research in hadron physics in joint 
efforts between experiment and phenomenological data analysis.

Studies of the excited nucleon ($N^*$) spectrum over the last decade have led to the discovery of new ``missing" resonances in the global 
multi-channel analysis of exclusive photoproduction data. Nine new $N^*$ and $\Delta^*$ resonances have been reported in the recent PDG 
edition with three- and four-star status~\cite{Tanabashi:2018oca}. Experimental studies of the pion and nucleon elastic and transition 
electromagnetic form factors, including nucleon resonance electroexcitation amplitudes, have allowed us to gain insight into the strong QCD 
dynamics underlying their mass generation~\cite{Burkert:2019bhp, Horn:2016rip, Aznauryan:2011qj, Mokeev:2018zxt, Mokeev:2019ron}. Contemporary 
knowledge on the ground state nucleon and pion structure has been extended considerably by the results on the proton and pion parton 
distribution functions (PDFs)~\cite{Accardi:2012qut, Aguilar:2019teb}.

The first results on the ground state nucleon structure in three-dimensions (3D) have started to emerge from semi-inclusive deep inelastic
scattering (SIDIS) in 3D momentum space~\cite{Bacchetta:2017gcc}. Analyses of the deeply virtual Compton scattering (DVCS) and deeply
virtual meson production (DVMP) exclusive electroproduction data have revealed the light-front structure of the 
ground state nucleon in the 3D space composed by the 1D space of the longitudinal parton momentum fraction $x$ of the nucleon and by the 
2D coordinates for the parton location in the plane transverse to the incoming photon~\cite{Dupre:2017}. Recent advances in DVCS studies 
have delivered a breakthrough result that provides insight into the parton pressure distribution in protons~\cite{Burkert:2018bqq}.

Extensive efforts on the exploration of the nucleon resonance electroexcitation amplitudes, mostly from CLAS exclusive meson electroproduction 
data, have provided unique information on many facets of strong QCD, especially connected with the generation of excited nucleon states of 
different quantum numbers with distinctively different structural features~\cite{Burkert:2019bhp, Aznauryan:2011qj}.

Impressive progress has been achieved during the past decade in relating hadron structure observables inferred from data to the strong QCD 
mechanisms underlying hadron generation. These efforts cover a broad research area from the {\it ab initio} QCD description of hadron structure 
to continuum QCD approaches capable of exposing emergent phenomena in hadron physics with a traceable connection to the QCD Lagrangian, as well 
as advances in quark models, including the hypercentral constituent quark model, relativistic approaches, light-front holography, and the chiral 
quark soliton model. Furthermore, progress in studies of the structure of atomic nuclei within symmetry-based symplectic models~\cite{Launey:2016fvy} 
opens up the opportunity to improve our understanding of the emergence of the structure of atomic nuclei from strong QCD~\cite{Draayer}.

The above sketch, which presents selected highlights from diverse areas of the research activity on the exploration of hadron structure with 
electromagnetic probes, demonstrates the promising potential of such studies to provide insights into the strong QCD dynamics underlying the 
generation of ground and excited state hadrons and to shed light on the emergence of the hadron properties from the QCD Lagrangian. In order 
to achieve these challenging objectives, synergistic efforts between experimentalists, phenomenologists, and hadron structure theorists across 
diverse research directions in hadron physics are especially critical.

\section{Executive Summary}

The impressive progress in charting the spectrum of hadrons and revealing their structure using data from experiments with electromagnetic probes 
and drawing heavily upon advances in theory has considerably extended our capabilities for gaining insight into the strong QCD dynamics underlying 
hadron generation. New and continuing efforts using these tools will address the most challenging open problems in the Standard Model, {\it e.g.} 
the emergence of hadron mass, the nature of quark-gluon confinement, and the role of dynamical chiral symmetry breaking (DCSB) in the generation of 
hadron mass and structure. It is anticipated that, with time, new and continuing studies will deliver explanations of the many facets of strong QCD 
dynamics that are expressed in hadron spectra and structure, beginning with only the QCD Lagrangian. This document provides an overview of the rapidly 
developing research paths that were highlighted by the workshop conveners as particularly promising avenues toward reaching these challenging goals.

Studies of the excited nucleon ($N^*$) spectrum over the last decade have delivered numerous breakthroughs. Several long-awaited missing resonances 
have been discovered in the global multi-channel analysis of exclusive photoproduction data with a decisive impact from the CLAS results on exclusive 
$K\Lambda$ and $K\Sigma$ photoproduction~\cite{Bur17}. Nine new $N^*$ and $\Delta^*$ resonances have been reported in the recent Particle Data Group
(PDG) edition with three- and four-star status~\cite{Tanabashi:2018oca}. In addition, a new $N'(1720)3/2^+$ resonance has been observed in combined 
studies of the CLAS $\pi^+\pi^-p$ photo- and electroproduction data~\cite{Mok20}. The first results on the electrocoupling of missing resonances have 
become available, allow for insight on the particular structural features of missing baryon states that make them so difficult to detect in experiments.

Investigations of meson and baryon spectra have highlighted the symmetries of the strong interaction that underlie the generation of hadrons. 
Completion of the excited nucleon and light meson spectrum using data from experiments with real and virtual photons will provide the ultimate 
experimental benchmarks for the theoretical description of the hadron spectrum starting from the QCD Lagrangian. Such studies are of particular
importance for understanding early Universe evolution in the transition between a deconfined mixture of gluons and quarks to a gas of hadrons. 
Highlights and prospects for mapping hadron spectra are presented in Sections~\ref{meson_spectr} and \ref{nstar_spectr}.

Experiments capitalizing on 12~GeV at JLab and studies foreseen with the U.S. Electron-Ion Collider (EIC) promise to considerably expand opportunities 
for the exploration of meson structure. The first results on the pion and kaon elastic electromagnetic form factors and their PDFs are expected in the 
still almost unexplored range of high photon virtualities $Q^2 > 5.0$~GeV$^2$. Results on pseudoscalar meson structure offer unique insights into the 
dual nature of these particles as a bound $q\bar{q}$ system and as the Goldstone boson whose emergence is driven by dynamical chiral symmetry breaking 
in the process of dressed quark mass generation. Furthermore, credible information on the pion Bethe-Salpeter amplitude constrained by experimental 
results on the pion elastic form factor and PDF offer clean insight into the dynamics of emergent hadronic mass. Comparative studies of pion and kaon 
structure will provide insight on the difference in the dressed quark mass generation for $u$, $d$, and $s$ quark flavors, which will a starting point 
for the exploration of the flavor dependence of hadron mass generation. The current status, prospects, and the impact on understanding of strong QCD 
from meson structure studies are presented in Section~\ref{Sec31}.

Consistent results on the strong QCD dynamics underlying hadron generation from independent studies of both meson and baryon structure are of particular 
importance. Contemporary knowledge of the structure of the ground state nucleon has been considerably extended by the results on their electromagnetic 
elastic form factors~\cite{Segovia:2014aza} and PDFs~\cite{Accardi:2012qut}. Analyses of the resonance electroexcitation 
amplitudes from CLAS data~\cite{Mokeev:2018zxt,Mokeev:2019ron} will extend our knowledge about the ground state nucleon PDF at large $x_B$ values in the 
resonance region. Experimental studies of the $d/u$ ratio in nucleon PDFs at $x_B$ values close to unity, which are in progress in Hall~A/C at JLab, 
offer a sensitive test for the QCD-rooted predictions of ground state nucleon structure. Data on inclusive and semi-inclusive electron scattering provide 
essential contributions to the store of detailed information on the flavor-separated quark/parton and gluon distributions in nucleons. Progress in lattice 
QCD and the development of novel quasi- and pseudo-PDF concepts pave the way for the description of nucleon PDFs starting from the QCD Lagrangian
\cite{Radyushkin:2017cyf,Chambers:2017dov,Ma:2014jla}. The results and prospects for the exploration of the structure of the ground state nucleon in 
1-dimension (1D) are presented in Section~\ref{nucleon_1d}.

The extensive efforts dedicated to the extraction of the nucleon resonance electroexcitation amplitudes, mostly from CLAS exclusive meson electroproduction 
data, have provided unique information on many facets of strong QCD as they appear in the generation of excited nucleon states of different quantum numbers 
with distinctively different structural features~\cite{Burkert:2019bhp,Aznauryan:2011qj}. During the last decade, the electroexcitation amplitudes for 
most excited nucleon states in the mass range up to 1.8~GeV have become available at photon virtualities $Q^2\leq\,$5.0~GeV$^2$
\cite{Mokeev:2018zxt,Mokeev:2019ron}. Analyses of these results within continuum QCD approaches and quark models have revealed the structure of nucleon 
resonances to emerge from a complex interplay between an inner core of three dressed quarks and an external meson-baryon cloud. A successful description 
of the electroexcitation amplitudes of the low-lying $N^*$ states achieved with continuum QCD approaches, which employ the same momentum dependence of the 
dressed quark mass inferred from the QCD Lagrangian~\cite{Segovia:2014aza,Segovia:2015hra,Mokeev:2015lda}, have demonstrated the capability of gaining 
insight into the dynamics of hadron mass generation from the combined studies of pion and nucleon electromagnetic form factors, including $N \to N^*$
electroexcitation amplitudes.

Verified predictions for meson electromagnetic form factors and Parton Distribution Functions (PDFs) with a dressed quark mass function constrained by 
data on the structure of ground and excited state nucleons will validate insight into this key ingredient of strong QCD in a nearly model-independent 
way. With the well-established dressed quark mass function and diquark correlation amplitudes from the studies of nucleon ground and excited states, 
the emergence of the shape of the nucleon in its intrinsic frame can be predicted within continuum QCD approaches, paving the way towards understanding 
the emergence of nuclear structure from strong QCD. The status and prospects for exploration of the structure of excited state nucleons are presented in 
Section~\ref{nstar_structure_intro}. Their impact on studies of the emergence of the structure of atomic nuclei from strong QCD is discussed in 
Section~\ref{sym}.

The first results on the ground state nucleon structure in 3D have started to emerge. Data from semi-inclusive meson and di-hadron
electroproduction experiments not only offer insight into the parton momentum distribution in the plane transverse to the virtual photon momentum, but 
have also provided detailed information on the correlations between the parton and nucleon spins, and between the partons' spin and their orbital 
momenta. Global analyses of these results have already produced 3D images of the parton distribution in the momentum space of the ground state
nucleons~\cite{Bacchetta:2017gcc}.

Analyses of DVCS and DVMP exclusive electroproduction data have revealed additional information about the light-front structure of the ground 
state nucleon in the 3D space composed by the 1D space of the longitudinal parton momentum fraction $x$ of the nucleon and the 2D coordinates for 
the parton location in the plane transverse to the incoming photon~\cite{Dupre:2017}. These two complementary perceptions of the ground state 
nucleon -- 3D in momentum and coordinate space -- offer new opportunities for insights into the dynamics driving the formation of the ground 
state nucleon. Such studies will drive future exploration of hadron structure at JLab in the 12~GeV era and at the future Electron-Ion Collider 
through partnerships between BNL and JLab.

Recent advances in DVCS studies have also delivered a breakthrough that reveals features of the parton pressure distribution in protons 
\cite{Burkert:2018bqq}. This information allows us to map out the components of the nucleon energy-momentum tensor underlying the balance of the 
explosive and implosive pressures within the nucleon. It opens a path toward understanding the strong QCD dynamics behind pressure generation, and 
may provide insight into the emergence of confinement.

Achievements and future potential of 3D imaging and derived insights into the energy-momentum tensor of the ground state nucleon via synergistic 
efforts using SIDIS, DVCS, and DVMP studies are discussed in Sections~\ref{semiinc} and \ref{tensor}.

Available and anticipated results on the spectrum and structure of ground and excited hadron states, and the novel potential of hadron 
femto-imaging in 3D space, offer qualitatively new opportunities for exploration of the strong QCD dynamics underlying the emergence of hadronic 
mass and gluon and quark confinement, and for developing insights into the facets of strong QCD that are responsible for generating the entire 
diverse array of hadrons with apparently distinct structural features. Strong theory support is essential in order to ensure full capitalization 
on the wealth of new data. Expanded efforts are needed in the development of reaction models and amplitude analysis approaches that are capable 
of extracting needed information from data, {\it i.e.} hadron elastic and transition electromagnetic form factors, and the 1D and 3D parton 
distributions. This activity is discussed in Sections~\ref{semiinc}, \ref{amp_electro}, and \ref{amp_photo}. The development of a multi-prong 
theoretical framework for the description of the hadron spectrum and structure and its emergence from QCD are presented in Sections~\ref{Sec37} 
and \ref{theory_out}.

Initial ideas for joint research activities between experimentalists, phenomenologists, and theorists that are aimed at drawing insights into strong 
QCD from studies of hadronic spectra and structure are outlined in the closing Section~\ref{outlook}. After further discussion within the hadron 
physics community, these ideas will be updated and developed at the next workshop in this strong QCD series, which will be hosted by the Institute 
for Nonperturbative Physics at Nanjing University in 2021. A bright future marked by rewarding collaboration on concrete new projects is anticipated.

\section{Hadron Spectra: From Strong QCD Symmetries to Cosmology}

\subsection{Highlights and Prospects in the Exploration of Meson Spectra}
\label{meson_spectr}

The turn of the century has seen a renaissance in meson spectroscopy. Unexpected results from older experiments (BELLE, BABAR, CLEO); new 
experiments specifically designed to hunt for exotica (COMPASS, GlueX); and analysis of new large datasets (LHCb, BESIII) have coincided 
to generate great excitement and opportunity in the exploration of the meson spectra.

The textbook description of hadrons postulates colorless three valence quark states as baryons with a quark-antiquark pair forming mesons. 
Quark models with these building blocks have historically done a remarkable job in reproducing the observed states. Typically such models 
take heavy spin $1/2$ constituent quarks, $m_{u,d}\simeq 1/3M_{proton}$, with an attractive potential mediated by single gluon exchange and 
a confining potential proportional to the distance between the quarks. Different couplings of the relative orbital angular momentum to the 
quark spins combined with the three light quark flavors ($u$, $d$, $s$) produces groups of meson nonets with specific $J^{PC}$ quantum numbers. 
Such constituent quark models provide an important benchmark with which to compare both experimental observations and sophisticated 
QCD-based calculations such as Lattice QCD or Light Front Holography as outlined in Section~\ref{theory_out}.

QCD however does not restrict states bound by the strong interaction to consist of only these configurations. Although it is remarkable that 
after fifty or so years of study we are still seeking to establish which types of hadrons may exist there are now significant experimental 
candidates that clearly do not fit the quark model picture alone. Such states are termed ``exotic" and their structure is likely to come 
from the following color neutral possibilities: meson-meson molecules, tetraquarks, glueballs, and hybrids. In molecules two color-singlet 
mesons are bound by virtual meson exchanges, similar to nucleons in a nucleus; while in tetraquarks a diquark and antidiquark are tightly
bound by the exchange of gluons similar to a quark and antiquark in a meson. Due to gluons being self-interacting it seems possible that 
they can form bound states with no valence quarks, giving rise to so called glueballs; while states consisting of mixed quark and gluon 
constituents are termed hybrids.

The trigger for the increased focus on meson spectroscopy was the unexpected discovery of the $X(3872)$, now known as $\chi_{c1}(3872)$ due 
to its $J^{PC} = 1^{++}$, by Belle in 2003~\cite{PhysRevLett.91.262001}. A number of experiments at this time were searching for missing 
quark model states above the open charm $D\bar{D}$ threshold and the Belle discovery was soon verified by others. Since then many more 
similar states have been found containing hidden charm $c\bar{c}$ and are commonly referred to as $X$, $Y$, and $Z$ mesons depending on 
their quantum numbers. However, none of these seem to match the expectations for the missing quark model states and so they are all regarded 
as candidates for exotic mesons. In particular the charmonium-like charged $Z_{c}(4430)$ state would require tetraquark or molecular 
compositions. While dynamic explanations, such as triangle singularities, for the observed structures seen in experiments also exist, it 
seems neither these nor specific exotic configurations can single handily explain the wealth of states now observed~\cite{Olsen2019}.
For a detailed review of the subject see Ref.~\cite{Brambilla2019}. 

In the light quark sector the lightest scalar meson nonet containing the $f_0(980)$ and $a_0(980)$ has been regarded as somewhat of a 
puzzle. Its masses are too low compared to quark model expectations and the mass ordering of the states does not follow the usual
isoscalar-isovector pattern. It is possible to explain these features via a molecular or tetraquark composition~\cite{PhysRevD.15.267}. 
Establishing a further nonet with non-exotic quantum numbers but without quark model predictions, by for example being too close in mass 
to a nonet with the same $J^{PC}$, would add further weight to this picture. For example the axial-vector states $f_{1}(1420)$ and 
$a_1(1420)$~\cite{PhysRevLett.115.082001} could be part of such a nonet as there are no free suitable quark model assignments.

Hybrid mesons can have exotic quantum numbers, that is quantum numbers that cannot be reached by straightforward quark models. Lattice 
QCD predicts 5 nonets consisting of exotic quantum numbers~\cite{Dudek:2011bn} and suggests a constituent gluon with mass around
$1-1.5$~GeV and $J^{PC}=1^{+-}$. Discovery of a nonet with quantum numbers $0^{+-}, 1^{-+}$, or $2^{+-}$ would provide conclusive 
evidence for the requirement of gluons as an underlying degree of freedom in the formation of the hadronic spectrum. To this end, the COMPASS 
experiment using a 190~GeV $\pi^-$ beam has provided strong experimental evidence for the existence of the exotic isovector $1^{-+}$. Early 
measurements with lower statistics already found indications of such a state in several final states: $3\pi$, $\eta' \pi$, $b_1\pi$, $f_1\pi$, 
for a full review see Ref.~\cite{Meyer:2015eta}. The new $3\pi$ data with 4.6M events allowed for analysis of unprecedented detail
\cite{KETZER2020103755}. In addition, the combination of the new $\eta \pi^-$ and $\eta' \pi^-$ final states allowed a sophisticated coupled 
channel analysis by JPAC to resolve two possible states, the $\pi_1(1400)$ and $\pi_1(1600)$, as coming from a single pole consistent with the 
$\pi_1(1600)$, and is presented in Section~\ref{amp_photo}. The mass of this single state is in good agreement with the Lattice calculation 
expectations for an exotic hybrid. 

A promising avenue for extending the search of exotic hybrids beyond the $1^{-+}$ isovector is to use photon beams and such is the plan at 
the upgraded Jefferson Lab with the flagship hybrid meson experiment GlueX and the MesonEx experiment with the CLAS12 detector. The spin-1 
photon acts as an effective beam of $\rho, \omega$, and $\phi$ mesons with isovector and isocalar components allowing production of states 
from all of the lightest five hybrid nonets. Charge exchange production mechanisms allow separation of isospin. In addition linear 
polarization of the photon acts as a filter for the naturality of the exchange particle giving more control over the production mechanism 
through, for example, separating $t$-channel pion and $\rho$ exchanges. 

For the already seen $\pi_1(1600)$ its decay modes to $\rho\pi$ and $\eta' \pi$ should be readily accessible and other decay modes such as 
$b_1\pi$ and $f_1\pi$ should be possible with large statistic datasets. Importantly, the isoscalar $\eta_1$ can also be measured with decays 
to $f_2 \eta$ and $a_2 \pi$, while the $\eta'_1$ can decay to $K^*K$. For the $2^{+-}$ sector decays of the $b_2$ to $\omega\pi$ and $\rho\eta$, 
and the $h_2$ to $\rho \pi$, could provide the first evidence for this nonet.

A possibility for investigating glueballs is to use the photoproduction of two identical pseudoscalar mesons, {\it i.e.} $\pi^0\pi^0$, $\eta \eta$, 
or $K_0K_0$, which filters out $f_{0,2,4}$ states to allow measurement of photocouplings and branching ratios of the scalar mesons 
\cite{Chandavar17}. 

The GlueX experiment started taking data in 2014. It features a linearly polarized tagged photon beam through coherent bremsstrahlung on a 
diamond radiator and a large acceptance detector providing coverage from $1$ to $120^{\circ}$. This allows it to measure many different 
final states in parallel and reconstruct mesons decaying to many particles. The linearly polarized portion of the beam covers photon 
energies from around 8 to 9~GeV, beyond where contributions from $s$-channel baryon resonances can provide significant background. 

The first results consist of photon beam asymmetry measurements of pseudoscalar mesons, using the linear polarization to determine the 
naturality of the exchange mechanism~\cite{PhysRevC.95.042201,PhysRevC.100.052201}. This provides a useful test of the reaction models 
that will be used to investigate the production of other mesons. In addition the photoproduction of the $J/\psi$ close to threshold
provided a means to hunt for s-channel resonances with hidden charm such as the LHCb pentaquark~\cite{PhysRevLett.123.072001}.

CLAS12 started production data taking in 2018 and first analyses are in progress~\cite{BURKERT2020163419}. In this experiment the production 
rate of mesons is increased through detecting electrons with small scattering angles in a forward calorimeter, thus tagging quasi-real photons 
with similar polarization and energy to the GlueX experiment~\cite{ACKER2020163475}. In this way similar final states to GlueX will be 
investigated giving complementary results.

\subsection{Highlights and Prospects in the Exploration of Baryon Spectra}
\label{nstar_spectr}

\begin{figure}[htp]
\includegraphics[width=0.9\columnwidth]{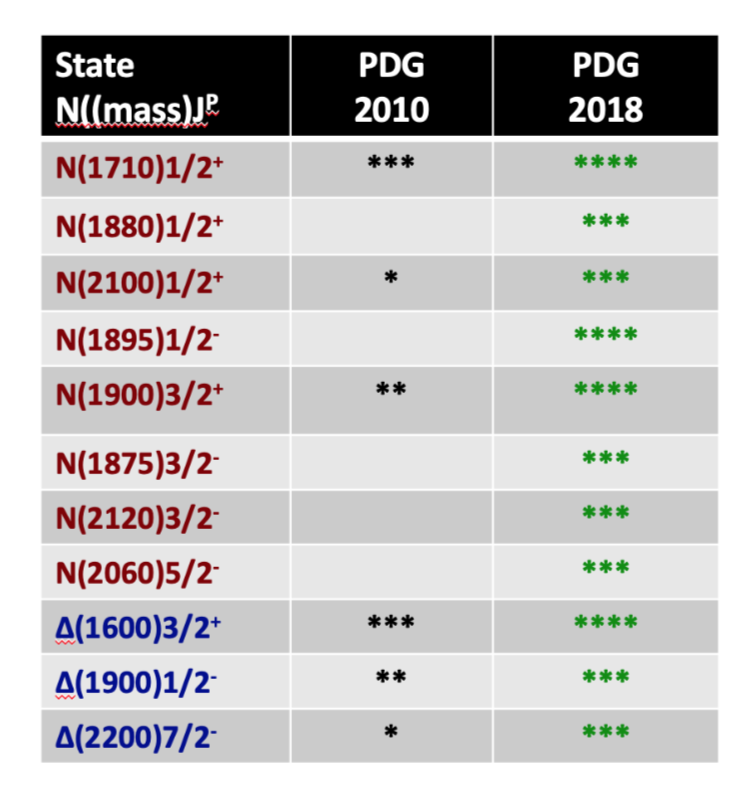}
\vspace{-0.25cm}
\caption{Evidence for 11 $N^*$ and $\Delta$ states in RPP 2010 compared with RPP 2018.}
\label{pdg2014} 
\end{figure}

Over the past six years eight light-flavored baryon states in the mass range from 1.85 to 2.15~GeV have been discovered or
evidence for the existence of states has been strengthened significantly, as visualized in the increase of the star ratings
assigned by the Particle Data Group (PDG) in their bi-annual Review of Particle Properties (RPP). To a large degree this has
been the result of adding very precise photoproduction data in open strangeness channels to the database that are included
in multi-channel partial-wave analyses. The possibility to measure polarization dependent observables with high statistics in
these processes has been critical. These discoveries shown in Fig.~\ref{pdg2014} and cataloged in the 2018 edition of the 
Review of Particle Properties~\cite{Tanabashi:2018oca}, while essential in completing the light-quark baryon spectrum, have so 
far not brought any surprises, as all the new states have quantum numbers that fit well in the spectrum predicted by 
SU(6)$\times$O(3) symmetry as well as in Lattice QCD with a pion mass of 400~MeV. Because of the large pion mass in LQCD and 
of constituent quark masses in the quark model, an agreement of mass predictions in LQCD with the physical resonance masses is 
not expected.

\begin{figure}[htp]
\includegraphics[width=0.9\columnwidth]{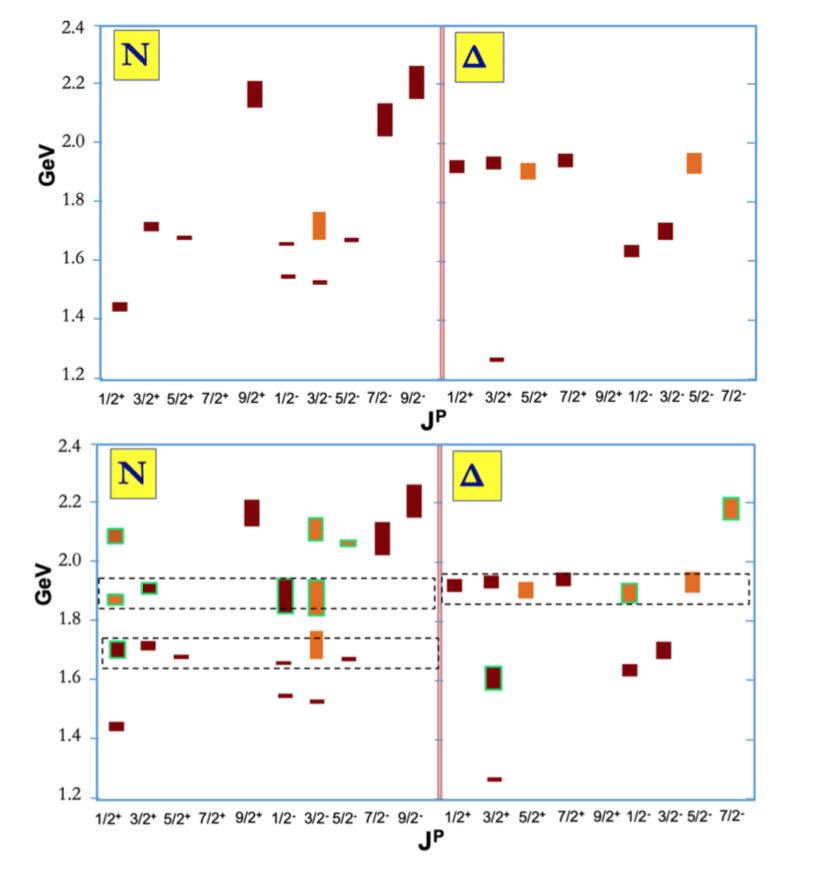}
\vspace{-0.3cm}
\caption{Top panel: Nucleon and $\Delta$ resonance spectrum below 2.2~GeV in the RPP 2010~\cite{Beringer:1900zz}.
Bottom panel: Nucleon and $\Delta$ resonance spectrum below 2.2~GeV in RPP 2018~\cite{Tanabashi:2018oca}.
The green frames highlight the new states and states with improved star ratings compared to 2010. The light brown color
indicate 3$^*$ states, the dark color indicates 4$^*$ states. The dashed frames indicate apparent mass degeneracy of states
with masses near 1.7~GeV and 1.9~GeV and different spin and parity. The Bonn-Gatchina analysis includes all of the
$K^+\Lambda$ and $K^+\Sigma^0$ photoproduction cross section and polarization data, as well as the pion photoproduction data.}
\label{pdg} 
\end{figure}

As shown in Fig.~\ref{pdg}, the resonant states appear to form mass-degenerate, spin-parity multiplets~\cite{Burkert:2019kxy}. At 
1900~MeV four of the newly established resonances in the $N^*$ sector (with green frames) form a nearly mass degenerate quartet. 
Also in the $\Delta$ sector six states are mass degenerate near 1950~MeV; five of them were known before, but a new state falls 
into the same mass window. Neither quark models nor current LQCD simulations with 400~MeV pion masses predict such a behavior.
Note that the same degeneracy is observed in the high-energy spectrum of light mesons and that some possible explanation has 
been proposed in Ref.~\cite{segovia-b662}, where the degeneracy is due to the screening of the linearly rising confinement 
potential due to the spontaneous creation of quark-antiquark pairs. It is possible that a similar mechanism could play a role in 
the baryon sector.

The more realistic tests of strong QCD will come when Lattice predictions of the light-quark baryon spectrum with physical pion
masses are undertaken, {\it i.e.} when resonances are allowed to decay. Forays in this arena have recently been undertaken in the meson
sector~\cite{Burkert:2019bhp}. Another critical aspect in testing strong QCD in baryon resonance production is the excitation
of hybrid baryons, {\it i.e.} excited states with a dominant component of glue in the wave function. In contrast to the meson sector,
gluonic baryons come with the same quantum numbers as ordinary quark model baryons, {\it i.e.} they are not exotic in the sense that
such states are not possible in the standard SU(6)$\times$O(3) quark model symmetry. We therefore will have to resort to other means
to uniquely distinguish hybrid baryons from ordinary baryons with the same quantum numbers. Searching for more states than
predicted within the quark model is one possibility, however not fully satisfying because of the remaining ambiguities.

\begin{figure}[htp]
\includegraphics[width=0.9\columnwidth]{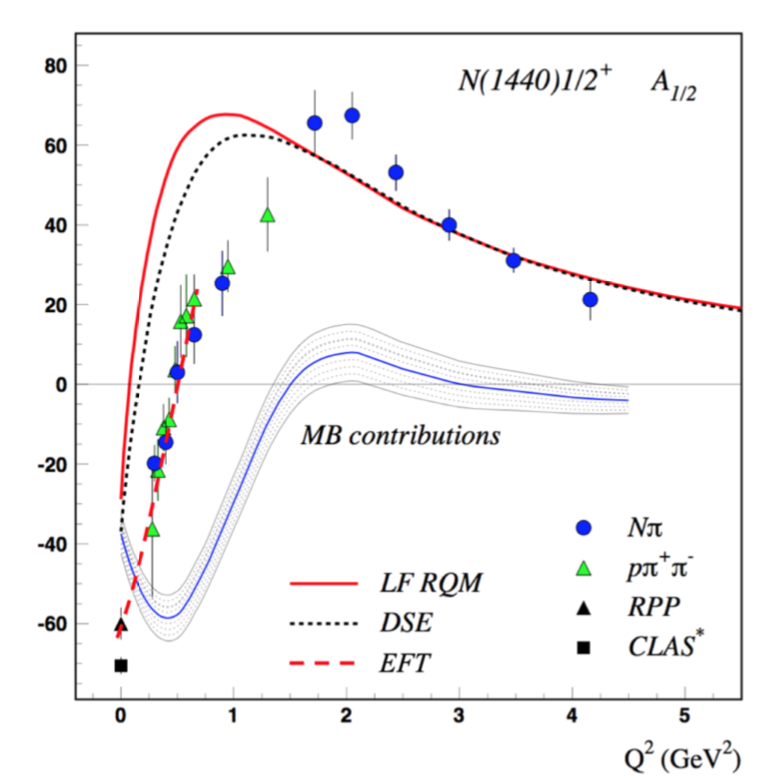}
\vspace{-0.25cm}
\caption{The transverse helicity amplitudes $A_{1/2}$ for the Roper resonance $N(1440)1/2^+$. Data are from CLAS compared to the LF-RQM 
with running quark masses (solid line), and with projections from the Dyson-Schwinger Equation (DSE/QCD) approach (dotted line). The shaded 
band indicates non-3-quark contributions inferred from the difference of the LF-RQM curve and the CLAS data. The red dashed line is the 
Effective Field Theory (EFT) calculation that describes the data at small $Q^2$.}
\label{p11} 
\end{figure}

Meson electroproduction is a possible tool that can provide separation of states with a different number of partonic degrees of
freedom, 3 in the case of quark excitations and 4 (3 quarks + 1 gluon) for gluonic excitations. The latter is expected to show a
faster drop of the transition form factor with $Q^2$, than the drop expected for regular 3-quark model states. For a realistic
separation it is essential to develop models for the transition form factor dependence of hybrid states. There is currently only
one calculation available, and that is for the Roper resonance~\cite{Li:1991yba}, which was originally considered a candidate for 
the lowest mass hybrid baryon in the nucleon sector~\cite{Aznauryan:2011qj}. This possibility has been eliminated based on the 
electroproduction data from CLAS. The transition amplitudes are in excellent agreement with the LF-RQM and with the DSE/QCD 
calculations that find that the Roper at its quark core is the first radial excitation of the nucleon~\cite{Burkert:2019bhp}. 
Figure~\ref{p11} shows the comparison of the CLAS data with the two calculations~\cite{Aznauryan:2018okk,Segovia:2015hra}, 
demonstrating good agreement of the data at $Q^2 > 2$~GeV$^2$ with the LF-RQM and DSE/QCD calculations. 

\begin{figure}[htp]
\includegraphics[width=0.9\columnwidth]{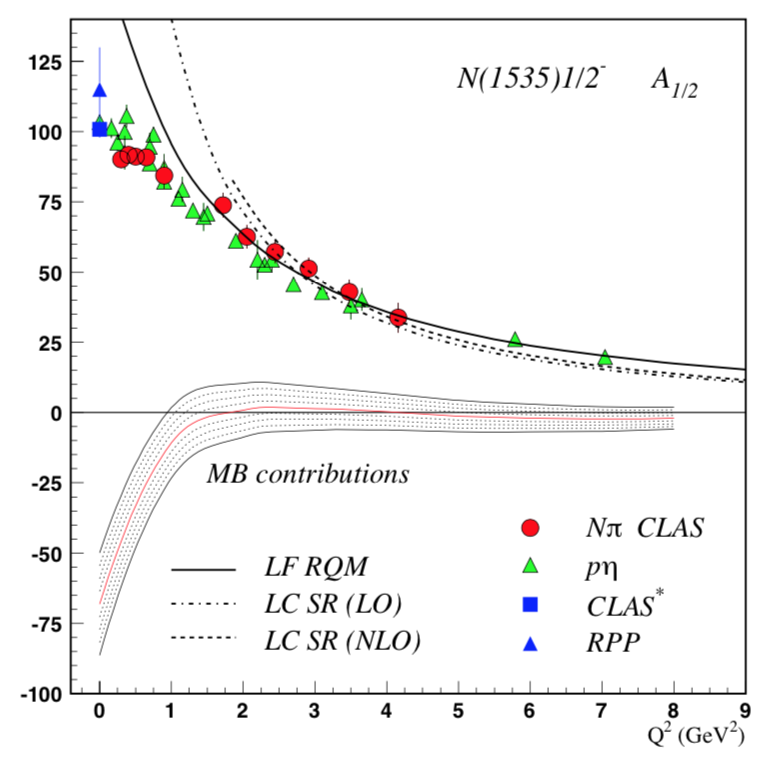}
\vspace{-0.25cm}
\caption{The transverse amplitude $A_{1/2}$ for the $N(1535)1/2^-$ resonance compared to LF-RQM calculations
(solid line) and QCD computations within the Light Cone Sum Rule approach.}
\label{s11} 
\end{figure}

Another example is the $N(1535)1/2^-$ state, where there is experimental information in the largest range of $Q^2$. In Fig.~\ref{s11} 
the data are compared with the light front relativistic quark model (LF-RQM) and the Light-Cone sum rule (SR) results in leading-order
(LO) and in next-to-leading-order (NLO)~\cite{Aznauryan:2017nkz,Anikin:2015ita}. There is obviously excellent agreement of the data with 
the LF-RQM at $Q^2 > 1.5$~GeV$^2$, and with the QCD-based results from Light-Cone LO and Light-Cone NLO at $Q^2 > 2$~GeV$^2$. The ongoing 
experiments with CLAS12 at the highest available energy of $\sim$11~GeV should allow for the extension of the range in $Q^2$ to beyond 
10~GeV$^2$ for both of these prominent excited states of the nucleon~\cite{Burkert:2018nvj}.

\subsection{Synergy Between Experiments with EM and Hadronic Probes in Hadron Spectra Exploration} 

Understanding the full nucleon spectrum, including gluonic excitations, within strong QCD will constitute a breakthrough
of similar importance as happened with the full understanding of the excitation spectrum of hydrogen and other atoms as a
consequence of the discovery of Quantum Electrodynamics (QED). Agreement between strong QCD projections and the experimental 
data of resonance masses (amplitude pole positions in the complex energy plane), width (imaginary part of the resonant
amplitude), should be commensurate with the uncertainties of the experimental data. In order to achieve such quantitative
agreement between strong QCD and experiment, much more experimental data, both in hadron scattering experiments and in
photo-/electroproduction must be produced and analyzed in multi-channel analyses. This is especially the case for multiple mesons
in the final state as in $\gamma p \to p\pi^-\pi^+$. The $N\pi \pi$ final state is the dominant final state coupling to the nucleon 
and $N^*$ resonances at masses from 1.65 to 1.9~GeV. There are very limited data in the hadron sector for this channel. 

\subsection{Strong QCD Symmetries from Hadron Spectra}

A full understanding of the excited baryon spectrum is very relevant in order to describe the transition of the microsecond old
universe from the quark-gluon plasma phase (perfect liquid) to the hadron phase of fully formed nucleons. It is now well understood
that this transition is not a simple first-order phase transition that occurs instantaneously when the boundary between two phases 
in the QCD phase diagram is crossed. 

\begin{figure}[htp]
\includegraphics[width=0.9\columnwidth]{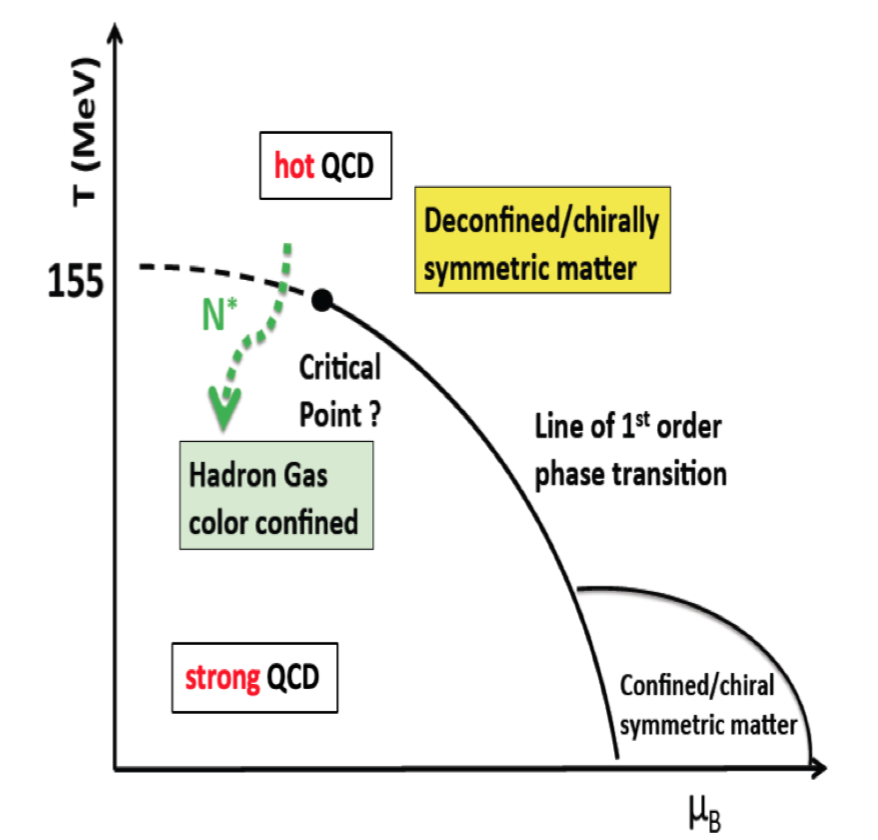}
\vspace{-0.25cm}
\caption{A generic phase diagram for the transition from the de-confined quark-gluon state to the confined hadron state.}
\label{universe} 
\end{figure}

\begin{figure}[htp]
\includegraphics[width=0.9\columnwidth]{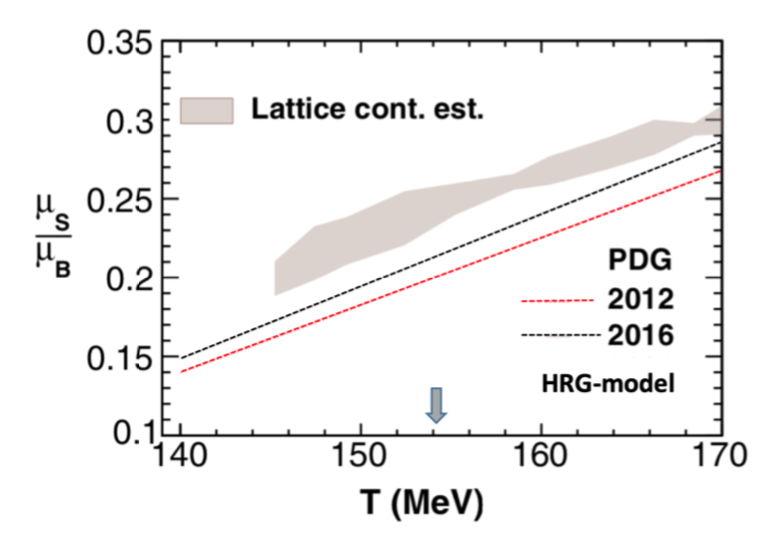}
\vspace{-0.25cm}
\caption{The ratio of baryon chemical potential of the strangeness versus all baryons for the RPP 2014 and RPP 2016. The hashed
gray area shows the LQCD calculation in ``hot QCD". The straight lines are calculations within a hadron gas model. The 2016
line, which includes more $N/\Delta$ baryon states, moves closer to the LQCD area. Note that only 3$^*$ and 4$^*$ states are
included. If the newly discovered states in RPP 2018 (seven states that are now at 3$^*$ or 4$^*$ status) were included, this line
would move even closer to the LQCD area.}
\label{chemical-potential}
\end{figure}

This is shown in Fig.~\ref{universe} where the temperature is plotted versus the baryon chemical potential. An analogy might be
water vapor transitioning to liquid water when the temperature versus pressure boundary is crossed. In our case a cross-over between the
two phases occurs that is governed and moderated by the excitation of baryons of all flavors. Figure~\ref{chemical-potential} shows
the ratio of the baryon chemical potential for strange baryons and all baryons using all baryons present in the editions of the PDG
2012 and 2016 for all baryons with ratings 3$^*$ (existence likely) and 4$^*$ (existence certain)~\cite{Chatterjee:2017yhp}. The lines
are hadron resonance gas (HRG) models compared to the latest hot Lattice QCD (LQCD) calculations. The HRG models undershoot the LQCD results,
indicating that many baryon resonances are still missing, and need to be searched for, both in the strangeness sector and in the
non-flavored sector. There are a lot more data already published or are currently under analysis that have not been included in 
multi-channel analyses, especially in vector meson photoproduction, including a significant amount of polarization data of all kinds.

An even more dramatic shortage of excited baryons exists in the charm sector, where the discrepancy in the HRG models in comparison with 
the LQCD band is much more dramatic compared to the strange- and the non-flavored baryon sector, {\it i.e.} excited hyperons and excited
nucleons~\cite{Bazavov:2014xya}. This should be an excellent motivation for a charm baryon spectroscopy program at the Electron-Ion Collider.

\section{Meson and Baryon Structure as a Window into Strong QCD Dynamics}

\subsection{Exposing the Emergence of Hadron Mass Through Studies of Pion and Kaon Structure}
\label{Sec31}

Continuum Schwinger function methods (CSMs) provide a systematic, symmetry preserving approach to solving problems in
QCD~\cite{Roberts:2015lja, Horn:2016rip, Binosi:2016rxz, Eichmann:2016yit, Burkert:2019bhp, Fischer:2018sdj}. Notably,
where fair comparisons can be drawn, predictions from continuum analyses are practically identical to those obtained via
numerical simulations of LQCD; and while lattice calculations maintain a tighter connection with the QCD
Lagrangian, the range of observables accessible to CSMs is greater. Evidently, the approaches are complementary and there
is real synergistic potential.

\begin{figure}[htp]
\includegraphics[width=0.9\columnwidth]{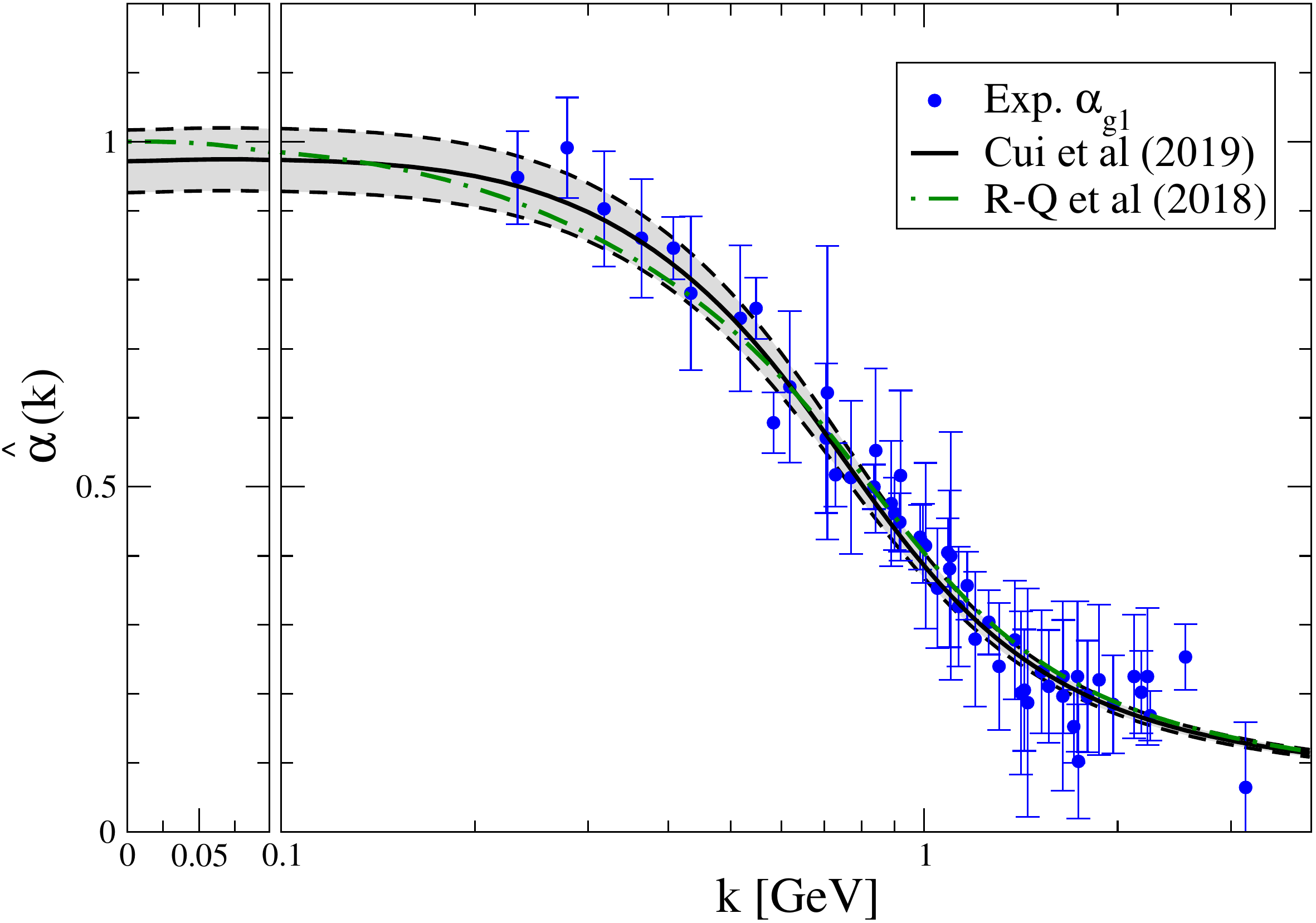}
\vspace{-0.25cm}
\caption{Solid black curve within gray band -- RGI PI running-coupling $\hat{\alpha}(k^2)/\pi$, obtained using lattice
  configurations for QCD generated with three domain-wall fermions at a physical pion mass~\cite{Cui:2019dwv}; and
  dot-dashed green curve -- earlier result~\cite{Rodriguez-Quintero:2018wma}. Also depicted, world data on the
  process-dependent effective coupling $\alpha_{g_1}$, defined via the Bjorken sum rule, the sources of which are listed
  elsewhere~\cite{Cui:2019dwv}. For additional details, see Refs.~\cite{Deur:2005cf,Deur:2008rf,Deur:2016tte,Binosi:2016nme}.
  The $k$-axis scale is linear to the left of the vertical dashed line and logarithmic otherwise.}
\label{FigwidehatalphaII}
\end{figure}

The merits of combining the strengths of the continuum and lattice QCD approaches are highlighted in Fig.~\ref{FigwidehatalphaII}, which 
depicts the process-independent effective charge of QCD~\cite{Binosi:2014aea, Binosi:2016nme, Rodriguez-Quintero:2018wma, Cui:2019dwv}. It 
was obtained by combining the best available results from continuum and lattice analyses of the gauge sector. Owing to the dynamical breaking 
of scale invariance, evident in the emergence of a gluon mass scale~\cite{Aguilar:2015bud}, $m_0= 0.43(1)$~GeV, this coupling saturates at 
infrared momenta: $\hat\alpha(0)/\pi=0.97(4)$. Among other things: $\hat\alpha(k^2)$ is almost identical to the process-dependent (PD) 
effective charge defined via the Bjorken sum rule~\cite{Deur:2005cf, Deur:2008rf, Deur:2016tte}; and also that PD charge which, employed in 
integrating the one-loop DGLAP evolution equations~\cite{Dokshitzer:1977sg, Gribov:1972, Lipatov:1974qm, Altarelli:1977zs}, delivers agreement 
between pion parton distribution functions computed at the hadronic scale and experiment
\cite{Ding:2019lwe, Ding:2019qlr, Rodriguez-Quintero:2019fyc, Cui:2019dwv}. The diversity of unifying roles played by $\hat\alpha(k^2)$ suggests 
that it is a strong candidate for that object that represents the interaction strength in QCD at any given momentum scale~\cite{Dokshitzer:1998nz}; 
and its properties support a conclusion that QCD is a mathematically well-defined quantum field theory in four dimensions.

The journey to this effective charge began in the late 1960s with the discovery of quarks. Much has since been learned about
the role of QCD's elementary excitations in building observable matter by studying hadron elastic and transition electromagnetic
form factors and also via DIS. The former provide empirical access, {\it e.g.} to the distributions
of charge and magnetization within hadrons; and the latter to momentum probability distributions. Crucially, some of the earliest
and most rigorous predictions of QCD relate to these observables; and new era detectors and facilities are expected to finally
bring their validation within reach.

Perhaps the best known QCD predictions are those made for the electromagnetic form factors of pseudoscalar mesons, {\it e.g.} the pion 
and kaon. Within the nuclear physics context, these states are abnormally light: they are pseudo-Nambu-Goldstone modes
\cite{Nambu:1960tm,Goldstone:1961eq}. Yet, modern theory predicts that their properties provide the cleanest window onto emergent 
hadronic mass (EHM) within the Standard Model~\cite{Roberts:2019ngp}. This connection is expressed in many ways; but, most forcefully, 
in the behavior of meson form factors at large momentum transfers. On this domain, QCD relates measurements simultaneously to low- and 
high-energy features of QCD, {\it viz.} to subtle features of meson wave functions and to the character of quark-quark scattering at
high-energy~\cite{Lepage:1979zb,Efremov:1979qk,Lepage:1980fj}. These relationships are expressed concretely in modern continuum analyses, 
with predictions that expose the crucial role of EHM~\cite{Chang:2013nia,Gao:2017mmp,Chen:2018rwz}. Hence, experiments focused in this 
area are of the greatest importance.

Exploiting such strengths as high-luminosity and high-energy, new era facilities have the potential to deliver precision results
on pion and kaon electromagnetic form factors at large-$Q^2$ using the Sullivan process~\cite{Sullivan:1971kd}. Contemporary
phenomenology and theory indicate that, with achievable kinematics, the proton's ``meson cloud'' can provide reliable access to
a meson target~\cite{Horn:2016rip,Qin:2017lcd}. An array of such experiments is already approved at JLab~\cite{E12-07-105,Carmignotto:2018uqj}; 
and could also be performed with an electron-ion collider (EIC) in the USA and, potentially, in China (EicC).

\begin{figure}[htp]
\includegraphics[width=0.7\columnwidth]{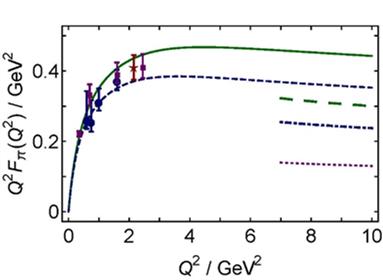}
\includegraphics[width=0.7\columnwidth]{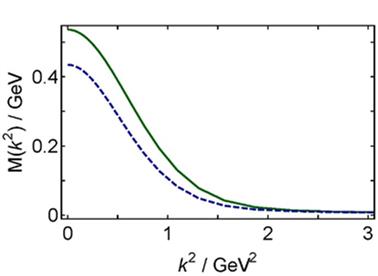}
\vspace{-0.25cm}
\caption{Top: $F_\pi(Q^2)$ obtained with the mass function in the lower panel: $r_\pi = 0.66$~fm with the solid
  green curve and $r_\pi = 0.73$~fm with the dashed blue curve. The long-dashed green and dot-dashed blue curves are
  predictions from the QCD hard-scattering formula, obtained with the related, computed pion parton distribution amplitudes
  (PDAs). The dotted purple curve is the result obtained from that formula if the asymptotic profile is used for the PDA:
  $\varphi(x)=6x(1-x)$. Bottom: Two dressed-quark mass functions distinguished by the amount of dynamical chiral
  symmetry breaking: emergent mass generation is 20\% stronger in the system characterized by the solid green curve, which
  describes the more realistic case.}
\label{figF9}
\end{figure}

To proceed experimentally, one needs to ensure that the Sullivan process is a valid tool for meson structure experiments. To
check whether these conditions are satisfied empirically, one can take data covering a range in $t$, particularly low $|t|$, and
compare with phenomenological and theoretical expectations. Furthermore, for the extraction of precision elastic form factors,
one must ensure that the virtual photon is longitudinally polarized. In the $W$, $Q^2$ regime accessible at JLab, the key to
such form factor extractions is longitudinal-transverse (L/T) separated cross sections and control over systematic uncertainties.
The latter include checking the consistency of the model used to extract the form factor from electroproduction data. This can
be done by extracting the form factor at two values of $t_{min}$ for fixed $Q^2$ and verifying that the pole diagram is the
dominant contribution to the reaction mechanism. Extensive studies~\cite{Horn:2016rip,Huber:2008id} over the last decades
have generated confidence in the reliability of this method and precision form factor data have been extracted to
$Q^2$=2.5~GeV$^2$~\cite{Horn:2006tm,Horn:2007ug} (see Fig.~\ref{figF9} -- upper panel).

The approved experiments at JLab~\cite{E12-09-011,E12-19-006} will provide a comprehensive set of L/T separated pion and
kaon electroproduction data up to $Q^2$ values of 8.5~GeV$^2$ (pion) and 5.5~GeV$^2$ (kaon), also enabling measurements of
the pion and kaon form factors. Beyond JLab, at the high $Q^2$, $W$ accessible with the EIC
\cite{Accardi:2012qut,Aguilar:2019teb} and possibly at EicC~\cite{EicCWP}, phenomenological models predict
$\sigma_L\gg \sigma_T$ at small $-t$. There, a practical method of isolating the longitudinal virtual photon is to use a
model to distinguish the dominant differential cross section, $d\sigma_L$/$dt$, from the measured, unseparated differential
cross section, $d\sigma$/$dt$. Focusing on the pion because the kaon is similar, one can then experimentally validate the
model, {\it i.e} the condition $\sigma_L\gg \sigma_T$, by using the $\pi^-/\pi^+$ ratio of charged pion data extracted
from electron-deuteron beam collisions in the same kinematics as charged pion data from electron-proton collisions.

The connection between pion and kaon form factors and mass generation in the Standard Model is illustrated in
Fig.~\ref{figF9}~\cite{Chen:2018rwz}. The lower panel depicts two similar but distinct dressed light-quark mass-functions,
$M(k^2)$, characterized by a different strength for EHM. The solid green curve was computed using a QCD effective
charge whose infrared value is consistent with modern continuum and lattice analyses of QCD's gauge sector~\cite{Cui:2019dwv},
whereas the dashed blue curve was obtained after reducing this value by 10\%. In a fully consistent calculation, such a
modification is transmitted to every element in the calculation; and subsequently to all observables. The resulting impact on
$F_\pi(Q^2)$ is depicted on the top: evidently, this experiment is a keen probe of the strength of EHM.
Figure~\ref{figF9}(top) also depicts results obtained using the QCD hard scattering formula derived for pseudoscalar
mesons~\cite{Lepage:1979zb,Efremov:1979qk,Lepage:1980fj}: at empirically accessible energy scales they, too, are sensitive
to the emergent mass scale in QCD~\cite{Chang:2013nia,Gao:2017mmp,Chen:2018rwz}.

At a similar level of rigor is the QCD prediction for the behavior of the meson structure functions; namely, the experimental
observables that probe the quark momentum distributions. More than forty years ago~\cite{Ezawa:1974wm,Farrar:1975yb,Berger:1979du}, 
it was shown that the momentum distributions of valence-quarks within the pion have the following behavior at large-$x$ ($x$ is the 
quark's fraction of the pion's light-front momentum): ${\mathpzc u}^{\pi}(x;\zeta =\zeta_H) \sim (1-x)^{2}$, where $\zeta_H$ is an 
energy scale characteristic of non-perturbative dynamics. Ongoing analyses are providing increasing support for the identification 
$\zeta_H = m_0$, {\it i.e.} the renormalization-group-invariant gluon-mass. $m_0$ is an essentially non-perturbative scale whose 
existence ensures that parton modes with $k^2 \lesssim m_0^2$ are screened from interactions. Hence, it appears as a natural boundary 
between soft and hard physics. The most recent measurements of ${\mathpzc u}^\pi(x;\zeta)$ are thirty years old
\cite{Badier:1980jq,Badier:1983mj,Betev:1985pg,Falciano:1986wk,Guanziroli:1987rp,Conway:1989fs}; and conclusions drawn from those 
experiments have proved controversial~\cite{Holt:2010vj}. For example, using a leading-order (LO) perturbative QCD (pQCD) analysis, 
Ref.~\cite{Conway:1989fs} (the E615 experiment) reported ($\zeta_5 = 5.2$~GeV):
${\mathpzc u}_{\rm E615}^{\pi}(x; \zeta_5) \sim (1-x)^{1}$, a striking contradiction of the QCD prediction. Subsequent CSM
calculations~\cite{Hecht:2000xa} confirmed the QCD prediction and eventually prompted reconsideration of the E615 analysis, with the 
result that, in a complete next-to-leading-order (NLO) study~\cite{Wijesooriya:2005ir,Aicher:2010cb}, the E615 data can be viewed as 
consistent with QCD. Notwithstanding these advances, uncertainty over ${\mathpzc u}^{\pi}(x)$ remains because more recent analyses of 
the E615 data have failed to consistently treat NLO effects and, crucially, modern data are lacking.

\begin{figure}[htp]
\centering
\includegraphics[width=0.9\columnwidth]{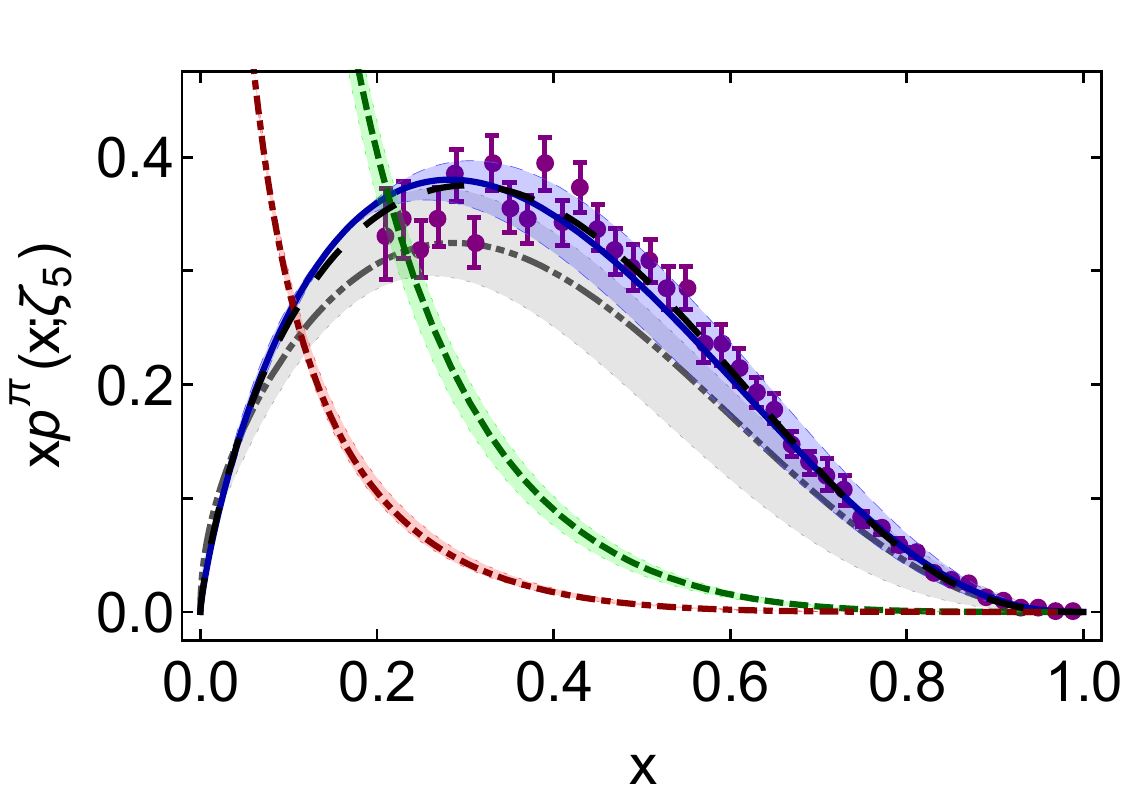}
\vspace{-0.25cm}
\caption{Pion valence-quark momentum distribution function, $x {\mathpzc u}^\pi(x;\zeta_5)$: dot-dot-dashed (gray) curve
  within shaded band -- LQCD result~\cite{Sufian:2019bol}; long-dashed (black) curve -- early continuum analyses~\cite{Hecht:2000xa}; 
  and solid (blue) curve embedded in shaded band -- modern, continuum calculation~\cite{Ding:2019lwe,Ding:2019qlr}. Gluon momentum 
  distribution in pion, $x g^\pi(x;\zeta_5)$ -- dashed (green) curve within shaded band; and sea-quark momentum distribution, 
  $x S^\pi(x;\zeta_5)$ -- dot-dashed (red) curve within shaded band. (In all cases, the shaded bands indicate the size of 
  calculation-specific uncertainties, as described elsewhere~\cite{Ding:2019lwe,Ding:2019qlr}.) Data (purple) from 
  Refs.~\cite{Conway:1989fs,Aicher:2010cb}.}
\label{figF12}
\end{figure}

Pressure is also being applied by modern advances in theory. Novel LQCD algorithms
\cite{Liu:1993cv,Ji:2013dva,Radyushkin:2016hsy,Radyushkin:2017cyf,Chambers:2017dov} are beginning to yield results for
the point-wise behavior of ${\mathpzc u}^\pi(x)$
\cite{Chen:2018fwa,Karthik:2018wmj,Karpie:2019eiq,Sufian:2019bol,Joo:2019bzr,Sufian:2020vzb}; and recent continuum
analyses are also yielding new insights. As displayed in Fig.~\ref{figF12}, the first parameter-free predictions of the valence,
glue and sea distributions within the pion~\cite{Ding:2019lwe,Ding:2019qlr} reveal that, like the pion's parton distribution
amplitude, ${\mathpzc u}^\pi(x)$ is hardened as a direct consequence of emergent mass. Very significantly, the new continuum
prediction for ${\mathpzc u}^\pi(x;\zeta_5)$ matches that obtained using lattice-QCD~\cite{Sufian:2019bol}. Consequently,
the Standard Model prediction: ${\mathpzc u}^{\pi}(x;\zeta =\zeta_H) \sim (1-x)^{2}$, is stronger than ever before; and an era
is approaching in which new experiments with novel design specifications will enable the ultimate validation
of this fundamental prediction.

To access the structure functions via experiment through the Sullivan process, one can work with the differential cross section, which is 
transverse in the Bjorken limit, and trust both ({\it i}) the phenomenology and theory, which predict that meson structure can reliably be 
extracted on a sizeable low-$|t|$ domain and ({\it ii}) comparisons with results from other experimental techniques on their common domain. 
The first data is expected from the tagged DIS program at JLab, including pion~\cite{Keppel:2015} and kaon~\cite{Keppel:2015B} structure 
functions. At the EIC, a large range in $x_{\pi}$ and $Q^2$, roughly covering down to $x_{\pi} = 10^{-3}$ at $Q^2= 1$~GeV$^2$ and up to 
$x_{\pi} = 1$ at $Q^2= 1000$~GeV$^2$, is accessible~\cite{Aguilar:2019teb}. Experiments that will deliver new pion and kaon Drell-Yan data 
are also proposed for the CERN M2 beam line by the AMBER collaboration~\cite{Denisov:2018unj}. Such data would constrain the separated valence 
and sea quark pion PDFs above $x_{\pi}= 0.2$. The previously published HERA results on the pion Sullivan process would continue to be used to 
constrain the pion PDFs on $x_{\pi}< 10^{-3}$ at $Q^2 = 1$~GeV$^2$.

The potential of such measurements to expose emergent mass is greatly enhanced if one includes similar kaon experiments because the combined 
power of continuum and lattice QCD analyses has revealed that strange-quark physics lies at the boundary between dominance of strong (emergent) 
mass generation and weak (Higgs-connected) mass~\cite{Roberts:2019ngp}. Hence, comparisons between distributions of truly light quarks and those 
describing strange quarks are well suited to exposing measurable signals of emergent mass in counterpoint to Higgs-driven effects. A striking 
example can be found in the contrast between the valence-quark PDFs of the pion and kaon. A significant disparity between these distributions 
would indicate a marked difference between the fractions of pion and kaon momentum carried by the other bound state participants, particularly 
gluons.

The measurement of cross sections from which one can extract GPDs and transverse-momentum-dependent parton distributions (TMDs); and, therefrom, 
3-D images of hadrons, presents a fascinating new frontier within the Standard Model, promising to deliver tomographic pictures of hadron structure. 
JLab and COMPASS have provided us with sketches; and EicC and EIC are being designed to complete the pictures~\cite{Accardi:2012qut,Aguilar:2019teb,EicCWP}.

The key to extracting GPDs from experiment are QCD factorization theorems. DVCS cross sections and polarized
asymmetries can provide detailed and precise information about GPDs, but are only sensitive to a particular flavor
combination. Exclusive meson production provides key additional information allowing the separation of different quark and
anti-quark flavors. To validate the meson factorization theorems and potentially extract flavor separated GPDs from
experiment, one must measure the separated longitudinal and transverse cross sections and their dependence on $(t,Q^2)$.
The experimental program at JLab takes advantage of the complementarity of four experimental halls and will provide a
comprehensive exclusive electroproduction data set~\cite{redmine:19}. Supposing, as accumulating evidence suggests,
that a material non-zero domain in $-t$ exists whereupon one can extract physical $\pi$ ($K$) information using the
Sullivan process, then within the projected EIC luminosity reach and detection capabilities, one can envision measurements
of the pion's GPD. Projected experimental results would be, at least, at the level achieved previously at HERA for the proton.
Moreover, should the data validate the assumption that reliable $\pi$ ($K$) structure data can be extracted for
$-t \leq 0.6\,(0.9)$~GeV$^2$, then one would even gain an order-of-magnitude in statistics as compared to HERA proton
data~\cite{Aguilar:2019teb}.

It should be recognized, however, that science faces many new challenges in extracting 3-D images from new generation
experiments. Phenomenological models of a wide variety of parton distribution functions will be crucial. They will provide
guidance on the size of the cross sections to be measured and the best means by which to analyze them~\cite{Berthou:2015oaw}.
On the other hand, as experience with meson structure functions has shown, in order to fully capitalize on such experiments,
one must use computational frameworks that can veraciously connect measurements with qualities of QCD. Complementing
LQCD, CSM analyses promise to fill this role because, as evidenced by calculations of meson
wave functions, GPDs, and TMDs~\cite{Mezrag:2014jka,Mezrag:2016hnp,Chouika:2017rzs,Xu:2018eii}, even simplified CSM
calculations can provide valuable insights.

\subsection{Highlights and Prospects for Nucleon Collinear Parton Distributions}
\label{nucleon_1d}

\subsubsection{Introduction}

There have been exciting developments recently in the study of nucleon structure at its most fundamental level in terms of
the quark and gluon (or generically parton) degrees of freedom of QCD. High-quality data from modern accelerator facilities,
such as the 12~GeV program at Jefferson Lab, COMPASS and LHC at CERN, and RHIC at BNL, along with plans for the future
EIC~\cite{Accardi:2012qut}, are pushing our knowledge of the momentum and spin distributions of the partons in the nucleon 
and nuclei.

With the accumulation of the new high-statistics, high-precision data comes the growing need for analysis methods that can
adequately analyze the results. This is particularly important when dealing with correlations between observables and their
inter-related dependence on the underlying PDFs. The standard method for obtaining information about PDFs has been through global
QCD analysis of various high-energy scattering observables, which because of QCD factorization can be described in terms of the same 
universal distributions~\cite{JMO13,Blumlein13,ForteWatt13}.

Much has been learned, but significant questions still remain, such as the nature of the $d/u$ valence quark distribution ratio
and the quark polarizations at large parton momentum fractions $x$, the flavor content of the proton's sea distributions, the
shape of the strange and antistrange PDFs, both unpolarized and polarized, as well as the possible modification of the PDFs in
nucleons bound inside nuclei, to name just a few. In the following we summarize recent developments in the study of PDFs, both
experimentally and theoretically, and outline the opportunities and challenges that are expected to arise in the near future.

\subsubsection{Modern Global QCD Analysis}

Traditional approaches to extracting PDFs have been based on the maximum likelihood method, which estimates an optimal set
of parameters for some chosen parametric form of the PDFs. A major difficulty with such single-fit analyses is that, when data
are scarce or the distributions are poorly known, these can introduce significant bias into the extraction. Moreover, since the
$\chi^2$ distribution is usually highly nonlinear in the parameters, it is common for local minima to exist in which a single fit 
can often get stuck.

To overcome these problems more sophisticated methods, such as Monte Carlo (MC) sampling, are necessary to thoroughly scan
the parameter space and take into account the multiple solutions. This methodology has recently been adopted by the Jefferson
Lab Angular Momentum (JAM) Collaboration~\cite{JAM-collab}, which has embarked on an ambitious program to {\it simultaneously}
constrain PDFs and parton to hadron fragmentation functions (FFs) in global QCD analyses. The analyses utilize inclusive
DIS, Drell-Yan lepton-pair production, and other high-energy scattering data that are sensitive to
PDFs, along with SIDIS data that depend on both PDFs and FFs. To better constrain the latter, data on
single-inclusive $e^+ e^-$ annihilation (SIA) are also used in the fits.

The JAM global analysis methodology implements MC algorithms that employ data resampling, with multi-step fitting strategies
tailored to the specific needs of the various extractions. Previous JAM studies of collinear distributions have included an
iterative MC analysis of spin-dependent PDFs of the nucleon~\cite{JAM15}, the first MC analysis of fragmentation functions from
$e^+ e^-$ SIA~\cite{JAMFF}, first simultaneous extraction of spin-dependent PDFs and FFs~\cite{JAM17}, simultaneous analysis
of unpolarized PDFs and FFs from DIS, SIDIS and SIA data~\cite{JAM19}, and a first MC analysis of pion PDFs~\cite{JAMpi}.
In the following we summarize some of these developments and how they relate to the upcoming experimental programs.

\subsubsection{Flavor Separation}

\begin{figure}[htp]
\includegraphics[width=0.9\columnwidth]{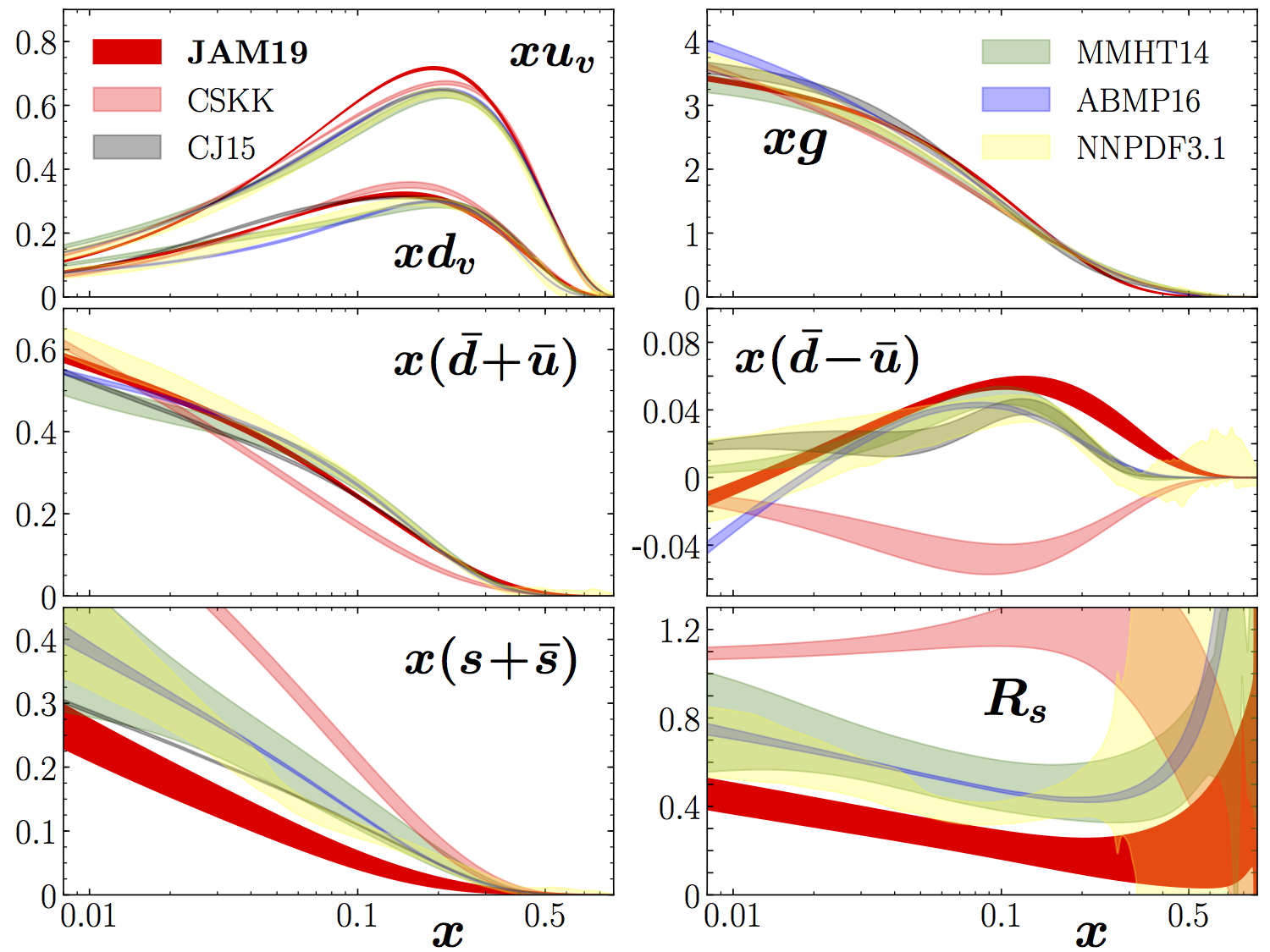}
\vspace{-0.25cm}
\caption{Comparison of spin-averaged PDFs from various recent global QCD analyses, at a common scale $Q^2=4$~GeV$^2$,
  including valence quark, antiquark, and gluon PDFs, and the strange to nonstrange ratio $R_s = (s+\bar s)/(\bar u+\bar d$)
  \cite{JAM19}.}
\label{fig:PDFs}
\end{figure}

The flavor and spin decomposition of the proton's valence and sea quark PDFs provides fascinating glimpses into the
non-perturbative QCD dynamics that govern quark and gluon interactions at long distances~\cite{MT96}. For spin-averaged PDFs,
in Fig.~\ref{fig:PDFs} we show various flavor combinations from several recent global QCD analyses. One of the important questions
addressed by the 12~GeV JLab physics program is what is the nature of the proton's $d$-quark PDF relative to the $u$-quark at
large~$x$, where most of the proton's light-cone momentum is carried by a single parton~\cite{Holt:2010vj}. The $d/u$ ratio has been
traditionally extracted from measurements of the neutron and proton structure functions at large~$x$, but the absence of free
neutron targets has required the use of deuterons as effective neutron targets, with subsequent complications due to nuclear
correction uncertainties at high $x$~\cite{MT96,Holt:2010vj,CJ15}.

\begin{figure}[htp]
\includegraphics[width=0.85\columnwidth]{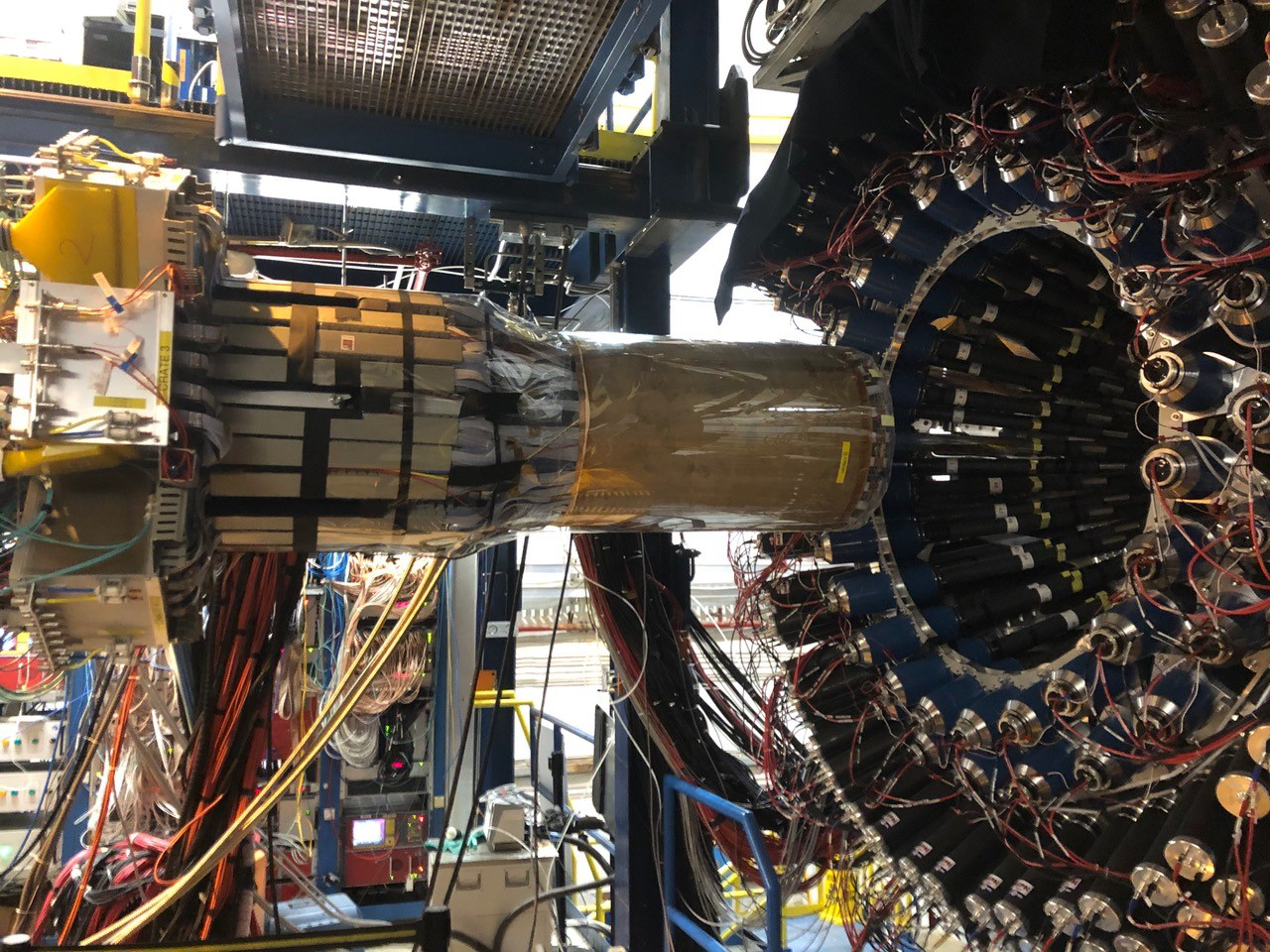}
\vspace{-0.25cm}
\caption{BONuS radial time projection chamber being inserted into the CLAS12 detector in Hall~B.}
\label{fig:BONuS12}
\end{figure}

The JLab 12~GeV program addresses this problem by focusing on the construction of more effective free-neutron targets, with
either spectator-tagging in electron scattering on a deuteron (BONuS in Hall~B~\cite{Tkachenko14}) or through comparison of
the isospin-mirror nuclei $^3$He and $^3$H (MARATHON in Hall~A~\cite{Afnan03}), both of which seek to minimize the nuclear
corrections. The MARATHON experiment ran in 2019, while the BONuS experiment started taking data in early 2020. The
experimental setup of the BONuS experiment is shown in Fig.~\ref{fig:BONuS12}, which shows the new detector built to measure
low-momentum spectator protons next to the CLAS12 detector that tags electrons scattered from a nearly free neutron. 

\begin{figure}[htp]
\includegraphics[width=0.95\columnwidth]{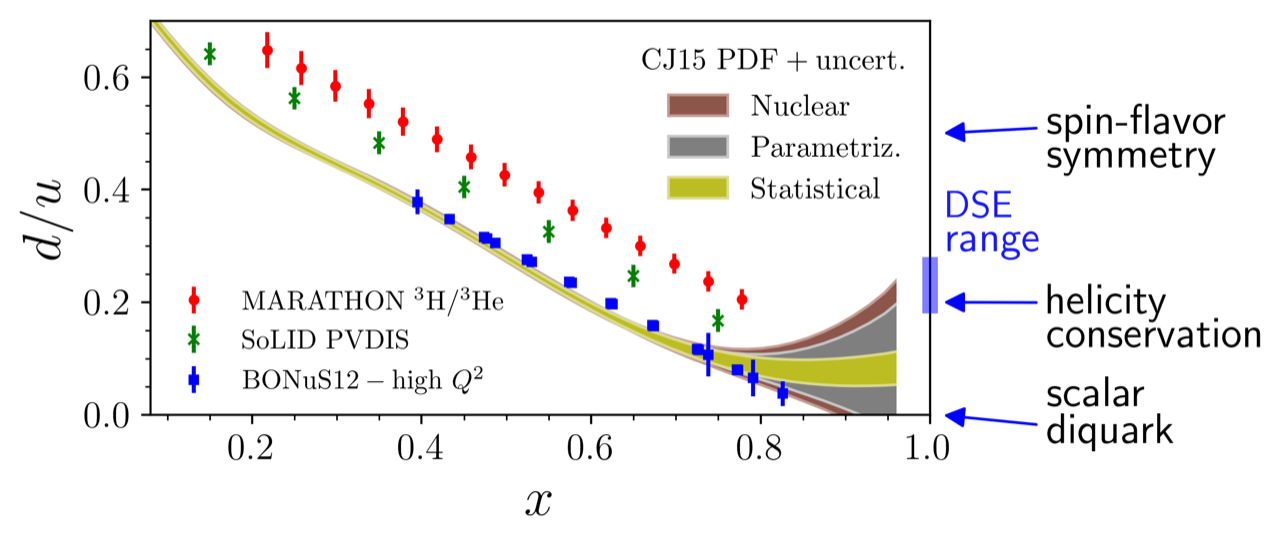}
\vspace{-0.25cm}
\caption{Ratio of $d$- to $u$-quark PDFs from the CJ15 fit~\cite{CJ15}, including uncertainties from various sources, compared
  with some projections from JLab 12~GeV experiments and theoretical predictions. The uncertainties on the experiment points
  do not include all of the systematic and model-dependent theoretical uncertainties.}
\label{fig:duratio}
\end{figure}

An alternative approach proposes measuring the $d$-quark PDF directly using parity-violating DIS (PVDIS) from the proton
\cite{Hobbs08}, which is being planned as part of the SoLID detector upgrade at JLab~\cite{SoLID:pCDR}. The projected impact of
all three experiments on the $d/u$ ratio is illustrated in Fig.~\ref{fig:duratio}, which compares the ratio extracted from the
CJ15 global analysis~\cite{CJ15} with the expected statistical uncertainties from the experiments. Note that the error bars on
the MARATHON $^3$He/$^3$H pseudo-data do not include systematic uncertainties from theoretical approximations, which may
underestimate the final uncertainties in a global QCD analysis. Nonetheless, the availability of data from all these experiments 
will be vital for obtaining more robust constraints on the $d$-quark PDF at large $x$.

In the strange-quark sector, the quantitative nature of the strange sea has remained even more elusive. Traditionally, the
strange-quark PDF has been constrained by inclusive charm meson production in charged current neutrino--nucleus DIS. Analyses
of $\nu$ and $\bar\nu$ cross sections from Fermilab and CERN experiments have generally found a strange to nonstrange sea
quark ratio \mbox{$R_s = (s+\bar s)/(\bar u + \bar d)$} $\sim 0.5$. Unfortunately, the interpretation of the $\nu A$ data 
suffers from uncertainties in the nuclear corrections associated with relating nuclear to free nucleon structure functions
\cite{Kalantarians17}, and in the treatment of charm energy loss and $D$ meson--nucleon interactions during hadronization inside
the nucleus~\cite{Accardi10, Majumder11}.

From another perspective, an independent source of information on strange PDFs is SIDIS with charged final state kaons as flavor
tags. A recent JAM analysis~\cite{JAM19} performed the first simultaneous extraction of unpolarized PDFs and FFs from a
combination of inclusive DIS, SIDIS, Drell-Yan, and SIA data. Inclusion of the SIDIS multiplicities in particular revealed a strong
suppression of the strange PDF at $x \gtrsim 0.01$, in contrast with earlier observation of enhanced strangeness by ATLAS at the
LHC~\cite{ATLAS-W}. This is illustrated in Fig.~\ref{fig:PDFs}, where the $x(s+\bar s)$ distribution and the related $R_s$ ratio
are seen to be rather smaller than other parameterizations at intermediate $x$ values.

The SIDIS production can in principle also discriminate between the $s$ and $\bar s$ PDFs through tagging of $K^+$ vs. $K^-$
mesons, although existing data are not precise enough to allow this discrimination. Future high-precision SIDIS data from Jefferson
Lab or EIC should allow more stringent determinations of the $s$ and $\bar s$ PDFs~\cite{Aschenauer19}, along with PVDIS, as
would $W$ + charm production in $pp$ collisions~\cite{ATLAS-Wc}.

\subsubsection{Spin-Dependent Structure}

\begin{figure}[htp]
\includegraphics[width=0.9\columnwidth]{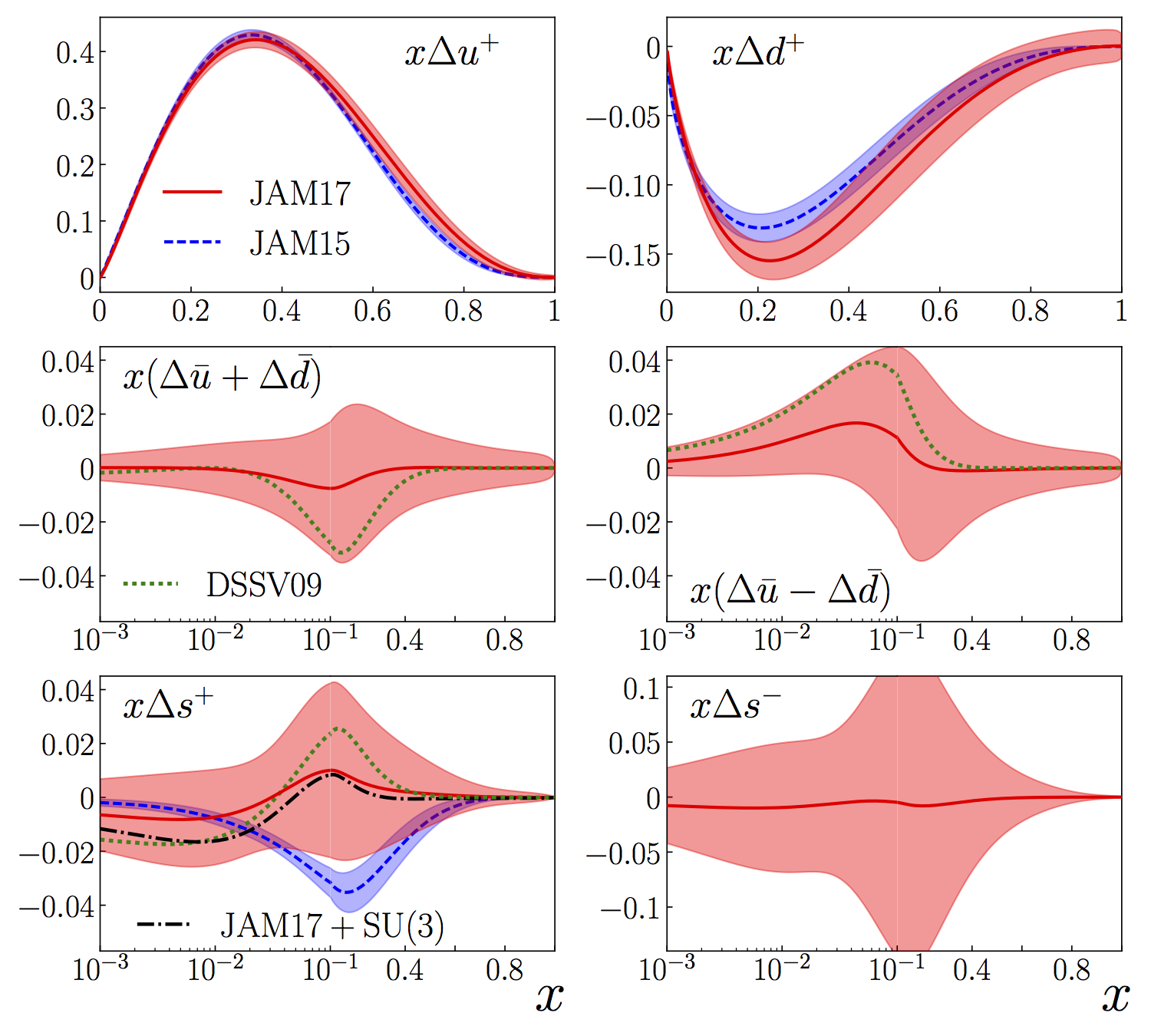}
\vspace{-0.25cm}
\caption{Comparison of spin-dependent PDFs from various recent global QCD analyses, at a common scale $Q^2=1$~GeV$^2$
  \cite{JAM17}.}
\label{fig:polPDFs}
\end{figure}

The decomposition of the proton's spin into its quark and gluon helicity and orbital angular momentum components has been one of
the driving problems in hadron physics for the past three decades~\cite{Aidala13}. While the behavior of the polarized $\Delta u$
and $\Delta d$ distributions has been reasonably well established at moderate $x$ values, as seen in Fig.~\ref{fig:polPDFs}, the
shape, and even sign, of the strange-quark polarization $\Delta s$ is much more uncertain. Moreover, even for the non-strange spin
distributions, there are significant uncertainties in kinematic regions, such as at large $x$, where data are scarce.

The latter observation is rather pertinent, given that polarized PDFs are quite sensitive to the details of non-perturbative
quark-gluon dynamics in the nucleon at high $x$~\cite{MT96}. To address these questions a comprehensive program of high-$x$
spin structure function measurements for the proton, deuteron and $^3$He has been initiated in JLab Halls B and C, the latter
taking data since Fall of 2019. These measurements will also improve the determination of higher twist contributions to structure
function moments, and test quark-hadron duality in the spin sector.

For strange quarks, the results of the combined analysis~\cite{JAM17} of polarized DIS, SIDIS and SIA data in
Fig.~\ref{fig:polPDFs}, which was the first performed {\it without} imposing flavor SU(3) symmetry, indicate that $\Delta s$ is
small across all $x$, within large uncertainties. This turns out to be largely due to the SIDIS $K^-$ production data, which are
most sensitive to strange quark spin. The $K^-$ polarization asymmetry computed with a negative $\Delta s$, for instance, gives
a significantly worse description of the data than the full result in Fig.~\ref{fig:polPDFs}. Of course, preference for a positive 
or negative $\Delta s$ depends on the FFs used in the evaluation of the asymmetry, so the solution is clearly to simultaneously
determine both PDFs and FFs, as was performed in Ref.~\cite{JAM17}. Upcoming SIDIS $K$ production data from JLab will be
helpful in further constraining the shape of $\Delta s$, and the future EIC offers the possibility of performing polarized PVDIS
to obtain unique sensitivity to strange quark polarization~\cite{Hobbs08,Anselmino94}.

\subsubsection{PDFs from Lattice QCD}

\begin{figure}[htp]
\includegraphics[width=0.9\columnwidth]{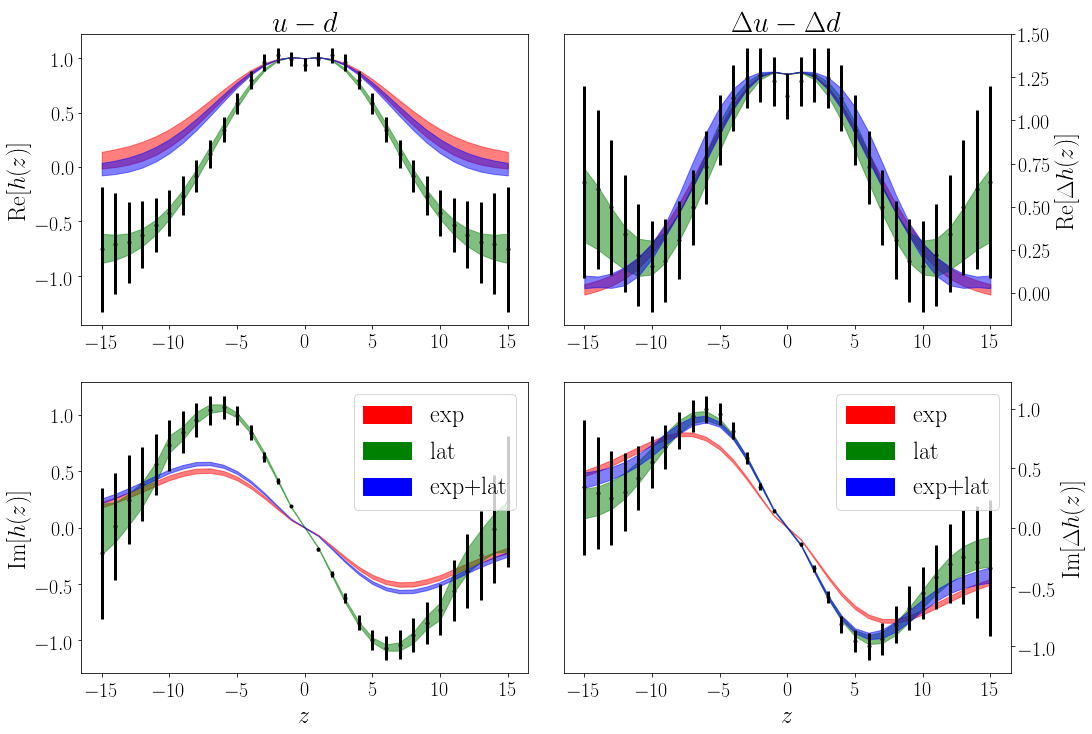}
\vspace{-0.25cm}
\caption{Lattice QCD simulations (green lines) of the real (top row) and imaginary (bottom row) parts of nonlocal matrix elements
  of twist-2 operators for unpolarized $u-d$ (left column) and polarized $\Delta u-\Delta d$ PDFs (right column), compared with
  phenomenological fits to experimental data (red lines) and fits to experimental + lattice data (blue lines)~\cite{JAMlat}. The 
  real and imaginary parts of the matrix elements are sensitive to different combinations of quark and antiquark PDFs.}
\label{fig:lattice}
\end{figure}

Complementing the experimental and phenomenological developments in PDF studies, recent progress in lattice QCD simulations of
parton quasi-distributions (qPDFs)~\cite{Ji:2013dva,Joo19} and pseudo-distributions~\cite{Alexandrou19} is paving the way towards the
study of the $x$ dependence of PDFs from first principles. The observables accessible in lattice calculations are matrix elements
of nonlocal operators, such as $h(z) = \left<{N\vert\bar\psi(0,z)\Gamma W_3(z)\psi(0,0)\vert N}\right>$, which can be related
to the PDFs via a Fourier transform and matching the off-light-cone qPDFs to the physical PDFs. An example of such simulations is
shown in Fig.~\ref{fig:lattice} for the isovector unpolarized $u-d$ and polarized $\Delta u-\Delta d$ combinations from the ETM
Collaboration~\cite{Alexandrou19}.

A new approach currently being developed~\cite{JAMlat} combines global QCD analysis of experimental data with lattice results
for $h(z)$, within the same analysis framework. The comparison in Fig.~\ref{fig:lattice} illustrates that the recent lattice data
\cite{Alexandrou19} for the spin-dependent $\Delta u-\Delta d$ PDFs are reasonably compatible with the empirical results,
while some systematic discrepancies exist for the more precise spin-averaged $u-d$ PDFs. The convergence of the
phenomenological and lattice results is currently an active area of research.

\subsection{Strong QCD Insights from $N^*$ Structure Studies with EM Probes}
\label{nstar_structure_intro}

Studies of the structure of excited nucleon states in terms of the $Q^2$-evolution of the $\gamma_vpN^*$ electrocouplings represents 
the only source of information on many facets of strong QCD that underlie the generation of these $N^*$ states with different quantum 
numbers and with distinctively different structural features. Continuum QCD approaches
\cite{Segovia:2019jdk,Roberts:2018hpf,Cui:2020rmu,Lu:2019bjs,Chen:2018nsg,Segovia:2016zyc} and most quark models
\cite{Burkert:2018oyl,Aznauryan:2012ec,Giannini:2015zia,Ramalho:2018wal,Ronniger:2012xp}, which reproduce the nucleon elastic form factors 
equally well, predict different behaviors for the $\gamma_vpN^*$ electrocouplings. Confronting theory expectations on the $\gamma_vpN^*$ 
electrocouplings with the data, allows us for insight into how resonances of different structure emerge from QCD. The data on the 
$\gamma_vpN^*$ electrocouplings are of particular importance for gaining insight into the strong QCD dynamics that underlie hadron mass 
generation. Advances in continuum QCD approaches~\cite{Binosi:2016wcx,Binosi:2014aea} make it possible to describe the momentum dependence 
of the dressed quark mass ({\it i.e.} the so-called dressed quark mass function), starting from the QCD Lagrangian.

Consistent results on the dressed quark mass function from independent analyses of the data on the nucleon elastic
form factors and the $\gamma_vpN^*$ electrocouplings of different resonances with a core of three dressed quarks
in different radial, spin-isospin-flip, and orbital excitations, will validate credible access to the dynamics responsible
for hadron mass generation. The $N^*$ program with the CLAS12 detector~\cite{Burkert:2018nvj,Az13} will allow us
to map out the dressed quark mass function at the distances where the transition between the almost bare, massless
QCD-quarks and fully dressed constituent quarks of $\approx$400~MeV mass is expected, addressing the key open
question of the Standard Model on the emergence of $>$98\% of hadron mass from QCD~\cite{Roberts:2019ngp}. 

\subsubsection{$\gamma_vpN^*$ Electrocouplings from Exclusive Meson Production and Their Impact on the
  Exploration of the Ground State Nucleon PDFs}
\label{extraction_incl}

The transverse $A_{1/2}(Q^2)$, $A_{3/2}(Q^2)$, and longitudinal $S_{1/2}(Q^2)$ $\gamma_vpN^*$ electrocouplings
have been extracted from the measured differential cross sections and polarization asymmetries for the exclusive
meson electroproduction channels by fitting the data within several different reaction models. A successful fit of
the data allows us to isolate the resonant contributions for the extraction of the resonance parameters. Consistent
results on the $\gamma_vpN^*$ electrocouplings from the independent analyses of the exclusive channels with
different non-resonant contributions validates the credible extraction of these quantities. The CLAS detector has
produced the dominant part of the available world-wide data on all of the relevant meson electroproduction channels
off the nucleon in the resonance region for $Q^2$ up to 5.0~GeV$^2$~\cite{Aznauryan:2011qj}. The data are stored
in the CLAS Physics Database~\cite{db}. So far, most of the results on the $\gamma_vpN^*$ electrocouplings have
been extracted from independent analyses of $\pi^+n$, $\pi^0p$, $\eta p$ and $\pi^+\pi^-p$ electroproduction
data off the proton
\cite{Aznauryan:2011qj,Aznauryan:2009mx,Park:2014yea,Denizli:2007tq,Mokeev:2012vsa,Mokeev:2015lda,Mokeev:2019ron}.
The parameterization of the $\gamma_vpN^*$ electrocouplings for $Q^2 < 5.0$~GeV$^2$ can be found in
Ref.~\cite{HillerBlin:2019hhz}.

The $\gamma_vpN^*$ electrocouplings from the $\pi N$ electroproduction data were determined from analyses of
a total of nearly 160,000 data points (d.p.) on unpolarized differential cross sections, longitudinally polarized beam
asymmetries, and longitudinal target and beam-target asymmetries within two conceptually different approaches: a
Unitary Isobar Model (UIM) and Dispersion Relations (DR)~\cite{Aznauryan:2009mx,Aznauryan:2011qj}. The
$\pi^+\pi^- p$ electroproduction data from CLAS~\cite{Mokeev:2019ron} provide information on nine independent
single differential and fully integrated cross sections binned in $W$ and $Q^2$ in the mass range $W < 2.0$~GeV
and for $Q^2$ from 0.25 -- 5.0~GeV$^2$. The $\gamma_vpN^*$ electrocouplings of most $N^*$ in the mass range
up to 1.8~GeV have become available from analyses of $\pi^+\pi^-p$ electroproduction data within the data driven
meson-baryon model JM~\cite{Mokeev:2008iw,Mokeev:2012vsa,Mokeev:2015lda,Golovatch:2018hjk,Mokeev:2018zxt}.
A good description of the $\pi^+\pi^- p$ differential cross sections for $W < 2.0$~GeV and $Q^2$ from 0.2 --
5.0~GeV$^2$ has been achieved with $\chi^2/d.p. < 3.0$ accounting only for the data statistical uncertainties
\cite{Mokeev:2019ron}.

\begin{figure}[htp]
\includegraphics[width=0.9\columnwidth]{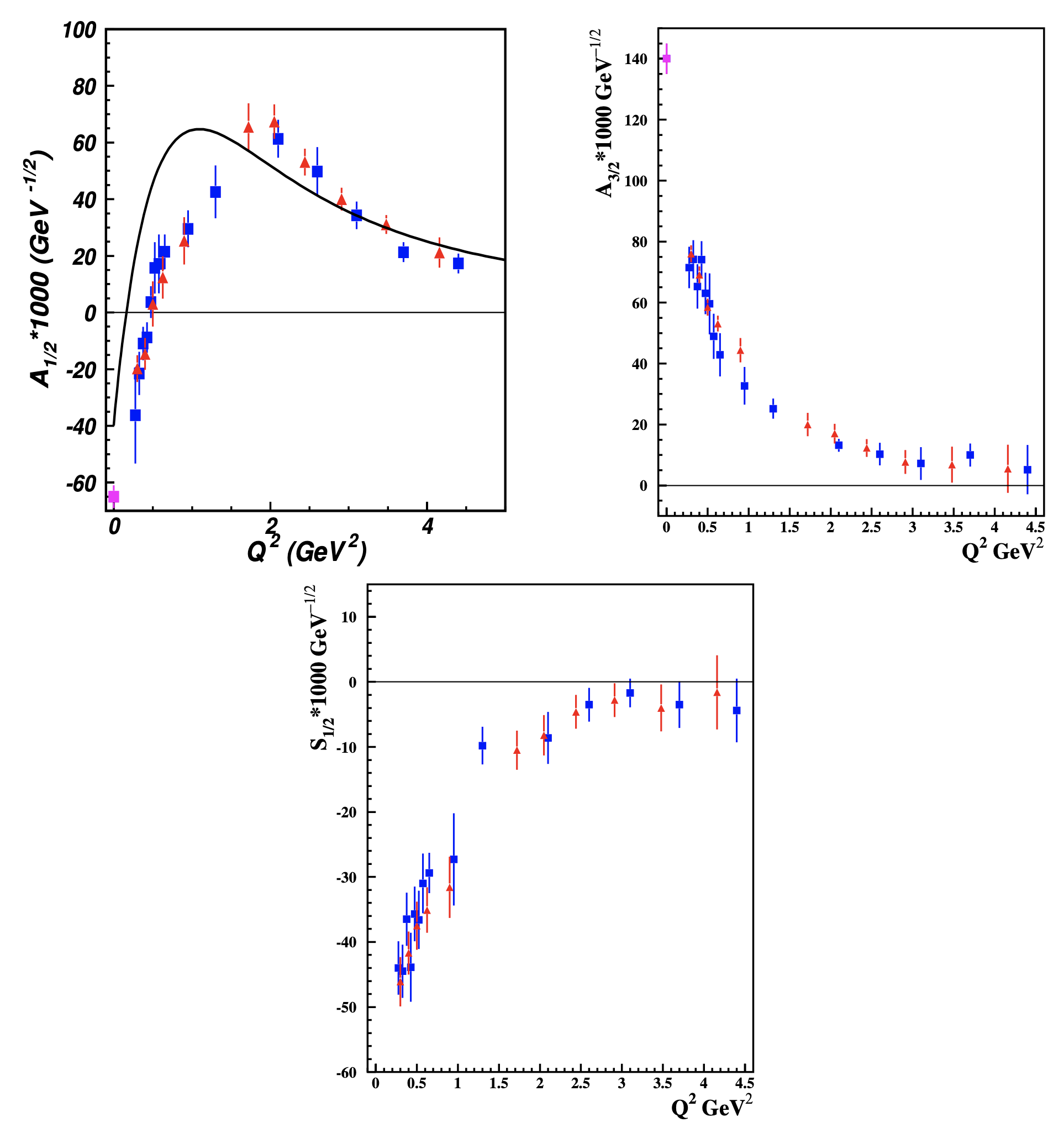}
\vspace{-4mm}
\caption{$\gamma_vpN^*$ electrocouplings $A_{1/2}$ of the $N(1440)1/2^+$ (UL), $A_{3/2}$ (UR), and
  $S_{1/2}$ (bottom) of $N(1520)3/2^-$ resonances vs. $Q^2$ from the studies of $\pi N$
  \cite{Aznauryan:2009mx,Park:2014yea} (red triangles) and $\pi^+\pi^-p$
  \cite{Mokeev:2012vsa,Mokeev:2015lda,Mokeev:2019ron} (blue squares) electroproduction data from CLAS. The
  $N(1440)1/2^+$ $A_{1/2}$ electrocouplings computed within the continuum QCD approach~\cite{Segovia:2015hra}
  starting from the QCD Lagrangian are shown by the curve in the UL plot.}
\label{electr_1pi2pi}
\end{figure}

Consistent results on the $\gamma_vpN^*$ electrocouplings for the $N(1440)1/2^+$ and $N(1520)3/2^-$ resonances,
which have been determined in independent analyses of the dominant meson electroproduction $\pi N$ and $\pi^+\pi^-p$
channels demonstrate that the extraction of these quantities is reliable (see Fig.~\ref{electr_1pi2pi}). The results 
on the $\gamma_vpN^*$ electrocouplings from the CLAS data analysis have been published in the recent PDG edition
\cite{Tanabashi:2018oca}.

For the first time, the resonance contributions for inclusive electron scattering has been evaluated based on the
experimental results in Ref.~\cite{HillerBlin:2019hhz}. A realistic accounting of these contributions is of particular
importance in order to gain insight into the parton distributions of the ground state nucleons at large values of the
Bjorken variable $x_B$. A gradual extension of the experimental results for the $\gamma_vpN^*$ electrocouplings
towards higher $W$ and $Q^2$ from CLAS/CLAS12, as well as from Halls~A/C on the transverse and longitudinal
inclusive cross sections will extend the knowledge of the nucleon parton distributions at large $x_B$. The novel
pseudo- and quasi-PDF concepts~\cite{Radyushkin:2019mye,Qiu:2019kyy}, which make it possible to relate the PDFs
to the QCD Lagrangian, are helping to drive the exploration of PDFs, including the resonant region at large $x_B$. The
studies of the resonant contributions in inclusive electron scattering represent an important step in the exploration
of 1D nucleon in synergistic efforts in Halls A, B, and C, JPAC, and the JLab Theory Center.

The experimental results on the $\gamma_vpN^*$ electrocouplings also offer important information for the development of 
the novel $N \to N^*$ transition GPD concepts. This is a promising new avenue in gaining insight into strong QCD from the 
data on the structure of excited nucleon states in 3D. The development of reaction models relating the transition GPDs to 
observables is needed in order to access this information from the CLAS12 data.

\subsubsection{From $\gamma_vpN^*$ Electrocouplings to Strong QCD Dynamics}
\label{sqcd}

\begin{figure}[htp]
\includegraphics[width=0.9\columnwidth]{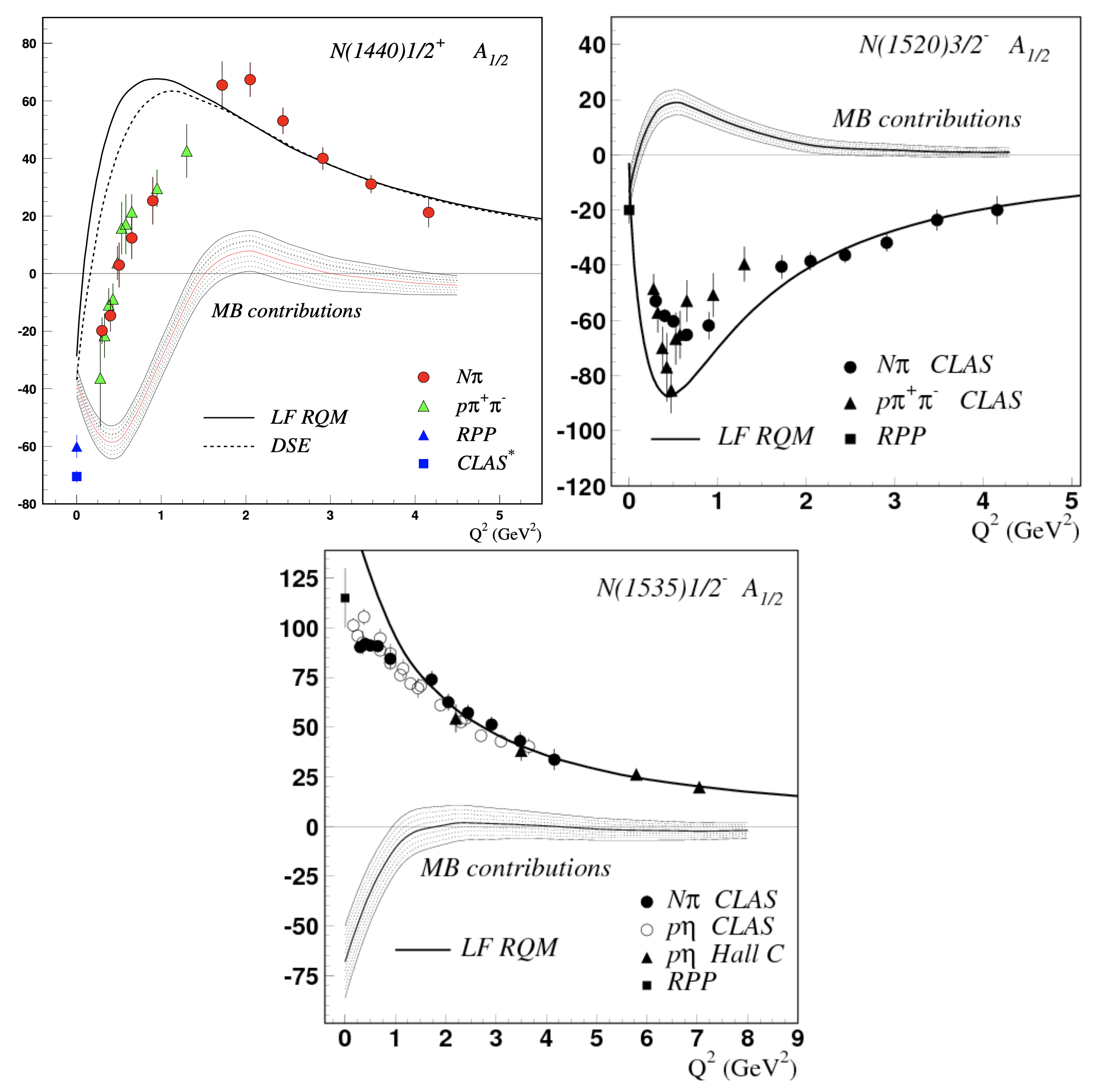}
\vspace{-2mm}
\caption{The quark core and meson-baryon cloud contributions to the $A_{1/2}$ electrocouplings of the $N(1440)1/2^+$
  (UL), $N(1520)3/2^-$ (UR), and $N(1535)1/2^-$ (bottom) resonances. The quark core contributions for the
  $N(1440)1/2^+$ were computed within the continuum QCD approach~\cite{Burkert:2019bhp} (dashed line). For each
  of the states the quark core contributions were obtained from the light-front quark model
  \cite{Aznauryan:2012ec,Aznauryan:2018okk} (solid line).}
\label{core_cloud}
\end{figure}

The analyses of the CLAS results on the $\gamma_vpN^*$ electrocouplings within the continuum QCD approach
\cite{Burkert:2019bhp,Cui:2020rmu,Lu:2019bjs,Chen:2018nsg,Segovia:2016zyc,Segovia:2015hra,Segovia:2014aza,Segovia:2013rca} and 
different quark models~\cite{Aznauryan:2012ec,Aznauryan:2018okk,Giannini:2015zia,Obukhovsky:2019xrs} have revealed $N^*$ structure
for $Q^2 < 5.0$~GeV$^2$ as a complex interplay between an inner core of three dressed quarks and an external
meson-baryon cloud. Representative examples for the contributions from the quark core and the meson-baryon cloud
to $N^*$ structure are shown in Fig.~\ref{core_cloud}. The relative contributions of the meson-baryon cloud depend
strongly on the quantum numbers of the excited nucleon state. For all resonances studied using CLAS data, this
contribution decreases with $Q^2$ in a gradual transition towards quark-core dominance for $Q^2 > 5.0$~GeV$^2$.
Understanding the emergence of the meson-baryon cloud from the quark core represents an important avenue in the
exploration of strong QCD. In order to address this problem, analyses of the data on the $\gamma_vpN^*$
electrocouplings at $Q^2$ where both the quark core and the meson-baryon cloud contribute is needed to bridge the
efforts between the multi-channel amplitude analyses of meson photo-, electro-, and hadroproduction
\cite{Ronchen:2015vfa,Kamano:2018sfb} and the $N^*$ quark models
\cite{Burkert:2018oyl,Aznauryan:2012ec,Giannini:2015zia,Ramalho:2018wal,Ronniger:2012xp}. Modeling of both
components of the $N^*$ structure has been forged by the advances in the AdS/CFT approaches that have provided a
successful description of the CLAS results on the $N(1440)1/2^+$ and $N(1535)1/2^-$ electrocouplings
\cite{Gutsche:2019yoo,Gutsche:2019jzh} for $Q^2 < 5.0$~GeV$^2$. The studies of the transition between the confined
quarks in the quark core and the deconfined mesons and baryons in the external cloud are of particular interest for
LQCD as it is the only approach with the potential to account for the effects of all relevant degrees of freedom in 
the $N^*$ structure. 

The electrocouplings of the $N(1440)1/2^+$ and $\Delta(1232)3/2^+$ resonances were evaluated starting from the
QCD Lagrangian within the continuum QCD approach~\cite{Segovia:2014aza,Segovia:2015hra}. The predictions for
the $N(1440)1/2^+$ state are shown in Fig.~\ref{electr_1pi2pi} (solid line). In the range of applicability,
$Q^2 > 2.0$~GeV$^2$, the approach~\cite{Segovia:2015hra} offers a good description of the experimental results
on the $N(1440)1/2^+$ electrocouplings, which was achieved with {\it exactly the same} dressed quark mass function
as that employed in the evaluations of the electromagnetic elastic nucleon form factors and the magnetic
$p \to \Delta(1232)3/2^+$ transition form factor~\cite{Segovia:2014aza}. This success conclusively demonstrates
the capability of gaining insight into the strong QCD dynamics underlying the dominant part of hadron mass generation
from the experimental results on the nucleon elastic form factors and the $\gamma_vpN^*$ electrocouplings.

In the upcoming experiments with the CLAS12 detector~\cite{Burkert:2018nvj}, the electrocouplings of all prominent
nucleon resonances will become available at the highest photon virtualities ($Q^2$ from 5 -- 12~GeV$^2$) ever
achieved in the studies of exclusive electroproduction, allowing us to map out the dressed quark mass function for
distances where the transition between the pQCD and strong QCD regimes is expected. Consistent results on dressed
quark mass function from independent studies of the electrocouplings of the resonances of distinctively different
structure will validate credible access to this key ingredient of strong QCD, addressing one of the most important
open questions of the Standard Model on the emergence of $>$98\% of hadron mass from QCD and quark-gluon confinement
\cite{Roberts:2019wov}. The results on the $\gamma_vpN^*$ electrocouplings for all prominent $N^*$ will offer insight 
into the $qq$ correlations in the resonances of different spin-parities. This information will enable us to predict 
the wave functions of the mesons consisting of $q\overline{q}$ with the same spin and opposite parity as for the $qq$
correlations from $N^*$ studies, emphasizing the need to combine the efforts in the exploration of both meson
\cite{Aguilar:2019teb} and baryon~\cite{Roberts:2019wov} structure.

With the dressed quark mass function and the $qq$-correlation amplitudes checked against the experimental results on the nucleon elastic 
form factors and the $\gamma_vpN^*$ electrocouplings, the continuum QCD approaches are capable of evaluating the Faddeev amplitude for 
the ground state nucleon state. The light-front wave function that contains the complete information on ground state nucleon structure 
can be computed from the Faddeev amplitude. Obtained in this way, the ground state nucleon light-front wave function can be used for the 
exploration of the intrinsic nucleon deformation. This will improve our understanding on the emergence of atomic nuclear structure from 
strong QCD by employing the symmetry-driven solution of the many-body nuclear physics problem~\cite{Draayer}. Decomposition of the nucleon 
light-front wave function over eigenvectors of the Elliot SU(3) irreducible representations will deliver full information on the nucleon 
shape. The intrinsic nucleon deformation can also be identified from the non-zero value of the computed TMD-pretzelosity that can be 
measured in semi-inclusive DIS experiments.

Continuum QCD approaches offer the prospect of evaluating all of the chiral-even and chiral-odd GPD structure
functions ($H$, $E$, $\tilde{H}$, $\tilde{E}$) from the ground state nucleon Faddeev amplitude~\cite{Roberts19}.
These GPD structure functions predicted from analyses of the data on the nucleon elastic form factors and the
$\gamma_vpN^*$ electrocouplings can be plugged into models that relate the GPDs to the DVCS and DVMP observables.
The predicted DVCS and DVMP cross sections, as well as the beam-, target-, and beam-target asymmetries for the
longitudinal and the transverse target polarizations, can be confronted with the experimental data that will be available
from the experiments with CLAS12 and in Halls A/C~\cite{Burkert:2018nvj}. Consistent results on the dressed quark
mass function and the different $qq$-correlations inferred from data analyses in the two independent and complementary
experimental areas, i) nucleon elastic form factors and $\gamma_vpN^*$ electrocouplings and ii) studies of the ground
state nucleon structure in 3D, will validate the credible insight into strong QCD underlying the generation of the ground
and excited state nucleons in a nearly model independent way.

Therefore, continuum QCD approaches~\cite{Burkert:2019bhp,Roberts:2018hpf,Binosi:2016wcx,Aguilar:2019teb}
offer a sound theoretical framework for the interpretation of the results on the ground and excited state nucleon
structure aimed at gaining insight into strong QCD dynamics underlying the generation of hadrons composed of light
$u$ and $d$ quarks with a traceable connection to QCD. These approaches also allow for the exploration of the
emergence of atomic nuclear structure from strong QCD. 

\begin{figure}[tp]
\includegraphics[width=0.95\columnwidth]{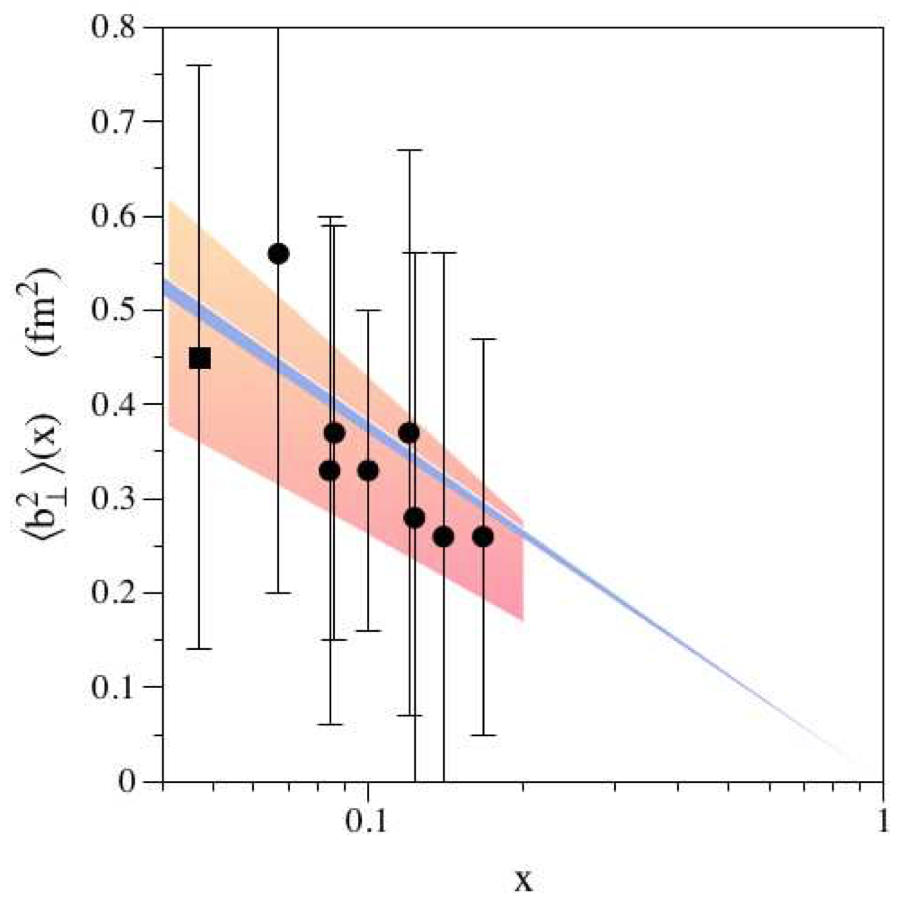}
\vspace{-0.35cm}
\caption{The impact parameter b$^2_\perp$ as a function of the quark momentum fraction $x$. The negative slope of the points 
is an indication of the decrease of the protons transverse size with increasing quark momentum $x$.}
\label{Proton-bT}
\end{figure}

\subsection{Insight into 3D Structure of the Ground State Nucleons in Impact Parameter Space from DVCS and DVMP}
\label{DVCS}

It is now well recognized~\cite{Burkardt:2002hr,Belitsky:2003nz,Belitsky:2001ns,Ji:1996nm} that deeply virtual Compton scattering, 
{\it i.e.} the process $ e p \to e' p' \gamma$, is most suitable for accessing the twist-2 vector GPDs $ H,~E$ and the axial GPDs 
${\tilde H},~{\tilde E}$ in $x$, $\xi$, and $t$. Here $\xi$ is the longitudinal momentum transfer to the struck quark, and $t$ the 
four-momentum transfer to the nucleon. Having access to a 3-dimensional image of the nucleon (two dimensions in transverse space, one 
dimension in longitudinal momentum) opens up completely new insights into the complex internal structure and dynamics of the nucleon 
that, eventually, QCD must reproduce. 
    
The beam helicity dependent cross section asymmetry is defined as $A_{LU} = {\Delta{\sigma_{LU}} / {2 \sigma}}$, where $\Delta{\sigma}_{LU}$ is
the cross section difference for electron spin parallel and spin antiparallel, respectively. In leading twist $\Delta{\sigma_{LU}}$ is given by 
3 GPDs $H$, $\tilde{H}$ and $E$ and two form factors as: $$ {\Delta\sigma_{LU} \propto \sin\phi[F_1H + \xi(F_1+F_2)\tilde{H}]d\phi~}, $$ where 
$\phi$ is the azimuthal angle between the electron scattering plane and the hadronic production plane. The kinematically suppressed term with 
GPD $E$ is omitted. $F_1(t)$ and $F_2(t)$ are the well-known Dirac and Pauli form factors. The asymmetry is mostly sensitive to the GPD 
$H(x,\xi,t)$. In a wide kinematics~\cite{Girod:2007aa,Jo:2015ema} the beam asymmetry $A_{LU}$ was measured at Jefferson Lab at modestly high 
$Q^2$, $\xi$, and $t$, and in a more limited kinematics~\cite{Camacho:2006qlk} the cross section difference $\Delta\sigma_{LU}$ was measured 
with high statistics. Moreover, the first measurements of the target asymmetry $A_{UL}=\Delta\sigma_{UL}/2\sigma$ were carried out
\cite{Chen:2006na,Pisano:2015iqa,Seder:2014cdc}, where 
    
\begin{equation}
\Delta\sigma_{UL} \propto \sin\phi[F_1\tilde{H} + \xi(F_1+F_2)H]~.
\end{equation}
The combination of $A_{LU}$ and $A_{UL}$ allows the separation of GPD $H(x=\xi,\xi,t)$ and $\tilde{H}(x=\xi,\xi,t)$. Using a transversely 
polarized target the asymmetry $A_{UT} = \Delta{\sigma_{UT}}/2\sigma$, with: 
$$\sigma_{UT} \propto \cos\phi\sin(\phi-\phi_s) \frac{t}{4M^2}(F_2H - F_1 E) $$ can be measured, where $\phi_s$ is the azimuthal 
angle of the target polarization vector relative to the electron scattering plane. $A_{UT}$ depends in leading order on GPD $E$. 
Measurement of $A_{UT}$ is the most efficient and most direct way to determine the GPD $E$, as $H$ will be well constrained by the 
other polarization observables $\Delta \sigma_{LU}$ and $\Delta\sigma_{UL}$. 
    
First DVCS experiments carried out at JLab~\cite{Stepanyan:2001sm,Girod:2007aa,Camacho:2006qlk} and DESY~\cite{Airapetian:2001yk} showed
results in terms of the applicability of the handbag mechanism to probe GPDs. The 12~GeV upgrade offers much improved possibilities for 
validating the dominance of the handbag mechanism, and accessing the GPDs. 
    
Measurements of all 3 asymmetries and the unpolarized cross section will enable the separation of GPDs $H$, $\tilde{H}$, and $E$ at some 
specified kinematics. To obtain a complete picture of the quark distribution in the nucleon, the GPDs need to be measured for both 
flavors $u$-quarks and $d$-quarks. This requires measurement of DVCS not only on the proton but also on the neutron. Experiment~\cite{e12-11-003} 
will measure the beam-spin asymmetry for the neutron. If the $t$-dependences are known, a Fourier transformation of, {\it e.g.} GPD $H^u(x=\xi,t)$ 
gives information about the $u$-quark distribution in impact parameter space. Similarly for GPD $\tilde{H}^d$ and GPD $E^d$. 
    
\begin{figure}[tp!]
\includegraphics[width=0.95\columnwidth]{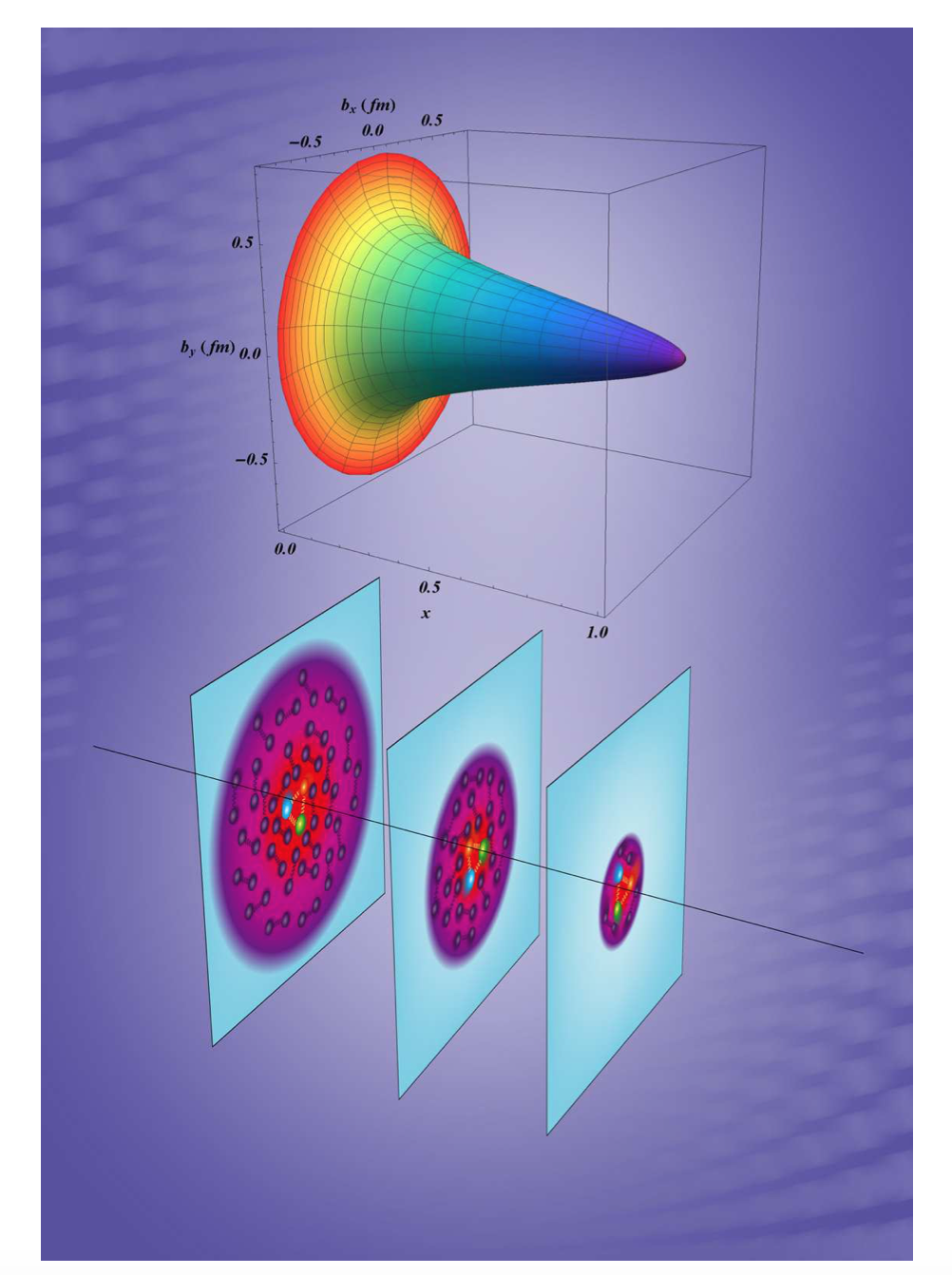}
\caption{The 3D image of the valence quarks in the proton versus quark fractional momentum $x$ using the data and fit parameters shown in 
Fig.~\ref{Proton-bT}.}
\label{Proton-3D}
\end{figure}
    
An essential aspect of DVCS measurements is the promise that 3D images of the nucleon may be extracted once high precision data are available 
covering a large part of the phase space in $x_B$, $t$, and $Q^2$, and to have these data collected with polarized lepton beams, with 
longitudinally and with transverse polarized nucleon targets. While some of the required measurements are currently underway, the information 
currently available is not sufficient for a full analysis with no or only minimal external constraints. The covered kinematics in $x_B$ and in 
$t$ is too limited for such an analysis. Nevertheless attempts have been made to get a first glimpse of the 3D structure of the proton
\cite{Dupre:2017hfs}. Results of this analysis, that uses all available DVCS data, are shown in Fig.~\ref{Proton-bT}. The impact parameter 
$b_\perp^2$ is plotted as a function of the quark momentum fraction $x$. The negative slope of the fit indicates a decrease in the 
transverse space the quarks occupy with increasing quark momentum.
    
This information can be employed to extract a 3D image of the proton's quark distribution a function of the quark $x$ momentum fraction, which 
is presented in Fig.~\ref{Proton-3D}. 
    
\begin{figure}[htp]
\centering
\includegraphics[width=0.98\columnwidth]{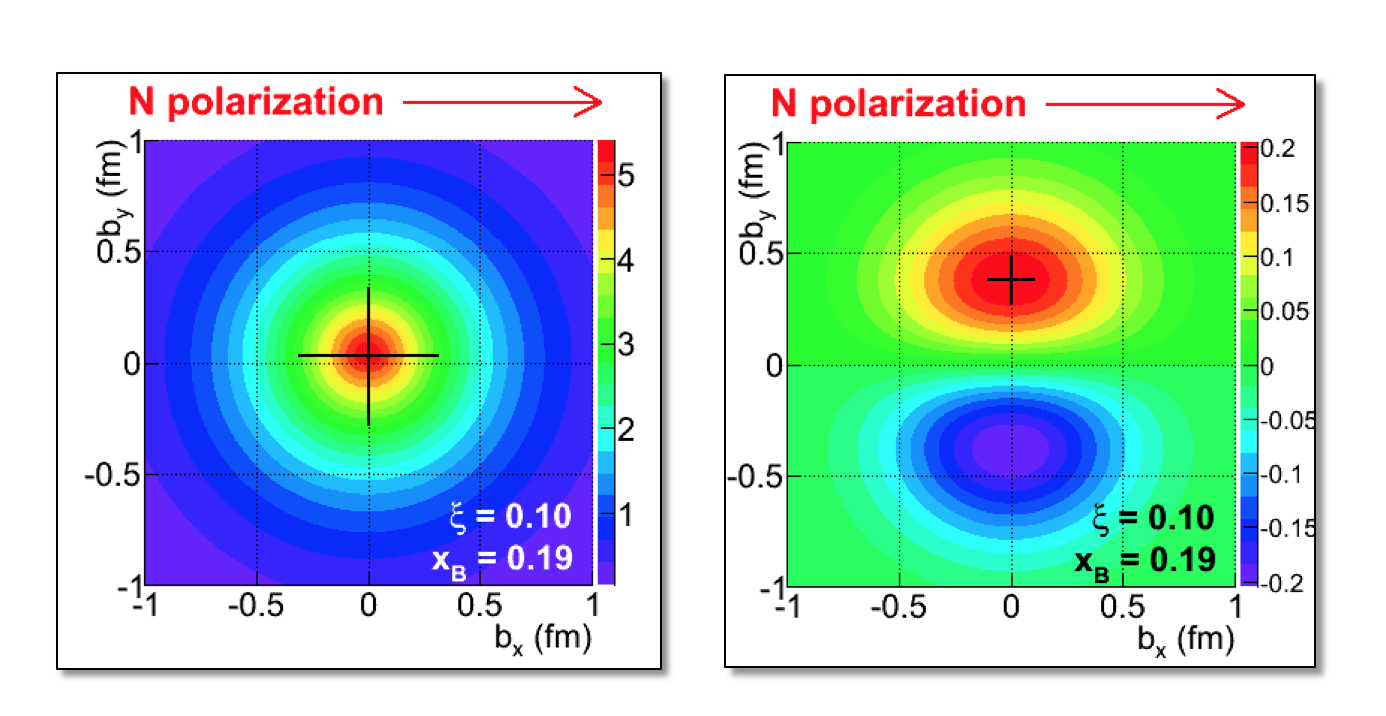}
\caption{The quark distributions in impact parameter space at fixed value of $x_B$ based on the CFF $\cal{H} + \cal{E}$ (left) for a transversely 
polarized target. The direction of the nucleon polarization is indicated by the arrow on top. The slight vertical shift of the center point is due 
to the non-zero value of $\cal{E}$. The right panel shows the quark distributions as projected from the CFF $\cal{E}$ separately, indicating an 
induced electric dipole moment due to the positive and negative charged quarks shifting in opposite directions. The graphs are extracted from 
simulated data weighted with the expected uncertainties for all projected measurements with polarized electron beams, longitudinally and transverse 
polarized targets.}
\label{H-E_GPD}
\end{figure}

Although these images are based on experimental data, the results include large uncertainties from insufficient kinematic coverage and incomplete 
data. It is therefore desirable to carry out analyses of generated data that include all expected future data with their coverage and expected 
statistical and systematic uncertainties. This has been done in a trial study~\cite{FXGirod}. In this study the Compton Form Factors 
${\cal H}(\xi, t)$ and ${\cal E}(\xi, t)$ were extracted from the simulated data. They are shown in Fig.~\ref{H-E_GPD}. 
    
GPDs are encoded in the off-forward matrix elements of two currents separated along the light cone. The formulation of lattice QCD in Euclidean 
space precludes their direct calculation, and therefore the approach of exploiting the operator product expansion to express the moments with respect 
to Bjorken $x$ as the matrix elements of local operators that are accessible to calculation on a Euclidean lattice, yielding Generalized Form
Factors~\cite{Hagler:2007xi,Bratt:2010jn,Alexandrou:2019ali,Bali:2018zgl}. These computations have provided important information on the 
three-dimensional imaging of the nucleon, notably in the decomposition of angular momentum within the nucleon~\cite{Hagler:2007xi,Bratt:2010jn}.
    
Recently, approaches have been adopted that enable the $x$-dependent distributions to be systematically related to quantities that are amenable to 
calculation on a Euclidean lattice; these approaches are described in detail in Section~\ref{theory_out}. These methods can be applied to the 
computation of GPDs~\cite{Radyushkin:2019owq}

While DVCS is regarded as the most promising exclusive channel for constraining the nucleon GPDs in leading order, DVMP has evolved into a dynamic 
field giving access to higher twists as well as to the GPD flavor decomposition. GPDs admit a particularly 
intuitive physical interpretation at zero skewness, where after a Fourier transform GPDs describe the spatial distribution of quarks with given 
longitudinal momentum in the transverse plane. There are eight independent twist-two GPDs, four correspond to parton helicity-conserving 
(chiral-even) processes, denoted by $H^q$, $\tilde H^q$, $E^q$ and $\tilde E^q$, and four correspond to parton helicity-flip (chiral-odd) processes, 
$H^q_T$, $\tilde H^q_T$, $E^q_T$, and $\tilde E^q_T$~\cite{Diehl:2003ny}. The chiral-odd GPDs are difficult to access since subprocesses with quark 
helicity-flip are usually strongly suppressed. By this reason very little is known about the chiral-odd GPDs. One of which, $H_T$, is related to the 
transversity distribution $H^q_T(x,\xi=0,t=0)=h^q_1(x)$ and the tensor charge $\kappa^q_T=\int{dx\,H^q_T(x,\xi,t=0)}$ while $\bar E_T$ is related to 
the anomalous tensor magnetic moment of the nucleon $\delta^q_T=\int{dx\,\bar E^q_T(x,\xi,t=0)}$.

The DVMP processes $\gamma^*N\to MN$ may include the exchange of vacuum quantum numbers in the t-channel and non-vacuum exchanges. To the first class we 
can assign the processes with $M=\rho^0,\omega,\phi,J/\psi,\Upsilon$, where we can access the gluon GPDs. The second class $M=\pi^0,\pi^+,\rho^+,K^+, 
K^*$  gives us access to the valence quark GPDs. The variety of reactions with mesons with different quark contents gives us the opportunity to make the 
GPD flavor decomposition. 

The special attention was attracted to the pseudoscalar meson production such as $\gamma^*(p/n)\to (\pi^0/\eta) (p/n)$. The early efforts to explain 
$\pi^0/\eta$ electroproduction focused on the chiral even (no helicity flip) GPDs, $\tilde H$ and $\tilde E$, as a means to parameterize only the 
longitudinal virtual photon amplitudes. However, these predictions were an order of magnitude lower than the recent experimental results
\cite{Bedlinskiy:2012be,Bedlinskiy:2014tvi,Bedlinskiy:2017yxe,Defurne:2016eiy,Mazouz:2017skh,Alexeev:2019qvd}. During the past few years, several 
approaches~\cite{Goloskokov:2009ia,Goloskokov:2011rd,Ahmad:2008hp,Goldstein:2010gu,Schweitzer:2016jmd} have been developed utilizing chiral odd GPDs 
in the calculation of pseudoscalar electroproduction. It was found that the the contribution from transverse polarized photons is strongly enhanced 
by the chiral condensate for pseudoscalar mesons. The role of the transversity GPDs, $H_T$ and $\bar E_T=2\tilde H_T+E_T$, has been examined and a 
strong increase of the $\pi^0/\eta$ cross section found. This conclusion was supported by the recent Rosenbluth separation of the structure functions 
$\sigma_L$ and $\sigma_T$~\cite{Defurne:2016eiy,Mazouz:2017skh} where it was found that the contribution of transverse $\sigma_T$ is dominant. The 
structure functions in the electroproduction of the pseudoscalar mesons read
$$
\frac{d\sigma_{T}}{dt} = \frac{4\pi\alpha}{2k Q^4} \left[ \left(1-\xi^2\right) \left| {\cal H}_T(\xi,t)\right|^2 - \frac{t'}{8m^2} \left|
 \bar{\cal E}_T(\xi, t)  \right|^2\right]
$$
$$
\frac{d\sigma_{TT}}{dt} = \frac{4\pi\alpha}{k Q^4}\frac{t'}{16m^2}\left|   \bar{\cal E}_T(\xi, t)   \right|^2
$$
\noindent 
The  generalized form factors ${\cal H}_T(\xi, t)$ and $\bar{\cal E}_T(\xi, t)$ denote the convolution of the elementary process $\gamma^*q\to q\pi^0$
with the GPDs $H_T$ and $\bar E_T$. As shown in the formulae above the form factors ${\cal H}_T(\xi, t)$ and $\bar{\cal E}_T(\xi, t)$ are directly 
accessible from the experimental observables $\sigma_T$ and $\sigma_{TT}$~\cite{Kubarovsky:2016yaa,Kubarovsky:2019nzr}.

In contrast to the form factors ${\cal H}_T(\xi, t)$ and $\bar{\cal E}_T(\xi, t)$  there is no direct way to get along without models for the GPDs 
in order to describe data on hard exclusive reactions. One such parameterization was presented in Ref.~\cite{Goloskokov:2011rd}. The proposed model 
is physically motivated on one hand by Regge phenomenology in the limit $x\to 0$ and, on the other hand, by the physical intuition gained in the 
impact parameter representation of GPDs. The main feature of this ansatz is an exponential $t$ behavior 
$\bar E^q(x,\xi=0,t)=N^qx^{-\alpha^q_0}(1-x)^{n^q}\exp[t(b^q-\alpha'^q\ln{x]}$. The $\xi$-dependence is generated by the double distribution 
representation of the GPDs~\cite{Radyushkin:1998bz}. There are 4 parameters for each quark flavor in the model. The $\bar E_T$ parameters for $u$ 
and $d$-quarks were determined in the global fit of all available data from CLAS, Hall~A, and COMPASS
\cite{Bedlinskiy:2012be,Bedlinskiy:2014tvi,Bedlinskiy:2017yxe,Defurne:2016eiy,Mazouz:2017skh,Alexeev:2019qvd} on the $\sigma_{TT}$ structure functions 
for the reactions $ep\to e'(\pi^0/\eta)p'$ and $en\to e'\pi^0n'$. According to Refs.~\cite{Diehl:2005jf,Diehl:2013xca}, the Fourier transform of $\bar E_T$ 
$$
 {\bar E_T}(x,\vec b)=\int\frac{d^2\vec \Delta}{(2\pi)^2}e^{-i\vec b\cdot \vec \Delta} \bar E_T(x,t=-\Delta^2).
$$
controls the density of polarized quarks in the unpolarized nucleon
$$
\delta(x,\vec b)=\frac{1}{2}[H(x,\vec b)-\frac{b_y}{m}\frac{\partial}{\partial b^2}\bar E_T(x,\vec b)].
$$

\begin{figure}[htp]
\centering
\includegraphics[width=0.98\columnwidth]{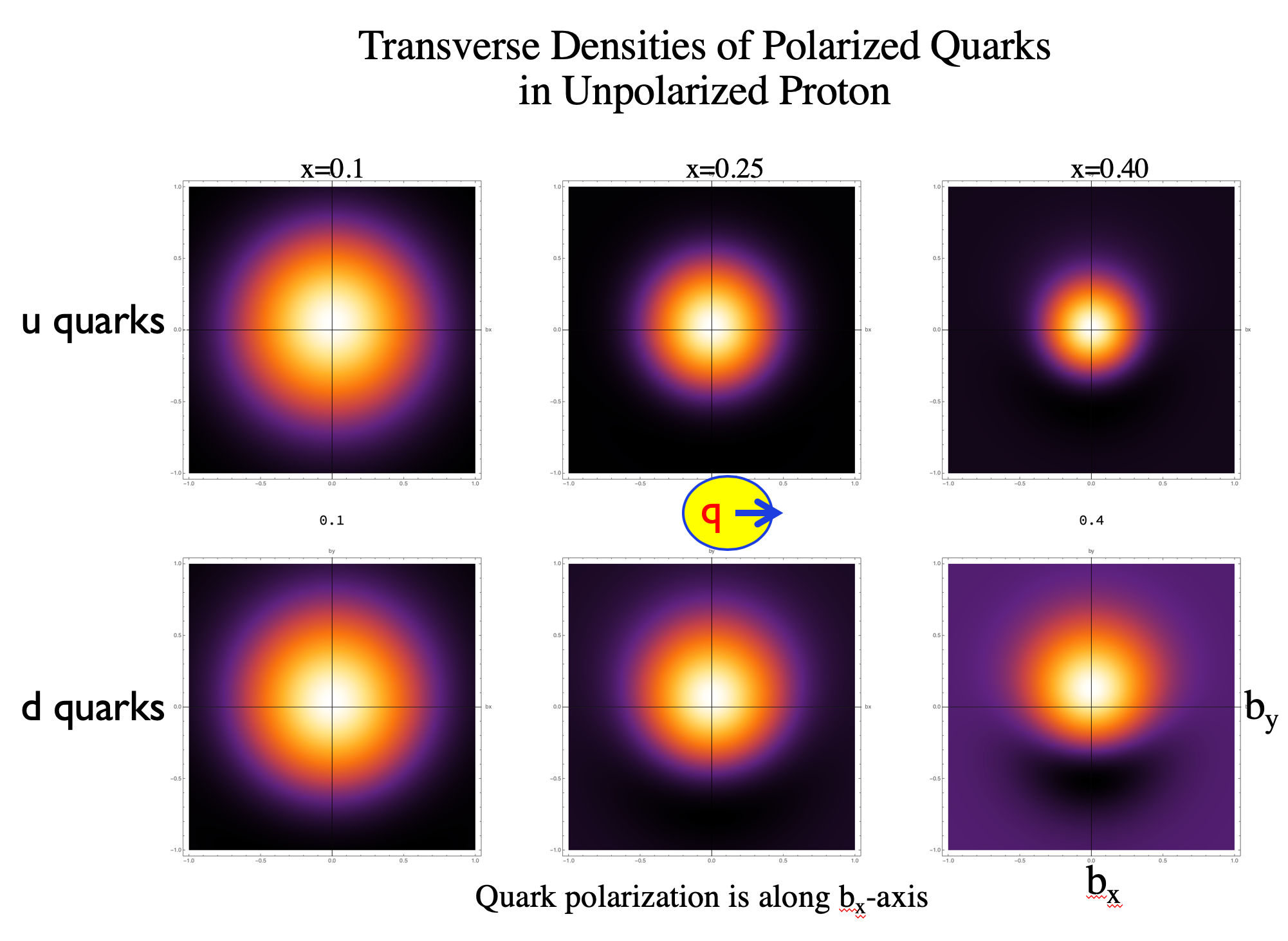}
\caption{The impact parameter density of the transversely polarized quarks along the $b_x$-axis in an unpolarized proton for Feynman $x=0.1$, $x=0.25$ 
and $x=0.40$, $u$-quarks distributions are in the top row and $d$-quarks in the bottom row. The quark polarization is indicated by the arrow in the 
middle of the figure. The density depends on $H(x,\vec b)$ and  $E_T(x,\vec b)$ as 
$\delta(x,\vec b)=\frac{1}{2}[H(x,\vec b)-\frac{b_y}{m}\frac{\partial}{\partial b^2}\bar E_T(x,\vec b)]$. The $E_T(x,\vec b)$ parameters were determined 
by the global fit of the world data on the $\pi^0/\eta$ electroproduction out of the proton and neutron. The GPD $H(x,\vec b)$ describes the density of 
unpolarized quarks and $\bar E_T (x,\vec b)$ is related to the distortion of the polarized quarks in the transverse plane. Note that the width of the 
quark cloud is diminished as $x\to1$. The visible vertical shift of the center point is due to $\bar E_T$.}
\label{density_3d}
\end{figure}

The transverse density of polarized quarks in an unpolarized proton, based on the global fit of the world data, is presented in Fig.~\ref{density_3d}
\cite{Kubarovsky:2019nzr}. This density depends on two GPDs $H(x,\vec b)$ and $E_T(x,\vec b)$. The GPD $H(x,\vec b)$ describes the density of unpolarized 
quarks and $\bar E_T (x,\vec b)$ is related to the distortion of the polarized quarks in the transverse plane. Note that the width of the quark cloud is
diminished as $x\to1$. This behavior is typical for the GPD models. The visible vertical shift in the positive $b_y$ direction of the center point is due
to  $\bar E_T$ that has the same sign for both $u$ and $d$-quarks.
 
The study of the pseudoscalar electroproduction is ongoing at Jefferson Lab using data sets that were collected during 2018-2020 years with higher 
statistics and kinematic reach. Extracting $d\sigma_L/dt$ and $d\sigma_T/dt$ and performing new measurements with transversely and longitudinally 
polarized targets would also be very useful, and are planned for the future Jefferson Lab at 12~GeV. The measurement of beam and target spin asymmetries 
can provide further constraints on the theoretical handbag models considered here. The physics community is actively planning DVMP experiments at the 
EIC that will definitely expand our understanding of the nucleon structure.
 
\subsection{Insight into 3D Structure of the Ground State Nucleons in Momentum Space from SIDIS}
\label{semiinc}

SIDIS data enables us to build 3D maps of the partonic structure of nucleons in momentum space. In fact, by 
measuring the transverse momentum $P_{hT}$ (with respect to the virtual photon direction) of an inclusively produced hadron $h$ we can reconstruct 
the transverse momentum $k_T$ of the confined parton, provided that a suitable factorization theorem holds. The necessary condition for this to happen 
is $P_{hT}/z \ll Q$, where $z$ is the fraction of the fragmenting parton energy carried by $h$, and $Q$ is the four-momentum of the virtual photon, the 
SIDIS hard scale~\cite{Ji:2004wu,Aybat:2011zv,Collins:2011zzd,Echevarria:2012pw}. Namely, factorization holds if SIDIS is a two-scale process. 

\begin{figure}[htp]
\includegraphics[width=0.95\columnwidth]{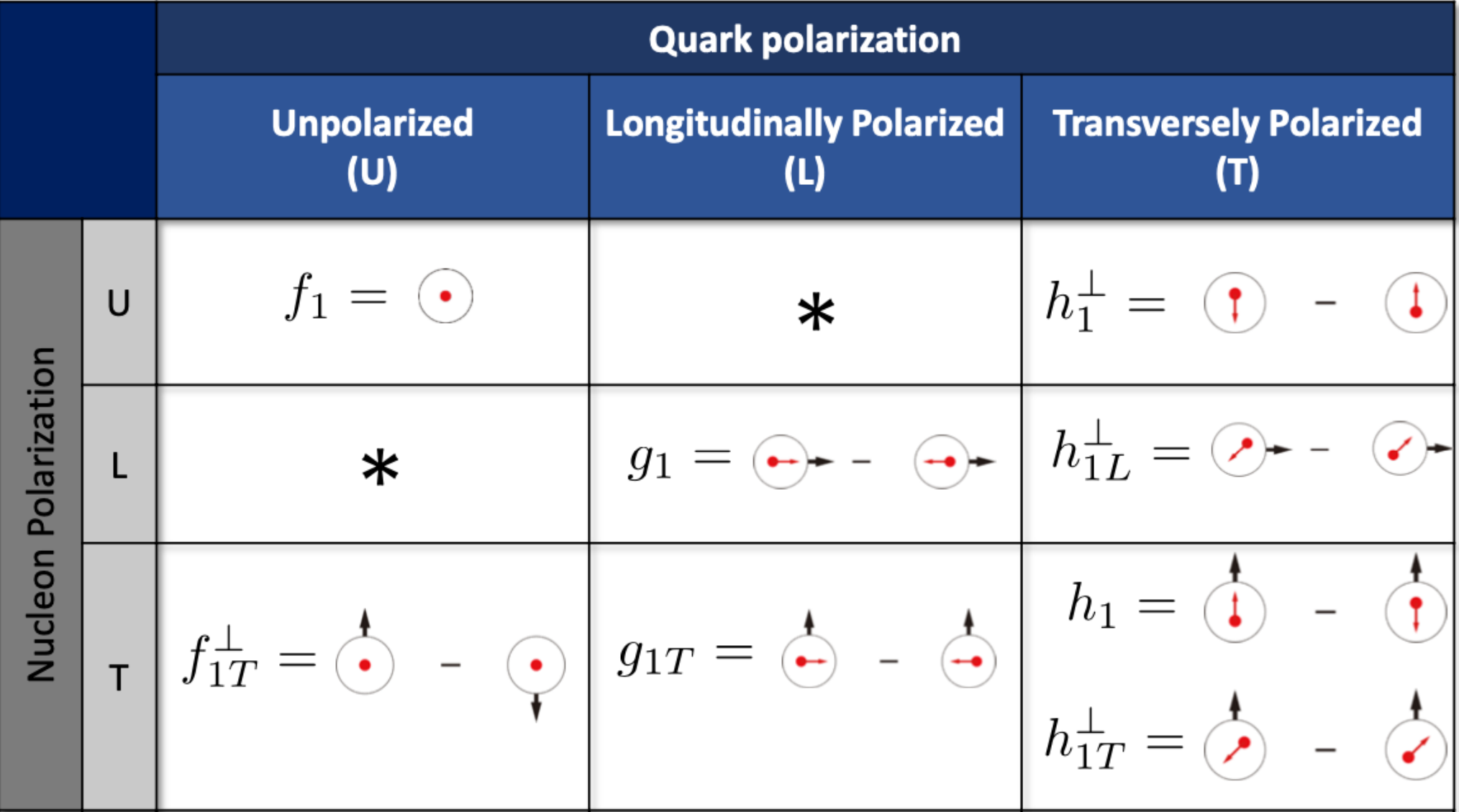}
\vspace{-0.25cm}
\caption{Table of all leading-twist TMD PDFs of a quark in a spin-1/2 hadron depending on quark polarization (columns: U - unpolarized, 
L - longitudinally polarized, T - transversely polarized) and hadron polarization (rows U, L, T, with same meaning). Horizontal (vertical 
or slanted) arrows indicate longitudinal (transverse) polarization. Red arrows for quark, black arrows for hadron. Asterisks denote 
combinations forbidden by invariance under parity transformations.}
\label{fig:TMDtable}
\end{figure}

In this case, the cross section can be factorized into a hard photon-absorption vertex and two non-perturbative objects, one 
describing the probability density of finding in the nucleon a parton with $k_T$ and longitudinal momentum fraction $x$, one 
describing the probability density for a parton to fragment into the observed hadron with $z$ and $P_{hT}$. These non-perturbative 
objects are called TMD PDFs and fragmentation functions (TMD FFs), respectively; 
in short TMDs. In Fig.~\ref{fig:TMDtable}, we list all the leading-twist TMD PDFs for quarks; an analogous table can be formed for TMD
FFs~\cite{Mulders:1995dh,Boer:1997nt}. Each entry in the table corresponds to a specific polarization status of the quark and of the
parent nucleon. The asterisks denote combinations that are forbidden by invariance under parity transformations. Building a 3D map
for each entry can expose details of spin-spin, spin-momentum (spin-orbit) partonic correlations, as well as correlations between the
partonic motion and nucleon spin. An even richer table can be made at subleading twist to explore quark-gluon dynamical correlations.
However, no factorization theorem is available yet at this level although conjectures towards it have been recently published
\cite{Bacchetta:2019qkv}. 

Factorization in the TMD framework holds also for other processes like $e^+e^-$ annihilation and production of Drell-Yan lepton pairs
(for a review, see Ref.~\cite{Collins:2011zzd}), making it possible to separately test the universality of TMD PDFs and TMD FFs.
Factorization is explicitly broken for inclusive hadron production in hadronic collisions~\cite{Rogers:2010dm}. An alternative approach
has been recently suggested to detect hadrons inside jets, because in this hybrid framework the cross section factorizes in terms of
the same universal TMD FF and a collinear PDF~\cite{Kang:2017glf}. 

Evolution of TMDs is more complicated than in the collinear framework. In the CSS scheme~\cite{Collins:2011zzd,Collins:1984kg}, the 
evolution operator contains a perturbative part and a non-perturbative part. The perturbative component of the operator resums all 
logarithms of $q_T / Q$ connected to soft gluon radiation, where $q_T = P_{hT}/z$. This component can be calculated nowadays at a 
level of sophistication comparable to the most refined phenomenological analyses at LHC~\cite{Bacchetta:2019sam,Scimemi:2019cmh}. But f
or very small parton $k_T$, the perturbative description breaks down. The non-perturbative component must take over, and a prescription 
for a smooth transition between the two regimes is needed. Both the non-perturbative part of the evolutor and the transition prescription 
can be described in an arbitrary way and depend on parameters to be fitted to data (the same holds also for other schemes
\cite{Laenen:2000de,Bozzi:2005wk,Echevarria:2012pw}). Moreover, matching the TMD framework at $q_T \ll Q$ to fixed-order calculations in 
collinear framework at $q_T \lesssim Q$ is still an open problem~\cite{Angeles-Martinez:2015sea,Collins:2016hqq,Gamberg:2017jha,Echevarria:2018qyi}.
Therefore, it is important to have experimental data sensitive to transverse momenta that span a large portion of phase space in $Q^2$ and 
also $x$. 

\begin{figure}[htp]
\includegraphics[width=0.95\columnwidth]{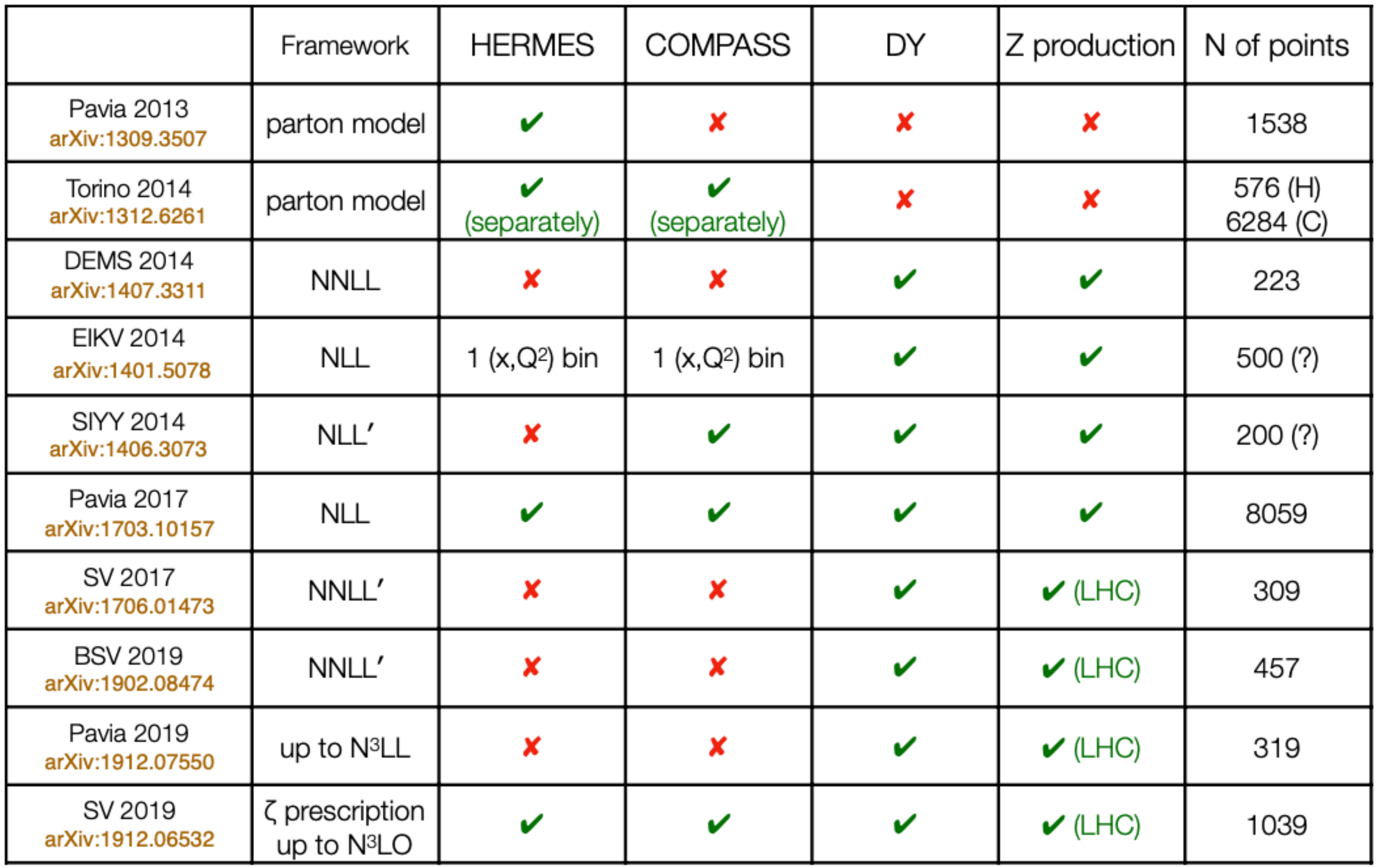}
\vspace{-0.25cm}
\caption{The most recent extractions of unpolarized TMD PDF $f_1 (x, k_T^2)$ from data on SIDIS unintegrated multiplicities provided by 
the HERMES (H) and COMPASS (C) collaborations, and/or cross sections for the production of Drell-Yan lepton pairs (DY) or $Z$-boson. In 
the second left column, the corresponding accuracy in the description of the perturbative component of TMD evolution is indicated: 
resummation of soft gluon radiation up to next-to-leading log precision (NLL), next-to-next-leading log (NNLL), next-to-next-to-next-leading 
log ($\mathrm{N}^3$LL), and slight variations of them indicated by NLL' and NNLL', respectively~\cite{Bacchetta:2019sam}. The last entry in 
the bottom box refers to the specific $\zeta$ prescription by the corresponding authors that amounts to a next-to-next-to-next-leading order 
($\mathrm{N}^3$LO) perturbative accuracy. The most recent four extractions use also LHC data on $Z$-boson production.}
\label{fig:f1fitlist}
\end{figure}

\begin{figure}[htp]
\includegraphics[width=0.9\columnwidth]{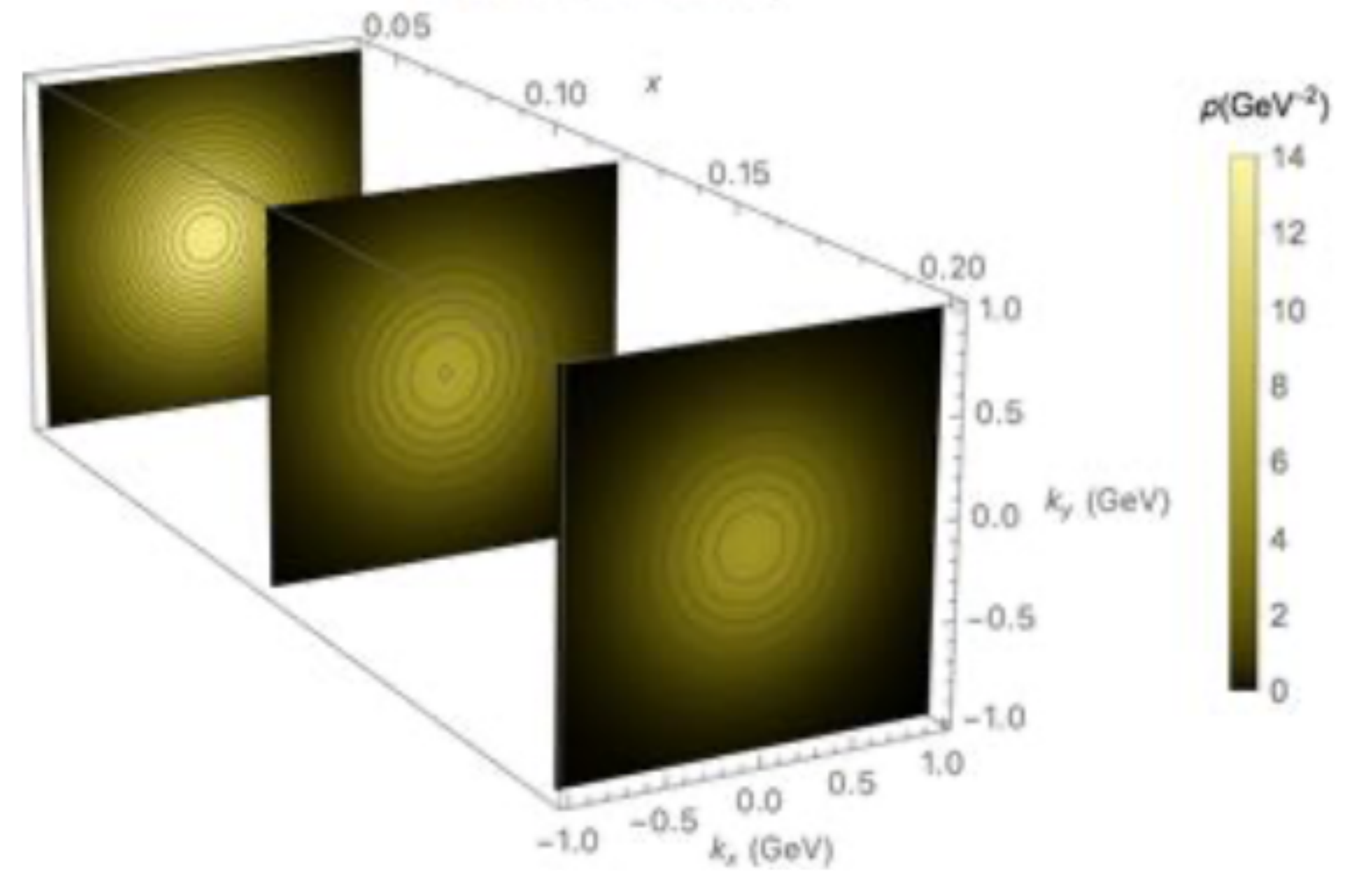}
\vspace{-0.25cm}
\caption{Tomography of unpolarized $f_1 (x, k_T^2)$ of quarks in the proton at $Q^2 = 1$~GeV$^2$~\cite{Bacchetta:2017gcc}.}
\label{fig:f1tomography}
\end{figure}

The best known leading-twist TMD PDF is the unpolarized $f_1 (x, k_T^2)$. In Fig.~\ref{fig:f1fitlist}, we list the most recent 
extractions of $f_1$ from data on SIDIS unintegrated multiplicities and/or cross sections for the production of Drell-Yan lepton pairs 
or $Z$-boson. The ``Pavia 2013" is the first (and unique, so far) analysis with an explicit flavor dependence in the fitting parameters 
of the functional form~\cite{Signori:2013mda}. The impact of this non-perturbative effect on the extraction of the $W$-boson mass $m_W$ at 
LHC has been explored using the template-fit technique, reaching the conclusion that the uncertainty $\Delta m_W$ induced by flavor sensitivity 
in quark $k_T$ distributions might be comparable with the error correlated to PDF uncertainties~\cite{Bacchetta:2018lna}. The ``Pavia 2017" is 
the first extraction from a global fit of 8059 points from SIDIS, Drell-Yan, and $Z$-boson data~\cite{Bacchetta:2017gcc}, reaching a 
$\chi^2/$d.o.f. $= 1.55 \pm 0.05$ with only 11 parameters. The main drawbacks of this analysis were the still marked anti-correlation between 
TMD PDF and TMD FF $k_T$ distributions (calling for an independent extraction of TMD FFs from $e^+ e^-$ data, which is still missing), and 
difficulties in reproducing the normalization of low-$Q^2$ SIDIS data (a similar finding is reported in other later works
\cite{Gonzalez-Hernandez:2018ipj,Boglione:2019nwk}). Very recently, a series of new analyses have been published where the details of perturbative 
resummation match the same accuracy of standard phenomenology at the LHC. In ``BSV 2019"~\cite{Bertone:2019nxa} and ``Pavia 2019"
\cite{Bacchetta:2019sam} the data set encompasses measurements from only Drell-Yan and $Z$-boson production including very precise ATLAS 
data points, implementing kinematical cuts on the final leptons and without {\it ad-hoc} adjusting the normalization~\cite{Bacchetta:2019sam}. 
The ``SV 2019" includes also SIDIS data~\cite{Scimemi:2019cmh} although the total number of fitted points is still far from the ``Pavia 2017" 
record. In all these cases, a large impact of LHC data was found on the behavior of the resulting $f_1 (x, k_T^2)$ at small $x$. The general 
outcome of results listed in Fig.~\ref{fig:f1fitlist} can be summarized in the tomography depicted in Fig.~\ref{fig:f1tomography}, which shows 
the $x$-dependence of the $k_T$-distribution of quarks inside Nucleons at $Q^2=1$~GeV$^2$. Through TMD evolution, we know also how this tomography 
evolves to higher scales and from ``Pavia 2013" we have some limited information on its flavor dependence~\cite{Signori:2013mda}. Nothing is known 
about the gluon TMD PDF. Several studies explored useful channels at RHIC or LHC with  $p-p$ collisions leading to $J/\psi+X$, or 
$J/\psi + \gamma + X$, or $\eta_c + X$~\cite{Dunnen:2014eta,Boer:2016fqd,DAlesio:2017rzj,DAlesio:2019gnu}. At the EIC, it would be possible to 
consider also SIDIS processes like $e-p$ collisions leading to $J/\psi + X$, $h_1 + h_2 + X$, jet + jet + $X$, $J/\psi$ + jet + $X$
\cite{Mukherjee:2016qxa,Boer:2016fqd,Rajesh:2018qks,Bacchetta:2018ivt}. 

\begin{figure}[htp]
\includegraphics[width=0.9\columnwidth]{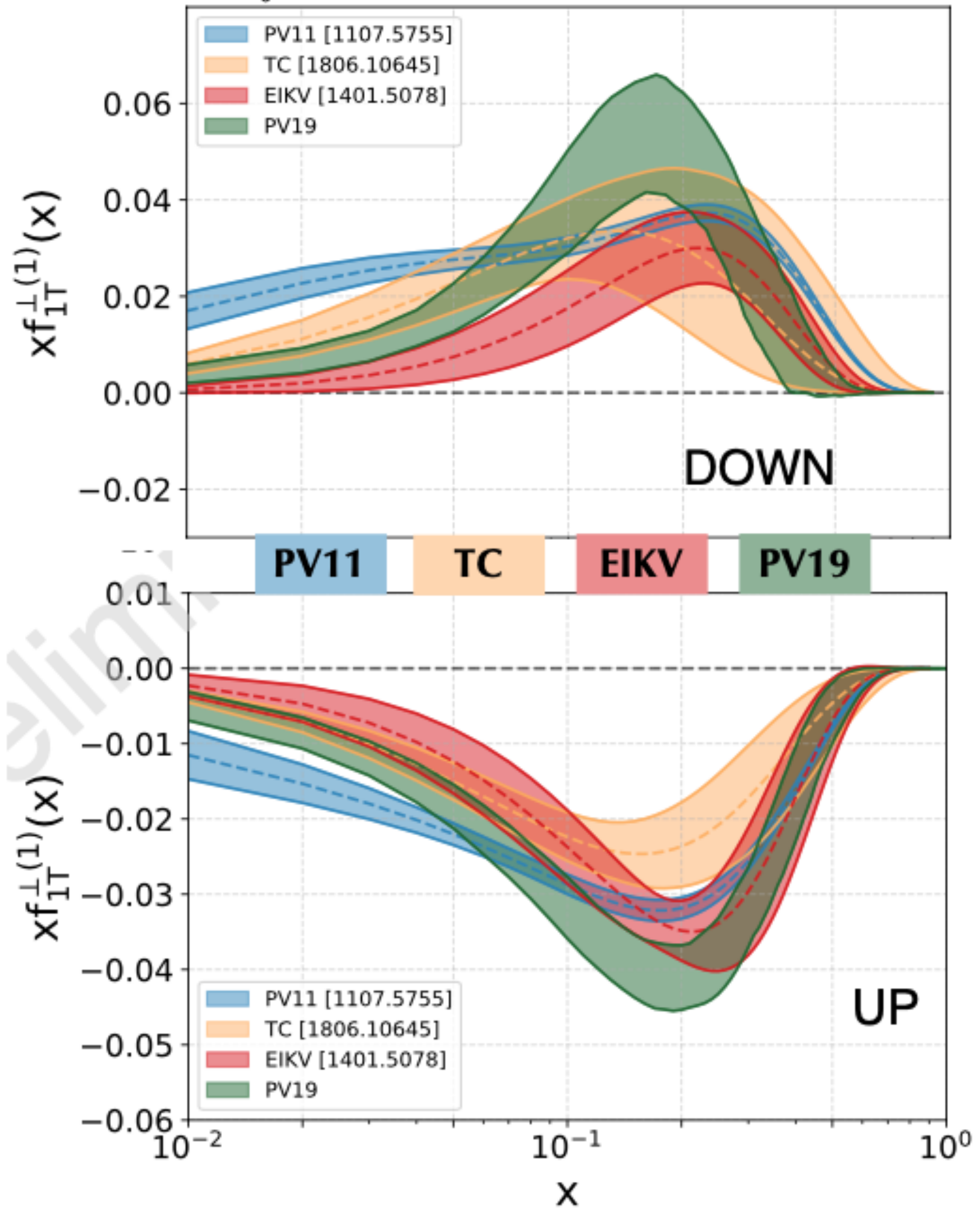}
\vspace{-0.25cm}
\caption{The $x f_{1T}^{\perp (1)} (x)$ (first $k_T$-moment of the Sivers function multiplied by $x$) as a function of $x$. Upper panel for
  down quark, lower for up quark. Parameterizations: ``PV11" at $Q^2 = 1$~GeV$^2$ from Ref.~\cite{Bacchetta:2011gx}; ``EIKV" at
  $Q^2 = 2.4$~GeV$^2$ from Ref.~\cite{Echevarria:2014xaa}; ``TC" at $Q^2 = 1.2$~GeV$^2$ from Ref.~\cite{Boglione:2018dqd};
  ``PV19" at $Q^2 = 1$~GeV$^2$ from Ref.~\cite{Bacchetta:2019xxx}.}
\label{fig:SiverskTmom}
\end{figure}

\begin{figure}[htp]
\includegraphics[width=0.9\columnwidth]{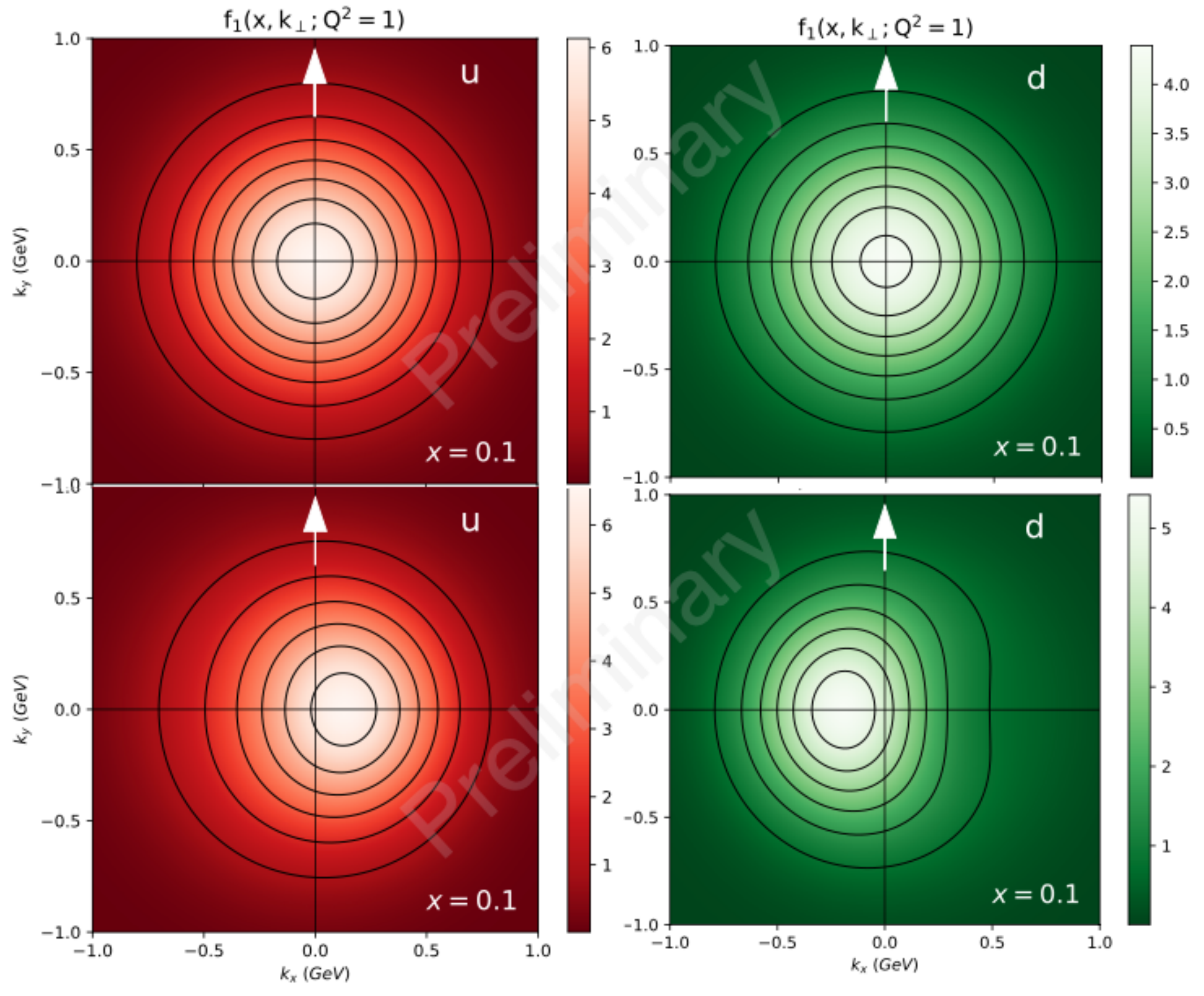}
\vspace{-0.25cm}
\caption{Upper panel: tomography of unpolarized $f_1 (x=0.1, k_T^2)$ at $Q^2 = 1$~GeV$^2$ as in Fig.~\ref{fig:f1tomography}.
  Lower panel: tomography of Sivers effect in same kinematic conditions from ``PV19" parameterization~\cite{Bacchetta:2019xxx}.
  Left panel for up quark, right panel for down quark.}
\label{fig:Siverstomography}
\end{figure}

In Fig.~\ref{fig:TMDtable}, another key entry in the table is the so-called Sivers function $f_{1T}^\perp$ appearing in the bottom left 
corner. It describes how the $k_T$ distribution of an unpolarized quark is distorted in a transversely polarized Nucleon~\cite{Sivers:1990cc}. 
Evidently, the $f_{1T}^\perp$ describes a spin-orbit effect at the partonic level. The Sivers function is representative of the class of na\"ive 
T-odd TMDs, namely of those TMDs that are not constrained by T-reversal invariance~\cite{Boer:1997nt}. Their universality is broken but in a 
calculable way. For example, the Sivers function extracted in a Drell-Yan process with a transversely polarized proton should turn out opposite 
to the one that is extracted in SIDIS. The $f_{1T}^\perp \vert_{\mathrm{DY}} = - f_{1T}^\perp \vert_{\mathrm{SIDIS}}$ prediction is based on 
very general assumptions, and it represents a fundamental test of QCD in the non-perturbative regime~\cite{Collins:2002kn}. Therefore, it is 
subject of intense experimental investigation. Preliminary results hint to statistically favor the prediction
\cite{Adamczyk:2015gyk,Aghasyan:2017jop,Anselmino:2016uie} although more precise data are needed to draw a sharp conclusion. Many parameterizations 
of the Sivers function are available on the market (for example, see
Refs.~\cite{Vogelsang:2005cs,Collins:2005ie,Bacchetta:2011gx,Anselmino:2012aa,Aybat:2011ta,Sun:2013hua,Boer:2013zca}). In Fig.~\ref{fig:SiverskTmom}, 
the first $k_T$-moment multiplied by $x$, {\it i.e.} $x f_{1T}^{\perp (1)} (x)$, is shown as a function of $x$ for the up and down quarks. Results 
from recent parameterizations (``PV11"~\cite{Bacchetta:2011gx}, ``EIKV"~\cite{Echevarria:2014xaa}, ``TC"~\cite{Boglione:2018dqd}, 
``PV19"~\cite{Bacchetta:2019xxx}) agree within statistical uncertainty. Figure~\ref{fig:Siverstomography} supplements the tomography of
Fig.~\ref{fig:f1tomography}: the upper panels correspond to the unpolarized $k_T$ distribution of that figure at $x=0.1$; the lower panels visualize 
how those $k_T$ distributions get distorted by the Sivers effect in an opposite way for up and down quarks. Both groups of panels show quark 
probability densities that are entirely based on real experimental data for Nucleons with or without transverse polarization. 

\begin{figure}[htp]
\includegraphics[width=0.8\columnwidth]{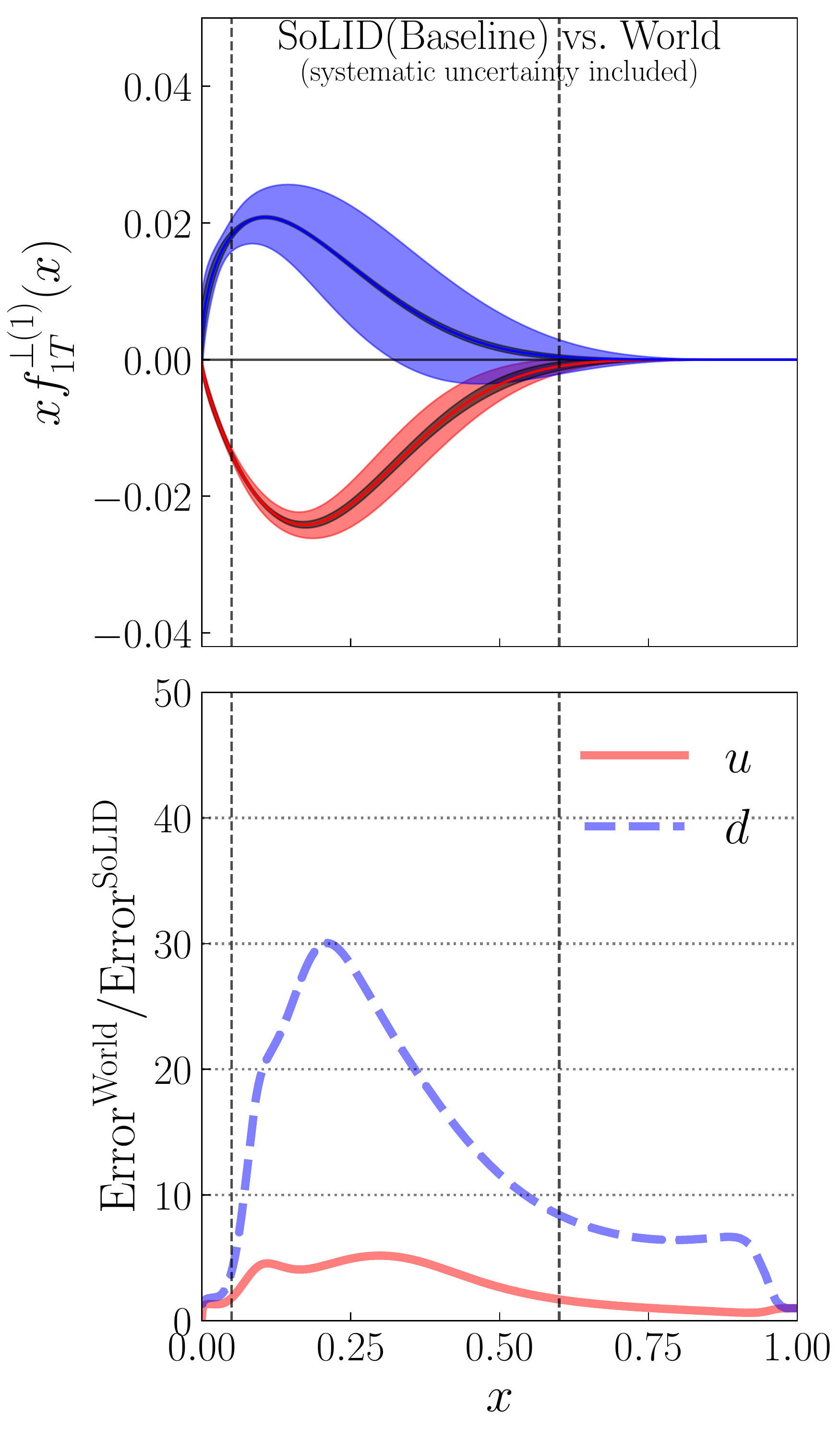}
\vspace{-0.25cm}
\caption{Upper panel: projections of uncertainty band of the Sivers function for up (red bands) and down (blue bands) quarks extracted
  from SoLID measurements (darker bands) comparing with those extracted from current world data (lighter bands) at
  $Q^2 = 2.4$~GeV$^2$. Lower panel: ratios of uncertainty for current world data over that for SoLID for up (red curve) and down
  (blue curve) quarks.}
\label{fig:SoLIDSivers}
\end{figure}

\begin{figure}[htp]
\includegraphics[width=0.9\columnwidth]{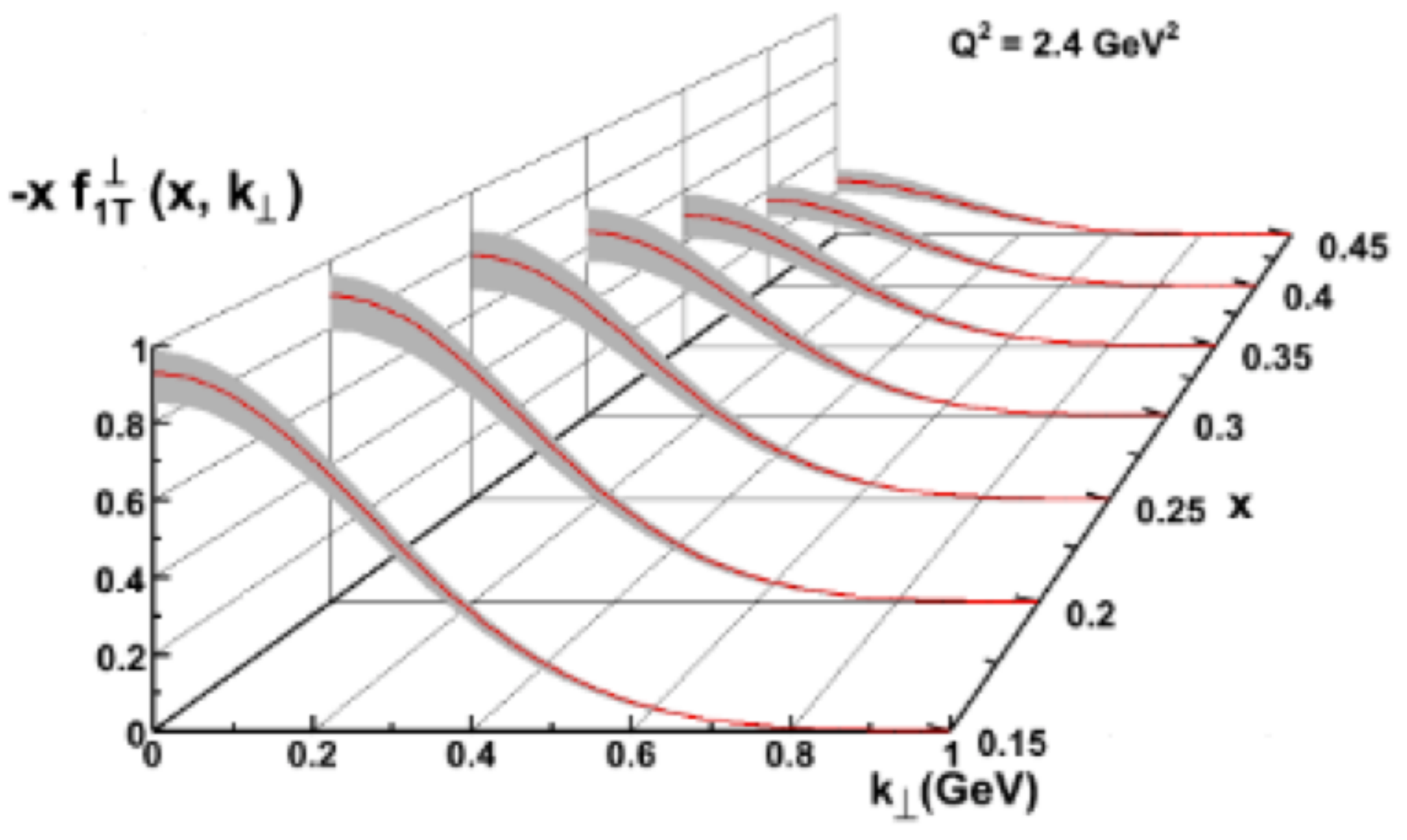}
\vspace{-0.25cm}
\caption{Sivers function of $x$ and $k_\perp$ for up quark at $Q^2 = 2.4$~GeV$^2$. Red band: uncertainty band for projection
  extracted from SoLID measurements. Gray band: uncertainty band extracted from current world data.}
\label{fig:SoLIDSivers2d}
\end{figure}

The precision of the current knowledge will be significantly improved with the planned measurements at Jefferson Lab after the recent 
12~GeV energy upgrade (JLab12) and with planned future EIC. The SoLID SIDIS experiments~\cite{Chen:2014psa,SoLID:SIDISn,SoLID:SIDISp} with 
JLab12 will provide the most precise measurements~\cite{SoLID:pCDR} of Sivers asymmetries in the valence quark region ($x > 0.05$).
Figure~\ref{fig:SoLIDSivers} shows the projected uncertainty bands of extracted Sivers function for up and down quarks with SoLID data 
comparing with the ones with current world data. Also shown at the bottom panel of the figure are the improvements (ratios of world data 
uncertainty over SoLID uncertainty). Figure~\ref{fig:SoLIDSivers2d} shows the 2D projection ($x$ and $k_\perp$) of the extracted Sivers 
function from SoLID measurements compared with that extracted from the current world data. 

\begin{figure}[htp]
\includegraphics[width=0.9\columnwidth]{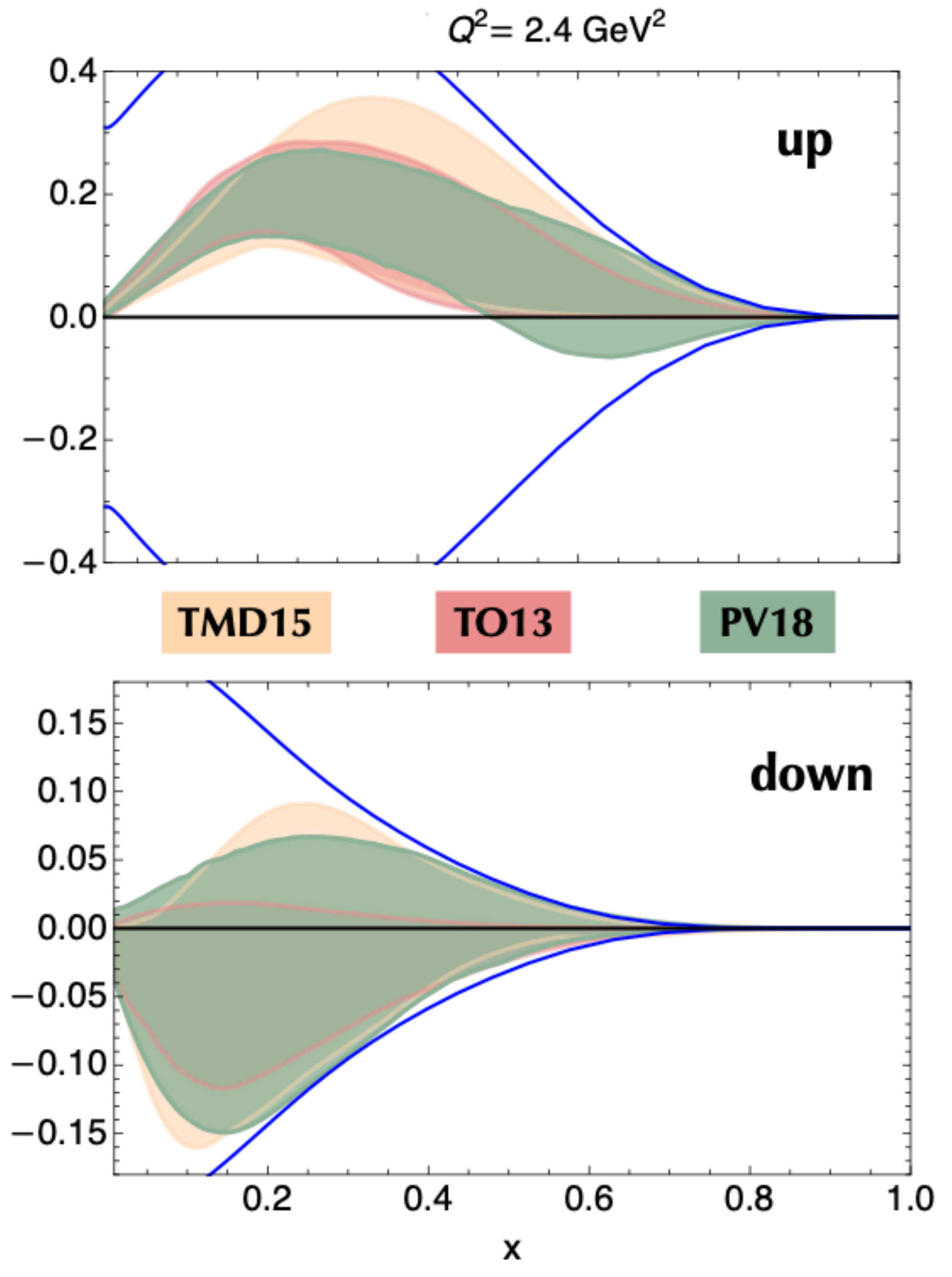}
\vspace{-0.25cm}
\caption{The $x h_1(x)$ at $Q^2 = 2.4$~GeV$^2$ for valence up quark (upper panel) and down quark (lower panel). Yellow band
  ``TMD15" from Ref.~\cite{Kang:2015msa}; red band ``TO13" from Ref.~\cite{Anselmino:2013vqa}; green band ``PV18" from
  Ref.~\cite{Radici:2018iag}. Blue solid lines for the Soffer bound at the same scale.}
\label{fig:xh1x}
\end{figure}

The right-most column of the table in Fig.~\ref{fig:TMDtable} contains the chiral-odd TMD PDFs, which are related to processes not 
conserving quark helicity. Each of them can appear in the cross section if it is associated to a chiral-odd partner~\cite{Jaffe:1991kp}. The 
best known chiral-odd TMD PDFs is the transversity $h_1$. It describes the correlation between the transverse polarizations of the nucleon 
and of its constituent quarks. Transversity survives in collinear framework: it is the third PDF needed to fully account for the partonic 
spin structure of the Nucleon at leading twist. The TMD PDF $h_1(x, k_T^2)$ can be extracted from SIDIS data for the so-called Collins
effect~\cite{Anselmino:2013vqa,Kang:2015msa,Lin:2017stx} (the chiral-odd partner being the Collins function $H_1^\perp$, simultaneously 
extracted from $e^+ e^-$ asymmetry data). As previously mentioned, the limitation of the TMD framework prevents this analysis to be extended 
to a global fit including data on the Collins effect in hadron-hadron collisions. This limitation can be overcome by considering the Collins 
effect for a hadron detected inside a jet~\cite{Adamczyk:2017ynk}, where the TMD FF $H_1^\perp$ is paired to the collinear PDF $h_1(x)$
\cite{Kang:2017btw}. Alternatively, in a fully collinear framework the PDF $h_1(x)$ can be extracted from data on the inclusive production of 
di-hadrons inside the same current jet and with small invariant mass~\cite{Jaffe:1998hf,Bianconi:1999cd,Radici:2001na,Bacchetta:2002ux,Bacchetta:2004it} 
(the chiral-odd partner being now the di-hadron function $H_1^\sphericalangle$, that can be extracted from asymmetry data on di-hadron pair 
production in $e^+ e^-$~\cite{Boer:2003ya,Bacchetta:2008wb,Courtoy:2012ry,Matevosyan:2018icf}). Because of the collinear framework, $h_1(x)$ 
could be extracted not only from SIDIS data~\cite{Bacchetta:2011ip,Bacchetta:2012ty,Radici:2015mwa} but also from a global fit including data 
on proton-proton collisions~\cite{Radici:2018iag}. In Fig.~\ref{fig:xh1x}, the results for $x h_1(x)$ and for valence up (upper panel) and down 
(lower panel) quarks extracted from Refs.~\cite{Anselmino:2013vqa,Kang:2015msa,Radici:2018iag} are compared at the same scale $Q^2 = 2.4$~GeV$^2$, 
showing a reasonable agreement. 

\begin{figure}[htp]
\includegraphics[width=0.8\columnwidth]{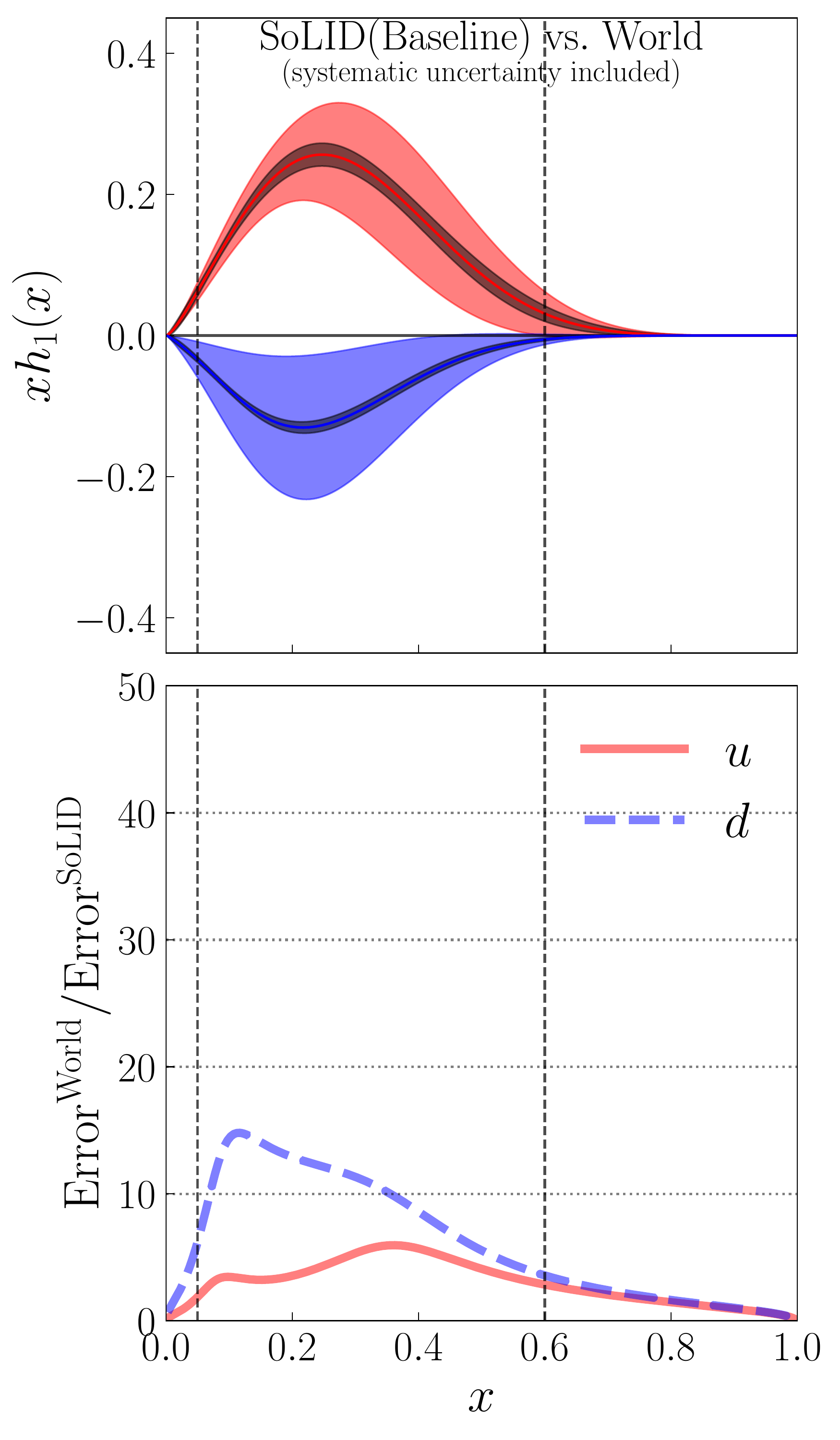}
\vspace{-0.25cm}
\caption{Same as in Fig.~\ref{fig:SoLIDSivers} but for the transversity distribution.}
\label{fig:SoLIDTransversity}
\end{figure}

\begin{figure}[htp]
\includegraphics[width=0.9\columnwidth]{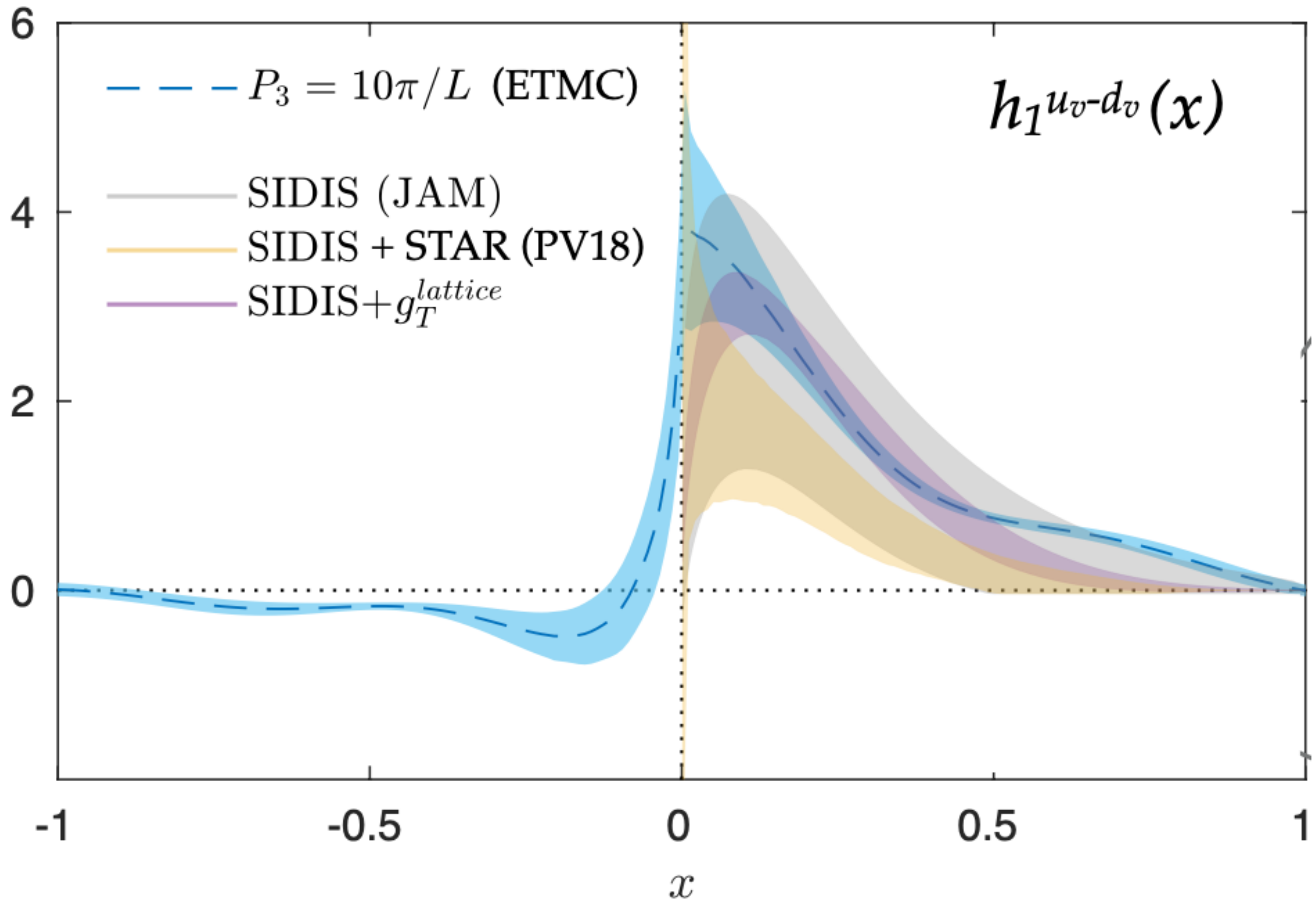}
\vspace{-0.25cm}
\caption{The isovector component of $h_1(x)$ at $Q^2 = 4$~GeV$^2$. Blue band ``ETMC" from lattice calculation of
  Ref.~\cite{Alexandrou19}. Gray band ``JAM" from extraction based on SIDIS data of Ref.~\cite{Lin:2017stx}, purple band
  when including constraint of reproducing lattice isovector tensor charge $g_T^{lattice}$. Yellow band ``PV18" from global fit of
  Ref.~\cite{Radici:2018iag}.}
\label{fig:h1PVlatt}
\end{figure}

\begin{figure}[htp]
\includegraphics[width=0.9\columnwidth]{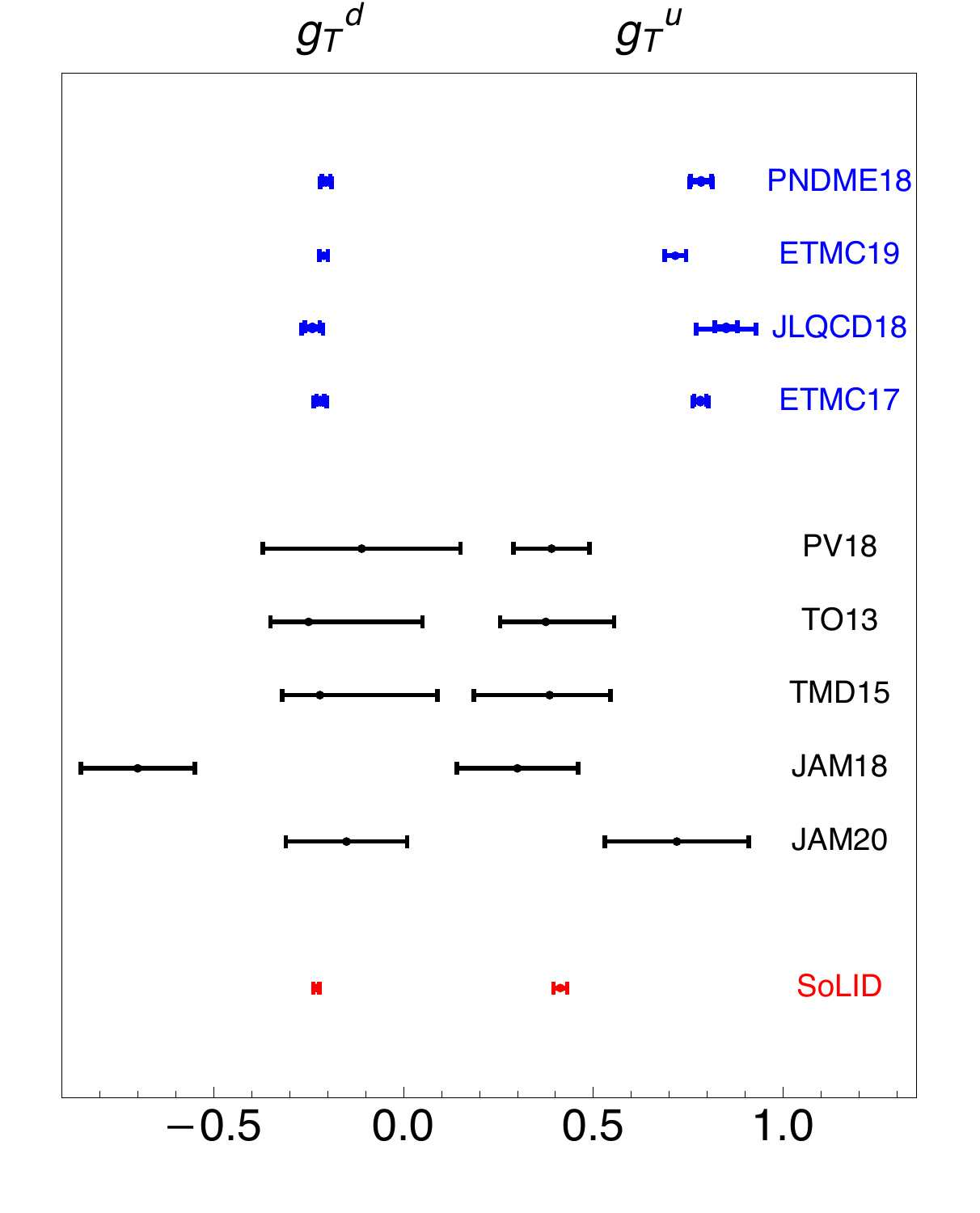}
\vspace{-0.25cm}
\caption{The tensor charge $g_T$ for down (left) and up quark (right). Blue points from lattice calculations (``PNDME18"
  \cite{Gupta:2018lvp}, ``ETMC19"~\cite{Alexandrou:2019brg}, ``JLQCD18"~\cite{Yamanaka:2018uud}, ``ETMC17"
  \cite{Alexandrou:2017qyt}). Black points from phenomenological extractions of transversity (``PV18"~\cite{Radici:2018iag},
  ``TO13"~\cite{Anselmino:2013vqa}, ``TMD15"~\cite{Kang:2015msa}, ``JAM18"~\cite{Lin:2017stx}, ``JAM20"
  \cite{Cammarota:2020qcw}). Red point from SoLID projection on ``TMD15" results~\cite{Ye:2016prn}.}
\label{fig:gTPDFlatt}
\end{figure}

The planned measurements at JLab12 will greatly improve our knowledge of transversity. The SoLID SIDIS experiments
\cite{Chen:2014psa,SoLID:SIDISn,SoLID:SIDISp} with JLab12 will provide the most precise measurements of Collins asymmetries~\cite{Ye:2016prn} 
in the valence quark region ($x > 0.05$). Figure~\ref{fig:SoLIDTransversity} shows the projected uncertainty bands of extracted transversity 
for up and down quarks with SoLID data comparing with the ones with current world data. Also shown in the bottom panel of the figure are the 
improvements (ratios of world data uncertainty over SoLID uncertainty). In Fig.~\ref{fig:h1PVlatt}, the global fit of Ref.~\cite{Radici:2018iag} 
and the Collins extraction of Ref.~\cite{Lin:2017stx} are compared at $Q^2 = 4$~GeV$^2$ with a lattice calculation of the corresponding
quasi-PDF~\cite{Alexandrou19}. The first Mellin moment of $h_1(x)$ is named tensor charge $g_T$ and has recently received increasing 
attention in searches of new physics Beyond Standard Model (BSM). For example, the tensor charge affects the contribution of quark electric dipole 
moments to the neutron electric dipole moment in searches of BSM sources of CP-violation~\cite{Dubbers:2011ns,Yamanaka:2017mef}. In 
Fig.~\ref{fig:gTPDFlatt}, the results for $g_T^u$ and $g_T^d$ from various phenomenological extractions of transversity are shown with black 
points and compared with recent lattice computations (shown with blue points). The red point shows the impact of JLab12 SoLID projection on the 
(SIDIS $+\, e^+ e^-$) analysis of the Collins effect~\cite{Ye:2016prn}. This would translate in an improvement of a factor 2 on the current limit 
of the up electric dipole moment~\cite{Gao:2017ade}.

Each TMD PDF in Fig.~\ref{fig:TMDtable} enters the SIDIS cross section with a specific dependence on azimuthal angles defined by the 
kinematics of the process. Hence, it can be extracted through measurements of azimuthal (spin) asymmetries. All relevant asymmetries 
corresponding to the entries in Fig.~\ref{fig:TMDtable} have been explored
\cite{Airapetian:2004tw,Airapetian:2009ae,Adolph:2012sn,Adolph:2012sp,Parsamyan:2013ug,Qian:2011py,Huang:2011bc,Zhang:2013dow} and will 
be precisely measured~\cite{Chen:2014psa,SoLID:SIDISn,SoLID:SIDISp,SoLID:WormGear}, providing crucial information on spin-orbit correlations 
and the orbital angular momentum of the confined quarks inside the Nucleon. 

In summary, studies of 3D maps of the partonic structure of Nucleons in momentum space are moving from an exploratory phase to the era of 
precision. Links to high-energy phenomenology are being established and inputs from non-perturbative effects encoded in TMDs could affect 
those analyses. Upcoming data from starting projects like JLab12 (in particular, the expected high precision data from SoLID) and future projects 
(LHCspin, ALICE, EIC) will significantly enlarge the available phase space, and will contribute in deepening our knowledge of parton dynamics, 
ultimately clarifying the mechanisms leading to confinement. 

\subsection{Mapping the Energy-Momentum Tensor of Ground State Nucleons from DVCS Data}
\label{tensor}

\begin{figure}[htp]
\includegraphics[width=0.9\columnwidth]{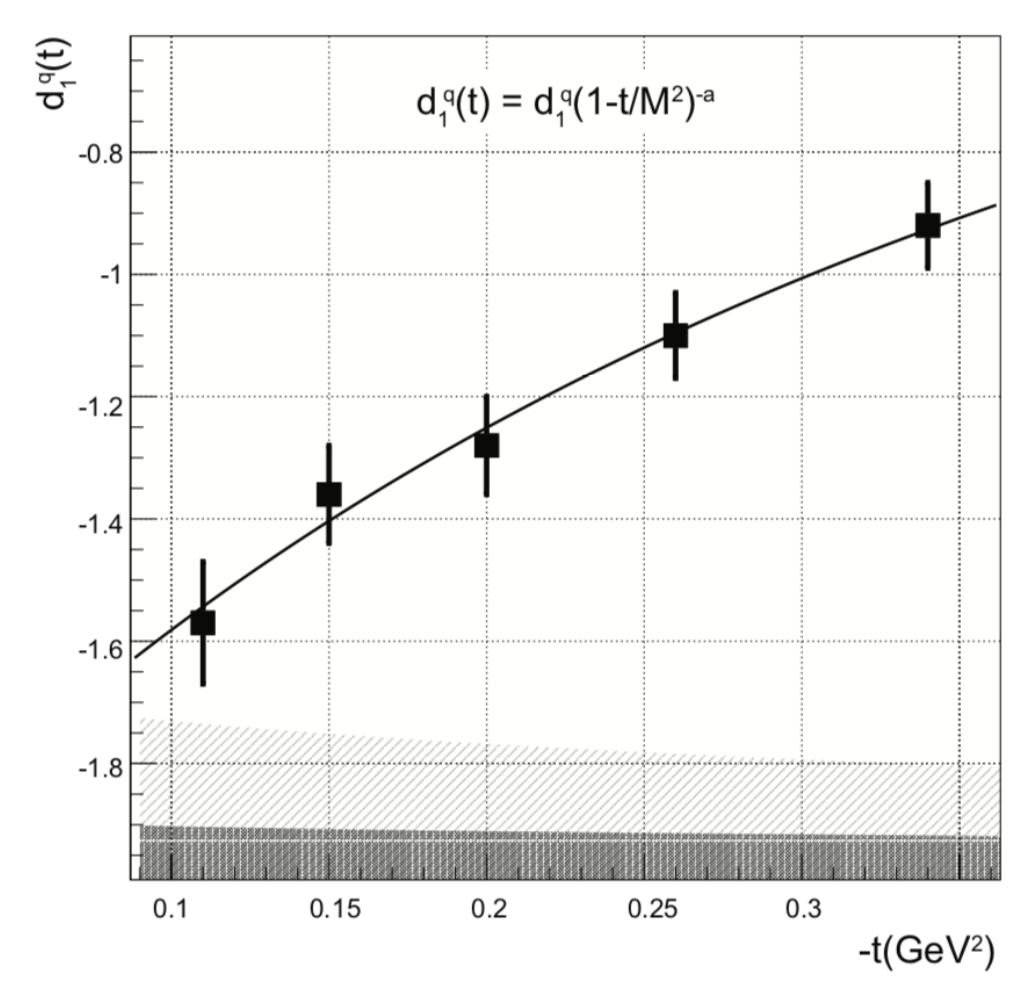}
\vspace{-0.25cm}
\caption{Example of a fit to $d_1(t)$. The error bars are from the fit to the cross sections at fixed value of $-t$
and correspond to 1 standard Gaussian deviation. The single-shaded area at the bottom corresponds to the uncertainties
from the extension of the fit into regions without data and is reflected in the green shaded area in Fig.~\ref{pressure}.
The double-shaded area corresponds with the projected uncertainties from future experiment as shown in Fig.~\ref{pressure} 
with the red shaded area.}
\label{fig:d1_fit} 
\end{figure}

The experimental study of the proton energy-momentum tensor is a novel tool to obtain information about the mechanical or
gravitational properties of particles. Russian theorists I. Yu. Kobzarev and L.B. Okun~\cite{Kobzarev:1962wt}, and American
theorist Heinz Pagels~\cite{Pagels:1966zza} were the first ones to explore, independently, the possibility of studying the
mechanical properties of subatomic particles. Pagels developed the framework of gravitational form factors in analogy to the
electromagnetic interaction. However, he concluded that {\it contrary to the case of electromagnetism, there is very little hope
of learning anything about the detailed mechanical structure of a particle, because of the extreme weakness of the gravitational
interaction}. The field remained dormant and was revived only with the discovery of the GPDs
\cite{Mueller:1994,Ji:1996ek,Radyushkin:1996nd} and the relationship of their second Mellin moments to $J(t)$, $M2(t)$, and $d1(t)$, 
which are the gravitational form factors (GFF) of the proton matrix element of the energy-momentum tensor. They relate to the 
distribution of angular momentum, the mass and energy, and the radial distribution of shear forces and pressure inside the proton
\cite{Polyakov:2002yz}. The first experimental application came in 2017 with the extraction of the GFF $d1(t)$, shown in 
Fig.~\ref{fig:d1_fit}, employing the two sets of DVCS data published by the CLAS Collaboration~\cite{Girod:2007aa,Jo:2015ema}, and 
the subsequent estimate of the pressure distribution inside the proton~\cite{Burkert:2018bqq}, which yielded a peak pressure near 
the proton center of 10$^{35}$~Pa, exceeding the gravitational pressure observed in neutron stars LIGO by an order of magnitude. 

\begin{figure}[htp]
\includegraphics[width=0.9\columnwidth]{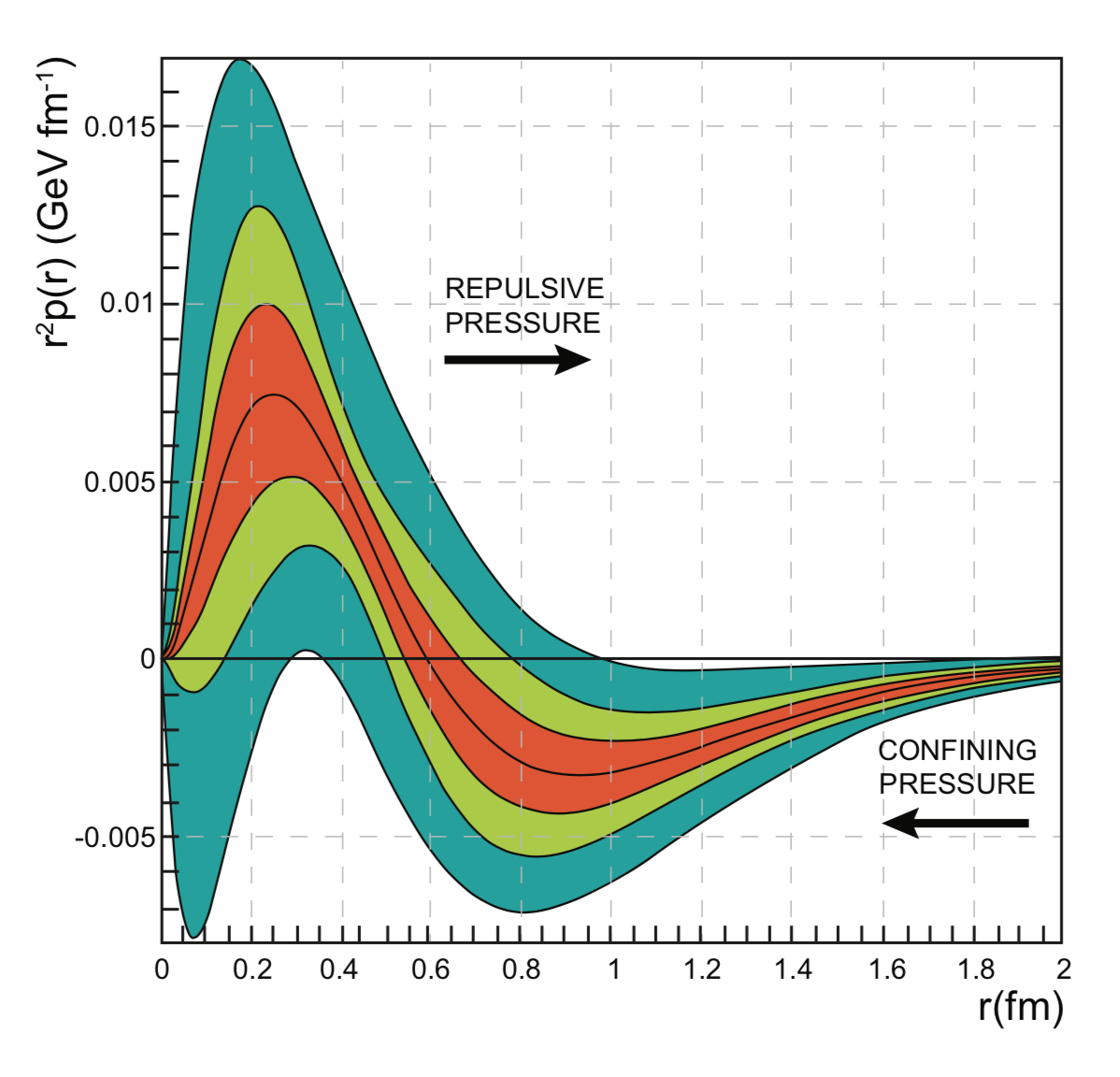}
\vspace{-0.25cm}
\caption{The radial pressure distribution in the proton. The figure shows the pressure distribution $r^2 p(r)$ resulting from
the interactions of the quarks in the proton versus the radial distance from the center in fm. The black central line
corresponds to the pressure extracted from the $D$-term parameters fitted to the published data at 6~GeV
\cite{Girod:2007aa,Jo:2015ema}. The corresponding estimated uncertainties are displayed as the shaded area shown in light
green. They correspond to 1 standard deviation. See the text for more details.}
\label{pressure} 
\end{figure}

The first extraction of the pressure distribution inside the proton has already generated over 70 citations, and inspired many 
theoretical papers that, among others, relate the pressure distribution to limits on deviations from general relativity at short
distance~\cite{Avelino:2019esh}, applications to computing the equation of state in neutron stars~\cite{Liuti:2018ccr}, Lattice 
calculations of gluon contributions to the pressure distribution~\cite{Shanahan:2018nnv}, studies of a two-scale structure of the 
proton related to LHC results~\cite{Troshin:2019ivz,Dremin:2019pza,Dremin:2019swd,Campos:2019hvw}, computations of the proton 
mechanical radius~\cite{Anikin:2019zul,Anikin:2019kwi}, a detailed review paper~\cite{Polyakov:2018zvc}, extensions of the concept of 
mechanical properties~\cite{Lorce:2018egm} and others.

In the near future, given a sufficient amount of DVCS data, including polarized beam asymmetries and cross section data, covering
a large range in photon virtuality $Q^2$, and in four-momentum transfer $t$ to the proton, the large systematic uncertainties of
these first data (green band) can be significantly reduced. The DVCS program at the JLab 12~GeV upgrade is now generating the
data that can result in significant advances of the field.

A significant extension of the JLab experimental program at 12~GeV that employs both electron and positron beams will directly 
access the real part of the Compton amplitude. This will allow for a much improved isolation of the Compton FF ${\cal H}(\xi, t)$ with 
much smaller systematic uncertainties. Estimates (see the proceedings of Positron workshop~\cite{Elouadrhiri:2009zz,Proceedings:2018umi}) 
have shown that the beam charge asymmetries are large and can directly access the real part of the Compton amplitude, which would be a 
major advance in the program to establish the 3D quark imaging of nucleons, and to determine one of the gravitational form factors, 
$d_1(t)$, in a larger $t$ range.

\subsection{Insights into Strong QCD from Combined Studies of Baryon Ground and Excited States}
\label{Sec37}

Contemporary strong interaction theory is beginning to reveal many consequences of EHM, as illustrated by
Figs.~\ref{FigwidehatalphaII} to \ref{figF12}. Numerous calculations connect these features with observable properties
of hadron ground states and the associated comparisons with experiment show that the ideas are viable. However, the ground
state is just one isolated member of a set of Hamiltonian eigenvectors with infinitely many elements: many Hamiltonians can
possess practically identical ground states and yet produce excited-state spectra that are vastly different. Moreover,
masses alone, being infrared-dominated quantities, contain relatively little information. Distinct Hamiltonians can satisfactorily
reproduce known hadron spectra; but these same Hamiltonians deliver predictions that disagree enormously when employed
to compute structural properties. Such properties -- like the $Q^2$-dependence of elastic and transition form factors --
possess the greatest discriminating power. Hence, modern theory must be deployed to compute these observables. With
CSMs, their reliable prediction requires the use of realistic wave functions and currents, the calculation of which is challenging;
and using LQCD, algorithmic and technical challenges must be surmounted in order to deliver simulations of realistic systems.

These remarks mean that results on the structure of nucleon resonances cannot be divorced from those obtained for
ground-state nucleons and mesons. In order to validate any description of features of the latter, one must elucidate the
properties of all systems that can be produced by the theoretical framework and insist that the approach provide a unifying
explanation. This is true for all sectors: mesons, baryons, hybrids and exotics. Consequently, any claim that progress has
been made toward understanding QCD should be tested by the requirement that its basis provides a simultaneous explanation
of, {\it inter alia}: how emergent mass is expressed in distinct bound states; and if there are differences between systems,
then how can they be understood?

Concerning baryons, an important aspect of EHM is expressed in the fact that any interaction capable of creating
pseudo--Nambu-Goldstone modes as bound-states of a light dressed-quark and -antiquark, and reproducing the measured value
of their leptonic decay constants, will necessarily, {\it inter alia}, also generate strong color-antitriplet correlations between
any two dressed-quarks contained within a nucleon~\cite{Segovia:2015ufa}.

The properties of such diquark correlations are known. As color-carrying correlations, diquarks are confined
\cite{Bender:1996bb,Bhagwat:2004hn}. Additionally, a diquark with spin-parity $J^P$ may be viewed as a partner to the analogous $J^{-P}$
meson~\cite{Cahill:1987qr}. Hence, the strongest diquark correlations in the nucleon are: scalar isospin-zero, $[ud]_{0^+}$; and pseudovector, 
isospin-one, $\{uu\}_{1^+}$, $\{ud\}_{1^+}$, $\{dd\}_{1^+}$. Moreover, while no pole-mass exists, the following mass-scales, which express 
the strength and range of the correlation, may be associated with these diquarks
\cite{Cahill:1987qr,Maris:2002yu,Eichmann:2008ef,Segovia:2015hra,Eichmann:2016hgl,Lu:2017cln,Chen:2017pse,Chen:2019fzn,Yin:2019bxe,Lu:2019bjs}
(in GeV):

\begin{equation}
m_{[ud]_{0^+}} \approx 0.7-0.8\,,\;
m_{\{uu\}_{1^+}} \approx 0.9-1.1 \,.
\end{equation}
With isospin symmetry, $m_{\{dd\}_{1^+}}=m_{\{ud\}_{1^+}} = m_{\{uu\}_{1^+}}$.

It should be stressed that contemporary applications of CSMs indicate that the ground-state nucleon necessarily contains both scalar-isoscalar 
and pseudo\-vec\-tor-isovector correlations: neither can be ignored and their presence has many observable consequences
\cite{Roberts:2013mja,Segovia:2013uga,Segovia:2016zyc,Mezrag:2017znp,Lu:2019bjs,Yin:2019bxe,Segarra:2019gbp}.
On the other hand, further amplifying the importance of excited states, odd-parity baryons are predicted to contain
pseudoscalar and vector diquarks and these correlations might also play a role in even-parity excited states of the nucleon
\cite{Eichmann:2016hgl,Lu:2017cln,Chen:2017pse}. Importantly, there is also strong evidence for the presence of diquark
correlations in baryons from LQCD~\cite{Alexandrou:2006cq,DeGrand:2007vu,Babich:2007ah,Bi:2015ifa}.

Realistic diquark correlations are soft and interacting. All carry charge, scatter electrons, and possess an electromagnetic size that is similar 
to that of the analogous mesonic system, {\it e.g.}~\cite{Maris:2004bp,Eichmann:2008ef,Roberts:2011wy}:
\begin{equation}
\label{qqradii}
r_{[ud]_{0^+}} \gtrsim r_\pi, \quad r_{\{uu\}_{1^+}} \gtrsim r_\rho,
\end{equation}
with $r_{\{uu\}_{1^+}} > r_{[ud]_{0^+}}$. As in the meson sector, these scales are set by that associated with EHM.

Here it should be emphasized that these fully dynamical diquark correlations are vastly different from the static, point-like ``diquarks'' 
introduced originally~\cite{Anselmino:1992vg} in an attempt to solve the so-called missing resonance problem~\cite{Aznauryan:2011ub}, {\it viz.} 
the fact that quark models predict many more baryon states than were observed in the previous millennium~\cite{Burkert:2004sk}. Moreover, their 
existence enforces certain distinct interaction patterns for the singly and doubly represented valence-quarks within the proton and its excited 
states, as exhibited elsewhere~\cite{Roberts:2013mja,Segovia:2014aza,Segovia:2016zyc,Chen:2018nsg}.

\begin{figure}[htp]
\includegraphics[width=0.98\columnwidth]{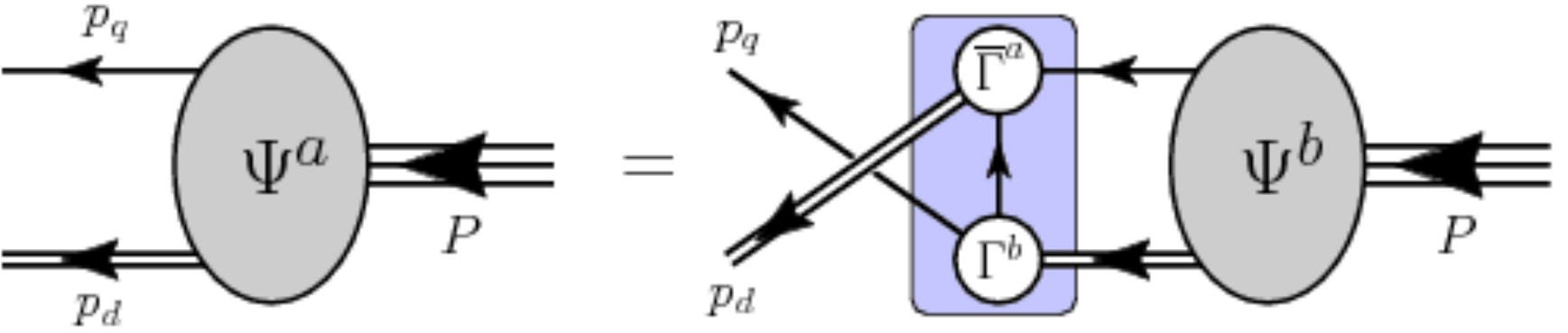}
\vspace{-0.25cm}
\caption{Poincar\'e covariant Faddeev equation: a linear integral equation for the matrix-valued function $\Psi$, being the
  Faddeev amplitude for a baryon of total momentum $P= p_q + p_d$, which expresses the relative momentum correlation
  between the dressed-quarks and -diquarks within the baryon. The shaded rectangle demarcates the kernel of the Faddeev
  equation: {\it single line}, dressed-quark propagator; $\Gamma$, diquark correlation amplitude; and {\it double line},
  diquark propagator.}
\label{figFHR4a}
\end{figure}

The existence of such tight correlations between two dressed quarks is the key to transforming the three valence-quark
scattering problem into a simpler Faddeev equation~\cite{Cahill:1988dx,Reinhardt:1989rw,Efimov:1990uz}; and this is
achieved without loss of dynamical information~\cite{Eichmann:2009qa}. The three-gluon vertex, a signature feature of
QCD's non-Abelian character, is not explicitly part of the bound-state kernel in this picture. Instead, one capitalizes on the
feature that phase-space factors materially enhance two-body interactions over $n\geq 3$-body interactions and exploits
the dominant role played by diquark correlations in the two-body subsystems. Then, while an explicit three-body term might
affect fine details of baryon structure, the dominant effect of non-Abelian multi-gluon vertices is expressed in the formation
of diquark correlations. Consequently, as depicted in Fig.~\ref{figFHR4a}, the active kernel describes binding within the
baryon through diquark breakup and reformation, which is mediated by exchange of a dressed-quark. Such a baryon is a
compound system whose properties and interactions are largely determined by its quark$+$diquark structure.

Importantly, the number of states in the spectrum of baryons obtained from the Faddeev equation
\cite{Eichmann:2016hgl,Lu:2017cln,Chen:2019fzn,Yin:2019bxe} is similar to that found in the three-constituent quark model.
This prediction is consistent with direct solutions of the Poincar\'e-covariant three-body bound-state equation
\cite{Eichmann:2016hgl, Qin:2019hgk} and LQCD results~\cite{Edwards:2011jj}; and, notably, modern data and recent analyses
have already reduced the number of missing resonances
\cite{Ripani:2002ss,Burkert:2012ee,Kamano:2013iva,Crede:2013sze,Mokeev:2015moa,Bur17}.

Furthermore, it is beginning to appear that diquark correlations also play a role in baryons involving one or more heavy quarks.
In these systems, owing to the dynamical character of the diquarks, it is typically the lightest allowed diquark correlation
that defines the most important component of a baryon's Faddeev amplitude~\cite{Yin:2019bxe}. This outcome challenges
the validity of phenomenological models that treat singly heavy baryons ($qq' Q, q, q' \in \{u, d, s\}, Q \in \{c, b\}$)
as two-body light-diquark+heavy-quark ($qq^\prime+Q$) bound states; similarly, those that treat doubly heavy baryons
($qQQ'$ as two-body light-quark+heavy-diquark bound-states, $q + QQ'$.

The increasing evidence in support of a role for two-body correlations in baryons has spurred analyses that seek to exploit diquarks 
in tetra- and pentaquark systems~\cite{Chen:2016qju,Liu:2019zoy} and others that search for different correlations in hybrid 
systems~\cite{Xu:2019sns}. The latter are sensitive to hitherto unexplored aspects of gluon-quark dynamics~\cite{Souza:2019ylx}.

Having explained the origin and sketched the features of diquark correlations, it is natural now to return to the central topic
of this subsection with a simple observation: QCD is not solved unless the Roper resonance is understood. In this connection,
the last twenty years have seen the acquisition and analysis of a vast amount of high-precision proton-target exclusive
electroproduction data with single- and double-pion final states on a large kinematic domain of energy and momentum-transfer;
development of a sophisticated dynamical reaction theory capable of simultaneously describing all partial waves extracted from
available, reliable data; and formulation and wide-ranging application of CSMs. Following these efforts, it is now widely accepted
that the Roper is, at heart, the first radial excitation of the nucleon, consisting of a well-defined dressed-quark core that is
augmented by a meson cloud, which both reduces the Roper's core mass by approximately 20\% and contributes materially to
the electroproduction transition form factors on $Q^2\lesssim 2 m_N^2$ ($m_N$ is the nucleon mass)~\cite{Burkert:2019bhp}.

The $\Delta(1232)$ also has a radial excitation and one should answer the question: ``Is this the $\Delta(1600)$?'' This is
because the spectrum of $\Delta$-baryons exhibits the same level-inversion seen in the nucleon spectrum, {\it viz.} the
lightest negative-parity state is heavier than the two lightest positive-parity states. The Roper experience has demonstrated
that it will only become possible to decide the character of the $\Delta(1600)$ after one has predictions for the associated
electroproduction form factors and experiments to test them. Fortunately, data exist~\cite{Trivedi:2018rgo,Burkert:2019opk}
and can be analyzed with this aim understood; and theoretical predictions are now available~\cite{Lu:2019bjs}.

Recall that Poincar\'e covariance demands that the proton possess intrinsic deformation. However,
quantum mechanics teaches that such deformation is not observable for $J=1/2$ systems. On the other hand, deformation
can be observed in proton\,$=p\to$ $\Delta(1232)$, $\Delta(1600)$ transitions.

Regarding $\gamma^* p \to \Delta(1232)$, on $Q^2 \gtrsim 0.5 m_N^2$, {\it i.e.} outside the meson cloud domain for
this process, the magnetic dipole and Coulomb quadrupole form factors reported in Ref.~\cite{Lu:2019bjs} agree well with
available data. Consistent with the data, too, the electric quadrupole form factor is very small in magnitude; hence, it is
particularly sensitive to the diquark content and quark+diquark angular-momentum structure of the baryons involved, and also
to meson-baryon final-state-interactions (MB\,FSIs) on a larger domain than the other form factors. These remarks are
supported by the following observations: the role played by higher partial waves in the wave functions increases with momentum
transfer (something also observed in meson form factors), here generating destructive interference; agreement with data on
$G_M^*$ is impossible without the higher partial waves; and the effect of such components is very large in $G_E^*$, with
the complete result for $G_E^*$ exhibiting a zero at $Q^2 \approx 4 m_N^2$, which is absent in the S-wave-only result(s).

\begin{figure}[htp]
\includegraphics[width=0.45\columnwidth]{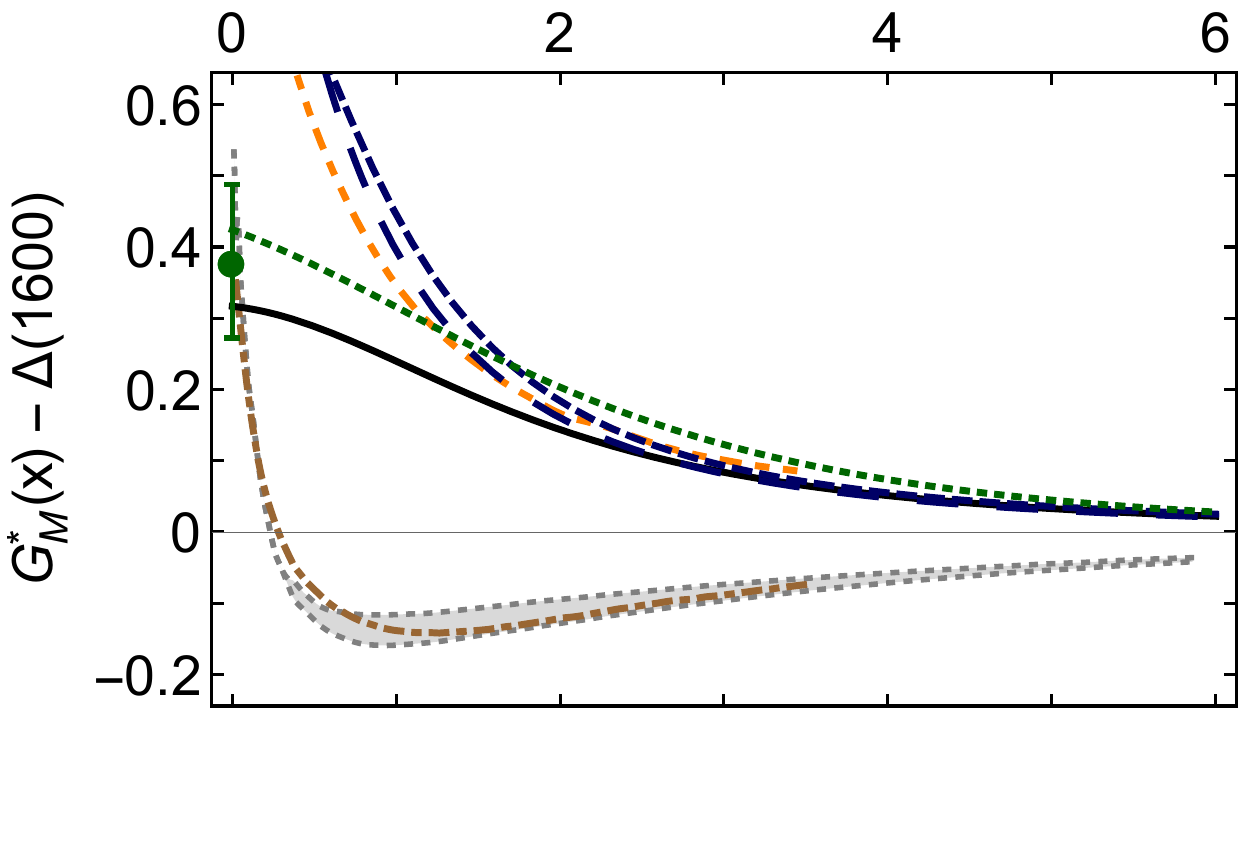}
\includegraphics[width=0.45\columnwidth]{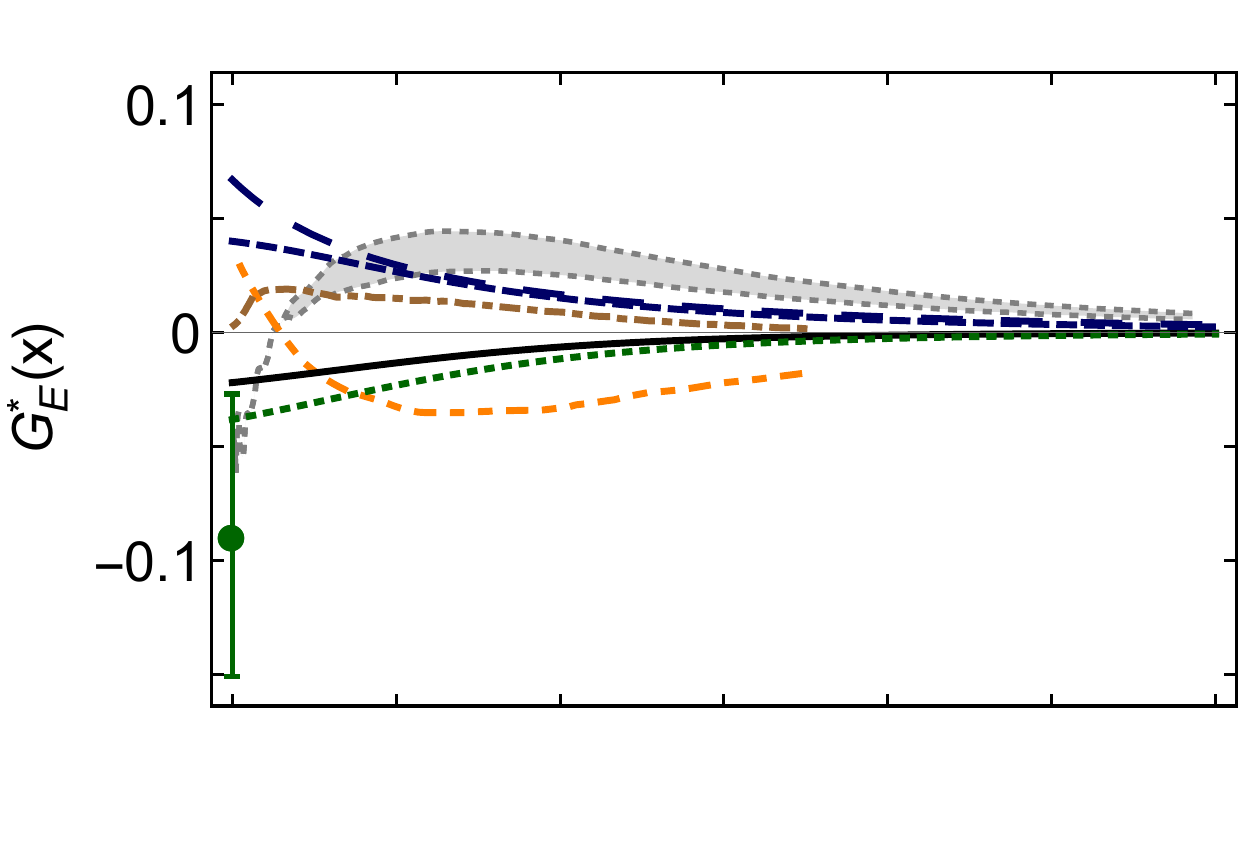}
\includegraphics[width=0.45\columnwidth]{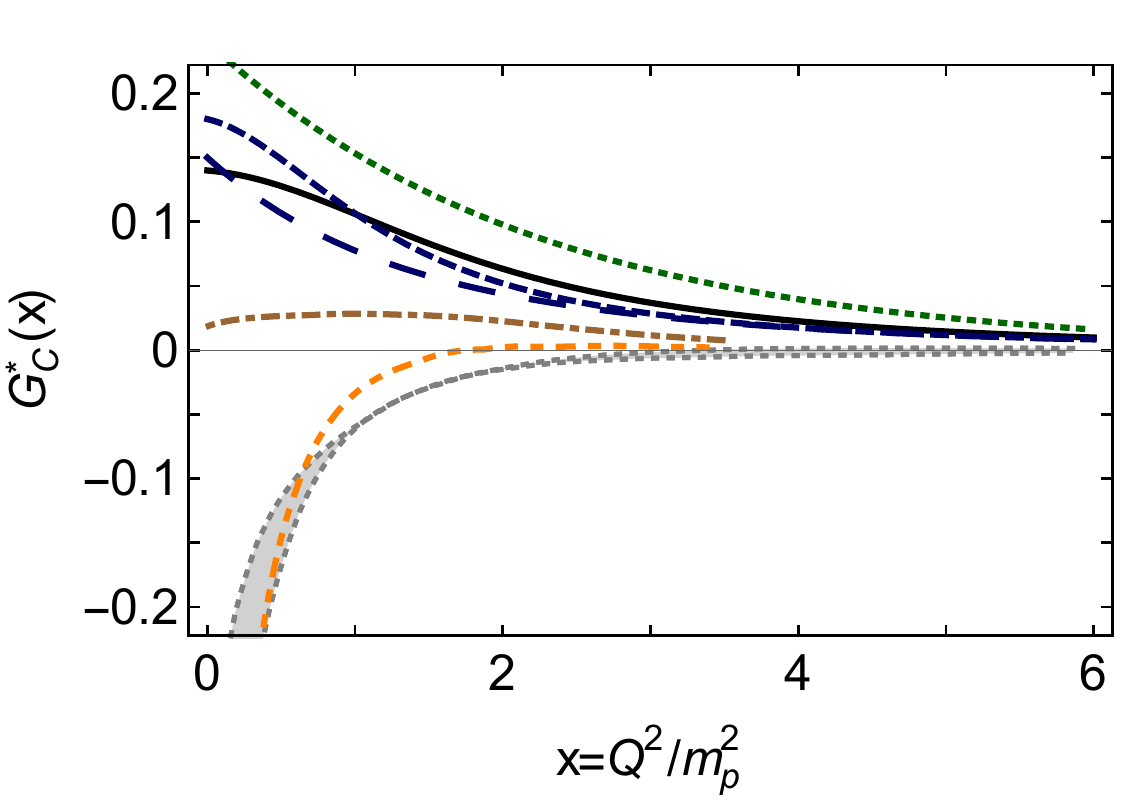}
\vspace{-0.25cm}
\caption{UL: Magnetic dipole $\gamma^* p \to \Delta(1600)$ form factor; UR: electric quadrupole; and Bottom: Coulomb quadrupole. Data 
from Ref.~\cite{Tanabashi:2018oca}; and the conventions of Ref.~\cite{Jones:1972ky} are employed. All panels: solid [black] curve, complete 
result; long-dashed [blue] curve, result obtained when $\Delta(1600)$ is reduced to $S$-wave state; dashed [blue] curve, both the proton and 
$\Delta(1600)$ are reduced to $S$-wave states; dotted [green] curve, obtained by enhancing proton's axial-vector diquark content; shaded [gray] 
band, light-front relativistic Hamiltonian dynamics (LFRHD)~\cite{Capstick:1994ne}; dot-dashed [brown] curve, light-front relativistic quark 
model (LF-RQM) with unmixed wave functions~\cite{Aznauryan:2015zta}; and dot-dot-dashed [orange] curve, LF-RQM with configuration mixing
\cite{Aznauryan:2016wwmO}.}
\label{D1600TFFs}
\end{figure}

Drawn from Ref.~\cite{Lu:2019bjs}, predictions for the $\gamma^* p\to \Delta(1600)$ transition form factors are displayed
in Fig.~\ref{D1600TFFs}. Here, empirical results are only available at the real-photon point: $G_M^*(Q^2=0)$,
$G_E^*(Q^2=0)$. Evidently, the quark model results -- [shaded gray band]~\cite{Capstick:1994ne}, dot-dashed (brown) curve
\cite{Aznauryan:2015zta}, and dot-dot-dashed [orange] curve~\cite{Aznauryan:2016wwmO}) -- are very sensitive to the wave
functions employed for the initial and final states. Furthermore, inclusion of relativistic effects has a sizeable impact on
transitions to positive-parity excited states~\cite{Capstick:1994ne}.

The quark+diquark Faddeev equation prediction is the solid [black] curve in each panel of Fig.~\ref{D1600TFFs}. In this
instance, every transition form factor is of unique sign on the domain displayed. Notably, the mismatches with the empirical
results for $G_M^*(0)$, $G_E^*(0)$ are commensurate in relative sizes with those in the $\Delta(1232)$ case, suggesting
that MB\,FSIs are of similar importance in both channels.

Axial-vector diquark contributions interfere constructively with MB FSIs~\cite{Hecht:2002ej}; hence, regarding form factors,
one can mimic some meson cloud effects by modifying the axial-vector diquark content of the participating hadrons. Accordingly,
to illustrate the potential impact of MB FSIs, the transition form factors were also computed using an enhanced axial-vector
diquark content in the proton. This was achieved by setting $m_{1^+} = m_{0^+} = 0.85$~GeV, values with which the proton's mass
is practically unchanged. The procedure produced the dotted [green] curves in Fig.~\ref{D1600TFFs}; better aligning the
$x\simeq 0$ results with experiment and suggesting thereby that meson-cloud effects will improve the Faddeev equation
predictions.

The short-dashed [blue] curve in Fig.~\ref{D1600TFFs} is the result obtained when only rest-frame $S$-wave components
are retained in the wave functions of the proton and $\Delta(1600)$; and the long-dashed [blue] curve is that computed with
a complete proton wave function and a $S$-wave-projected $\Delta(1600)$. Once again, the higher partial-waves have a visible
impact on all form factors, with $G_E^*$ being most affected: the higher waves produce a change in sign. This reemphasizes one 
of the conclusions from the quark model studies, {\it viz.} data on the $\gamma^* p\to \Delta(1600)$ transition form factors 
will be sensitive to the structure of the $\Delta(1600)$.

These observations emphasize that it is crucial to study the complete spectrum of states and expose the internal structure
of characterizing members of each level if one wishes to arrive at answers to numerous fundamental questions in hadron physics.
For instance, is relativity important to hadron structure; and if so, when is it expressed, how, and what can be learned from
seemingly good predictions made with nonrelativistic models? Furthermore, are there two-body subclusters (correlations) in
baryons and other hadrons; if so, what sort of correlations, in which channels, how are they manifested in observables, and
what do they reveal about EHM? In answering such questions, tools will be developed that are necessary to chart the
distributions of energy, momentum and mass within QCD bound states, and learn how they are influenced by the hadronic and
nuclear environments.

\subsection{Exploring the Emergence of Nuclear Structure from Strong QCD}
\label{sym}

Systematic studies of monopole, quadrupole, and rotational collective modes in atomic nuclei suggest that approximate symmetries
play an important, if not dominant role in determining their low-energy structure. Indeed, to leading order - at least for light 
nuclei - it has been shown that calculated eigenstates using realistic $NN$ interactions correlate strongly with eigenvectors of 
the symplectic [Sp(3,R)] group~\cite{Rowe_1985,Dytrych_2008,Launey:2016fvy}.

From a microscopic perspective the latter should not come as a surprise as the 21 generators of Sp(3,R) consist of the
particle-symmetric independent quadratic forms in the momenta and spatial coordinates of the constituent particles. These
in turn can be re-grouped into four physically important subset modalities: six generators of monopole and quadrupole moments,
six generators of monopole and quadrupole deformation, three generators of rotation, and another set of six generators
associated with quadrupole flow. For purposes of this contribution, which is to probe the linkage between low-energy nuclear
physics (as seemingly governed by the strong interaction between nucleons and mesons), and high-energy nuclear physics (as
governed by the strong interaction between quarks and gluons), it is important to realize that Sp(3,R) can also be characterized
as a multi-shell generalization of the single-shell Elliott SU(3) model. The Elliott SU(3) symmetry enters as a subgroup of Sp(3,R)
in a chain that ends with SO(3), the symmetry group of the orbital angular momentum. Specifically, if one removes from Sp(3,R)
the generators of U(3) - all of which act only within single major shells of the oscillator (for SU(3) plus 1 for U(1) that counts 
the total number of oscillator quanta within a shell), the remaining 12 generators are 6 raising operators each of which adds two
quanta to a configuration [a $L$=0 (monopole, or '$s$') mode, and another five $L$=2 (quadrupole, or '$d$') modes, with the other 6
lowering operators that are simply conjugates of the raising operators].

The latter features, when understood within the context of a many-particle shell-model theory that takes into account Pauli
statistics between like particles, provides a linkage to the more common Slater determinant characterization of shell-model
basis states, as well as our ability to understand the microscopic $np$-$nh$ structures~\cite{Dytrych_2008,Launey:2016fvy}
that are also everywhere apparent as clustering degrees of freedom in atomic nuclei. The good news is that within the past few
years a so-called symmetry-adapted no-core shell-model (SA-NCSM) code~\cite{Dytrych:2013cca} has been developed that takes all 
of these features into account, including the ability to start from realistic interactions that are considered to be part
of what is now commonly referred to as {\it ab initio} shell-model theories.

As the above implies, the Sp(3,R) model can also be used to expose the microscopic underpinnings of clustering in nuclei; 
especially, but not only, when the total number of nucleons is a simple multiple of the $\alpha$-particle, which is the case for 
the $\alpha$-particle ($^{4}$He) itself, $^{8}$Be, $^{12}$C, $^{16}$O, $^{24}$Mg, and so on. In particular, in contrast with the 
simplest of $\alpha$-cluster models, the reach-back to {\it ab initio} features in nuclei afforded by the Sp(3,R) model suggests 
that the Hoyle state in $^{12}$C is considerably more subtle than simply geometrical configurations of three $\alpha$ particles 
suggest. Indeed, SA-NCSM calculations have demonstrated the relevance of Sp(3,R) for light $p$-shell nuclei through the $sd$-shell 
and even into the intermediate mass region of the $pf$ shell. Beyond this, additional considerations are needed due to a breaking 
of this simple picture as evidenced in a growing of neutron excesses over the proton count. But even for strongly deformed nuclei 
of the rare earth and actinide regions~\cite{Launey:2016fvy}, the prominence of deformation and clustering is everywhere apparent, 
especially in fission fragments. In short, nuclear deformation and shape coexistence dominate the entire nuclear landscape. However, 
while advanced tools of the SA-NCSM type await further developments for heavy nuclei, for the current focus of examining the bridging 
between nuclear structure and strong QCD, a focus on light nuclei is best; and for this, the existing SA-NCSM should suffice.

\begin{figure}[htp]
\includegraphics[width=0.75\columnwidth]{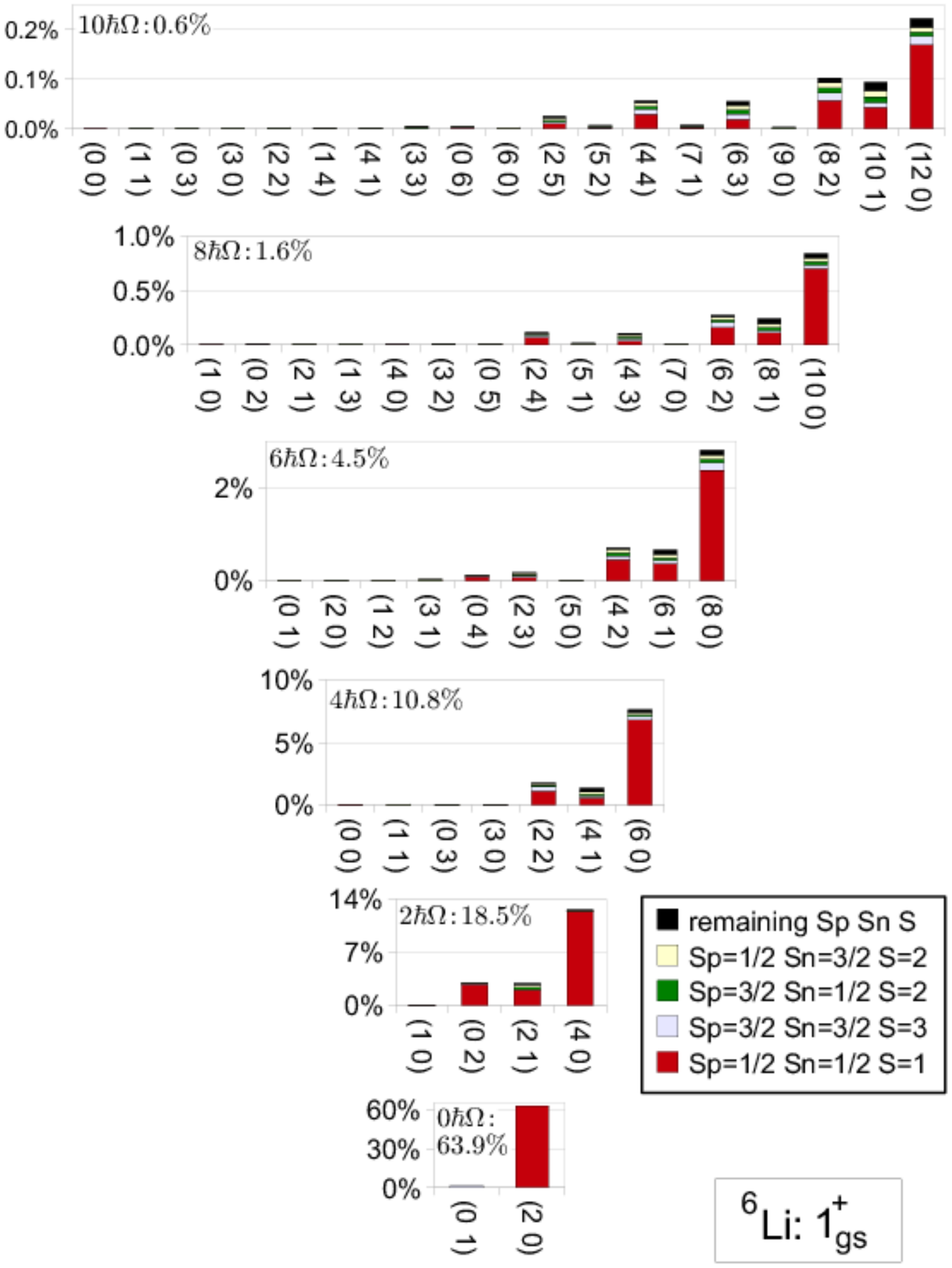}
\includegraphics[width=0.75\columnwidth]{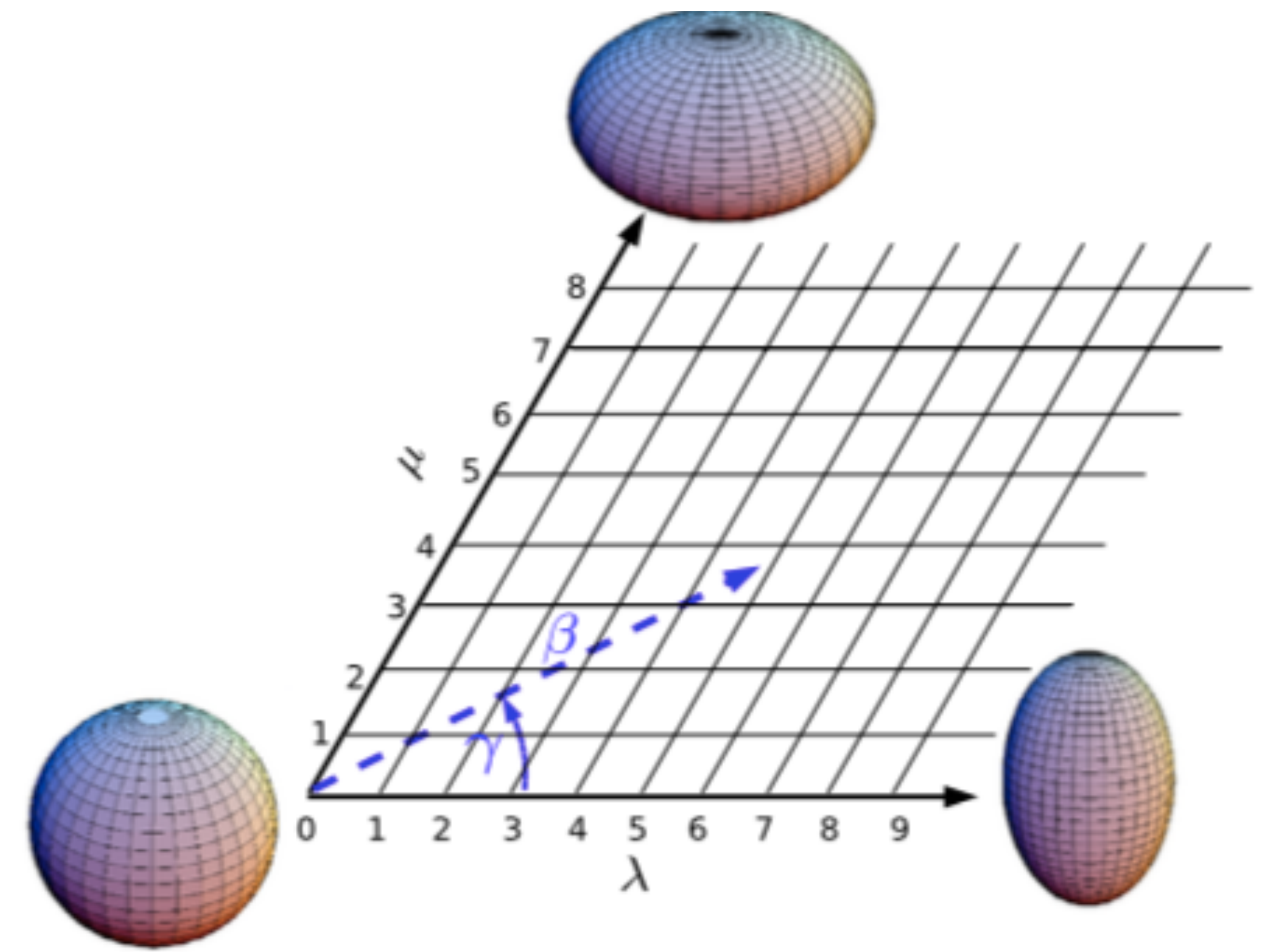}
\vspace{-0.25cm}
\caption{(Top) The symplectic configurations in the wave function of the ground state of $^6$Li computed within the SA-NCSM
  \cite{Dytrych:2013cca}. Less than 1\% of the included configurations account for greater than 90\% of the physics encoded in
  the wave function. (Bottom) The correspondence between the Elliott SU(3) group quantum numbers ($\lambda, \mu$) and the
  parameters $\beta$ and $\gamma$ that describe the shape of atomic nuclei.}
\label{li6}
\end{figure}

Before pushing ahead with addressing low/high energy aspects of the nuclei/nucleon interface, it seems useful to also recall that the
simplest Bohr-Mottelson liquid-drop picture of a nucleus -- which predated the work of Elliott -- was used to successfully describe in
the simplest of terms the dominance of rotational and vibrational modes in light as well as heavy nuclei. Under the assumption of
deformation (quadrupole) dominance described in terms of two shape variables, $\beta$ and $\gamma$, many observed features of
nuclei -- light and heavy -- could be reasonably reproduced. Interestingly, the corresponding quantities in the Sp(3,R) picture are the
($\lambda,\mu$) representation labels of the lowest-weight SU(3) configuration in a Sp(3,R) representation that uniquely defines its
entire structure. In fact, up to an overall scaling, the rotational invariants of SU(3) can be put in one-to-one correspondence with
$\beta$ and $\gamma$ of the Bohr-Mottelson theory. In short, $\beta$ is a radial measure of the prolate-oblate character of a
nucleus and $\gamma$ is an angular measure of its triaxiality. This relationship is depicted schematically in Fig.~\ref{li6} (bottom)
where the continuous $\beta$ and $\gamma$ values are shown by a vector, while the discrete ($\lambda, \mu$) quantum numbers
of the Elliott SU(3) group are depicted by a mesh overlaying the same space. Each node corresponds to a certain shape (spherical,
prolate, oblate, or triaxial) as determined by the $\beta$ and $\gamma$ parameters of each node. In summary, the basis states of
a specific Sp(3,R) symmetry should be considered to be a coherent combination of the well-defined shapes.

The above background on symmetry informed advances in {\it ab initio} approaches for studying the structure of atomic nuclei
\cite{Launey:2016fvy}, coupled with progress in gaining a description of the ground state structure of nucleons from a strong QCD
perspective within continuum QCD approaches~\cite{Eichmann:2016yit,Segovia:2015ufa,Chen:2018nsg,Chen:2019fzn,Lu:2019bjs}
discussed below, opens up a promising new avenue to explore at a deeper level how deformation of atomic nuclei emerges from strong
QCD. These studies address two important questions: ({\it a}) whether the ground state of a nucleon in its intrinsic frame is round or
deformed, and ({\it b}) how the interactions between nucleons within nuclei are driven by strong QCD and the role it plays in the
generation of dynamic deformation found in atomic nuclei? Clearly the scales are different -- MeV for nuclei and GeV for the strong
interaction between the dressed quarks and gluons in the strong QCD regime; yet, the deuteron is bound and deformed while
di-neutrons and di-protons are unbound, and the $\alpha$-particle ($^4$He) is the most tightly bound of all light nuclei, and also
deformed!

\begin{figure}[htp]
\includegraphics[width=0.95\columnwidth]{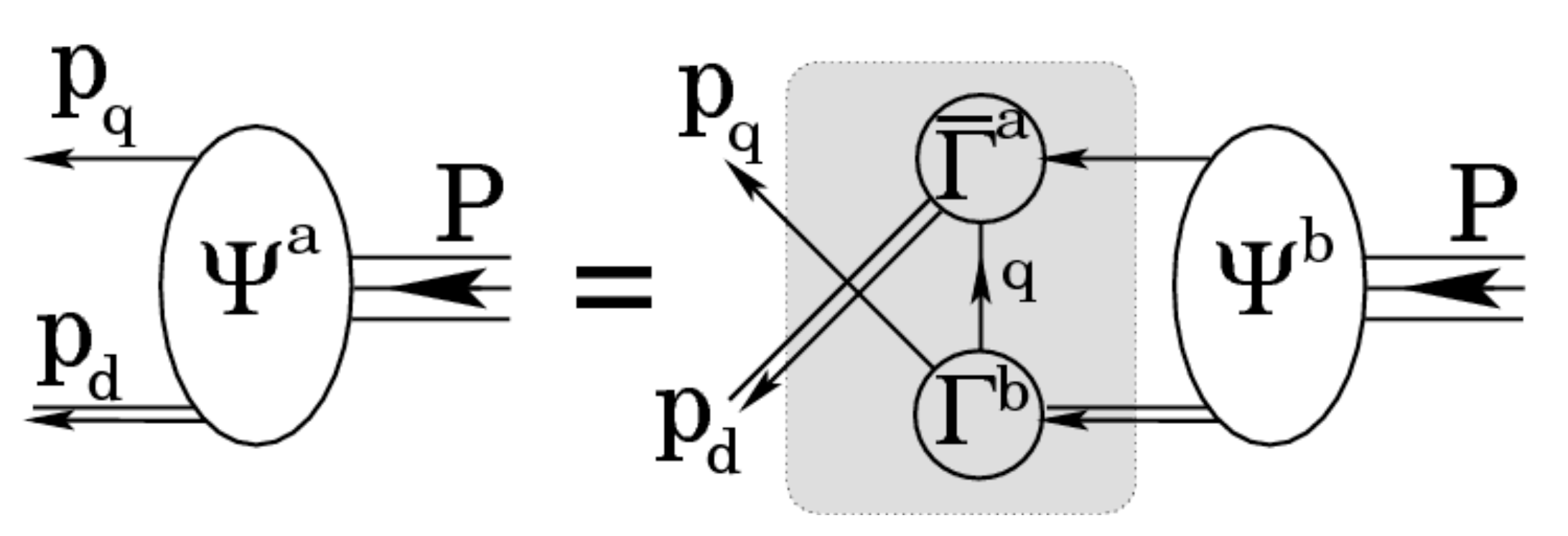}
\vspace{-0.25cm}
\caption{Nucleon = quark+diquark Faddeev equation. This is a linear integral equation for the Poincar\'e-covariant matrix-valued
  function $\Psi$, the Faddeev amplitude for a state with total momentum $P= p_q + p_d$. It describes the relative momentum
  correlation between the dressed-quarks and -diquarks. Legend. Shaded rectangle -- kernel of the Faddeev equation;
  {\it single line} -- dressed-quark propagator; $\Gamma$ -- diquark correlation amplitude; and {\it double line} -- diquark
  propagator. Ground-state nucleons ($n$ - neutron, $p$ - proton) contain both isoscalar-scalar diquarks, $[ud]\in (n,p)$, and
  isovector-pseudovector diquarks $\{dd\}\in n$, $\{ud\}\in (n,p)$, $\{uu\}\in p$.}
\label{figFaddeev}
\end{figure}

\subsection{Nucleon Shape from Continuum QCD}
\label{dseqcd}

Substantial progress has been achieved in the exploration of ground state nucleon structure within continuum QCD approaches with
a traceable connection to the QCD Lagrangian~\cite{Eichmann:2016yit,Segovia:2015ufa,Chen:2018nsg,Chen:2019fzn,Lu:2019bjs}.
Within continuum QCD approaches, full information on the ground state nucleon structure is encoded in the Faddeev amplitude. This
amplitude comes from the solution of the integral equation depicted in Fig.~\ref{figFaddeev}. The ground state nucleon Faddeev
amplitude is given by the residue at the position of the pole of minimal mass in the partial wave $J^{P}=1/2^+$. The kernel in
Fig.~\ref{figFaddeev} implies two sources for the dressed constituent quark binding within the nucleon: ({\it a}) tight, dynamical
color-antitriplet quark-quark correlations in the scalar-isoscalar and pseudovector-isotriplet channels and ({\it b}) exchange
associated with diquark breakup and reformation, which is required in order to ensure that each valence-quark participates in all
diquark correlations to the complete extent allowed by its quantum numbers.

\begin{figure}[htp]
\includegraphics[width=0.9\columnwidth]{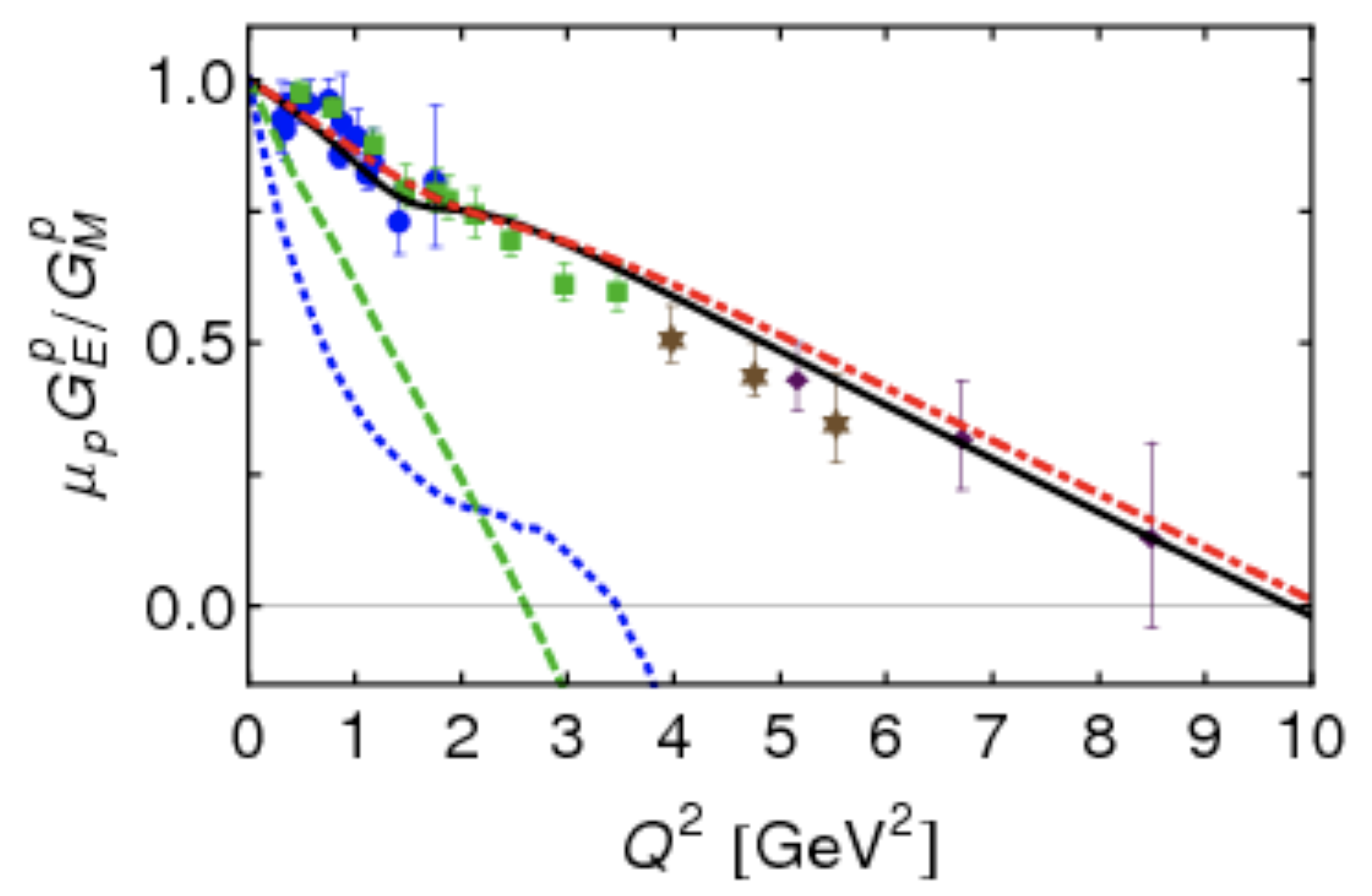}
\vspace{-0.25cm}
\caption{Continuum QCD prediction for the ratio $\mu_{P}G^{P}_{E}/G^{P}_{M}$~\cite{Segovia:2015ufa}: solid (black) curve -- full result;
  dot-dashed (red) -- momentum-dependence of the scalar-diquark contribution, including both $L=0,1$ nucleon rest-frame quark+diquark
  orbital angular momentum components. The contribution from configurations of quark+diquark orbital angular momentum $L$=0 resulting
  in a spherical nucleon shape is shown by the dashed (green) curve, while the dashed (blue) curve shows the contribution from the
  pseudovector diquark, which can generate a deformation in the intrinsic frame of the nucleon. All partial contributions have been
  renormalized to produce unity at $Q^2=0$. Data: circles (blue)~\cite{Gayou:2001qt}; squares (green)~\cite{Punjabi:2005wq};
  asterisks (brown)~\cite{Puckett:2010ac}; and diamonds (purple)~\cite{Puckett:2011xg}.}
\label{grnust}
\end{figure}

In Fig.~\ref{grnust} we compare the experimental data on the ratio of electric to magnetic proton form factors,
$\mu_{P}G^{P}_{E}/G^{P}_{M}$, with results computed using the Faddeev equation solution outlined above~\cite{Segovia:2015ufa}. 
Here we also show (green dashed curve) the contribution from the configuration corresponding to the orbital angular momentum 
between the quark and scalar diquark in the proton rest frame $L$=0 ($S$-wave). This contribution gives rise to the spherically 
symmetric shape of the nucleon in its intrinsic frame. However, the pseudovector diquark correlations (shown by blue dashed line) 
also contributes to the ground state nucleon structure. Owing to the presence of the selected direction in the intrinsic nucleon 
frame defined by the pseudovector, it becomes possible to generate a deformed shape of the nucleon in its intrinsic frame. 

\begin{figure}[htp] 
\includegraphics[width=0.9\columnwidth]{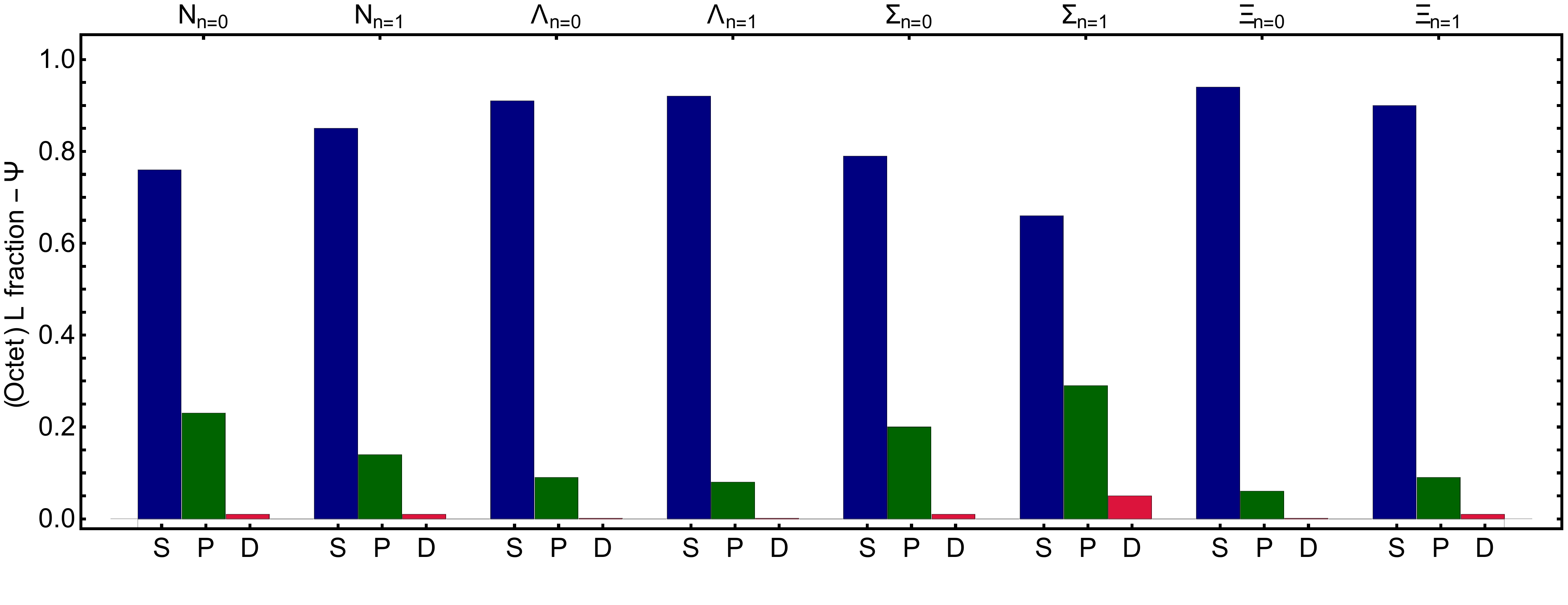}
\includegraphics[width=0.9\columnwidth]{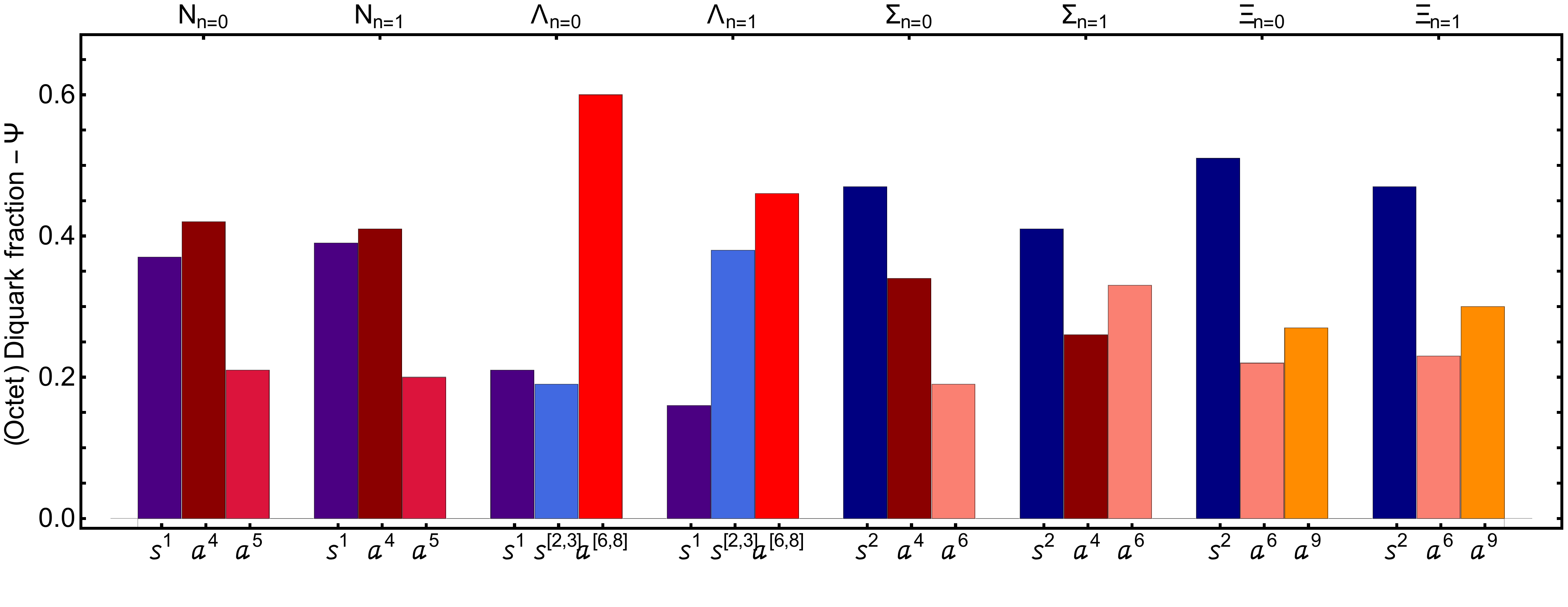}
\vspace{-0.25cm}
\caption{Octet baryon ground states ($n=0$) and their first positive-parity excitations ($n=1$). Upper panel
  -- Baryon rest-frame quark-diquark orbital angular momentum fractions: $L=0$ ($S$), $L=1$ ($P$) and $L=2$ ($D$).
  Lower panel -- Relative strengths of various diquark components within the indicated baryon's Faddeev
  amplitude: scalar diquark ($s$) and pseudovector diquarks ($a$, $a$). The superscripts here indicate quark flavor content.
  This is listed for the neutron and proton in Fig.~\ref{figFaddeev}. (Details provided in Ref.~\cite{Chen:2019fzn}.)}
\label{orbdiq}
\end{figure}

In Fig.~\ref{orbdiq} the relative contributions of the configurations of different quark+diquark orbital angular momenta in the
nucleon rest frame, as well as the contributions from scalar and pseudovector diquark correlations computed using the methods of
continuum QCD~\cite{Chen:2019fzn} are shown. Based on these expectations for the ground state nucleon structure, we conclude
that the contributions from pseudovector diquark correlations, together with the admixture of the configurations with non-zero
quark-diquark orbital angular momenta, may generate a nucleon deformation in its intrinsic frame.

Confirming the continuum QCD expectations in comparison with the experimental data on ground and excited state nucleons, as well
as atomic nuclear structure opens up a new avenue in the joint research efforts between nuclear and hadron physics.

\subsubsection{Validating Theory Predictions on the Nucleon Shape}
\label{shape}
The Faddeev equation depicted in Fig.~\ref{figFaddeev} includes: ({\it a}) the dressed quark propagator with the momentum-dependent 
dressed quark mass, ({\it b}) the diquark correlation amplitude $\Gamma$, and ({\it c}) the diquark propagator. The CLAS results on 
the nucleon resonance electroexcitation amplitudes ($\gamma_vpN^*$ photo-/electrocouplings) offer an excellent opportunity to validate 
the continuum QCD evaluations for the momentum dependence of the dressed quark mass, the so-called dressed quark mass function. The s
tructure of all excited nucleon states is encoded in their respective Faddeev amplitudes $\psi_{N^*}$, which are the solution of the 
similar Faddeev equations for the partial waves and with the diquark correlations allowed and determined by the resonance quantum numbers.

The consistent results on the dressed quark mass function available from independent continuum QCD studies of nucleon elastic form
factors and electroexcitation amplitudes for resonances of distinctively different structure, $\Delta(1232)3/2^+$ and $N(1440)1/2^+$,
strongly support the credible evaluation of the dressed quark mass function and the dressed quark propagator within continuum QCD
\cite{Roberts:2019wov}. Comparison between the continuum QCD expectations on the $\gamma_vpN^*$ photo-/electrocouplings for
other excited nucleon states~\cite{Lu:2019bjs} with the experimental results from CLAS~\cite{Mokeev:2019ron} will further validate
the continuum QCD evaluation of the dressed quark propagator starting from the QCD Lagrangian.

The spin-parities for the diquark correlations contributing to $N^*$ structure are unambiguously determined by the resonance
spin-parity. Therefore, studies of the $\gamma_{r,v}pN^*$ photo-/electrocouplings for all prominent nucleon resonances will allow us
to explore all relevant types of diquark correlations within nucleon structure. In this way, the experimental results on the
$\gamma_{r,v}pN^*$ photo-/electrocouplings will provide for the adequate description of the diquark correlation amplitudes $\Gamma$
and the diquark propagators obtained within the continuum QCD approaches. The description of diquark correlations in the nucleon
structure can also be checked by making predictions for the meson$^\prime$\,$\to$\,meson time-like transition form factors. These
form factors should be computed by employing the Bethe-Salpeter amplitudes associated with the meson partners of a given diquark
correlation, ranging over all diquarks relevant for the structure of all prominent $N^*$ states. The meson-transition time-like form
factors predicted in this way can be confronted with the future experimental results, thus bridging the efforts between studies of
meson and baryon structure with the outcome of particular interest for the exploration of nucleon shape in connection with the nucleon
deformation impact on the structure of atomic nuclei.

\subsubsection{From Nucleon Structure to the Structure of Atomic Nuclei}
\label{nucleon_nuclei}

In order to gain insight on the shape of the nucleon in its intrinsic frame, the ground state nucleon Faddeev amplitude should be
transformed into the light-front wave function~\cite{Mezrag:2017znp} to allow a probabilistic interpretation for the matter distribution
within the nucleon. A special approach should be developed to allow us to express the ground state nucleon light-front wave function in
terms of the superposition of the Elliott SU(3) eigenvectors projected into the light front. Each of these vectors ($\lambda, \mu$)
corresponds to the contribution of a certain and well-known shape. Therefore, the decomposition of the ground state nucleon light-front
wave function over the set of ($\lambda, \mu$) SU(3) eigenvectors will provide the information on the co-existence of different shapes
that generate the resulting nucleon shape in its intrinsic frame.

An alternative pathway in the search for the intrinsic nucleon deformation is the computation of the pretzelosity $h_{1T}$ TMD
structure function. Its non-zero values will offer evidence for the intrinsic nucleon deformation. The presence of the nucleon
deformation is implied by the tensor structure of the projection operator for the pretzelosity~\cite{Miller:2008sq}. Comparison of
the experimental results on $h_{1T}$ from the SIDIS data with the expectations from continuum QCD, with all elements of the
Faddeev equation kernel checked against the results on the nucleon elastic and transition form factors, will bridge the efforts between
the $N^*$ and DIS physics in the exploration of hadron structure, in particular, shedding light on the nucleon's shape.

In order to reveal the impact of the nucleon shape on the structure of atomic nuclei, the nucleons should be treated in the many-body
nuclear physics computations as space-extended objects of the certain shape, in contract with most previous studies of the nuclear
structure where the nucleons were considered as point-like objects. As a starting point, the structure of the simplest nuclei $^2$H,
$^3$H, $^3$He, and $^4$He should be described accounting for the space-extension of the nucleons, paving the way for description
of complex nuclei as bound systems of spatially extended nucleons~\cite{Draayer}. Besides the shape of the nucleon, nuclear
structure is also determined by nucleon interactions within atomic nuclei. Therefore, the emergence of nucleon interactions from strong
QCD should also be elucidated. As a first step, the $NN$ interaction amplitudes should be related to the dressed quark and the dressed
gluon interactions within nucleons and pions. This work is in progress within the continuum QCD approaches~\cite{Roberts19}.

Eventually, the emergence of nuclear structure from the interactions between the dressed quarks and gluons confined within the
nucleons and mesons will be explored. The success in this challenging adventure will certainly open up a new chapter in the experimental
studies of nuclear structure in nuclear-nuclear, photon-nucleus, and electron-nucleus collisions.

\section{Phenomenology for the Extraction of QCD Interpretable Hadron Parameters}

\subsection{Advances in Multi-Channel Amplitude Analyses}

Partial-wave analysis provides the link between large-scale experimental programs and theory approaches that focus
on the intermediate-energy region, where quark confinement manifests itself in a rich spectrum of resonances and
where the key to our understanding of the strong interaction will be found. Improved and extended analysis techniques
are necessary to further our understanding of baryon structure and, in particular, help resolve the missing-resonance
problem, formulated as a priority in the 2015 Long-Range Plan of the National Science Advisory Council (NSAC).

The intermediate-energy region holds the key to our understanding of confinement and its manifestation in a rich phenomenology of 
resonances. Partial-wave analysis (PWA) of experimental data reveals the spectrum of these resonances. It, thus, provides the bridge 
to compare experiments with theories such as Quark Models~\cite{Aznauryan:2012ec,Isgur:1978xj,Capstick:1986bm,Capstick:1992uc,
Capstick:1993kb,Ronniger:2011td,Ramalho:2011ae,Golli:2013uha,Jayalath:2011uc}, Dynamical Dyson-Schwinger approaches
\cite{Roberts:1994dr,Roberts:2007ji,Wilson:2011aa,Chen:2012qr,Eichmann:2016yit}, Chiral Unitary calculations
\cite{Meissner:1999vr,Kolomeitsev:2003kt,GarciaRecio:2003ks,Doring:2005bx,Doring:2007rz,Doring:2010fw,Doring:2009qr,Borasoy:2006sr,
Mai:2012wy,Gasparyan:2010xz,Garzon:2012np,Khemchandani:2013hoa,Wu:2010vk}, and Lattice QCD simulations~\cite{Durr:2008zz,Burch:2006cc,
Alexandrou:2008tn,Bulava:2010yg,Menadue:2011pd,Melnitchouk:2002eg,Edwards:2011jj,Edwards:2012fx, Engel:2013ig, Lang:2012db, 
Pelissier:2012pi, Alexandrou:2013ata}. Note that, in the latter two approaches, one can directly compare theory to partial waves 
assuming that finite-volume, discretization, and quark mass effects are under control in lattice QCD calculations (see, e.g
Refs.~\cite{Doring:2016bdr,Mai:2017bge,Mai:2018djl}).

Our knowledge of the baryon spectrum, as determined from analyses of experimental data, has advanced rapidly~\cite{Tanabashi:2018oca} 
over the past decade. The progress has been most significant for non-strange baryons, due largely to the wealth of new and more precise 
measurements made at electron accelerators worldwide. The majority of these new measurements have been performed at Jefferson Lab (using 
the CLAS and Hall A detectors), with the MAMI accelerator in Mainz (the Crystal Ball/TAPS detector being particularly well suited for the 
measurement of neutral final states), and with the Crystal Barrel detector at ELSA in Bonn.

While most of the early progress~\cite{Hoehler,Cutkosky:1979zv,Arndt:2006bf,Shrestha:2012ep} in baryon spectroscopy
was based on the analysis of meson-nucleon scattering data, particularly pion-nucleon scattering ($\pi N\to \pi N$,
$\pi N\to \pi \pi N$), photon-nucleon interactions offer the possibility of detecting unstable intermediate states with small
branchings to the $\pi N$ channel. Many groups have performed either single-channel or multi-channel analyses of these
photo-induced reactions. In the more recent single-channel analyses, fits have typically used isobar models
\cite{Drechsel:2007if,Aznauryan:2002gd} with unitarity constraints at the lower energies, $K$-matrix-based formalisms,
having built-in cuts associated with opening inelastic channels~\cite{Workman:2012jf}, and dispersion-relation constraints
\cite{Aznauryan:2002gd, Hanstein:1997tp}. Multi-channel fits have analyzed data (or, in some cases, amplitudes) from
hadronic scattering data together with the photon-induced channels. These approaches have utilized unitarity more directly.
The most active programs are being carried out by the Bonn-Gatchina~\cite{Anisovich:2011fc}, J\"ulich-Bonn
(JuBo)~\cite{Ronchen:2015vfa}, ANL-Osaka~\cite{Kamano:2013iva}, Kent State~\cite{Shrestha:2012ep}, JPAC
\cite{Nys:2016vjz}, and Giessen~\cite{Shklyar:2006xw} groups. At low energies the chiral MAID analysis provides a
comprehensive description of photo and electroproduction data~\cite{Hilt:2013fda}.

\subsubsection{From Photo to Electroproduction}
\label{amp_electro}

Our knowledge of the baryon spectrum has rapidly evolved over the last decade, due largely to the refinement of dynamical
and phenomenological coupled-channel approaches for the analysis of pseudoscalar-meson photoproduction reactions. The
electroproduction process is closely related but, so far, no unified coupled-channel analysis of photo and electroproduction
experiments exists that simultaneously studies the $\pi N$, $\eta N$ and $\Lambda K$ final states, where the $Q^2$
variation of resonance couplings is expected to provide a link between perturbative QCD and the region where quark
confinement sets in. 

Going from photo- to electroproduction of pseudo-scalar meson, the number of helicity amplitudes increases from four to
six, requiring more measurements for the analogous `complete experiment'~\cite{Tiator:2017cde,Dmitrasinovic:1987pc},
with a multipole decomposition adding longitudinal components to the transverse elements anchored by photoproduction
analyses at $Q^2=0$. Variation of resonance couplings with $Q^2$ is expected to provide a link between perturbative
QCD and a region where quark confinement requires the use of lattice QCD, ChPT, or more phenomenological approaches.
Exactly where this transition occurs is not precisely known. The well-known prediction~\cite{Carlson:1985mm} of an E2/M1
ratio, for the $\Delta(1232)$ state, approaching unity shows no sign of occurring, remaining essentially flat at a small
negative value. In contrast, other clear resonances, such as the $N(1520)$, do show rapid behavior in the low-$Q^2$ region,
followed by a smooth transition to higher values of $Q^2$.

The reliable determination of helicity amplitudes for $Q^2>0$ is also relevant for neutrino physics. Neutrino oscillation
experiments are currently evolving from the discovery to the precision era. At this stage, systematic errors will become
comparable to the statistical ones. A deeper understanding and realistic modeling of neutrino interactions with the detector
target is therefore required. At the future DUNE experiment, most of the interaction events will be inelastic, demanding an
accurate description of weak meson (typically pion) production. The multipole decomposition of the electroproduction
amplitude provides a powerful framework incorporating all the valuable experimental information available to constrain the
corresponding weak processes. Furthermore, methods to extract information about the axial nucleon inelastic current from
measurements in detector materials containing hydrogen are being developed. The possibility of performing new
measurements of neutrino cross sections on elementary targets is also being considered. In such a scenario, our formalism is
particularly well suited to extract useful information about the axial properties of the inelastic nucleon current at various
kinematics. See also Ref.~\cite{Nakamura:2015rta} for progress in this direction.

Recent advances in the baryon spectrum, based upon real-photon data analyzed via dynamical and phenomenological
coupled-channel models, have not yet been fully reproduced in the realm of electroproduction. Electroproduction experiments,
e.g., by the CLAS collaboration at JLab, are producing a wealth of data that, in many cases, still await a detailed analysis, 
in pion electroproduction~\cite{Bosted:2016spx,Bosted:2016hwk,Zheng:2016ezf,Park:2014yea} (JLab),
\cite{Stajner:2017fmh} (A1 collaboration at MAMI), $\eta$ electroproduction~\cite{Bedlinskiy:2017yxe} (JLab),
\cite{Merkel:2007ig} (A1 collaboration at MAMI), and kaon electroproduction~\cite{Achenbach:2017pse} (A2 collaboration),
\cite{Nasseripour:2008aa} (JLab). Analyses at JLab continue; for example in the near future data on $\pi^0p$ differential
cross sections at $0.3<Q^2<1.0$~GeV$^2$ and $1.1<W<1.8$~GeV will be released~\cite{Mokeevprivate}.

\begin{table}[htp]
\begin{tabular}{l|l|l|l|l}
Reaction		& Observable	& $Q^2$~[GeV]	& $W$ [GeV]	& Ref. \\ \hline \hline
\multirow{3}{*}{$ep\to e'p'\eta$}	& $\sigma_U,\,\sigma_{LT}, \,\sigma_{TT}$ & $1.6-4.6$ & $2.0-3.0$ & \cite{Bedlinskiy:2017yxe}	\\
& $\sigma_U,\,\sigma_{LT},\,\sigma_{TT}$ & $0.13-3.3$ & $1.5-2.3$ & \cite{Denizli:2007tq} \\
& $d\sigma/d\Omega$	& $0.25-1.5$ & $1.5-1.86$ & \cite{Thompson:2000by} \\ \hline
\multirow{5}{*}{$ep\to e'K^+\Lambda$}	& $P^0_N$ & $0.8-3.2$ & $1.6-2.7$ & \cite{Gabrielyan:2014zun} \\
& $\sigma_U,\,\sigma_{LT},\,\sigma_{TT},\,\sigma_{LT'}$ & $1.4-3.9$ & $1.6-2.6$ & \cite{Carman:2012qj} \\
& $P_x',\,P_z'$	& $0.7-5.4$ & $1.6-2.6$ & \cite{Carman:2009fi} \\
& $\sigma_T,\,\sigma_L,\,\sigma_{LT},\,\sigma_{TT}$ & $0.5-2.8$ & $1.6-2.4$ & \cite{Ambrozewicz:2006zj} \\
& $P_x',\,P_z'$	& $0.3-1.5$	& $1.6-2.15$ & \cite{Carman:2002se} \\\hline
\end{tabular}
\caption{Overview of $\eta p$ and $K^+\Lambda$ electroproduction data measured at CLAS for different photon
  virtualities $Q^2$ and total energy $W$. Based on material provided by courtesy of D.S. Carman (JLab) and
  I. Strakovsky (GW).}
\label{tab:clasdata}
\end{table}

In Table~\ref{tab:clasdata}, CLAS data on $\eta$ and kaon electroproduction are summarized. The much longer list for
pion electroproduction is omitted here for brevity. Many pion electroproduction data are already included in the SAID
database~\cite{onlinesaid}. It should also be stressed that pion and kaon electroproduction experiments with the new 
CLAS12 detector at the 12~GeV upgrade of Jefferson Lab~\cite{clas12,exclusive_KY,more_exclusive} will provide many
data that require a timely analysis.

The ANL-Osaka group is currently extending its dynamical coupled-channel analysis of pion electroproduction
\cite{JuliaDiaz:2009ww} to higher $Q^2$-values~\cite{Kamano:2018sfb}. Plots of the $\Delta(1232)$ amplitudes at the
resonance pole position (yielding a complex amplitude) also seem to qualitatively reproduce results found for the MAID and
SAID analyses~\cite{Tiator:2016btt}. However, results have generally been restricted to the low-energy $\Delta(1232)$
and $N(1440)$ states.

The most widely used single-pion electroproduction analyses, covering the resonance region, have been performed by the
Mainz (MAID)~\cite{Drechsel:2007if} and Jefferson Lab~\cite{Aznauryan:2009mx,Az13} groups. An
extensive single-pion electroproduction database, and a $K$-matrix based SAID fit, is also available~\cite{Arndt:2001si}.
Eta electroproduction has been analyzed in the Eta MAID framework~\cite{Chiang:2001as}, and kaon electroproduction by
the Ghent group~\cite{Corthals:2007kc}. These fits have generally utilized a Regge~\cite{Vanderhaeghen:1997ts} or
Regge-plus-resonance approach~\cite{Vrancx:2014pwa} at high to medium energies. (We mention here parenthetically that
the Ghent Regge approach can be improved by correctly implementing the local gauge-invariance constraints
\cite{HWH2015}.) Effective Lagrangian and isobar models have also been used~\cite{Mart:2002gn,Maxwell:2014txa}, with
some of these available via the MAID website, for both kaon and eta electroproduction~\cite{Chiang:2001as}.

These are all single-channel analyses with approaches similar to the associated real-photon fits, but generally, with the
exception of the MAID and SAID analyses, not including the real-photon data as a constraint at $Q^2=0$. Both the MAID
and Jefferson Lab groups have made fits using a Breit-Wigner plus background models with resonance couplings extended
to include a $Q^2$ dependence. In the Jefferson Lab analyses~\cite{Aznauryan:2009mx}, a further fit was again based on
satisfying fixed-t dispersion relation. It should be mentioned that two-pion electroproduction fits have also been performed,
and compared to the single-pion results, at Jefferson Lab
\cite{Aznauryan:2005tp,Aznauryan:2011qj,Isupov:2017lnd,Mokeev:2015lda}. See Ref.~\cite{Aznauryan:2011qj} for a review.

We emphasize that the electromagnetic resonance properties are encoded in the helicity couplings. In the past, they have
often been defined as Breit-Wigner couplings, {\it i.e.}, a Breit-Wigner plus background term was fitted to the energy-dependent
multipoles or helicity amplitudes that are a superposition of multipoles. The couplings are real by construction, but not
unambiguously defined. Instead, a reaction-independent definition of helicity couplings is only possible by utilizing residues 
of the resonance poles, since it is these singularities in the complex energy plane and their properties that lead to resonance
shapes for real physical scattering energies. See Ref.~\cite{Tiator:2016btt} for recent work of our group towards this goal.
Indeed, such a definition is reasonable because many hadronic models can only compare to the transition form factors at the
pole, as in the unitarized versions of Chiral Perturbation Theory~\cite{Gail:2005gz,Dorati:2007pv,Doring:2010rd,Jido:2007sm} or 
perturbative calculations using the complex mass scheme~\cite{Hilt:2017iup,Bauer:2014cqa}. Transition form factors at the complex 
pole position are complex themselves. which adds a new and independent piece of information compared to the reaction-dependent 
definition in terms of Breit-Wigner parameters~\cite{Tiator:2016btt,Hilt:2017iup}. 

\subsubsection{Recent Progress with the Juelich-Bonn and SAID Analysis Frameworks}

The Juelich-Bonn analysis is currently upgraded to simultaneously analyze pion, eta, and kaon photo- and electroproduction
reactions. A few recent developments in the joint analysis effort of the SAID and Juelich-Bonn group are highlighted in the
following.

Data from the CLAS collaboration at Jefferson Lab have been incorporated into the SAID website. These have included
neutron-target (deuteron) measurements of pion photoproduction cross sections and the beam-target polarization observable
($E$). Data were fitted via multipole analyses that were included in the associated publications of these datasets
\cite{Mattione:2017fxc,Ho:2017kca}. Similarly, the beam asymmetry $\Sigma$ in $\eta$ photoproduction was analyzed
using the JuBo approach and published with the CLAS collaboration~\cite{Collins:2017sgu}; an intriguing structure in the
data, close to the position of a supposed pentaquark, could be conventionally explained in terms of interference effects. 
The JuBo approach reached a major milestone by analyzing the world database of $K^+\Lambda$ photoproduction
\cite{Ronchen:2018ury}. This reaction is particularly interesting due to the richness of polarization data available from
experiments at Jefferson Lab, ELSA, and MAMI. Two resonances claimed by other groups could be confirmed and properties
of known states could be determined with greater precision. 

The SAID, JuBo, and other groups compared their analysis frameworks and the impact of new high-precision data in a common
effort~\cite{Beck:2016hcy}. This synergistic study revealed that the new data brought the analyses significantly closer
together although differences still remain.

In anticipation of future electroproduction data from JLab, the SAID group has analyzed baryon transition form factors at
the resonance pole~\cite{Tiator:2016btt} in collaboration with Lothar Tiator and Alfred \v Svarc. For the $\Delta(1232)$,
existing SAID and MAID transition form factors showed qualitative agreement, apart from $R_{SM}(Q^2)$ at intermediate
values of $Q^2$.

In collaboration with Tiator (Mainz) and Wunderlich (Bonn), we have also re-examined the connections between complete
experiments and truncated multipole analyses for both pion photoproduction and pion electroproduction. New relationships
were discovered~\cite{Tiator:2017cde,Workman:2016irf}. Studies of the phase-ambiguity problem, associated with
amplitude analyses, were conducted in collaboration with the Mainz, Bonn, and Zagreb groups
\cite{Wunderlich:2017dby,Svarc:2017yuo}. These studies have more clearly defined the amplitude information that can
be extracted from data with minimal input from reaction models.

\subsection{Extension of Amplitude Analyses, Development of Reaction Models for the Extraction of Hadron Parameters from Data}
\label{amp_photo}

The recent JLab 12~GeV upgrade, the proposed EIC, and the continued precision experimental 
efforts at accelerator facilities around the world all aim to explore the dynamics of strong QCD and hadron structure. These experiments 
necessitate close collaboration between experimentalists, phenomenologists, theorists to untangle the underlying physics from 
measurement. The mission of the Joint Physics Analysis Center (JPAC) at Indiana University and Jefferson Lab is to facilitate this 
collaboration by developing tools for precision data and amplitude analysis as well as training a new generation of practitioners of 
hadron/QCD phenomenology.

Amplitude analysis, {\it i.e.} the construction of models satisfying general quantum mechanical scattering principles ($S$-matrix
theory) to describe physical measurement is a vital step in connecting experiment to the theory of strong interactions. 
Here we summarize recent JPAC efforts in the this regard with a special focus on amplitude analysis in photo- and
electro-production processes. 

\begin{figure}[htp]
\includegraphics[width=0.95\columnwidth]{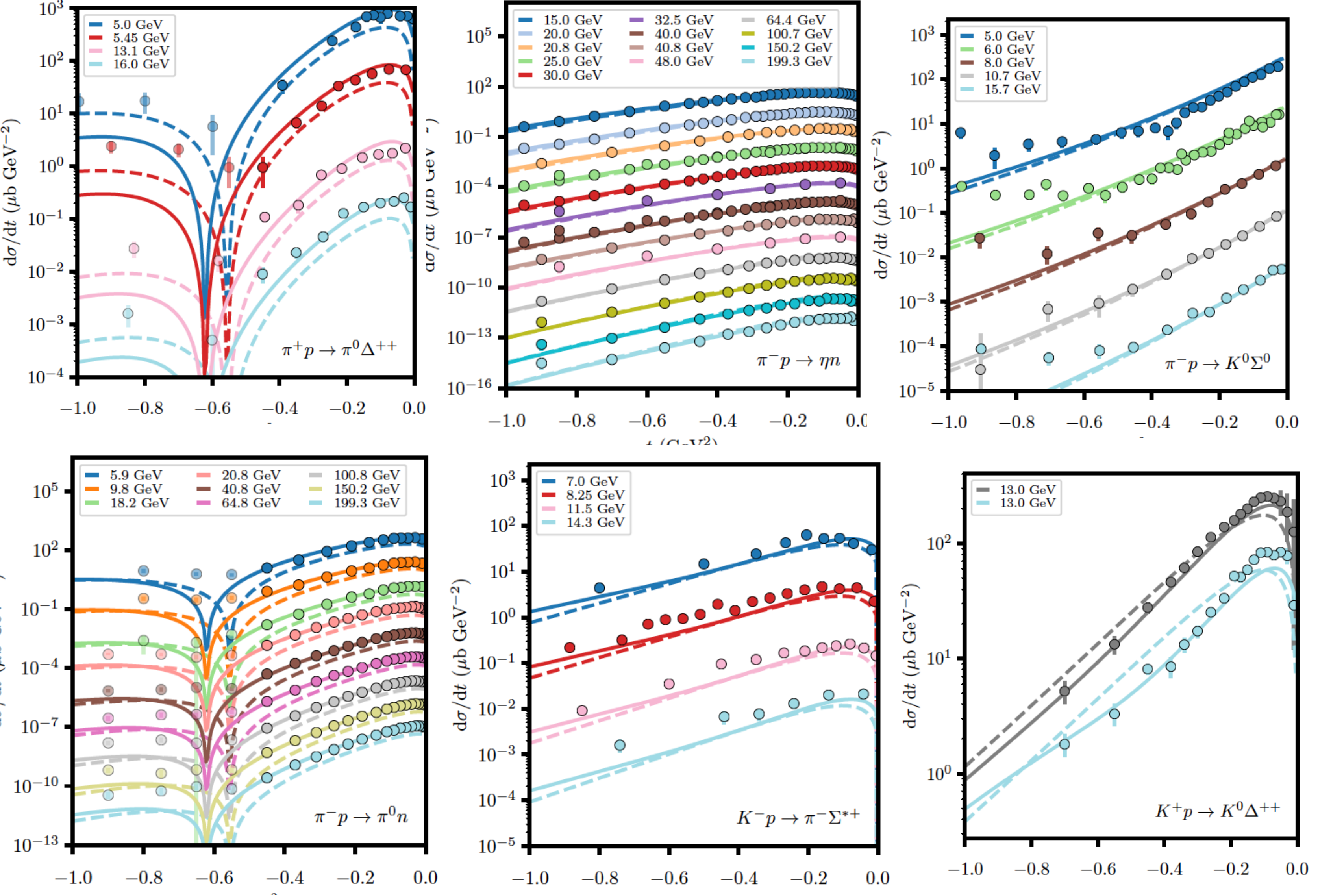}
\vspace{-0.25cm}
\caption{Sample of results from a global fit of world's data on quasi-elastic charge exchange reactions using Regge poles only.}
\label{adamfig:1}
\end{figure}

Many ongoing experiments, {\it e.g.} the CLAS12 and GlueX experiments both at JLab, rely on diffractive production of mesons
recoiling against (excited) nucleons to probe the properties of hadrons. These processes are primarily studied within the
framework of Regge phenomenology that provides a rich set of theoretical tools for amplitude analysis. The increased
statistics of next-generation experiments necessitate a deeper understanding of the validity of the tools used to extract
physical quantities. In particular, Regge theory predicts the high-energy diffractive production to be factorizable, {\it i.e.}
describe by independent fragmentation of the beam and target. Dynamics of the production arise from the presence of
resonances in the exchange channels, so called ``reggeon" exchanges. Photon induced, and in particular charge exchanged
processes allow the possibility of non-Pomeron reggeons to be exchanges and thus require identifying the dominant
contributions in terms of the singularity structure of the amplitude, {\it i.e.} contribution from leading and daughter Regge poles
or Regge cuts. The validity of this Regge factorization hypothesis was recently assessed by a global fit to charge and strange
exchange quasi-two body reactions in Ref~\cite{nys2018}. In particular kinematic domains where Regge pole models faithfully
reproduce observation are identified for reactions dominated by different exchanges (see Fig.~\ref{adamfig:1}).

\begin{figure}[htp]
\includegraphics[width=0.95\columnwidth]{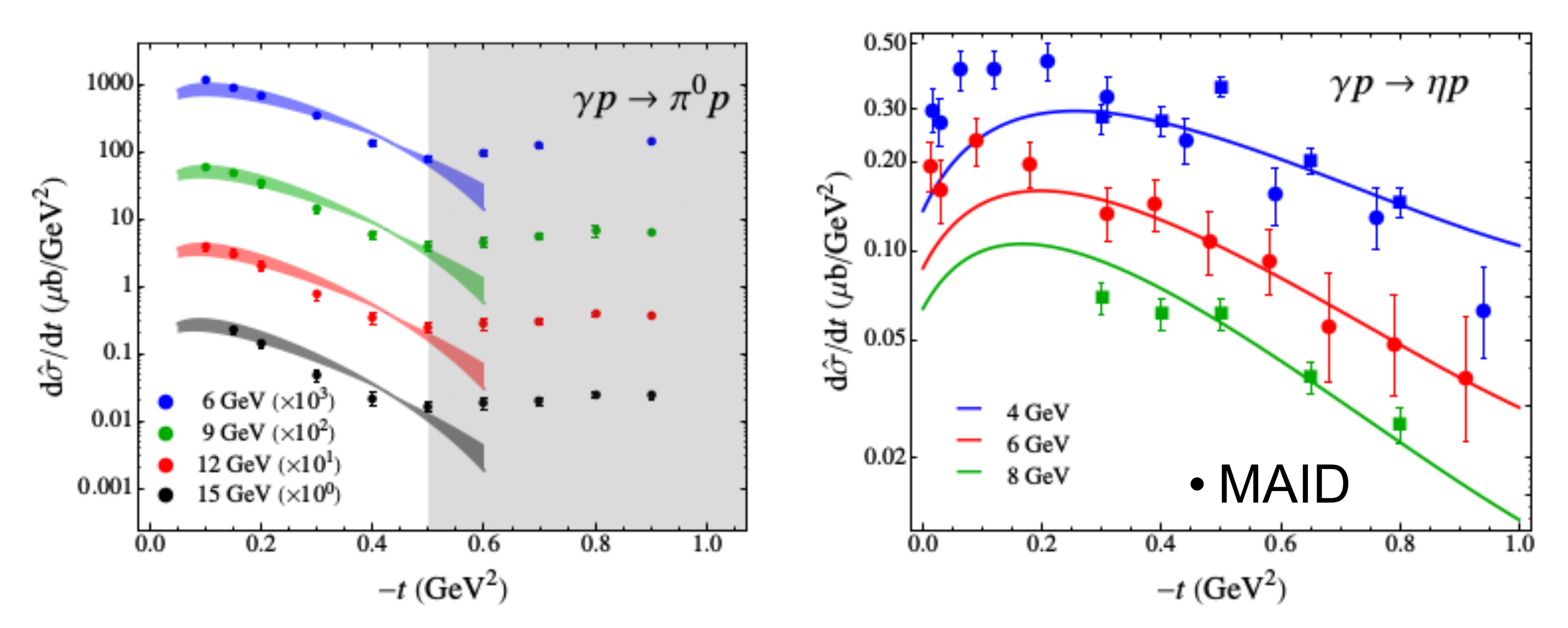}
\vspace{-0.25cm}
\caption{Finite Energy Sum Rule predictions from pion (left) and eta (right) low-energy partial wave analyses }
\label{adamfig:2}
\end{figure}

This framework was then applied to $\pi^0$ and $\eta$ photoproduction off a nucleon in Ref.~\cite{Mathieu2018,Nys:2016vjz}. Here,
analyticity in the form of finite-energy sum rules are used to relate the different energy regimes ({\it i.e.} the high-energy
diffractive regime and the low-energy resonance regime). This provides a rigid constraint on model amplitudes and connects
the baryon spectrum dominating the low-energy regime with the mesonic exchanges of Regge dominated peripheral scattering
(see Fig.~\ref{adamfig:2}).
 
\begin{figure}[htp]
\includegraphics[width=0.95\columnwidth]{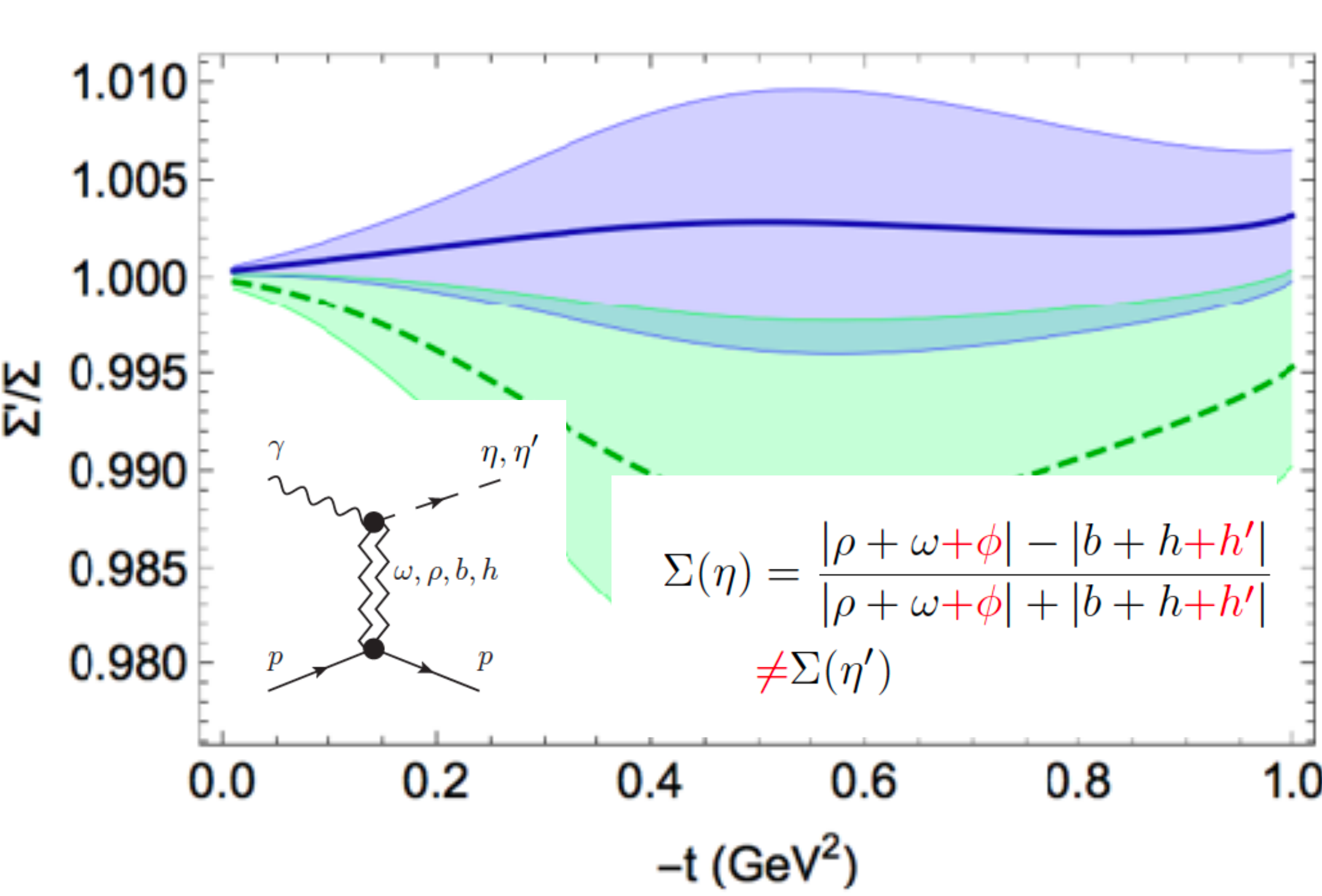}
\vspace{-0.25cm}
\caption{Regge pole prediction for the ratio of $\eta'$ to $\eta$ beam asymmetries.}
\label{adamfig:3}
\end{figure}

Similar Regge models are considered in $\eta/\eta'$ photoproduction at JLab energies in Ref.~\cite{Mathieu2018}. The contributions of 
hidden-strange exchanges are estimated to give predictions for the ratio of beam asymmetries between the $\eta$ and $\eta'$. An observed 
deviation from prediction may indicate significant sub-leading contributions (see Fig.~\ref{adamfig:3}).

Subsequent measurement from GlueX indicate agreement with prediction~\cite{PhysRevC.95.042201}. Generalizations for the beam
asymmetries and moments of angular distribution for $\eta\pi^0$ photoproduction have also been recently made
\cite{Mathieu2019}.

\begin{figure}[htp]
\includegraphics[width=0.95\columnwidth]{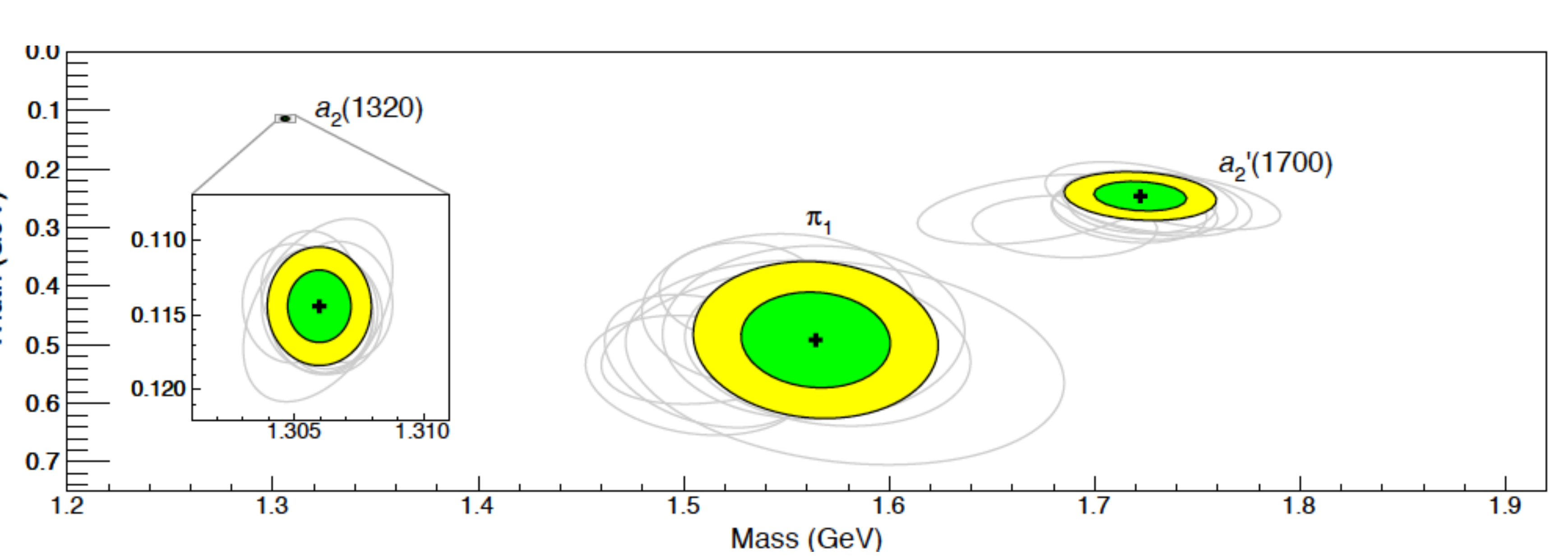}
\vspace{-0.25cm}
\caption{Pole position of (from left to right) the $a_2$, $\pi_1$, $a'_2$ resonances form analytical fits to the COMPASS
  partial waves analysis.}
\label{adamfig:4}
\end{figure}

JPAC is also extensively involved in the application of amplitude analysis to search for exotic hadrons. The possibility of
structures beyond the usual $q\bar{q}$ or $qqq$ have long been predicted to exist. In particular, hybrid mesons, {\it i.e.}
mesons with an excited gluonic degree of freedom, are expected to be accessible in a variety of reactions and uncovering
their properties ({\it e.g.} masses, widths, decays) provides a window into the role of glue at low-energies. The lowest lying
hybrid states are anticipated to exist with exotic quantum numbers $J^{PC} = 1^{-+}$ near 2~GeV. These exotic states, called
$\pi_1$ states have lead to controversies with experiments reporting the possible existence of two different states,
$\pi_1(1400)$ and $\pi_1(1600)$, coupling separately to $\eta\pi$ and $\eta^\prime\pi$ final states. JPAC provided a
robust extraction of resonant pole parameters from available COMPASS experiment data that indicates the existence
of only one exotic pole that couples to both $\eta^{(\prime)}\pi$ channels~\cite{Rodas2018} (see Fig.~\ref{adamfig:4}). 

The new data coming from dedicated photoproduction experiments at JLab will prove valuable in testing additional hypotheses. 

In addition to hybrid states, the observation of signals with pentaquark-like quantum numbers by the LHCb collaboration
\cite{LHCbcollaboration2019} spurred immediate interest in searching for these state with photoproduction. JPAC provided
first estimates for the branching fraction of these states that serve as a benchmark for the experimental reach of the
CLAS12 detector~\cite{Blin2016}. In an updated analysis incorporating data from the GlueX experiment, the branching
fraction estimates and sensitivity of polarization observables accessible at Hall A of JLab are given~\cite{Winney}. 

\section{Multi-Prong Hadron Structure Theory for Exploration of Strong QCD Emergence}
\label{theory_out}

\subsection{Description of Hadron Spectrum and Structure Within LQCD}

\begin{figure}[htp]
\centering
\includegraphics[width=0.7\columnwidth]{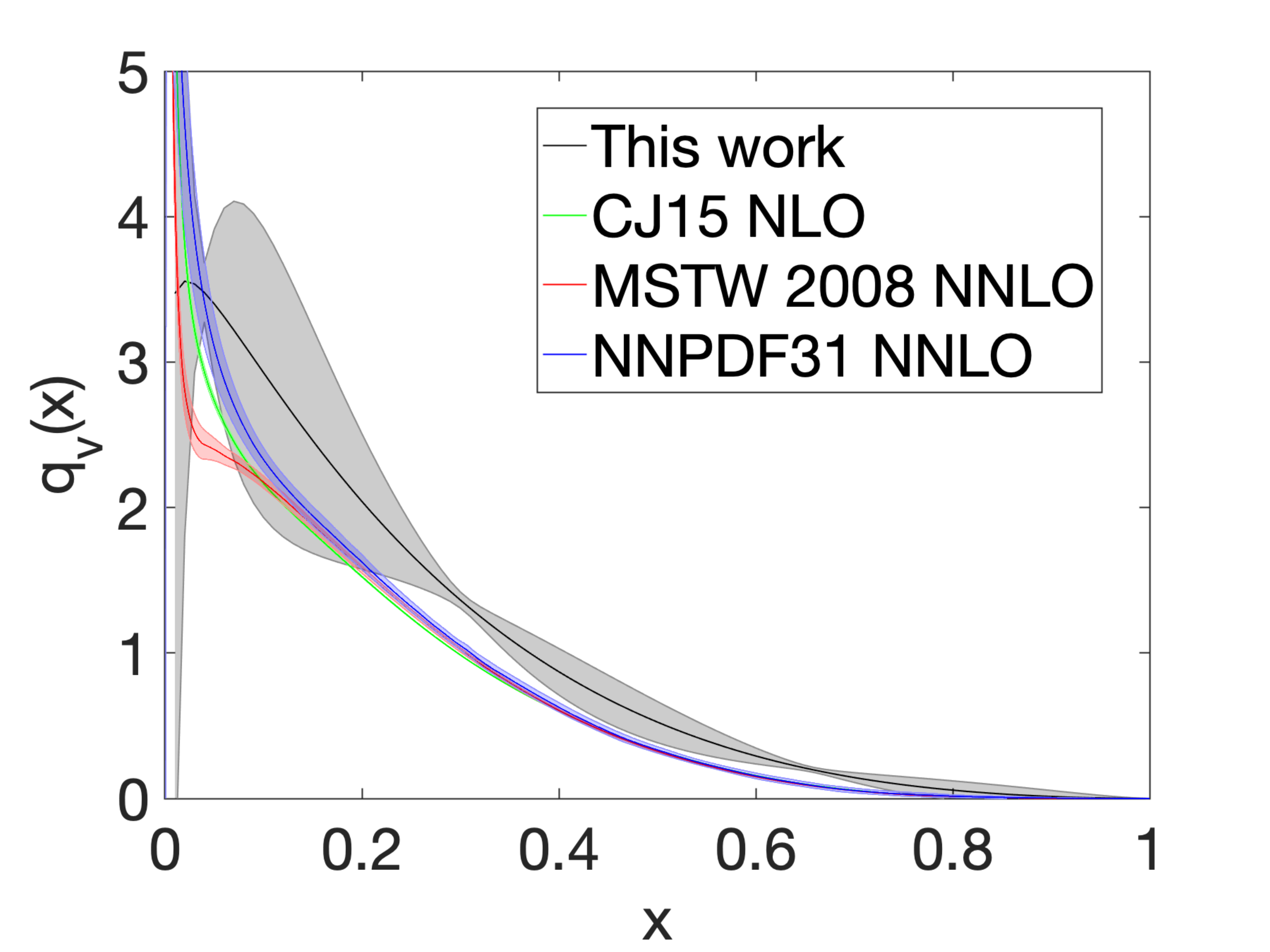}
\includegraphics[width=0.7\columnwidth]{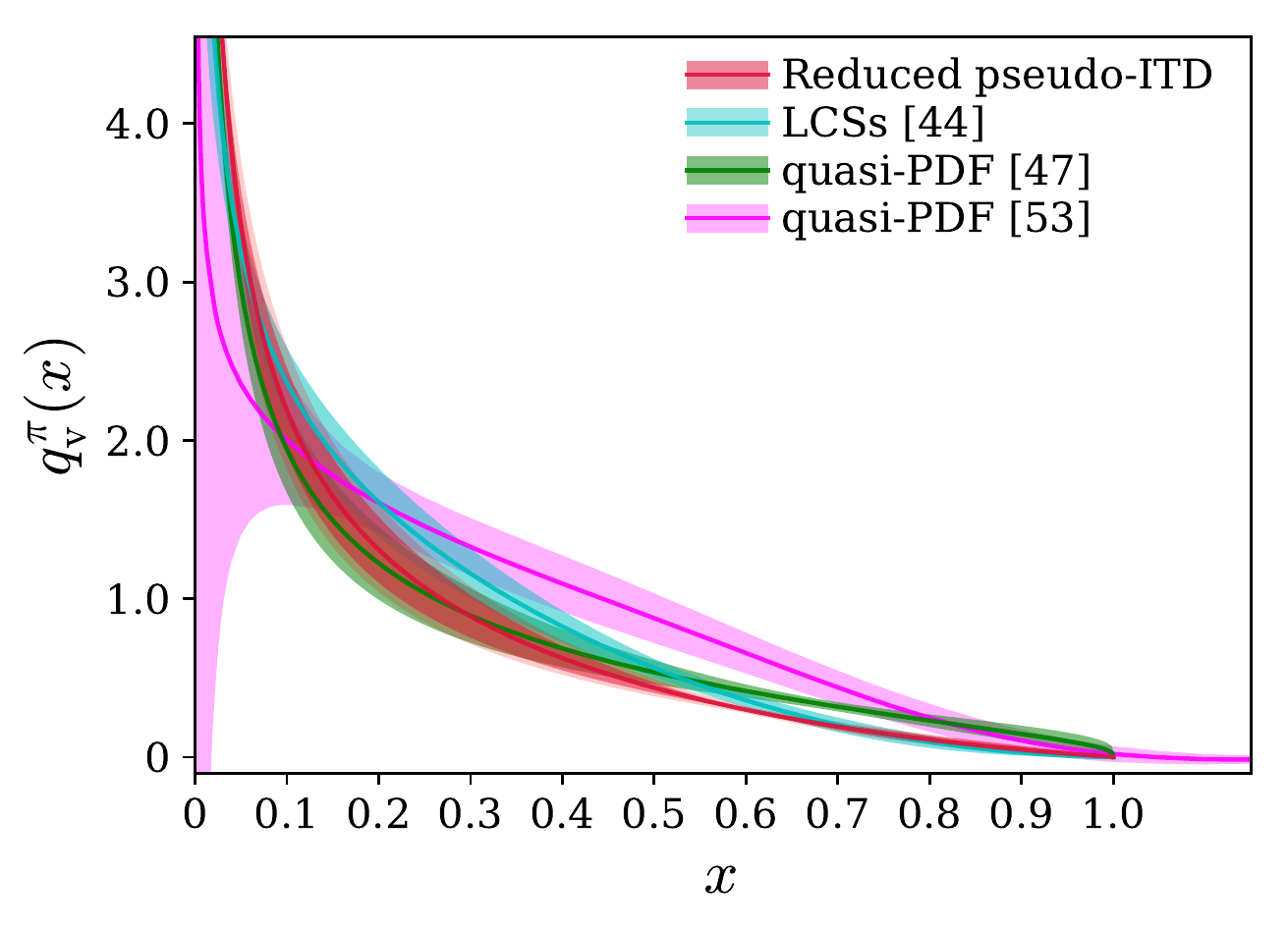}
\vspace{-0.25cm}
\caption{The top panel shows the unpolarized, isovector PDF of the nucleon using the pseudo-PDF approach~\cite{Joo:2019jct}, obtained on 
a lattice with $m_\pi \simeq 415$~MeV and spacing $a = 0.127$~fm. The bottom panel shows the valence-quark distribution of the pion using 
the pseudo-PDF approach~\cite{Joo:2019bzr}, compared with calculations using the quasi-PDF~\cite{Chen:2018fwa,Izubuchi:2019lyk} and 
current-current correlator approaches~\cite{Sufian:2019bol}.}
\label{fig5.1_1}
\end{figure}

The study of the structure of the nucleon and the pion has long been the subject of an intense lattice-QCD effort. Those studies
have focused on the different nucleon charges, the electromagnetic and axial-vector form factors; and on the Bjorken-$x$ moments
of parton distribution functions, and of the GPDs and TMDs that encode three-dimensional properties. However, the formulation of 
lattice QCD in Euclidean space
precluded the direct calculation of $x$-dependent PDFs, GPDs, and TMDs since these are matrix elements of operators separated
along the light-cone. The last few years have witnessed a series of advances that have circumvented this restriction
\cite{Ji:2013dva,Ji:2014gla,Ma:2014jla,Liu:2016djw,Orginos:2017kos,Chambers:2017dov,Radyushkin:2017cyf,Ma:2017pxb},
beginning with the Large Momentum Effective Theory (LaMET)~\cite{Ji:2013dva,Ji:2014gla}, through the introduction of
pseudo-PDFs~\cite{Radyushkin:2017cyf}, to the calculation of the matrix elements of gauge-invariant current-current correlators
\cite{Ma:2017pxb}. Each involves the computation of matrix elements of operators separated in space, with differences in the
renormalization prescription and kernel relating the lattice matrix elements to the target light-cone distributions. Since the
introduction of these new ideas, there has been an increasing body of calculations of the $x$-dependent parton distributions both 
of the nucleon and of the pion, exemplified in Fig.~\ref{fig5.1_1}.

\begin{figure}[htp]
\includegraphics[width=0.7\columnwidth]{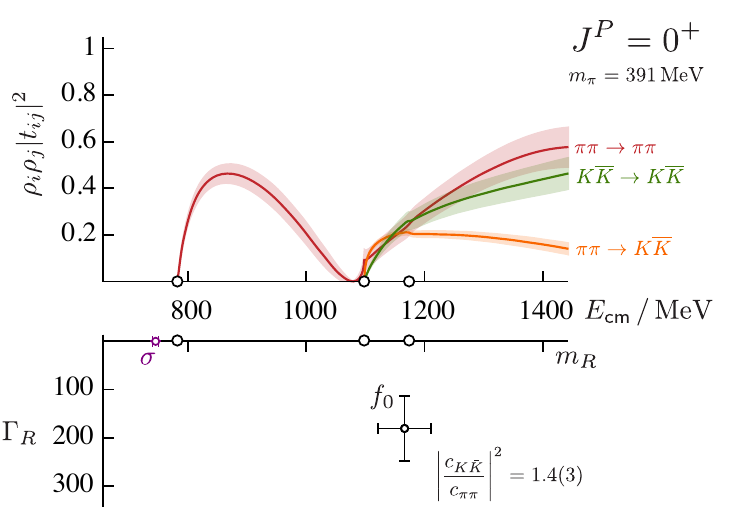}\vspace{3mm}
\includegraphics[width=0.7\columnwidth]{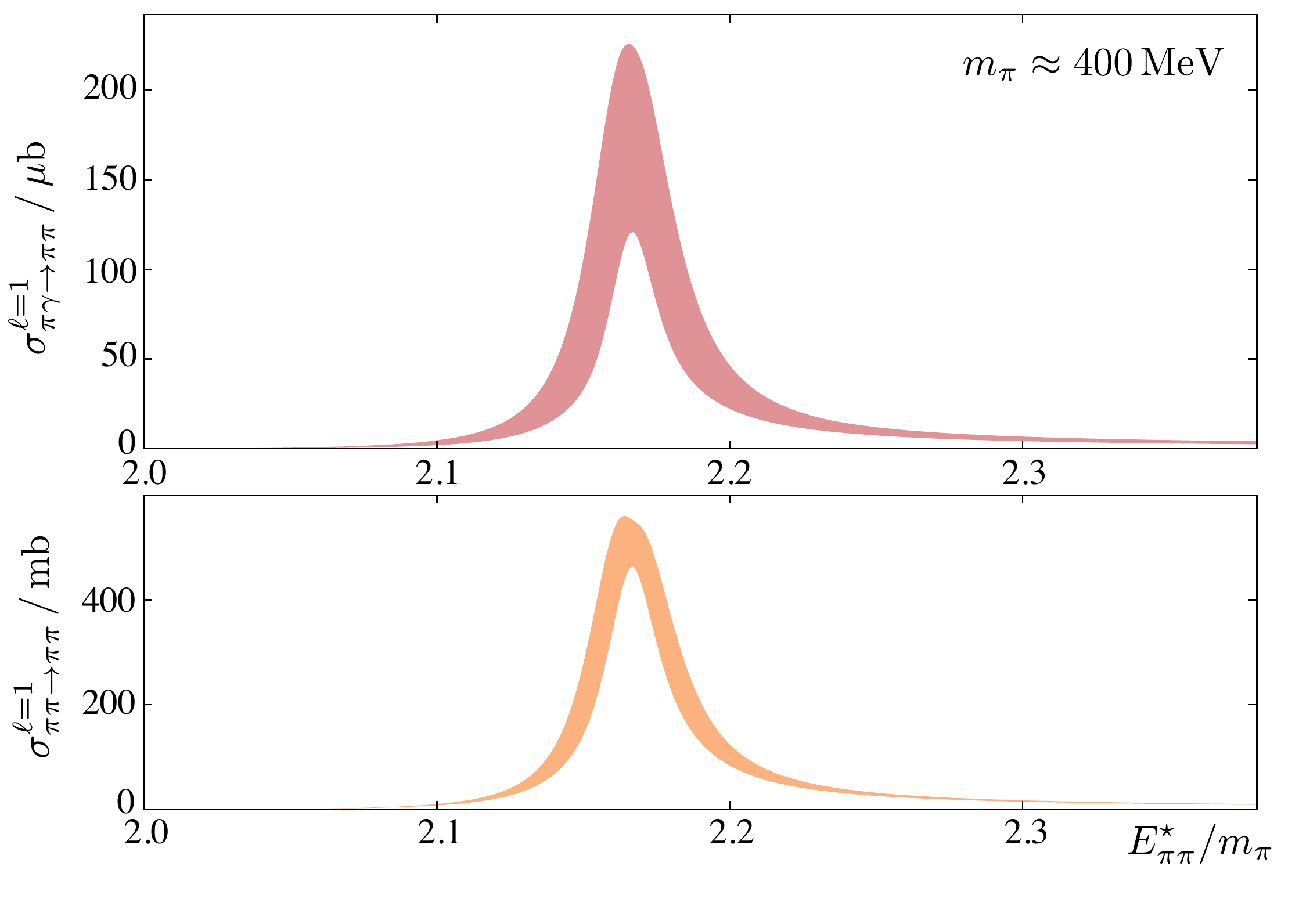}
\vspace{-0.25cm}
\caption{The top panel shows the coupled $\pi\pi$ $K\bar{K}$ amplitudes in the scalar sector, where the open circles denote
    the $\pi\pi$, $K\bar{K}$, and $\eta\eta$ thresholds~\cite{Briceno:2017qmb}. The bottom panel shows the
    $\pi^+ \gamma \to \pi^+ \pi^0$ cross section, together with the elastic $\pi\pi$ scattering cross sections showing the $\rho$
    resonance~\cite{Briceno:2016kkp}.}
\label{fig5.1_2}
\end{figure}

Our ability to explore the resonance spectrum of QCD has likewise been transformed through the application of the so-called
L\"{u}scher method~\cite{Luscher:1990ux} and its extensions, whereby infinite-volume scattering amplitudes can be related to
energy-shifts on a Euclidean lattice of finite spatial volume. The current state-of-the-art is to explore multi-channel and inelastic
scattering, for which the formulation has recently developed~\cite{Hansen:2012tf,Briceno:2012rv}. Together these calculations
are providing new insights into the resonance spectrum of QCD, most recently for the $\sigma, f_0$ and $f_2$ mesons, illustrated
in the top plot of Fig.~\ref{fig5.1_2}. With the formalism to explore the resonance spectrum of QCD established, the next
challenge is to calculate their structure and transitions through interactions with external currents, where the formalism has also
been developed~\cite{Briceno:2015tza}, and applied to the calculation of the $\gamma^* \pi \to \pi\pi$ transition, shown as the
bottom plot of Fig.~\ref{fig5.1_2}.

The advent of the exa-scale era in leadership-class computing provides an opportunity for lattice QCD to truly capitalize on both these
advances, and to perform {\it ab initio} calculations that can both extend experimental and QCD-inspired studies, and compute key
quantities that are either inaccessible to experiment, or at least highly model dependent. The success at understanding the resonance
spectrum for mesons will be extended to provide a first-principles understanding of the computationally more demanding baryons, where 
few calculations have been performed~\cite{Andersen:2017una}. Precision studies of the pion will be performed without the need to
extrapolate to the pion pole, and form factors of resonances such as the $\Delta$, inaccessible to experiment, will be computed can be
computed from first principles. Finally, the contribution of gluons to the structure of hadrons will be explored, in anticipation
of the future Electron-Ion Collider. Indeed, the potential of lattice QCD to advance of our understanding of hadrons over the next
five has been explored in detail in Refs.~\cite{Detmold:2019ghl} and~\cite{Joo:2019byq}.

\subsection{Continuum Strong QCD: Achievements and Prospects}
\label{cqcd_prosp}

The Faddeev equation was introduced almost sixty years ago~\cite{Faddeev:1960su}. It treats the quantum mechanical problem
of three-bodies interacting via pairwise potentials by reducing it to a sum of three terms, each of which describes a solvable
scattering problem in distinct two-body subsystems. An analogous approach to the three-valence-quark (baryon) bound-state
problem in QCD was explained thirty years ago~\cite{Cahill:1988dx,Reinhardt:1989rw,Efimov:1990uz}.
In this case, owing to EHM and the importance of symmetries~\cite{Binosi:2016rxz}, a Poincar\'e-covariant quantum field theory
generalization of the Faddeev equation is required. Like the meson Bethe-Salpeter equation, it is natural to consider analyses
using such a Faddeev equation within the class of CSMs.

\begin{figure}[htp]
\includegraphics[width=0.9\columnwidth]{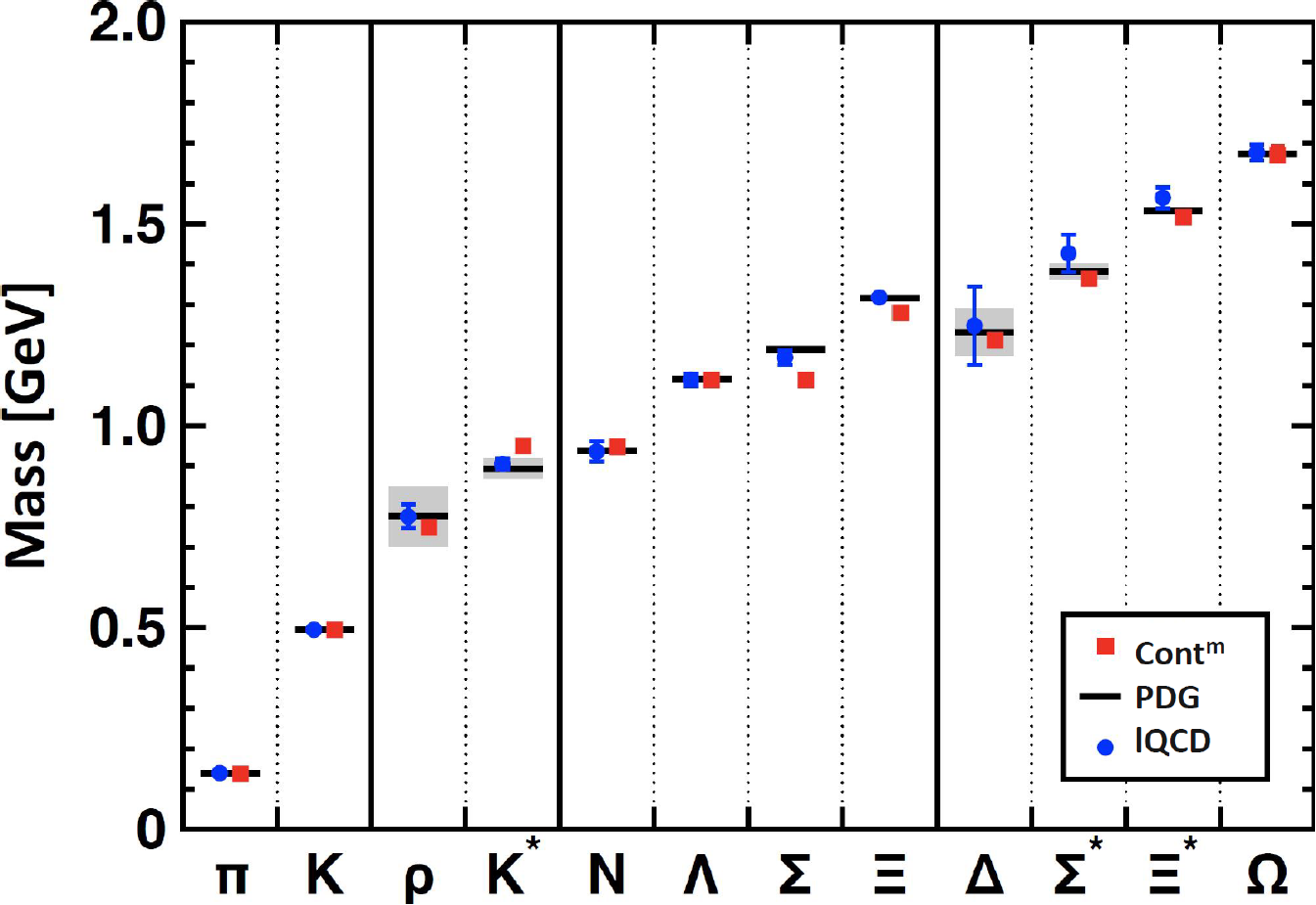}
\vspace{-0.25cm}
\caption{Masses of pseudoscalar and vector mesons, and ground-state positive-parity octet and decuplet baryons calculated using
  continuum (squares, red)~\cite{Qin:2019hgk} and lattice~\cite{Durr:2008zz} methods in QCD compared with experiment
  (PDG -- black bars, with decay-widths of unstable states shaded in gray)~\cite{Tanabashi:2018oca}. The
  continuum study did not include isospin symmetry breaking effects, which are evidently small, as highlighted by the empirically
  determined $\Sigma$-$\Lambda$ mass difference ($\sim 7\%$).}
\label{810MassComparison}
\end{figure}

The first direct treatment of the nucleon Faddeev equation is described in Ref.~\cite{Eichmann:2009qa}. Following that
approach, Refs.~\cite{Qin:2018dqp,Qin:2019hgk} calculated the spectrum of ground-state $J=1/2^+$, $3/2^+$
$(qq' q'')$-baryons, where $q, q', q'' \in \{u,d,s,c,b\}$, their first positive-parity excitations
and parity partners. Introducing two parameters, to compensate for deficiencies of the leading-order truncation when used for
excited hadrons~\cite{Qin:2011xq}, a description of the known spectrum of 39 such states was obtained, with a
mean-absolute-relative-difference between calculation and experiment of $3.6\,(2.7)$\%. This is exemplified in
Fig.~\ref{810MassComparison}. The framework was subsequently used to predict the masses of 90 states not yet seen
empirically.

The success of this treatment of the spectrum supports a view that the states are constituted of dynamically dressed quarks
bound by exchanges of gluons, which are themselves dressed. Being dressed, the gluons also have a momentum-dependent
mass-function as a consequence of EHM; and this function is large at infrared momenta, $m_0= 0.43(1)$~GeV~\cite{Cui:2019dwv}.
Given their role in the Faddeev equation, there is a sense in which the dressed-quarks, whose properties can be and are calculated
in QCD~\cite{Bhagwat:2003vw,Bowman:2005vx,Bhagwat:2006tu}, serve as Nature's embodiment of the constituent-quarks used
effectively in bringing order to hadron physics~\cite{GellMann:1964nj, Zweig:1981pd}. A key advantage of modern CSMs is their
manifestly Poincar\'e covariant formulation, essential to any treatment of systems involving light quarks, it ensures that wave
functions and currents can be used to calculate form factors to arbitrarily large $Q^2$.

\begin{figure}[htp]
\includegraphics[width=0.9\columnwidth]{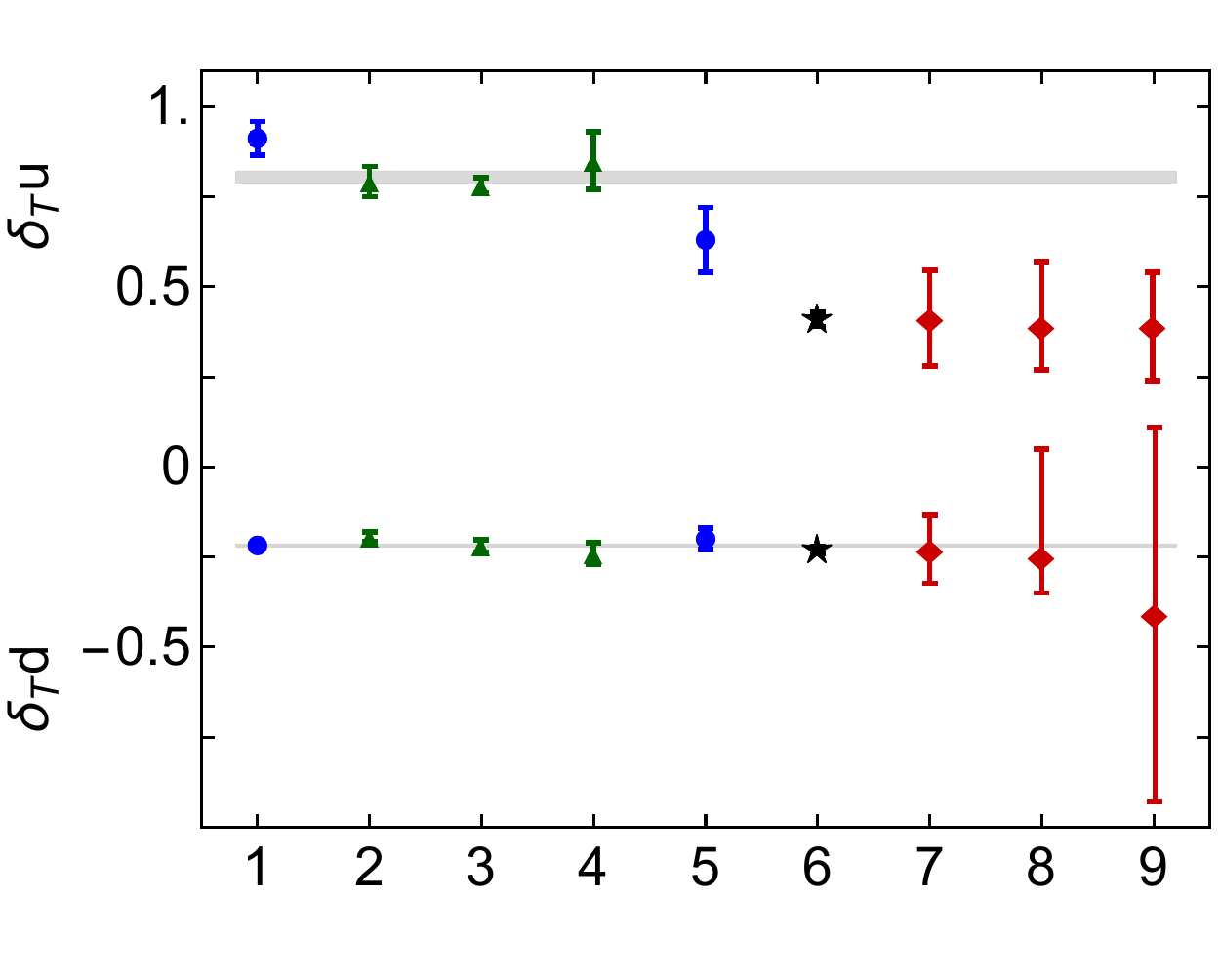}
\vspace{-0.35cm}
\caption{Comparison of the CSF prediction for the proton's tensor charges, position 1~\cite{Wang:2018kto}, with those obtained
  using: LQCD (2 -- \cite{Bhattacharya:2016zcn}, 3 -- \cite{Alexandrou:2017qyt}, 4 -- \cite{Yamanaka:2018uud}); and a
  contact-interaction Faddeev equation (5 -- \cite{Xu:2015kta}). The renormalization scale is $\zeta^2 = 4$~GeV$^2$ in all
  these cases; and the gray bands depict the averages in Eq.~\eqref{DSElQCD}. Position 6 -- projected errors achievable at
  JLab12 with the Solenoidal Large Intensity Device (SoLID)~\cite{Gao:2017ade}, using Eq.~\eqref{DefineTensorCharge} and
  anticipated transversity distribution data. The central values are chosen to match those estimated elsewhere
  \cite{Ye:2016prn} (7, $\zeta^2=2.4$~GeV$^2$) following an analysis of extant transversity distribution data. Earlier estimates
  from transversity distribution data are also depicted (8 -- \cite{Anselmino:2013vqa}, $\zeta^2=2.4$~GeV$^2$, and 9 --
  \cite{Radici:2015mwa}, $\zeta^2=1$~GeV$^2$.)}
\label{TCcompare}
\end{figure}

At the other extreme, the symmetry-preserving character of CSMs means that low-$Q^2$ observables can also be computed
and the associated truncation error quantified. An example is provided by the proton's tensor charges $\delta_T q$
\cite{Wang:2018kto}:
\begin{equation}
\label{DefineTensorCharge}
\delta_T q = \int_{-1}^1 dx\, h^q_{1T}(x) = \int_0^1 dx\, \left[ h_{1T}^q(x) - h_{1T}^{\bar q}(x)\right]\,,
\end{equation}
expressed here in terms of the quark transversity distributions, $h^q_{1T}(x)$. $\delta_T q$ measures the light-front
number-density of valence-$q$ quarks with transverse polarization parallel to that of the proton minus that of such quarks
with antiparallel polarization; namely, it measures any bias in quark transverse polarization induced by a polarization of the
parent proton. These charges are analogs of the nucleon flavor-separated axial-charges, which measure the difference
between the light-front number-density of quarks with helicity parallel to that of the proton and the density of quarks with
helicity antiparallel. In nonrelativistic systems, the helicity and transversity distributions are identical because boosts and
rotations commute with the Hamiltonian. Hence, tensor charges are a basic property of the nucleon and may be judged to
measure, {\it inter alia}, the importance of Poincar\'e-covariance in treatments of the nucleon bound state.

\begin{figure}[htp]
\includegraphics[width=0.9\columnwidth]{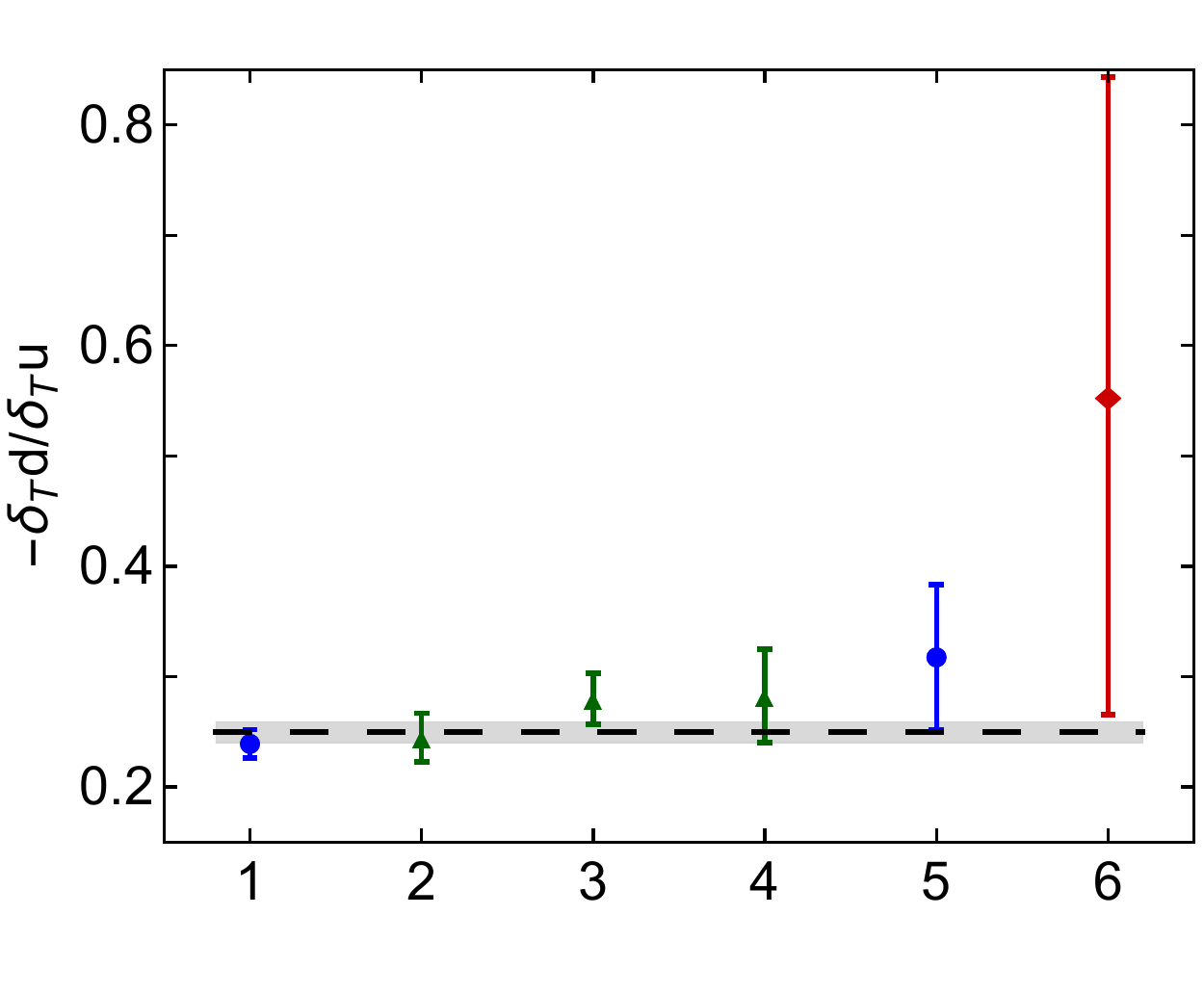}
\vspace{-0.35cm}
\caption{Ratio $(-\delta_T d/\delta_T u)$. Position 1: CSM result~\cite{Wang:2018kto}; LQCD: 2 --~\cite{Bhattacharya:2016zcn},
  3 -- \cite{Alexandrou:2017qyt} and 4 -- \cite{Yamanaka:2018uud}; contact-interaction Faddeev equation: 5 -- \cite{Xu:2015kta}.
  The gray band depicts the weighted average in Eq.~\eqref{DSElQCDratio}; and the dashed horizontal line is the quark model
  result $(-\delta_T d/\delta_T u)=1/4$~\cite{He:1994gz}. Position 6 -- estimate obtained using Eq.~\eqref{DefineTensorCharge}
  and extant transversity distribution data~\cite{Ye:2016prn}.}
\label{ratiocompare}
\end{figure}

Figure~\ref{TCcompare} compares the most recent CSF predictions for the proton's tensor charges with those obtained using
LQCD~\cite{Bhattacharya:2016zcn,Alexandrou:2017qyt,Yamanaka:2018uud} and an earlier contact-interaction Faddeev equation
study~\cite{Xu:2015kta}. A weighted combination of the CSF result and recent LQCD values
\cite{Bhattacharya:2016zcn,Alexandrou:2017qyt,Yamanaka:2018uud} yields the following estimates (gray bands in
Fig.~\ref{TCcompare}):
\begin{equation}
\label{DSElQCD}
\overline{\delta_T} u = 0.805(17)\,,\; \overline{\delta_T} d = - 0.216(4)\,.
\end{equation}
It is evident from the figure that CSF predictions are consistent with recent LQCD values; and produce results for $\delta_T u$
that differ markedly from those obtained via Eq.~\eqref{DefineTensorCharge} using extant transversity distribution data.

The mismatch between theory and phenomenology is also apparent in the scale independent 
ratio $(-\delta_T d/\delta_T u)$, Fig.~\ref{ratiocompare}. In this case, the weighted average of theoretical predictions is
\begin{equation}
\label{DSElQCDratio}
-\overline{\frac{\delta_T d}{\delta_T u}} = 0.25(1)\,,
\end{equation}
\noindent
illustrated by the gray band in the figure. Using a simple non-relativistic quark model spin-flavor wave function, this ratio is
$0.25$. It is practically the same in the MIT bag model~\cite{He:1994gz}; but, in both cases, the individual tensor charges
are measurably larger in magnitude than the modern theory predictions.

\begin{figure}[htp]
\includegraphics[width=0.95\columnwidth]{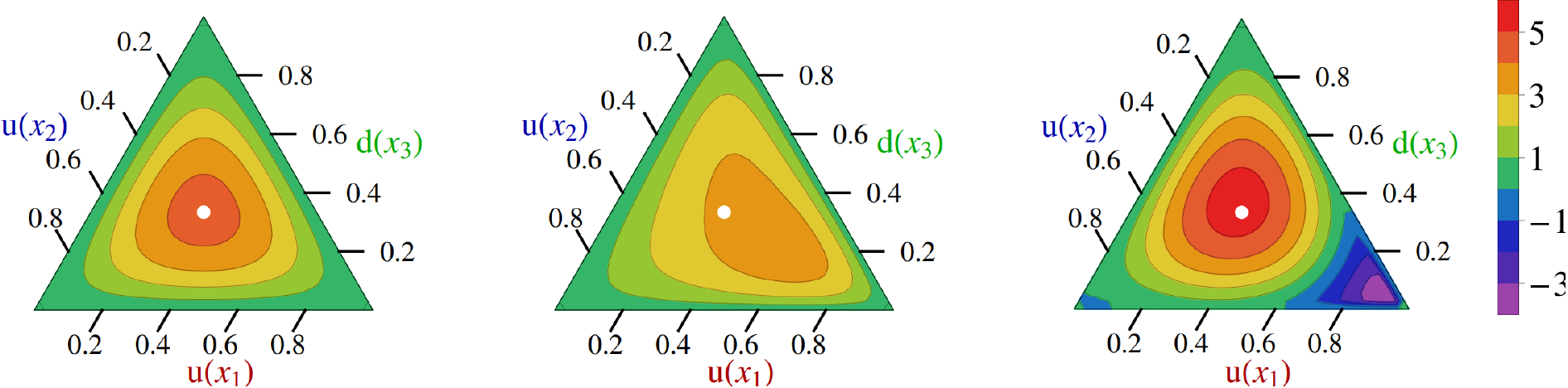}
\vspace{-0.25cm}
\caption{(Left) Barycentric plots: asymptotic profile, baryon PDA, $\varphi_N^{\rm asy}([x])=120 x_1 x_2 x_3$;
  (Middle) computed proton PDA evolved to $\zeta=2$~GeV, which peaks at $([x])=(0.55,0.23,0.22)$; and
  (Right) Roper resonance PDA. The white circle in each panel marks the center of mass for
  $\varphi_N^{\rm asy}([x])$, whose peak lies at $([x])=(1/3,1/3,1/3)$.}
\label{PlotPDAs}
\end{figure}

The agreement between CSM and LQCD predictions for the proton's tensor charges increases tension between theory and
phenomenology, {\it viz.} while there is agreement on $\delta_T d$, direct computations of the tensor-charge matrix
element produce a value of $\delta_T u$ that is approximately twice as large as that obtained via analyses of extant transversity
distribution data. Evidently, it is very important to complete a sound analysis of new data on the proton's transversity distribution.

While a complete treatment of the Poincar\'e-covariant Faddeev equation is now possible using modern hardware, it remains
a complex task. Hence, as remarked in Sec.~\ref{Sec37}, for the development of insights in a wide array of baryon problems,
it remains common to treat the equation in a quark+diquark approximation, where the diquark correlations are non-point-like and
dynamical~\cite{Segovia:2015ufa}.

One of the predictions of this approach is the $x\to 1$ behavior of the ratio of neutron and proton structure functions,
$F_2^n/F_2^p$~\cite{Roberts:2013mja}. This ratio is a clean probe of the proton's valence-quark structure, exposing crucial
features of its wave function. However, for roughly 30 years, theoretical predictions have produced answers over a very wide
range, anything between $1/4$ and $2/3$. Conflicting with common sense, the former may be interpreted as indicating that
there are NO valence $d$-quarks in the proton. In an important step toward a solution of puzzle, a new analysis of the world's
data on this ratio~\cite{Segarra:2019gbp}, accounting for the role of short-range correlations, produces a result in agreement
with the quark+diquark CSM picture, {\it viz.} $\lim_{x\to 1}F_2^n(x)/F_2^p(x) \approx 0.45$. This value is consistent with
axial-vector diquarks being a significant part of the nucleon wave function.

The surest path to a picture of the proton with a clean probability interpretation is via projection of the bound-state's Faddeev
amplitude onto the light-front. Following this route, the proton's leading-twist dressed-quark PDA was calculated
\cite{Mezrag:2017znp}. The result is depicted in Fig.~\ref{PlotPDAs}.

Table\,\ref{interpolation} lists the four lowest-order moments of the proton PDA. They reveal valuable insights, {\it e.g.}
when the proton is drawn as solely a quark+sca\-lar-diquark correlation, $\langle x_2 \rangle_u=\langle x_3 \rangle_d$, because
these are the two participants of the scalar quark+quark correlation; and the system is very skewed, with the PDA's peak being
shifted markedly in favor of $\langle x_1 \rangle_u > \langle x_2 \rangle_u$. This outcome conflicts with LQCD results
\cite{Braun:2014wpa, Bali:2015ykx}. On the other hand, realistic Faddeev equation calculations indicate that pseudovector
diquark correlations are an essential part of the proton's wave function. When these $uu$ and $ud$ correlations are
included, momentum is shared more evenly, shifting from the bystander $u(x_1)$ quark into $u(x_2)$, $d(x_3)$. Adding these
correlations with the known weighting, the PDA's peak moves back toward the center, locating at $([x])=(0.55,0.23,0.22)$; and
the computed values of the first moments align with those obtained using LQCD. This confluence delivers a more complete
understanding of the LQCD simulations, which are thereby seen to validate a picture of the proton as a bound-state with both
strong scalar {\it and} pseudovector diquark correlations, in which the scalar diquarks are responsible for $\approx 65$\%
of the Faddeev amplitude's canonical normalization.

\begin{table}[htp]
\caption{Computed values of the first four moments of the proton and Roper-resonance PDAs. The error on $f_N$, a
    dynamically determined quantity that measures the proton's ``wave function at the origin'', reflects a nucleon scalar diquark
    content of $65\pm 5$\%; and values in rows marked with ``$\not\supset \mbox{av}$'' were obtained assuming the baryon is
    constituted solely from a scalar diquark.
(All results listed at a renormalization scale $\zeta=2\,$GeV.)
\label{interpolation}
}

\begin{tabular*}
{\hsize}
{
l|@{\extracolsep{0ptplus1fil}}
c|@{\extracolsep{0ptplus1fil}}
c|@{\extracolsep{0ptplus1fil}}
c|@{\extracolsep{0ptplus1fil}}
c@{\extracolsep{0ptplus1fil}}}\hline
 & $10^3 f_N/\mbox{GeV}^2$ & $\langle x_1\rangle_u$ & $\langle x_2\rangle_u$ & $\langle x_3\rangle_d$ \\\hline
asymptotic PDA & & $0.333\phantom{(99)}$ & $0.333\phantom{(9)}$ & $0.333\phantom{(9)}$ \\\hline
LQCD \mbox{\cite{Braun:2014wpa}} & $2.84(33)$ & $0.372(7)\phantom{9}$ & 0.314(3) & 0.314(7) \\
LQCD \mbox{\cite{Bali:2015ykx}} & $3.60(6)\phantom{9}$ & $0.358(6)\phantom{9}$ & $0.319(4)$ & 0.323(6) \\\hline
CSM proton \mbox{\cite{Mezrag:2017znp}}& $3.78(14)$ & $0.379(4)\phantom{9}$ & 0.302(1) & 0.319(3) \\
CSM proton $\not\supset \mbox{av}$ & $2.97\phantom{(17)}$ & $0.412\phantom{(17)}$ & $0.295\phantom{(7)}$ & $0.293\phantom{(7)}$ \\\hline\hline
CSM Roper \mbox{\cite{Mezrag:2017znp}} & $5.17(32)$ & $0.245(13)$ & $0.363(6)$ & $0.392(6)$ \\
CSM Roper $\not\supset \mbox{av}$ & $2.63\phantom{(14)}$ & $0.010\phantom{(19)}$ & $0.490\phantom{(9)}$ & $0.500\phantom{(9)}$ \\\hline
\end{tabular*}
\end{table}

Like ground-state $S$-wave mesons~\cite{Chang:2013pq,Shi:2015esa,Braun:2015axa,Gao:2016jka,Zhang:2017bzy,Chen:2017gck},
the leading-twist PDA of the ground-state nucleon is both broa\-der than $\varphi_N^{\rm asy}([x])$ (defined by
Table~\ref{interpolation}, Row~1) and decreases monotonically away from its maximum in all directions, {\it i.e.} the PDAs of
these ground-state $S$-wave systems possess endpoint enhancements, but neither humps nor bumps.

As with meson elastic form factors, the veracity of this result for the proton PDA can be tested in future experiments. For
instance, it can be used to provide a realistic assessments of the scale at which exclusive experiments involving proton targets
may properly be compared with predictions based on pQCD hard scattering formulae~\cite{Rodriguez-Quintero:2018wma}.
Analogous to the meson case, the value of this mass-scale is an empirical manifestation of EHM, here within the
three-valence-quark proton bound-state.

Additionally, {\it e.g.} DVMP is sensitive to the wave functions of both the target hadron
and final-state meson. Consequently, the predictions being made for these objects using CSMs can be useful for both planning
experiments and understanding their outcomes; and progress toward these goals might be achieved by exploiting the predictions
via the PARTONs software~\cite{Berthou:2015oaw}.

As noted above, the quark+diquark Faddeev equation has also been used to elucidate properties of the Roper-resonance
\cite{Segovia:2015hra,Chen:2017pse,Chen:2018nsg}. Working with this input, Ref.~\cite{Mezrag:2017znp} delivered the
associated leading-twist PDA, depicted in the rightmost panel of Fig.~\ref{PlotPDAs} and whose first four moments are listed
in Table\,\ref{interpolation}. The prediction reveals some curious features, {\it e.g.}: the excitation's PDA is not positive
definite and there is a prominent locus of zeros in the lower-right corner of the barycentric plot, both of which echo aspects of
the wave function for the first radial excitation of a quantum mechanical system and have also been seen in the leading-twist PDAs
of radially excited mesons~\cite{Li:2016dzv,Li:2016mah}; and here the impact of pseudovector correlations is opposite to that
in the ground-state, {\it viz.} they shift momentum into $u(x_1)$ from $u(x_2)$, $d(x_3)$.

The character of EHM and the diquark correlations it induces within baryons are also visible in baryon elastic and transition
form factors. Particular examples of contemporary significance are neutron and proton elastic form factors, which provide
vital information about the structure and composition of the most basic elements of nuclear physics. They are a measurable and
physical manifestation of the nature of the nucleons' constituents and the dynamics that binds them together; and new, accurate
form factor data are driving paradigmatic shifts in our understanding of these things.

This is particularly clear in analyses of experimental data acquired in the past twenty years, which have imposed a new ideal.
Namely, despite its simple valence-quark content, the internal structure of the nucleon is very complex, with marked differences
between the distributions of total charge and magnetization~\cite{Jones:1999rz,Gayou:2001qd,Punjabi:2005wq,Puckett:2010ac,Puckett:2011xg} 
and also between the distributions carried by the different quark flavors~\cite{Cates:2011pz,Segovia:2014aza,Segovia:2016zyc}.

Data available before the year 1999 led to a view that
\begin{equation}
\label{GEeqGM}
\left. \mu_p \frac{G_E^p(Q^2)}{G_M^p(Q^2)} \right|_{\rm Rosenbluth} \approx 1,
\end{equation}
$G_M^p(Q^2=0)/\mu_p=1$, where $G_{E,M}^p$ are the proton's electric and magnetic form factors; and hence a conclusion that
the distributions of charge and magnetization within the proton are approximately identical~\cite{Perdrisat:2006hj}.
Significantly, this outcome is consistent with the simple pictures of the proton's internal structure popular at the time, in which,
{\it e.g.} correlations between quarks and attendant orbital angular momentum play little role.

The situation changed dramatically when the combination of high-energy, -current and -polarization at JLab enabled a new type
of experiment to be performed, {\it viz.} polarization-transfer reactions~\cite{Jones:1999rz}, which are directly proportional
to $G_E(Q^2)/G_M(Q^2)$~\cite{Akhiezer:1974em,Arnold:1980zj}. A series of these experiments
\cite{Jones:1999rz,Gayou:2001qd,Punjabi:2005wq,Puckett:2010ac,Puckett:2011xg} has determined that
\begin{equation}
\label{GEGMJLab}
\left. \mu_p \frac{G_E^p(Q^2)}{G_M^p(Q^2)} \right|_{\rm JLab\,PT} \approx 1 - {\rm constant} \times Q^2,
\end{equation}
where the constant is such that the ratio might become negative for $Q^2 = 9.8 m_N^2 = 8.7\,$GeV$^2$. This behavior
contrasts starkly with Eq.~\eqref{GEeqGM}; and since the proton's magnetic form factor is reliably known on a space-like
domain that extends to $Q^2 \approx 30$~GeV$^2$~\cite{Kelly:2004hm,Bradford:2006yz}, the $Q^2$-dependence of
$G_E^p/G_M^p$ exposes novel features of the proton's charge distribution, as expressed in $G_E^p(Q^2)$.

Numerous analyses have sought to explain the behavior of $G_E^p(Q^2)/G_M^p(Q^2)$; and insights deriving from CSMs are
described in Refs.~\cite{Wilson:2011aa, Segovia:2014aza}.

More recently, a new, indirect approach to the problem has been explored~\cite{Xu:2019ilh}. Consider that the electric form
factor of a positively charged vector meson decreases with increasing $x=Q^2/m_{\mathpzc V}^2$, where $m_{\mathpzc V}$ is
the vector meson's mass. However, setting it apart from that of a pseudoscalar meson, which is positive-definite, the large-$x$
prediction from Refs.\,\cite{Brodsky:1992px,Haberzettl:2019qpa} suggests that $G_E^{\mathpzc V}(x)$ may possess a zero at
$x \sim 6$. This was the outcome in Ref.~\cite{Roberts:2011wy}, which used a symmetry preserving regularization of a contact
interaction and was thus able to compute form factors to arbitrarily large $x$.

In exhibiting a zero crossing, the vector mesons electric form factor, $G_E^{\mathpzc V}$, can serve as a surrogate for the
proton's electric form factor because the reason for the potential appearance of a zero is similar in both cases. For the proton,
a zero can be produced by destructive interference between the Dirac and Pauli form factors, and will appear if the transition
between the strong and perturbative domains of QCD is pushed to a sufficiently large value of $Q^2$
\cite{Wilson:2011aa,Segovia:2014aza}. In the vector meson case, there are three elastic form factors: $F_1^{\mathpzc V}$ is
Dirac-like; $F_2^{\mathpzc V}$, Pauli; and $F_3^{\mathpzc V}$ is quadrupole-like. Here, $F_{1,3}^{\mathpzc V}$ are positive and
$F_2^{\mathpzc V}$ is negative; and if the magnetic form factor, $F_2^{\mathpzc V}$, is removed, then the vector meson's electric
form factor is positive-definite at space-like momenta.

The merit of using vector meson studies to locate and explain a zero in the electric form factor of a $J\neq 0$ hadron is the
relative simplicity of the two-body continuum bound-state problem as compared to the analogous three-body problem.
Ref.~\cite{Xu:2019ilh} calculated the electric charge form factors of $u\bar d$ vector mesons, varying the current-quark
masses of the valence quarks: realistic, $s$-quark, $c$-quark. A zero is found in each case; and importantly, as the
current-mass of the system's valence-quarks is increased, the $x=Q^2/m_{\mathpzc V}^2$-location of the zero, $x_{\mathpzc z}$,
moves toward $x=0$:
\begin{equation}
\label{GEzero}
\begin{array}{l|c|c|c}
{\mathpzc V} & \rho & \rho_s & \rho_c\\\hline
\rule{0em}{3ex} x_{\mathpzc z} & 10.6(3) & 10.1^{(9)}_{(7)}& 4.5^{(2.5)}_{(1.0)}
\end{array}
\end{equation}
The shift is initially slow; but the pace increases as one leaves the domain upon which emergent mass is dominant and enters
into that for which explicit (Higgs-connected) mass generation overwhelms effects deriving from strong-QCD dynamics.
Reverting to $Q^2$, the location of the zero in $G_E^{\mathpzc V}$ moves to larger values with increasing current-quark mass.

Focusing on the $\rho$-meson case, because the $\rho$ is made from the same valence-quarks as the proton, if one replaces
$m_{\mathpzc V}^2$ by $m_N^2$,
\begin{equation}
x = 10.6(3) \to Q^2 = 9.4(3)\,{\rm GeV}^2.
\end{equation}
Using a quark+diquark Faddeev equation to describe the proton, Ref.~\cite{Segovia:2014aza} predicted that $G_E^p$ possesses a zero at 
$Q^2 = 9.5$~GeV$^2$. No symmetry protects $G_E^p$. It can be negative and very likely exhibits a zero in the neighborhood of this point 
because dressed-quarks have large anomalous magnetic moments that run slowly to zero~\cite{Singh:1985sg,Bicudo:1998qb,Chang:2010hb}. Once 
again, the existence of a zero is a consequence of EHM, driven by strong-QCD dynamics.

The existence of a zero in vector meson form factors has another important corollary; namely, single-pole vector meson dominance (VMD), 
{\it viz.} $G_E^{\mathpzc V}(x) \approx 1/(1+x)$, can only be a useful tool for approximating (off-shell) vector meson properties within a 
limited $x$-domain. The vector-meson electric form factor presents the best case for a VMD model because it necessarily agrees with the 
computed result in some neighborhood of $x=-1$ and, by charge conservation, also in the vicinity of $x=0$. The analysis in 
Ref.~\cite{Xu:2019ilh} reveals that the discrepancy is less-than 20\% within the following regions:
\begin{equation}
\label{VMDregions}
\begin{array}{lccc}
\rho_{\phantom{s}}\,: &-1 < x < 0.81\,, & & \\
\rho_s\,: &-1 < x < 0.60\,, & & \\
\rho_c\,: &\; -1 < x < -0.96 & \mbox{\&} & -0.15 < x < 0.24\,,
\end{array}
\end{equation}
where the subscript indicates the size of the current-mass for the valence-quarks defining the bound state. Evidently, a
single-pole VMD approximation is a fair assumption on a reasonable domain for light-quark systems. However, it is poor for
states in which the Higgs-mechanism of mass generation is dominant, {\it i.e.} $c \bar c $ and more massive systems.
In fact, without the $x=0$ constraint imposed by current conservation, a VMD approximation for the $c \bar c $ system
becomes quantitatively unreliable once bound-state virtuality exceeds 4\%. Hence, as highlighted by the analysis in
Ref.~\cite{Wu:2019adv}, it is likely that VMD estimates of heavy-quarkonia photo- and electroproduction cross sections are
both quantitatively and qualitatively unsound. This question is currently the focus of detailed analysis because, {\it e.g.}
it bears upon flagship experiments aimed at exposing EHM~\cite{PhysRevLett.123.072001,Wang:2019mza}.

\begin{figure}[htp]
\includegraphics[width=0.7\columnwidth]{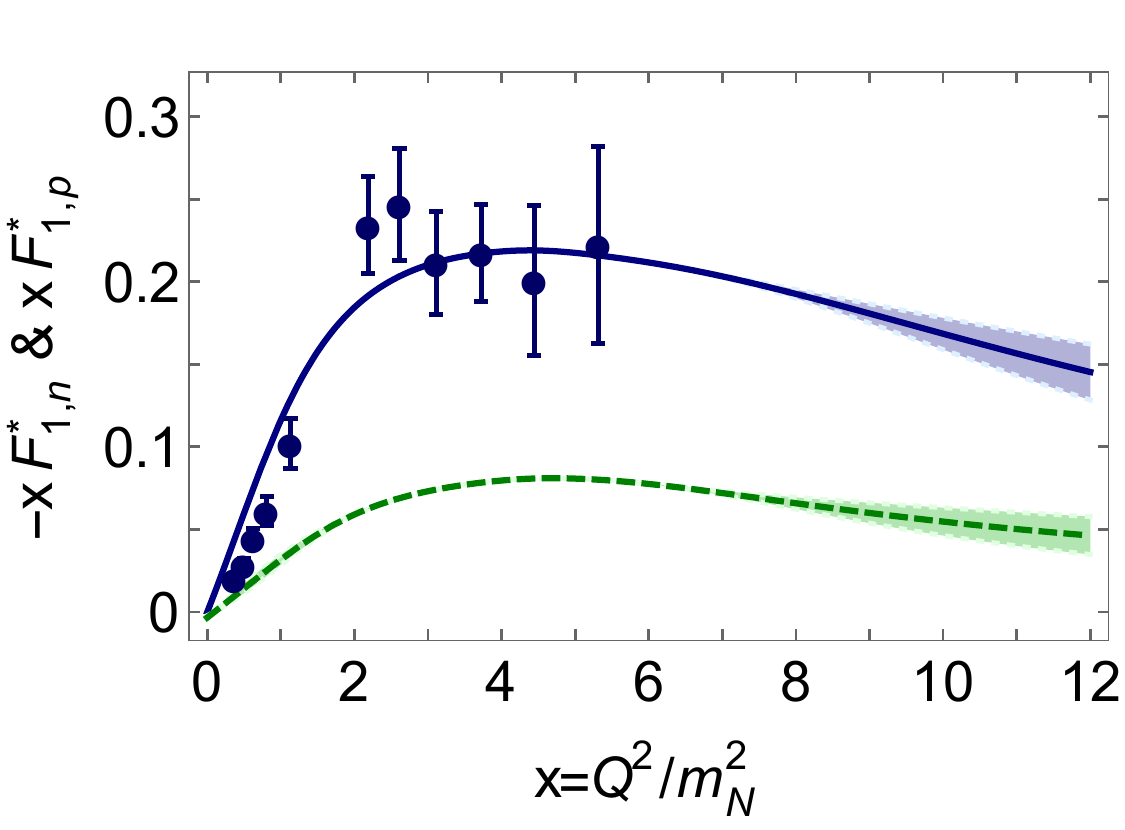}
\includegraphics[width=0.7\columnwidth]{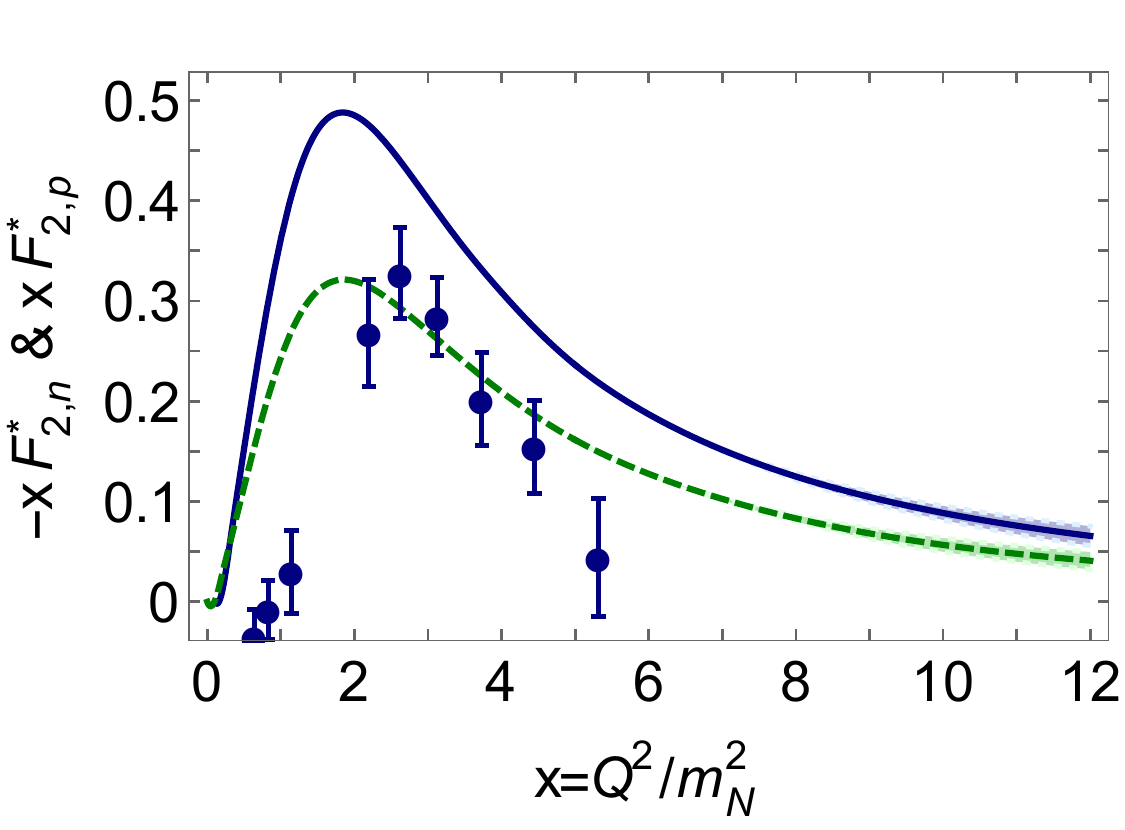}
\vspace{-0.25cm}
\caption{$x$-weighted Dirac (top) and Pauli (bottom) transition form factors for the reactions
  $\gamma^* p \to R^+$ (solid blue curves) and $\gamma^* n\to R^0$ (dashed green curves) reported in
  Ref.~\cite{Chen:2018nsg}. In all cases, the results on $x\in [6,12]$ are projections, obtained via extrapolation of analytic
  approximations to results on $x\in [0,6]$: at each $x$, the width of the band associated with a given curve indicates the
  confidence-level for the extrapolated value. (See Ref.~\cite{Chen:2018nsg} for details.) Data in both panels are for the
  charged-channel transitions, $F_{1,p}^*$ and $F_{2,p}^*$: circles (blue)~\cite{Aznauryan:2009mx}. No data currently
  exist for the neutral channel, but could be obtained using a deuteron to provide the neutron target. In this case, there are
  indications that the quality of the cross section data should be comparable to that for charged-Roper production off the free
  proton~\cite{Tian:2018siq}.}
\label{figLargeQ2}
\end{figure}

Key to completing the analysis in Ref.~\cite{Xu:2019ilh} is the recent development of new numerical techniques for extending
the domain of $Q^2$ upon which CSMs can provide reliable form factor predictions. The approach, which adds a powerful
statistical aspect to the Shlessinger point method (SPM)~\cite{Schlessinger:1966zz,PhysRev.167.1411,Tripolt:2016cya,Binosi:2019ecz}, 
was first tested in Ref.~\cite{Chen:2018nsg}. In anticipation of new data from the CLAS12 detector at JLab that will reach to 
unprecedented photon virtuality, Ref.~\cite{Chen:2018nsg} employed a standard quark+diquark approximation to the three valence-quark 
bound-state problem to compute $\gamma^* p \to R^+$ and $\gamma^* n \to R^0$ transition form factors on $Q^2/m_N^2 \in [0,12]$, 
greatly extending the reach of existing theory predictions~\cite{Burkert:2019bhp}. The results are illustrated by the curves in
Fig.~\ref{figLargeQ2}. The precision of the projections in Fig.~\ref{figLargeQ2} can be exemplified by quoting the form factor 
values at the upper bound of the extrapolation domain, $x_{12}=12$:
\begin{subequations}
\begin{align}
 F^*_{1,p} & = 0.0121(14), \; & -F^*_{1,n} & = 0.0039(10),\\
x_{12}F^*_{1,p} & = 0.145(17), \; & -x_{12}F^*_{1,n} &= 0.046(11),
\end{align}
\end{subequations}
\vspace*{-7ex}

\begin{subequations}
\begin{align}
F^*_{2,p} & = 0.0055(8), \; & -F^*_{2,n} &= 0.0034(7),\\
x_{12}F^*_{2,p} & = 0.066(10), \; & -x_{12}F^*_{2,n} &= 0.041(9).
\end{align}
\end{subequations}

The $x$-weighted form factors were drawn in Fig.~\ref{figLargeQ2} so as to accentuate, but not over-magnify, the larger-$x$
behavior of the form factors. On the domain depicted, there is no indication of the scaling behavior expected of the transition
form factors: $F^*_1 \sim 1/x^2$, $F^*_2 \sim 1/x^3$. Since each dressed-quark in the baryons must roughly share the
impulse momentum, $Q$, one should expect that such behavior will only become evident on $x\gtrsim 20$~\cite{Maris:1998hc}.

Similar techniques are currently being employed to update the predictions in Ref.~\cite{Segovia:2014aza}, reanalyzing the
brute-force results described therein and delivering predictions for proton and neutron elastic form factors on $x\lesssim 15$.

\subsection{Further Developments of Quark Models for the Description of the Hadron Spectrum, Structure, and GPDs}

\textit{Why do we use mean-field approaches to describe a baryon?} In 1979, E. Witten proposed an ingenious approach to describe
a baryon based on large $N_c$ QCD in which a baryon can be viewed as $N_c$ valence quarks bound by the meson mean fields, since
the mass of the baryon is proportional to $N_c$ whereas the meson-loop contributions are suppressed by $1/N_c$
\cite{Witten:1979kh,Witten:1983tx}. He showed explicitly that in two-dimensional QCD the baryon arises as a bound state of $N_c$
valence quarks by the meson mean fields. It resembles a well-known Hartree approximation. The presence of the $N_c$ valence
quarks polarize the Dirac sea or the QCD vacuum, which in turn affects self-consistently the $N_c$ valence quarks. The chiral
quark-soliton model ($\chi$QSM) was constructed such that this self-consistent Hartree picture is realized
\cite{Diakonov:1987ty, Christov:1995vm, Diakonov:1997sj}. As a result, the baryon emerges as a soliton that consists of the $N_c$
valence quarks. We can summarize the features of the $\chi$QSM.
\begin{itemize}
\item A meson mean-field theory
\item A model based on relativistic quantum field theory 
\item A model that incorporates chiral symmetry and its spontaneous breakdown 
\item A very constraint model with all parameters fixed in the mesonic sector
\item It is related to strong or non-perturbative QCD via the instanton vacuum 
\end{itemize}

The $\chi$QSM starts from the effective chiral action (E$\chi$A), which contains the dynamical quarks and the pseudo-Goldstone
boson fields: 
\begin{align}
S_{\mathrm{eff}} = -N_c\mathrm{Tr}\log\left[ i\rlap{/}{\partial} + i
  \sqrt{M(i\partial)} U^{\gamma_5}\sqrt{M(i\partial)} + i \hat{m}\right], 
\label{eq:1} 
\end{align}
where $M$ denotes the dynamical quark mass and $U^{\gamma_5}$ the pseudo-Goldstone boson field. $\hat{m}$ represents the mass
matrix of the current quarks in flavor space $\hat{m}=\mathrm{diag}(m_{\mathrm{u}},\,m_{\mathrm{d}}, m_{\mathrm{s}})$. In fact, 
chiral symmetry and its spontaneous breakdown determines the explicit form of the E$\chi$A given in Eq.~\eqref{eq:1}. A specific
quark-gluon dynamics is encoded in the dynamical quark mass. However, local models such as the NJL model and the chiral quark model
do not provide any information on the internal quark-gluon dynamics. On the other hand, if one considers the QCD instanton vacuum,
which was investigated by D. Diakonov and V. Petrov~\cite{Diakonov:1985eg}, the dynamical quark mass arises as a momentum-dependent
one. Moreover, the instanton vacuum determines not only the explicit form of the momentum-dependent quark mass but also its value at
the zero virtuality by the saddle-point approximation that is justified in the large $N_c$ limit. In the $\chi$QSM constructed from the 
instanton vacuum, the renormalization point inheres naturally~\cite{Diakonov:1996sr}. This is indeed very important, when one compares
the results from the $\chi$QSM. Since the E$\chi$A~\eqref{eq:1} contains all orders of the effective chiral Lagrangian, the model is
distinguished from the Skyrme model, which was constructed by the truncated effective chiral Lagrangian. Moreover, if one turns off
the self-consistent interaction between the mean fields and valence quarks, the model produces the results of the well-known
nonrelativistic quark model (NRQM). Thus, the model interpolates between the Skyrme model (large soliton size) and the NRQM (zero
soliton size)~\cite{Christov:1995vm}.

Since the momentum-dependence of the quark mass causes additional complexity, the quark mass is often taken to be constant, which
brings about a regularization to tame divergences arising from quark loops. The $\chi$QSM has been successfully applied to the
description of properties of the lowest-lying light baryons over decades~\cite{Christov:1995vm}. Recently, the model has been utilized
to explain singly heavy baryons~\cite{Yang:2016qdz,Diakonov:2010tf}. Taking the limit of the infinitely heavy mass of the heavy quark
($m_Q\to\infty$), one can view a singly heavy baryon as $N_c-1$ light valence quarks bound by the pion mean field. In this limit, the
spin of the heavy quark is conserved, which makes the spin of the light degrees of freedom
\cite{Isgur:1989vq,Isgur:1991wq, Georgi:1990um, Manohar:2000dt}. By exactly the same self-consistent procedure, a singly heavy
baryon comes about again as a chiral soliton but a bosonic and colored one~\cite{Yang:2016qdz,Kim:2018xlc,Kim:2018cxv,Kim:2019rcx}.
Combining it with the heavy quark that is considered to be a mere static color source, we can construct a singly heavy baryon. Thus, 
the $\chi$QM has a virtue of explaining the light and singly heavy baryons on an equal footing. 

\textit{What have we studied for the structures of the baryons} Within the framework of the $\chi$QSM, various observables of the 
baryons have been examined: we list some of them as follows:
\begin{itemize}
\item Mass spectra of the SU(3) baryons~\cite{Blotz:1992pw,Yang:2010fm}
\item Electromagnetic form factors of the baryon octet and decuplet~\cite{Kim:1995mr,Silva:1999nz,Ledwig:2008es,Kim:2019gka}
\item Transition magnetic moments of the SU(3) baryons~\cite{Kim:2005gz}
\item Axial-vector form factors of the baryon octet and decuplet~\cite{Silva:2005fa,Ledwig:2008rw}
\item Parity violation and Strange vector form factors of the nucleon~\cite{Silva:2001st,Silva:2005qm,Silva:2013laa} 
\item Tensor charges and tensor form factors of the nucleon~\cite{Kim:1995bq,Kim:1996vk,Ledwig:2010tu,Ledwig:2010zq,Ledwig:2011qw} 
\item Gravitational form factors of the nucleon~\cite{Goeke:2007fp}
\item Semileptonic hyperon decays~\cite{Kim:1999uf,Ledwig:2008ku,Yang:2015era} 
\item Unpolarized quark distributions (twist-2) of the nucleon~\cite{Diakonov:1996sr,Diakonov:1997vc} 
\item Transversity distributions of the nucleon~\cite{Efremov:2004qs}
\item Antiquark distributions and unpolarized flavor asymmetry~\cite{Efremov:2000za} 
\item Quasi-parton distributions of the nucleon~\cite{Son:2019ghf} 
\item Electromagnetic form factors of singly heavy baryons~\cite{Kim:2018nqf,Kim:2019wbg} 
\item Excited $\Omega_c$s as exotic baryons~\cite{Kim:2017jpx, Kim:2017khv}
\item Generalized parton distributions~\cite{Goeke:2001tz} 
\end{itemize}

\textit{Possible extensions of the $\chi$QSM} The present version of the $\chi$QSM is mainly applicable to the lowest-lying baryons.
In order to explain the mass spectra and properties of both the excited light and heavy baryons, we have to extend the mean-field
approximation. To do that, we need to take into account at least four important physics:
\begin{itemize}
\item Inclusion of the vector, axial-vector, and tensor mean fields~\cite{Diakonov:2013qta} 
\item Confining background fields
\item $1/N_c$ meson-loop fluctuations
\item Momentum-dependent dynamical quark mass~\cite{Golli:1998rf}
\end{itemize}
So far, the $\chi$QSM contains only the pion mean field, which is represented by the single pion profile function. Once we introduce the
vector, axial-vector, and tensor structures, we have to determine the twelve different profile functions~\cite{Diakonov:2013qta}, which
is technically not at all an easy task. Nevertheless, they should be included, since some of excited baryons can decay into vector mesons
and lower baryon states. In general, a valence quark or quarks in a excited baryon should be excited to the next level. It implies that
certain confining effects should be taken into account to keep the valence quarks stay in the excited baryon. Unfortunately, the
confining background field cannot be derived or determined self-consistently. Moreover, it should be saturated at a certain scale such
that the confining string is broken or a meson is created. Unfortunately, there is no consistent and rigorous theory on the quark
confinement. We can only resort to phenomenological methods to incorporate it. 

When a valence quark is excited in the baryon, it is natural that there must be a quark-antiquark correlation, which corresponds to the 
$1/N_c$ meson-loop fluctuations. Considering these $1/N_c$ meson loops means that we have to go beyond the mean-field approximation.
It bears resemblance to the RPA approximation in nuclear physics but it requires more involved investigation. So far, we have turned off
the momentum dependence of the dynamical quark mass. However, the momentum-dependent quark mass embraces quark-gluon dynamics
arising from the instanton vacuum. In particular, when it comes to the quark distribution amplitudes and generalized parton distributions,
it is essential to take into account the momentum-dependent quark mass. 

\textit{Future investigation on the baryon properties} Once the $\chi$QSM is successfully extended, one can perform the
investigations on the structures of excited light and heavy baryons. We can present some examples as follows:
\begin{itemize}
\item Mass spectra of excited light and heavy baryons
\item Strong decays of the excited baryons
\item Various form factors of the excited baryons
\item Decays of heavy baryons to light baryons
\item Light-cone distribution amplitudes of the baryons and form factors at higher $Q^2$ regions and in the time-like region
\end{itemize}
In particular, the transitions from the heavy baryons to the light baryons require additional theoretical efforts. Since the heavy and
light baryons have different meson mean fields, one has to scrutinize the transitions between different meson mean fields. 

\textit{Conclusion} The chiral quark-soliton model is a robust and viable model for the description of the structure of baryons. It has
the great virtue that explains the light and heavy baryons on an equal footing. Moreover, since it is a model based on relativistic quantum
field theory, it can describe not only the static properties and form factors of the baryons but also the internal quark structure of the
baryons such as the quark distributions, generalized parton distributions, light-cone distribution amplitudes, and so on. Since the
renormalization point of the model is well defined from the instanton vacuum, results from the model can be directly compared
with both the experimental data and those from lattice QCD data. 

\subsection{Light-Front Holography and Supersymmetric Conformal Algebra: A Novel Approach to Hadron Spectroscopy, Structure,
  and Dynamics}

Recent insights into the non-perturbative structure of QCD based on the gauge/gravity correspondence and light-front (LF) quantization,
{\it light-front holography} for short~\cite{deTeramond:2008ht}, have lead to effective semiclassical bound state equations for mesons
and baryons where the confinement potential is determined by an underlying superconformal algebraic structure
\cite{Brodsky:2013ar, deTeramond:2014asa, Dosch:2015nwa}. The formalism provides a remarkably first approximation to QCD, including its
hidden supersymmetric hadronic features. The resulting light-front wave equation allows the familiar tools and insights of
Schr\"odinger's nonrelativistic quantum mechanics and the Hamiltonian formalism to be applied to relativistic hadronic physics
\cite{Brodsky:1997de, Brodsky:2014yha, Zou:2018eam}. It should be noted that supersymmetry in this approach is supersymmetric quantum
mechanics~\cite{Witten:1981nf} and refers to bound state wave functions and not to elementary quantum fields.

Our work in this area can be traced back to the original article of Polchinski and Strassler~\cite{Polchinski:2001tt}, where the
exclusive hard-scattering counting rules~\cite{Brodsky:1973kr, Matveev:ra} were derived from the warped geometry of Maldacena's
five-dimensional anti-de Sitter AdS$_5$ space: The hadron in elastic scattering at high momentum transfer shrinks to a small size 
near the AdS boundary at $z = 0$ where the dual space is conformal ($z$ is the fifth coordinate of AdS space). Hadron form factors 
(FFs) look very different in AdS space~\cite{Polchinski:2002jw} or in physical spacetime~\cite{Drell:1969km,West:1970av}: One can 
show, however, that a precise mapping can be carried out at Dirac's fixed light-front time~\cite{Dirac:1949cp} for an arbitrary 
number of partons~\cite{Brodsky:2006uqa}. As a result, the impact parameter generalized parton distributions
\cite{Soper:1976jc, Burkardt:2000za} are expressed in terms of the square of AdS eigenmodes, provided that the invariant transverse
impact variable $\zeta$ for the $n$-parton bound state is identified with the holographic variable $z$. For a two-parton system, 
$\zeta^2 = x(1-x) {\bf b}^2_\perp$, the AdS modes are mapped directly to the square of effective light-front wave functions (LFWFs) 
that incorporate the non-perturbative pole structure of FFs~\cite{Brodsky:2006uqa}. Similar results follow from the mapping of the 
matrix elements of the energy-momentum tensor~\cite{Brodsky:2008pf}.

A semi-classical approximation to light-front QCD follows from the LF Hamiltonian equation
$P_\mu P^\mu \vert \psi \rangle = M^2 \vert \psi \rangle$ with $P = \left(P^-, P^+, {\bf P}_\perp\right)$. In the limit 
$m_q \to 0$ the LF Hamiltonian for a $q \bar q$ bound state can be systematically reduced to a wave equation in the variable
$\zeta$~\cite{deTeramond:2008ht}
\begin{equation}
\left(-\frac{d^2}{d\zeta^2} 
- \frac{1 - 4L^2}{4\zeta^2}+ U(\zeta) \right) \phi(\zeta) = M^2 \phi(\zeta) ,
\end{equation}
where the effective potential $U$ includes all interactions, including those from higher Fock states. The orbital angular momentum 
$L = 0$ corresponds to the lowest possible solution. The LF equation has similar structure of wave equations in AdS, and can be 
embedded in AdS space provided that $\zeta = z$~\cite{deTeramond:2008ht}. The precise mapping allows us to write the LF confinement
potential $U$ in terms of the dilaton profile that modifies AdS~\cite{deTeramond:2010ge}.

The separation of kinematic and dynamic components can be extended to arbitrary integer-spin $J$ by starting from a dilaton-modified
AdS action for a rank-$J$ symmetric tensor field $\Phi_{N_1 \dots N_J}$. Variation of the AdS action leads to a general wave equation
plus kinematical constraints to eliminate lower spin from the symmetric tensor~\cite{{deTeramond:2013it}}. LF mapping allows to determine
the mass function in the AdS action in terms of physical kinematic quantities consistent with the AdS stability bound
\cite{Breitenlohner:1982jf}. Similar derivation for arbitrary half-integral spin follows for Rarita-Schwinger spinors in AdS
\cite{deTeramond:2013it}. In this case, however, the dilaton term does not lead to an interaction~\cite{Kirsch:2006he} and an effective
Yukawa-type interaction has to be introduced instead~\cite{Abidin:2009hr}. Embedding light-front physics in a higher dimension gravity
theory leads to important insights into the non-perturbative structure of bound state equations in QCD for arbitrary spin, but does not
answer how the effective confinement dynamics is determined and how it can be related to the symmetries of QCD itself?

\begin{figure}[htp]
\includegraphics[width=0.9\columnwidth]{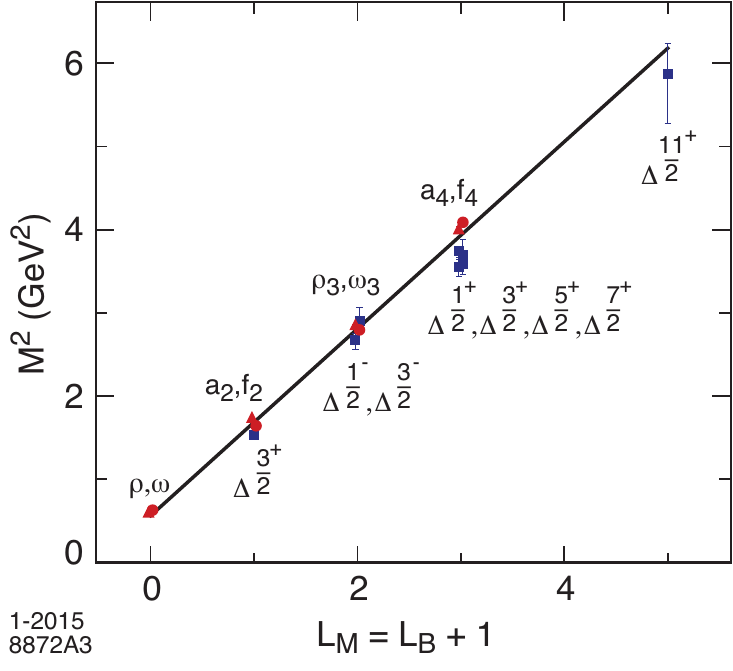}
\vspace{-0.25cm}
\caption{Supersymmetric vector meson and $\Delta$ partners from Ref.~\cite{Dosch:2015nwa}. The experimental values of $M^2$ for confirmed
states~\cite{Tanabashi:2018oca} are plotted vs $L_M = L_B+1$. The solid line corresponds to $\sqrt \lambda= 0.53$~GeV. The $\rho$ and 
$\omega$ mesons have no baryonic partner, since it would imply a negative value of $L_B$.}
\label{rho-delta}
\end{figure}

Conformal algebra underlies in LF holography the scale invariance of the QCD Lagrangian~\cite{Brodsky:2013ar}. It leads to the
introduction of a scale $\lambda = \kappa^2$ and harmonic confinement, $U \sim \lambda \zeta^2$, maintaining the action conformal
invariant~\cite{Brodsky:2013ar, deAlfaro:1976je}. The oscillator potential corresponds to a quadratic dilaton profile and thus to linear
Regge trajectories~\cite{Karch:2006pv}. Extension to superconformal algebra leads to a specific connection between mesons and
baryons~\cite{Dosch:2015nwa} underlying the $SU(3)_C$ representation properties, since a diquark cluster can be in the same color
representation as an antiquark, namely $\bar 3 \in 3 \times 3$. We follow~\cite{Fubini:1984hf} and define the fermionic generator
$R_\lambda = Q + \lambda S$ with anticommutation relations $\{R_\lambda,R_\lambda\} = \{R_\lambda^\dagger, R_\lambda^\dagger\} = 0$. 
It generates a new Hamiltonian $G_\lambda = \{R_\lambda, R_\lambda^\dagger\}$ that closes under the graded algebra 
$ [R_\lambda, G_\lambda] = [R_\lambda^\dagger, G_\lambda] = 0$. The generators $Q$ and $S$ are related to the generator of time
translation $H = \{ Q,Q^\dagger \}$~\cite{Witten:1981nf} and special conformal transformations $K = \{S,S^\dagger\}$: together with the
generator of dilations $D$ they satisfy the conformal algebra. The new Hamiltonian $G_\lambda$ is an element of the superconformal
(graded) algebra and uniquely determines the bound-state equations for both mesons and baryons~\cite{deTeramond:2014asa,Dosch:2015nwa}
\begin{eqnarray}
 \left(-\frac{d^2}{d\zeta^2} + \frac{4 L_M^2 -1}{4 \zeta^2} + V_M(\zeta) \right)\phi_M &=& M^2 \, \phi_M, \label{M}\\
\left(-\frac{d^2}{d\zeta^2} + \, \frac{4 L_B^2 -1}{4 \zeta^2} + V_B(\zeta) \right)\phi_B &=& M^2 \, \phi_B , \label{B}
\end{eqnarray}
including essential constant terms in the effective confinement potential
$ V_{M,B}(\zeta) = \lambda_{M,B}^2\, \zeta^2 + 2 \lambda_{M,B} (L_{M,B} \mp 1)$, with $\lambda_M = \lambda_B \equiv \lambda$ 
(equality of Regge slopes) and $L_M = L_B + 1$~\cite{Note1}. This is shown in Fig.~\ref{rho-delta}. The mass spectrum from
Eqs.(\ref{M}-\ref{B}) is $M^2_M = 4 \lambda (n+ L_M )$ and $M^2_B = 4 \kappa^2 (n+ L_B+1)$ with the same slope in $L$ and $n$, the radial
quantum number. Since $[R_\lambda^\dagger, G_\lambda] = 0$, it follows that the state $\vert M, L \rangle$ and
$R_\lambda^\dagger \vert M, L \rangle = \vert B, L - 1\rangle$ have identical eigenvalues $M^2$, thus $R_\lambda^\dagger$ is 
interpreted as the transformation operator of a single constituent antiquark (quark) into a diquark cluster with quarks (antiquarks) 
in the conjugate color representation. The pion, however, has a special role as the unique state of zero mass that is annihilated
by $R_\lambda^\dagger$, $R_\lambda^\dagger \vert M, L = 0 \rangle = 0$: The pion has not a baryon partner and thus breaks the
supersymmetry. 

\begin{figure}[htp]
\includegraphics[width=0.9\columnwidth]{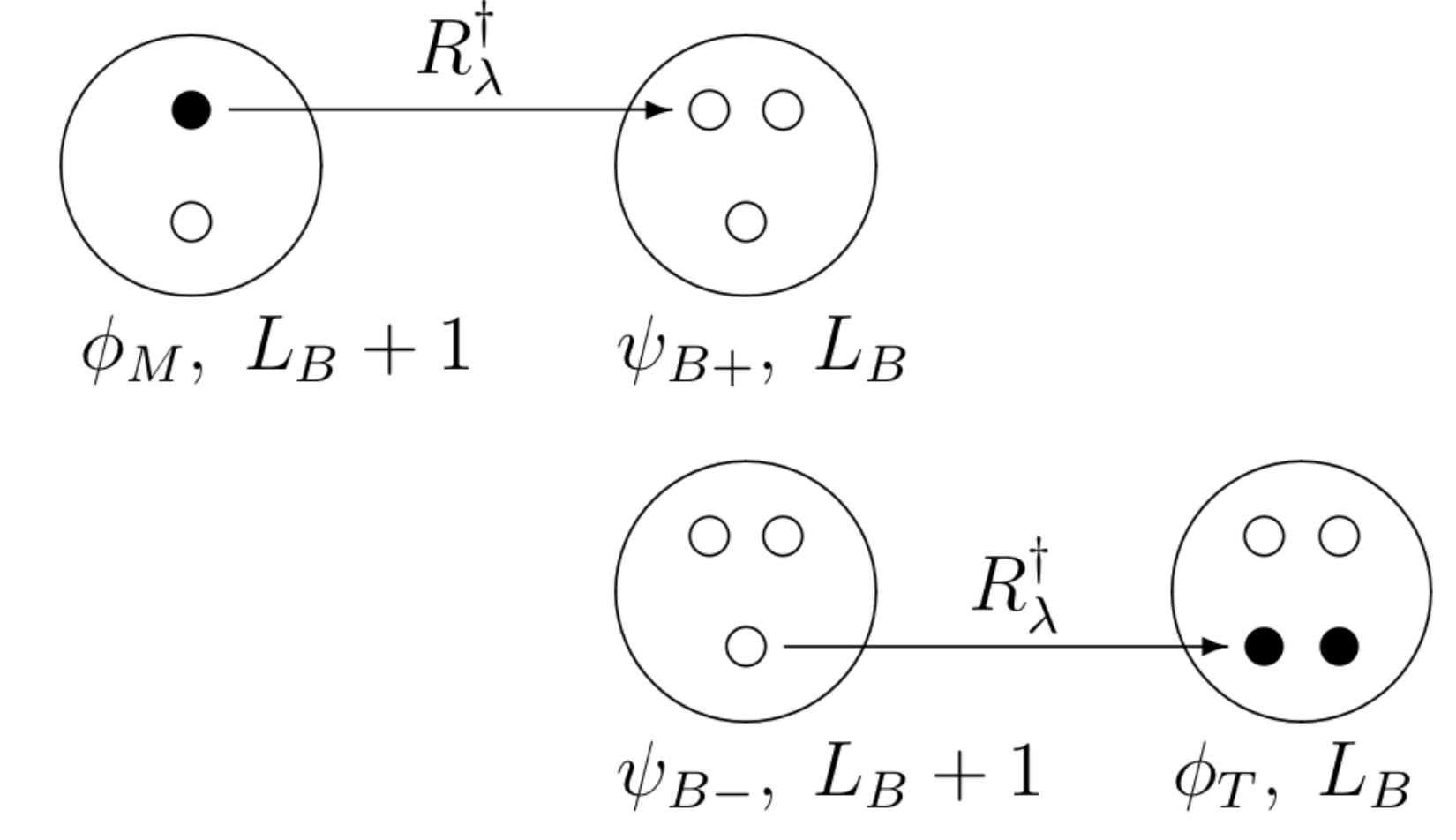}
\vspace{-0.25cm}
\caption{Supersymmetric 4-plet representation of same-mass and parity hadronic states $\{\phi_M, \psi_{B+}, \psi_{B-},\phi_T\}$
\cite{Brodsky:2016yod}. Mesons are interpreted as $ q \bar q$ bound states, baryons as quark-antidiquark bound states and tetraquarks as 
diquark-antidiquark bound states. The fermion ladder operator $R^\dagger_\lambda$ connects antiquark (quark) and diquark (anti-diquark) 
cluster of the same color. The baryons have two chirality components with orbital angular momentum $L$ and $L+1$.}
\label{4-plet}
\end{figure}

Embedding in AdS is also useful to extend the superconformal Hamiltonian to include the spin-spin interaction: From the spin dependence
of mesons~\cite{deTeramond:2013it} one concludes that $G_\lambda = \{R_\lambda, R_\lambda^\dagger\} + 2 \lambda s$, with $s = 0, 1$ 
the total internal spin of the meson or the spin of the diquark cluster of the baryon partner~\cite{Brodsky:2016yod}. The lowest 
mass state of the vector meson family, the $\rho$ (or the $\omega$) is also annihilated by the operator $R^\dagger$, and has no 
baryon partner: The effect of the spin term is an overall shift of the quadratic mass scale without a modification of the LFWF as
depicted in Fig.~\ref{rho-delta}. The analysis was consistently applied to the radial and orbital excitation spectra of the 
$\pi, \rho, K, K^*$ and $\phi$ meson families, as well as to the $N, \Delta, \Lambda, \Sigma, \Sigma^*, \Xi$ and $\Xi^*$ in the 
baryon sector, giving the value $\kappa = \sqrt{\lambda} = 0.523 \pm 0.024$~GeV from the light hadron spectrum~\cite{Brodsky:2016yod}. 
Contribution of quark masses~\cite{Brodsky:2008pg} are included via the Feynman-Hellman theorem, 
$\Delta M^2 = \langle \sum_q m_q^2/x_q \rangle$, with the effective values $m_u = m_d = 46$~MeV and $m_s = 357$~MeV~\cite{Brodsky:2014yha}. 
The complete multiplet is obtained by applying the fermion operator $R_\lambda^\dagger$ to the negative-chirality component baryon wave
function~\cite{deTeramond:2014asa, Brodsky:2014yha} $\phi_B = \left\{\psi_{+}(L_B), \psi_{-}(L_B+ 1)\right\}$ leading to a tetraquark 
bosonic partner, $R_\lambda^\dagger\, \psi_{-} =\phi_T$, a bound state of diquark and anti-diquark clusters with angular momentum $L_T=L_B$
\cite{Brodsky:2016yod}: The full supermultiplet (see Fig.~\ref{4-plet}) contain mesons, baryons, and tetraquarks~\cite{Note2}. A systematic 
analysis of the isoscalar bosonic sector was also performed using the framework described here; the $\eta'-\eta$ mass difference is correctly
reproduced~\cite{Zou:2019tpo}.

\begin{figure}[htp]
\includegraphics[width=0.9\columnwidth]{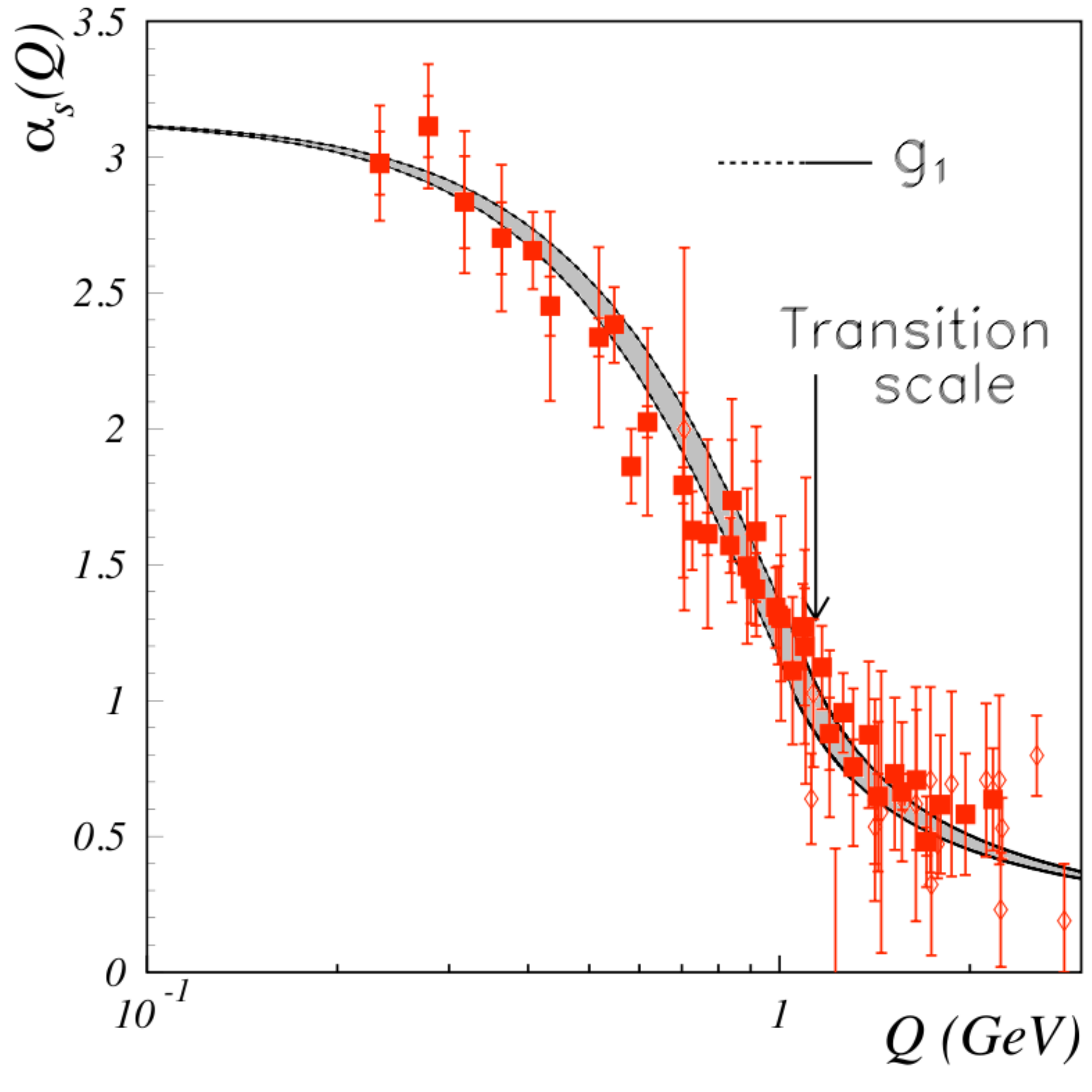}
\vspace{-0.25cm}
\caption{Matching the non-perturbative and perturbative couplings regimes at 5-loop $\beta$-function in the $\bar{MS}$ renormalization 
scheme and comparison with $\alpha_s$ measurements from the Bjorken sum rule. For $\sqrt{\lambda} = 0.523 \pm 0.024$~GeV we obtain 
$\Lambda_{\bar{MS}}=0.339 \pm 0.019$~GeV compared with the world average $\Lambda_{\bar{MS}}= 0.332 \pm 0.017$~GeV~\cite{Deur:2016opc}.}
\label{alphaIR}
\end{figure}

We have shown in Ref.~\cite{Dosch:2015bca} that the basic underlying hadronic supersymmetry still holds and gives remarkable connections
across the entire spectrum of light and heavy-light hadrons even if quark masses break the conformal invariance. In particular, the
$L = 0$ lowest mass meson defining the $K, K^*, \eta', \phi, D, D^*, D_s, D_s^*, B, B^*,B_s$ and $B^*_s$ families examined in Ref.~\cite{Dosch:2015bca} 
has in effect no baryon partner, conforming to the SUSY mechanism found for the light hadrons. The analysis was extended in
\cite{Dosch:2016zdv} by showing that the embedding of the light-front wave equations in AdS space nevertheless determines the form of
the confining potential in the LF Hamiltonian to be harmonic, provided that: a) the longitudinal and transverse dynamics can be factored
out to a first approximation and b) the heavy quark mass dependence determines the increasing value of the Regge slope according to Heavy
Quark Effective Theory (HQET)~\cite{Isgur:1991wq}. This model has been confronted with data in the detailed analysis performed in
\cite{Nielsen:2018uyn} including tetraquarks with one charm or one bottom quark as illustrated in Tables~\ref{c} and \ref{b}. The
double-heavy hadronic spectrum, including mesons, baryons and tetraquarks and their connections was examined in Ref.~\cite{Nielsen:2018ytt}
confirming the validity of the supersymmetric approach applied to this sector. The lowest mass meson of each family, the $\eta_c$,
$J/\psi$, $\eta_B$, and $Y$ have no hadronic partner and the increase in the Regge slope qualitatively agrees with the HQET prediction. 
Embedding LF dynamics in AdS allow us to study the infrared (IR) behavior of the strong coupling. In fact, it is possible to establish a 
connection between the short-distance behavior of the QCD coupling $\alpha_s$ with JLab long-distance measurements of $\alpha_s$ from the 
Bjorken sum rule~\cite{Brodsky:2010ur, Deur:2014qfa, Deur:2016cxb, Deur:2016opc}. In light-front holography the IR strong coupling is
$\alpha_s^{IR}(Q^2) = \alpha_s^{IR}(0) e^{- Q^2/4 \lambda}$. One can obtain $\Lambda_{QCD}$ from matching the perturbative (5-loop) and
non-perturbative couplings at the transition scale $Q_0$ as shown in Fig.~\ref{alphaIR}. For $\sqrt{\lambda} = 0.523 \pm 0.024$~GeV we
find $\Lambda_{\bar{MS}}=0.339 \pm 0.019$~GeV compared with the world average $\Lambda_{\bar{MS}}= 0.332 \pm 0.017$~GeV and
$Q_0^2 \simeq 1$~GeV$^2$. Therefore, one can establish a connection between the proton mass $M^2_p = 4 \lambda$ and the perturbative QCD
scale $\Lambda_{QCD}$ in any renormalization scheme.

An extensive study of FFs~\cite{Sufian:2016hwn} and parton distributions~\cite{deTeramond:2018ecg, Liu:2019vsn} has been
carried out recently using an extended model based on the gauge-gravity correspondence, light-front holography, and the generalized
Veneziano model~\cite{Veneziano:1968yb,Ademollo:1969wd,Landshoff:1970ce}. The non-perturbative strange and charm sea content of the nucleon
has been studied by also incorporating constraints from lattice QCD~\cite{Sufian:2018cpj, Sufian:2020coz}. Meson~\cite{Brodsky:2011xx} and
nucleon transition form factors, such as the proton to Roper $N(1440)1/2^+$ transition, can also be described within the light-front
holographic framework~\cite{deTeramond:2011qp, Ramalho:2017pyc} and extended to other nucleon transitions~\cite{Gutsche:2019yoo}, such as
the transition to the $\Delta(1232)3/2^+$, $N(1520)3/2^-$, $N(1535)1/2^-$, $\Delta(1600)3/2^+$, and $\Delta(1620)1/2^-$ states measured at
CLAS~\cite{Mokeev:2015lda}.

Hadron FFs in the light-front holographic approach are a sum from the Fock expansion of states $F(t) = \sum_\tau c_\tau F_\tau(t)$,
where the $c_\tau$ are spin-flavor coefficients and $F_\tau(t)$ has the Euler's Beta form structure
\cite{Veneziano:1968yb,Ademollo:1969wd,Landshoff:1970ce} 
\begin{equation}
F_\tau(t) = \frac{1}{N_\tau} B\big(\tau-1, 1 - \alpha(t) \big),
\end{equation}
where $\alpha(t)$ is the Regge trajectory of the vector meson that couples to the quark current in the hadron. For twist $\tau= N$, the
number of constituents in a Fock component, the FF is an $N-1$ product of poles 
\begin{equation}
F_{\tau}( Q^2) = \frac{1}{\big(1 + \frac{Q^2}{M^2_{n=0}}\big) \big(1 + \frac{Q^2}{M^2_{n=1}} \big) \cdots 
\big(1 + \frac{Q^2}{M^2_{n =\tau - 2}} \big)}, 
\end{equation}
located at $ - Q^2 = M^2_n = (n + 1 - \alpha(0)) / \alpha'$, which generates the radial excitation spectrum of the exchanged particles in
the $t$-channel~\cite{Brodsky:2014yha, Zou:2018eam}. The trajectory $\alpha(t)$ can be computed within the superconformal framework and
its intercept $\alpha(0)$ incorporates the quark masses~\cite{Sufian:2018cpj}.

\begin{figure}[htp]
\includegraphics[width=0.9\columnwidth]{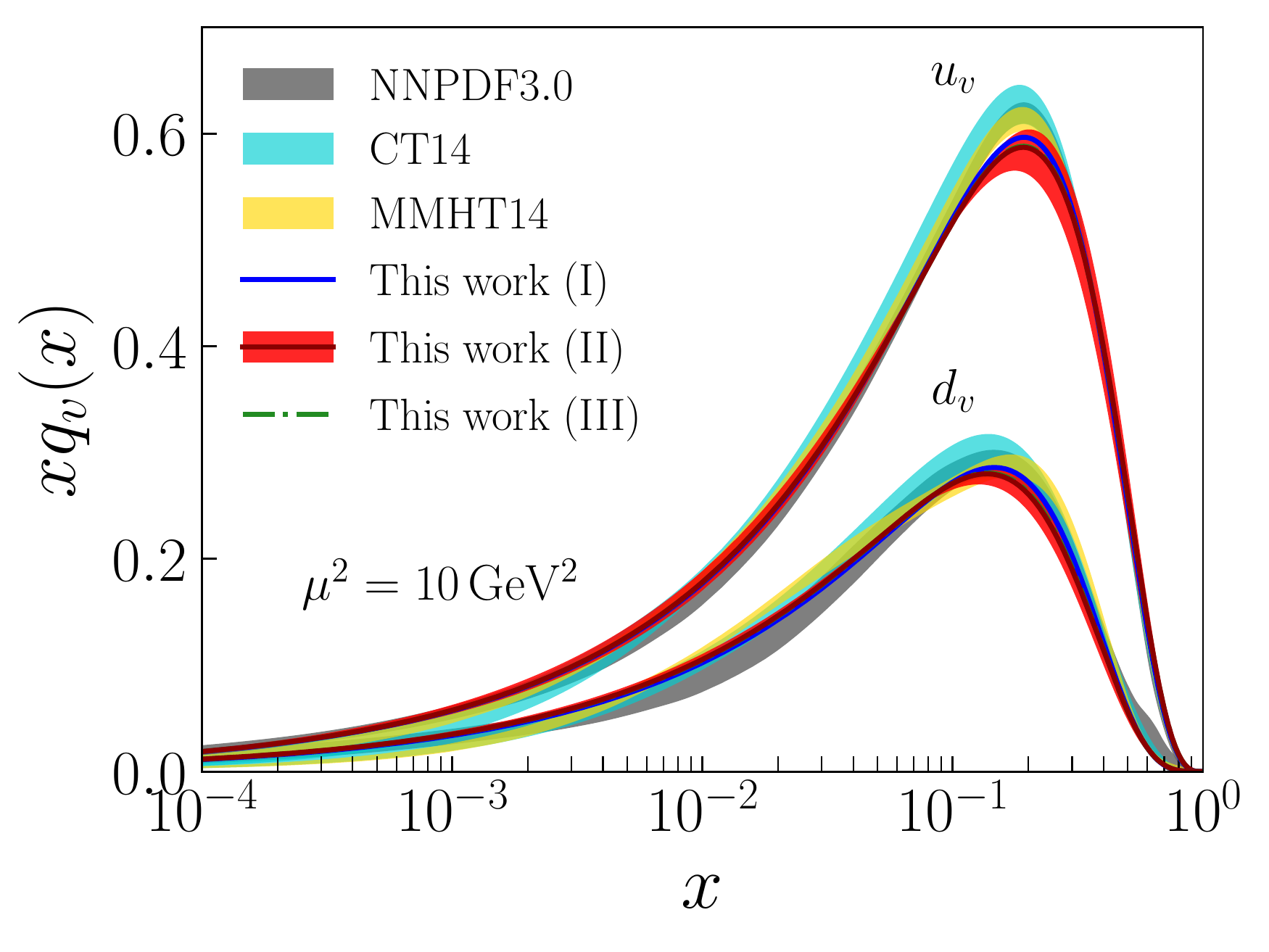}
\vspace{-0.25cm}
\caption{Comparison of $x q(x)$ in the proton from LF holographic QCD~\cite{deTeramond:2018ecg} with global
 fits~\cite{Ball:2014uwa, Harland-Lang:2014zoa, Dulat:2015mca} for models I, II and III in ~\cite{Liu:2019vsn}. The results are evolved
 from the initial scale $\mu_0 = 1.06 \pm 0.15$~GeV~\cite{Deur:2016opc}.}
\label{unpolPDFs}
\end{figure}

Using the integral representation of the Beta function, the form factor is expressed in a reparameterization invariant form 
\begin{equation}
F(t)_\tau = \frac{1}{N_\tau} \int_0^1 dx w'(x) w(x)^{-\alpha(t)} \left[1 - w(x) \right]^{\tau -2} ,
\end{equation}
with $w(0) = 0, \quad w(1) = 1, \quad w'(x) \ge 0$. The flavor FF is given in terms of the valence GPD at zero skewness
$F^q_\tau(t) = \int_0^1 dx \, q_\tau(x) \exp[t f(x)]$ with the profile function $f(x)$ and PDF $q(x)$ determined by $w(x)$
\begin{eqnarray}
f(x)&=&\frac{1}{4\lambda}\log\Big(\frac{1}{w(x)}\Big) , \\ 
q_\tau(x)&=&\frac{1}{N_\tau}[1-w(x)]^{\tau-2}w(x)^{-\frac{1}{2}}w'(x).
\end{eqnarray}
Boundary conditions at $x \to 0$ follow from the expected Regge behavior, $w(x) \sim x$, and at $x \to 1$ from the inclusive-exclusive
counting rules~\cite{Drell:1969km} $q_\tau(x) \sim (1-x)^{2 \tau - 3}$ that imply $w'(1) = 0 $. These physical conditions, together 
with the constraints written above, basically determine the form of $w(x)$. If the universal function $w(x)$ is fixed by the nucleon PDFs 
then the pion PDF is a prediction~\cite{deTeramond:2018ecg}. The unpolarized PDFs for the nucleon are compared with global fits in
Fig.~\ref{unpolPDFs}.

\begin{figure}[htp]
\includegraphics[width=0.9\columnwidth]{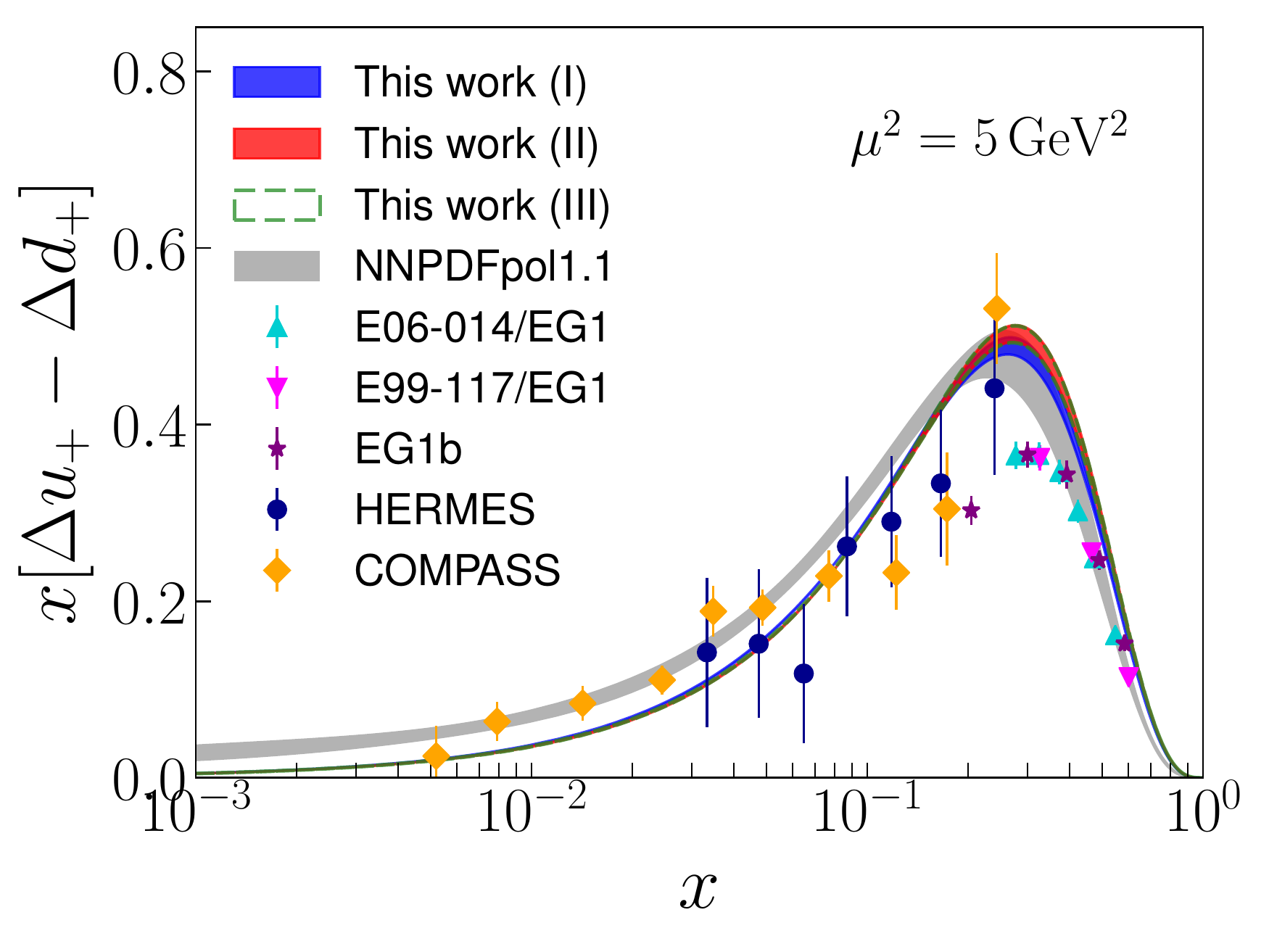}
\vspace{-0.25cm}
\caption{Polarized distributions of the isovector combination $x[\Delta u_+(x)-\Delta d_+(x)]$ in comparison with NNPDF
  global fit~\cite{Nocera:2014gqa} and experimental data
  \cite{Zheng:2003un,Zheng:2004ce,Parno:2014xzb,Dharmawardane:2006zd,Airapetian:2003ct,Airapetian:2004zf,Alekseev:2010ub}. Three sets of
  parameters are determined from the Dirac form factor and unpolarized valence distributions.}
\label{poldist}
\end{figure}

\begin{figure}[htp]
\includegraphics[width=0.9\columnwidth]{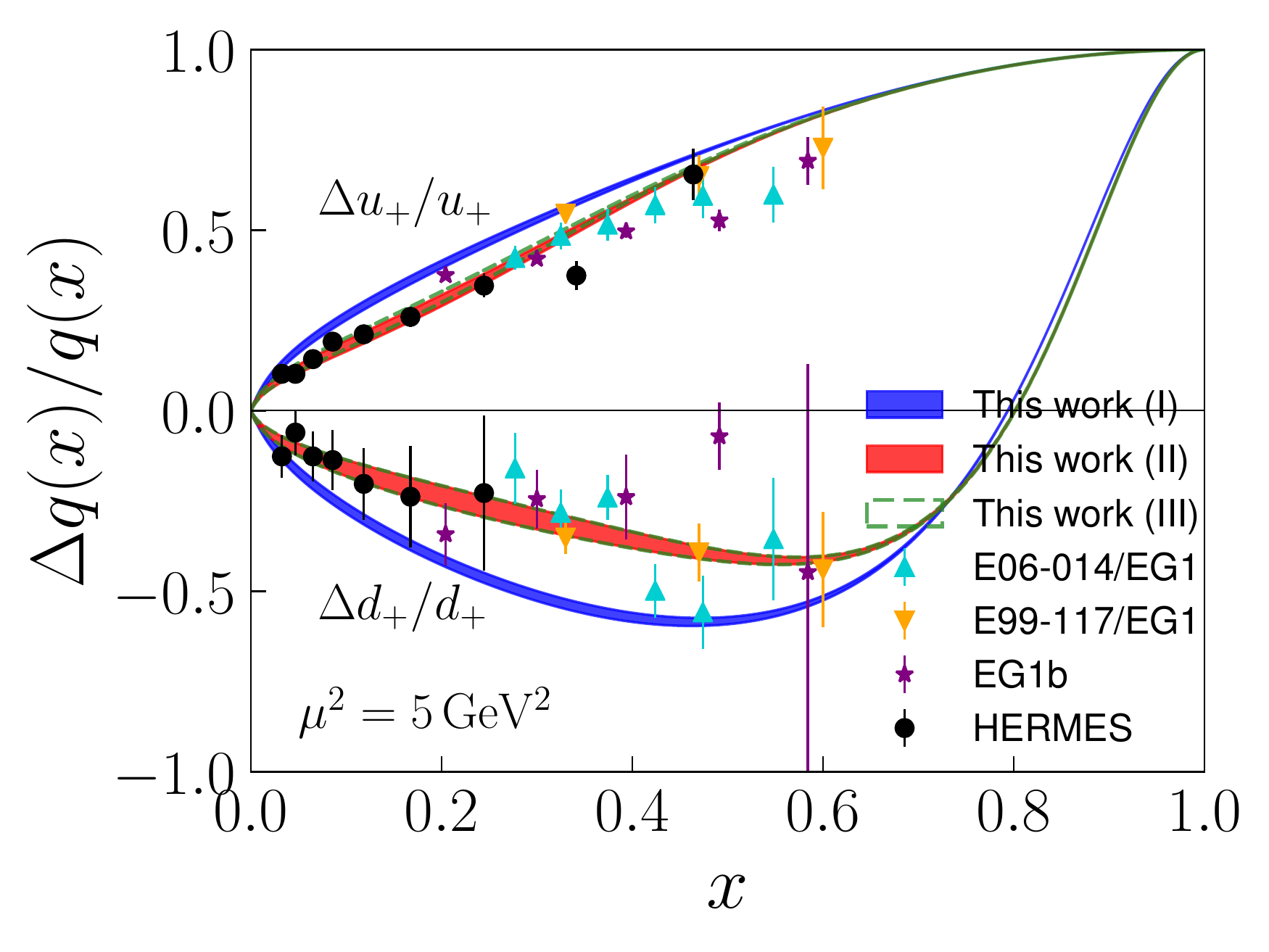}
\vspace{-0.25cm}
\caption{Helicity asymmetries of $u + \bar{u}$ and $d + \bar{d}$. Symbols as in Fig.~\ref{poldist}.}
\label{Dqoq}
\end{figure}

\begin{figure}[htp]
\includegraphics[width=0.9\columnwidth]{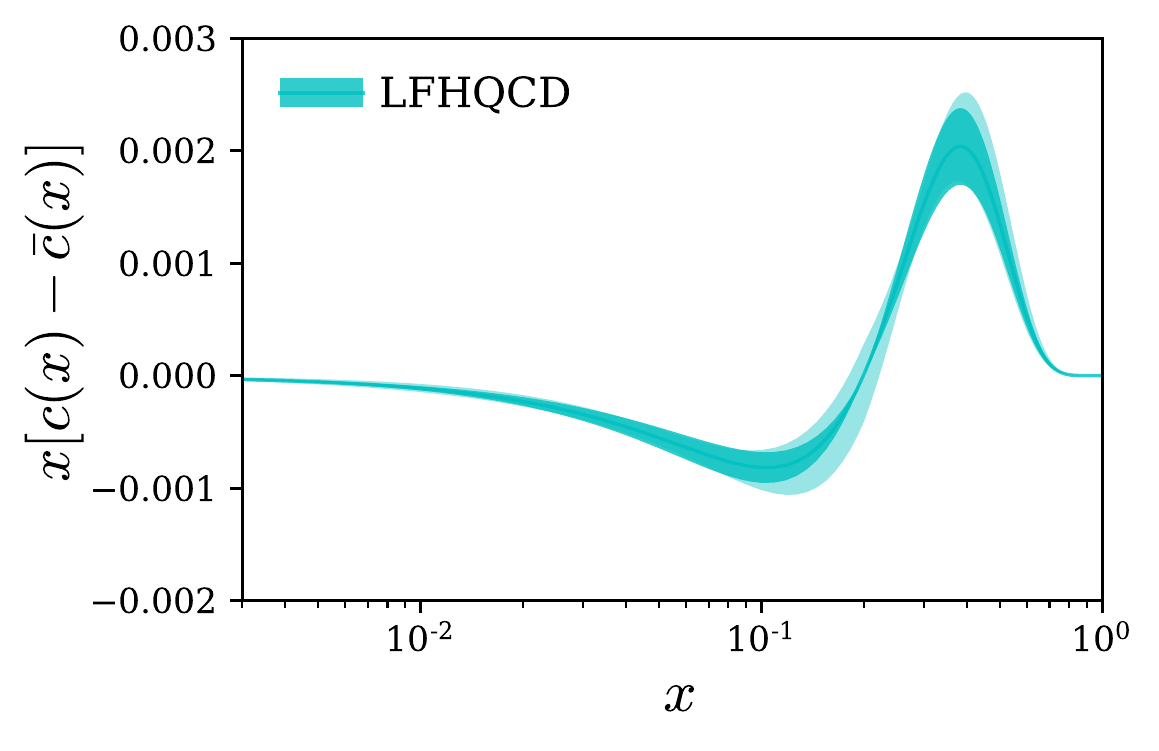}
\vspace{-0.25cm}
\caption{The distribution function $x[c(x) - \bar{c}(x)]$ computed in the light-front holographic framework using lattice
  QCD input of the charm electromagnetic form factors~\cite{Sufian:2020coz}.}
\label{cbarc}
\end{figure}

To study the polarized GPDs and PDFs we perform a separation of chiralities in the AdS action: It allows the computation of the
matrix elements of the axial current - including the correct normalization, once the coefficients $c_\tau$ are fixed for the
vector current~\cite{Liu:2019vsn}. The formalism incorporates the helicity retention between the leading quark at large $x$ and the
parent hadron: $\lim_{x \to 1} ~ \frac{\Delta q(x)}{q(x)} = 1$, a perturbative QCD result~\cite{Farrar:1975yb}. It also predicts no-spin
correlation with the parent hadron at low $x$: $\lim_{x \to 0} ~ \frac{\Delta q(x)}{q(x)} = 0$. We compare our predictions with available
data for spin-dependent PDFs in Fig.~\ref{poldist} and for the ratio $\Delta q(x)/q(x)$ in Fig~\ref{Dqoq}. The first lattice QCD
computation of the the charm quark contribution to the electromagnetic form factors of the nucleon with three gauge ensembles (one at the
physical pion mass) was performed in Ref.~\cite{Sufian:2020coz}. It gives the necessary constraints to compute the non-perturbative intrinsic
charm-anticharm asymmetry $c(x) - \bar c(x)$ using the light-front holography approach. The results are shown in Fig.~\ref{cbarc}
($q_+ \equiv q + \bar{q}$).

We have shown how the classical equations of motion for hadrons of arbitrary spin derived from the 5-dimensional gravity theory have
the same form of the semiclassical bound-state equations for massless constituents in LF quantization. The implementation of
superconformal algebra determines uniquely the form of the confining interaction. This new approach to hadron physics incorporates
basic non-perturbative properties that are not apparent from the chiral QCD Lagrangian, such as the emergence of a mass scale and the
connection between mesons and baryons. In particular, the prediction of a massless pion in the chiral limit is a consequence of the
superconformal algebraic structure and not of the Goldstone mechanism. The structural framework of LF holography also provides nontrivial
connections between the structure of form factors and polarized and unpolarized quark distributions with non-perturbative results such as
Regge theory and the Veneziano model. 
 
Specific key results, such as the prediction of the ratio $\Delta q(x)/ q(x)$ at large $x$ will be tested very soon in upcoming
experiments at JLab~\cite{E12-06-110,E12-06-122}. The strange-antistrange asymmetry could be explored in semi-inclusive $\phi$
electroproduction with CLAS 12. Our study of the nucleon to Roper transition form factor will be extended up to $Q^2 = 12$~GeV$^2$
for comparison with new CLAS data. The prediction of hadron states within superconformal multiplets of meson-baryon-tetraquarks (for
example the multiplets shown in Tables~\ref{c} and \ref{b}) can motivate the search for new tetraquark states. Many other important
applications to hadron physics based on the holographic framework have been studied in addition to the new developments described here;
unfortunately it is not possible to review them in this short overview and we apologize to the authors in advance.
 
\begin{table*}[htp]
\caption{\label{c} Quantum number assignment of different meson families with quarks: $q=u,~d,~s$ and one charm quark $c$ and their
    supersymmetric baryon and tetraquark partners from Ref.~\cite{Nielsen:2018uyn}. Each family is separated by a horizontal line. For
    baryons multiplets with same $L_B$ and $S_D$ only the state with the highest possible value for $J$ is included. Diquarks clusters
    are represented by $[~]$ have total spin $S_D=0$, and $(~)$ represents $S_D=1$. The quantum numbers $J^P=1^+$ and $J^P=2^-$ are assigned
    to the states $D(2550)$ and $D_J(2600)$, but their quantum numbers have not yet been determined. States with a question mark (?) are the
    predicted ones. The lowest meson bound state of each family has no baryon or tetraquark partner and breaks the supersymmetry.}
 \begin{ruledtabular}
 \begin{tabular}{ccc|ccc|ccc}
\multicolumn{3}{c|}{Meson} & \multicolumn{3}{c|}{Baryon} & \multicolumn{3}{c}{Tetraquark}\\
$q$-cont&$J^{P(C)}$ & Name & $q$-cont & $J^{P}$ &Name &$q$-cont &$J^{P(C)}$& Name \\
\hline
$\bar{q}c$&$0^{-}$&$D(1870)$& --- &---&---& --- & --- &--- \\
$\bar{q}c$&$1^{+}$&${D}_1(2420)$& $[ud]c$ & $(1/2)^+$&$\Lambda_c(2290)$ & $[ud][\bar{c}\bar{q}]$ & $0^{+}$ &$\bar{D}_0^*(2400)$ \\
$\bar{q}c$&$2^{-}$&$D_J(2600)$ & $[ud]c$ & $(3/2)^-$ & $\Lambda_c(2625)$ & $[ud][\bar{c}\bar{q}]$ & $1^{-}$ & --- \\
\hline
$\bar{c}q$&$0^{-}$&$\bar{D}(1870)$& --- &---&---& --- & --- &--- \\
$\bar{c}q$&$1^{+}$&$\bar{D}_1(2420)$& $[cq]q$ & $(1/2)^+$&$\Sigma_c(2455)$ & $[cq][\bar{u}\bar{d}]$ & $0^{+}$ &${D}_0^*(2400)$ \\
\hline
$\bar{q}c$&$1^{-}$&$D^*(2010)$& --- &---&---& --- & --- &--- \\
$\bar{q}c$&$2^{+}$&$D_2^*(2460)$& $(qq)c$ & $(3/2)^+$&$\Sigma_c^*(2520)$ & $(qq)[\bar{c}\bar{q}]$ & $1^{+}$ &${D}(2550)$ \\
$\bar{q}c$&$3^{-}$&$D_3^*(2750)$ & $(qq)c$ & $(3/2)^-$ & $\Sigma_c(2800)$ & $(qq)[\bar{c}\bar{q}]$ & --- & --- \\
 \hline
$\bar{s}c$&$0^{-}$&$D_s(1968)$& --- &---&---& --- & --- &--- \\
$\bar{s}c$&$1^{+}$&${D}_{s1}(2460)$& $[sq]c$ & $(1/2)^+$&$\Xi_c(2470)$ & $[sq][\bar{c}\bar{q}]$ & $0^{+}$ &$\bar{D}_{s0}^*(2317)$ \\
$\bar{s}c$&$2^{-}$&$D_{s2}(\sim2830)$? & $[sq]c$ & $(3/2)^-$ & $\Xi_c(2815)$ & $[sq][\bar{c}\bar{q}]$ & $1^{-}$ & --- \\
\hline
$\bar{s}c$&$1^{-}$&$D_s^*(2110)$& --- &---&---& --- & --- &--- \\
$\bar{s}c$&$2^{+}$&$D_{s2}^*(2573)$& $(sq)c$ & $(3/2)^+$&$\Xi_c^*(2645)$ & $(sq)[\bar{c}\bar{q}]$ & $1^{+}$ &${D}_{s1}(2536)$ \\
$\bar{s}c$&$3^{-}$&$D_{s3}^*(2860)$ & $(sq)c$ & $(1/2)^-$ & $\Xi_c(2930)$ & $(sq)[\bar{c}\bar{q}]$ & --- & --- \\
\hline
$\bar{c}s$&$1^{+}$&$\bar{D}_{s1}(\sim2700)$?& $[cs]s$ & $(1/2)^+$&$\Omega_c(2695)$ & $[cs][\bar{s}\bar{q}]$ & $0^{+}$ & ?? \\
\hline
$\bar{s}c$&$2^{+}$&$D_{s2}^*(\sim2750)$?& $(ss)c$ & $(3/2)^+$&$\Omega_c(2770)$ & $(ss)[\bar{c}\bar{s}]$ & $1^{+}$ & ?? 
\end{tabular}
\end{ruledtabular}
\end{table*}

\begin{table*}[htp]
\caption{\label{b} Same as Table~\ref{c} but for mesons containing one bottom quark $b$ from Ref.~\cite{Nielsen:2018uyn}. The quantum numbers
    $J^P=1^+,~J^P=0^+ $ and $J^P=2^-$ are assigned to the states $B_J(5732)$, $B_{J}^*(5840)$ and $B_J(5970)$, but their quantum numbers
    have not yet been determined. States with a question mark (?) are the predicted ones. The lowest meson of each family has no baryon
    or tetraquark partner and breaks the supersymmetry.}
 \begin{ruledtabular}
\begin{tabular}{ccc|ccc|ccc}
\multicolumn{3}{c|}{Meson} & \multicolumn{3}{c|}{Baryon} & \multicolumn{3}{c}{Tetraquark}\\
$q$-cont&$J^{P(C)}$ & Name & $q$-cont & $J^{P}$ &Name &$q$-cont &$J^{P(C)}$& Name \\
\hline
$\bar{q}b$&$0^-$&$\bar{B}(5280)$& --- &---&---& --- & --- &--- \\
$\bar{q}b$&$1^+$&$\bar{B}_1(5720)$& $[ud]b$ & $(1/2)^+$&$\Lambda_b(5620)$ & $[ud][\bar{b}\bar{q}]$ & $0^{+}$ &${B}_J(5732)$ \\
$\bar{q}b$&$2^{-}$&$\bar{B}_J(5970)$ & $[ud]b$ & $(3/2)^-$ & $\Lambda_b(5920)$ & $[ud][\bar{b}\bar{q}]$ &$1^{-}$ & --- \\
\hline
$\bar{b}q$&$0^{-}$&${B}(5280)$& --- &---&---& --- & --- &--- \\
$\bar{b}q$&$1^{+}$&${B}_1(5720)$& $[bq]q$ & $(1/2)^+$&$\Sigma_b(5815)$ & $[bq][\bar{u}\bar{d}]$ & $0^{+}$ &$\bar{B}_J(5732)$ \\
\hline
$\bar{q}b$&$1^{-}$&$B^*(5325)$& --- &---&---& --- & --- &--- \\
$\bar{q}b$&$2^{+}$&$B_2^*(5747)$& $(qq)b$ & $(3/2)^+$&$\Sigma_b^*(5835)$ & $(qq)[\bar{b}\bar{q}]$ & $1^{+}$ &${B}_J(5840)$ \\
  \hline
$\bar{s}b$&$0^{-}$&$B_s(5365)$& --- &---&---& --- & --- &--- \\
$\bar{s}b$&$1^{+}$&${B}_{s1}(5830)$& $[qs]b$ & $(1/2)^+$&$\Xi_b(5790)$ & $[qs][\bar{b}\bar{q}]$ & $0^{+}$ &$\bar{B}_{s0}^*(\sim5800)$? \\
\hline
$\bar{s}b$&$1^{-}$&$B_s^*(5415)$& --- &---&---& --- & --- &--- \\
$\bar{s}b$&$2^{+}$&$B_{s2}^*(5840)$& $(sq)b$ & $(3/2)^+$&$\Xi_b^*(5950)$ & $(sq)[\bar{b}\bar{q}]$ & $1^{+}$ &${B}_{s1}(\sim5900)$? \\
\hline
$\bar{b}s$&$1^{+}$&${B}_{s1}(\sim6000)$?& $[bs]s$ & $(1/2)^+$&$\Omega_b(6045)$ & $[bs][\bar{s}\bar{q}]$ & $0^{+}$ & ?? 
\end{tabular}
\end{ruledtabular}
\end{table*}

\section{Contributions to the USA EIC Physics Program}

With the recent decision by the Department of Energy (DOE), the prospects for the construction the Electron-Ion Collider have
been increased significantly. The nucleon 3-D imaging program with exclusive and semi-inclusive processes is of high priority. The
expected availability both of highly polarized electrons as well as of polarized protons, either longitudinally (along the beam) or
transversely polarized (either in the $ep$ scattering plane or perpendicular to it), will give access not only to the CFF 
${\cal{H}}(\xi,t)$ but will directly access the CFF ${\cal{E}}(\xi,t)$, which is related to the gravitational spin form factor $J(t)$, 
and can therefore help solve the proton's still unresolved spin puzzle of the proton. When comparing the 12~GeV kinematics with the 
EIC kinematics we have to take into account the much lower luminosity for the EIC in case of unpolarized $ep$ scattering. 

However, in the case of polarized proton targets the effective luminosity, {\it i.e.} taking into account the dilution factor for polarized
NH$_3$, approximately equal to 0.17, the situation is less dramatic. Moreover, for transverse polarization, {\it e.g.} using a polarized HD 
target, the expected acceptable effective proton luminosity is $L \approx 0.3\times10^{34}$~cm$^{-2}$s$^{-1}$, more comparable with 
the EIC luminosity on polarized protons. The process $\gamma p \to p e^+ e^-$, which is the time-reversed process to DVCS, also accesses 
the GPDs and Compton Form Factors, but provides direct access to the real part of the Compton amplitude, {\it i.e.} it complement DVCS, which, 
in polarized beam experiments directly accesses the imaginary part of the Compton amplitude. 

Owing to the high center-of-mass energy, the EIC offers the possibility to study the equivalent 3D and gravitational properties of the glue. 
Here processes such as $e+p \to e+p+J/\psi$ and $e+p \to e+p+\phi$ may be explored, and in greater detail. Last by not least, the EIC, in 
particular in its lower energy implementation, can be an excellent tool for spectroscopy especially in the light-flavor and strangeness area,
as well as in the $X$, $Y$, $Z$ exotic states. In the latter, the nature of the already found states can be explored, for example 4-quarks 
states versus meson-meson molecules, and dynamical processes near threshold. The slope of the $t$-dependence in the production cross 
section can be an effective tool in discriminating among such possibilities. 

A critical parameter of such an EIC machine is its operating luminosity as all of these exclusive processes are quite rare, and require high 
luminosity and long beam times to collect sufficient statistics in a multi-dimensional binning. Especially the $\xi$ and $t$ dependence of 
DVCS cross sections and of the beam and target spin asymmetries are required for the analysis and the Fourier transform of the gravitational 
form factors in order to extract mechanical properties of the proton and light nuclei. 

\section{Recommendations on Future Joint Activities Between Experiment, Phenomenology, and Theory}
\label{outlook}

The achievements described in this document, which summarizes the exploration of the spectra and structure of hadrons using experiments with 
electromagnetic probes and attendant advances in theory for the description of hadron properties, suggests good prospects for gaining insight 
into the strong QCD dynamics that underlie the generation of the ground and excited state hadrons. In this section we introduce possible future 
research activities covering a wide field of hadron physics efforts that can all benefit from synergistic engagement between experimentalists, 
phenomenologists, and theorists.

\subsection{Hadron Spectroscopy}

Observations of new missing resonances extend our knowledge about the spectrum of excited nucleon states. They support the relevance of SU(6) 
spin-flavor approximate symmetry in the generation of the excited nucleon spectrum. This symmetry, however, predicts many resonances in addition 
to those already discovered within the mass range from 1.7~GeV to 2.5~GeV. Experiments on the studies of exclusive meson photoproduction collecting 
data on differential cross sections, and single-, double-, and triple-polarization asymmetries, supported by results from experiments with hadron 
beams, will provide the nearly complete set of data needed to solve the fundamental several-decade-old problem of how the spectrum of nucleons arises 
from QCD. The amplitude or reaction model analyses of the experimental photoproduction data for the individual exclusive channels as well as within 
global coupled-channel approaches are critical in order to achieve this objective. Joint research activity between experimentalists in the area of 
exclusive meson photo-/hadroproduction and the experts in phenomenological data analyses will provide the final results on the spectrum of the 
excited nucleons as they are seen from the data.

This activity should be extended by the combined studies of exclusive meson photo- and electroproduction data aimed at searches for new resonances. 
New resonances established in photoproduction can be seen in the electroproduction data too. A successful description of both photo- and 
electroproduction data within a substantial range of $Q^2$, with $Q^2$-independent resonance masses, and total and partial hadronic decay widths, 
will validate the resonance existence in a nearly model-independent way. The efficiency of this strategy was demonstrated recently by observation 
of a new $N'(1720)3/2^+$ nucleon resonance from combined studies of CLAS $\pi^+\pi^-p$ photo- and electroproduction data~\cite{Mok20}. These studies 
will benefit from new CLAS12 experimental data. Crucially, the amplitude analyses and reaction models, used successfully in the discovery of new 
baryon states from photoproduction data, must be extended to electroproduction as part of these efforts.

Searches for new states of baryon matter with glue as an active structural component, the so-called hybrid baryons, are in progress with the CLAS12 
detector. Contemporary theory suggests that hybrid baryons may be seen as missing resonances in the $W$ range from 2.0~GeV to 2.5~GeV in the $KY$ and 
$\pi^+\pi^-p$ exclusive photo-/electroproduction channels. In order to identify the hybrid nature of these resonances, it is necessary to study the 
$Q^2$ evolution of the resonance electrocouplings with $Q^2$-independent hadronic parameters. In regular baryons, the three valence quarks are in 
a color-singlet configuration, while in hybrid-baryons they are in a color octet configuration. This difference can result in peculiar features of 
the $Q^2$-evolution of hybrid baryon electroexcitation amplitudes in comparison with those for three-quark systems. Modeling of the evolution of 
the hybrid-baryon electroexcitation amplitudes is critically needed for identification of the hybrid baryon states.

Establishing the spectrum of nucleon resonances is important for elucidating the emergence of strong QCD phenomena from the QCD Lagrangian. 
Coupled-channel analyses of exclusive meson photo- and hadroproduction data have demonstrated that the $N^*$ spectrum is determined by combined 
contributions from both a dressed-quark core and meson-baryon cloud. Currently, LQCD has offered the potential to describe the $N^*$ spectrum starting 
from the QCD Lagrangian and accounting for all active components in the resonance structure with an impact on the generation of the $N^*$ spectrum. 
Therefore, extension of the results on the $N^*$ spectrum are of particular interest for LQCD to advance toward a description of this spectrum using 
quark masses that approach physical values and accounting for all relevant multi-particle configurations. Approaching the physical masses of the 
$N^*$ states requires accounting for the resonance hadronic decays. This can be done in joint efforts between LQCD and amplitude analysis theory.

Establishing the resonance spectrum is a key element in understanding the inner workings of QCD, which is the key objective of the various efforts 
described in this document. Specifically, understanding the nature of color confinement rests in quantitative determination of the role of gluons in 
the spectrum. Besides baryons, as described above, precision studies of the meson spectrum have and will continue, benefiting from new precision data 
collected in collider and fixed-target experiments. For example, the GlueX and CLAS12 programs have dedicated experiments to search for light-quark 
meson resonances in order to identify states that do not fit the conventional quark model. The $J^{PC}=1^{-+}$ exotic hybrid meson is expected to 
have mass in the $1.5-2$~GeV range with a width that is comparable to that of other resonances with a similar mass~\cite{Meyer:2015eta}. The analysis 
of the COMPASS data on the $\eta-\pi$ and $\eta'-\pi$ spectrum~\cite{Adolph:2014rpp} indicates the existence of a pole in the $P$-wave partial wave 
amplitude consistent with existence of just such a resonance~\cite{Rodas:2018owy}. In addition, phenomenological models~\cite{Guo:2008yz}, lattice
QCD~\cite{Dudek:2011bn}, and continuum studies~\cite{Xu:2019sns} imply the existence of other non-$q\bar q$ states in the same mass range, possibly 
containing significant gluonic contributions and filling a quadruplet with $J^{PC} = (0^{-+},1^{-+},1^{--},2^{-+})$ quantum numbers. 

The spectrum of mesons with heavy flavors appears to be much richer than predicted by the naive quark model. Some of the $XYZ$ states observed, for 
example, in $e^+ e^-$ annihilation or $B$-meson decays are candidates for tetraquarks, di-meson molecules, or alternatively could be a kinematic 
effect associated with exchange forces~\cite{Esposito:2016noz}. To reduce this range of possibilities and identify the key mechanisms for producing 
these novel phenomena, much higher statistics are needed along with access to similar states in the bottomonium spectrum. This could be achieved at 
the EIC provided the luminosity is high enough.

The resonance spectrum is determined from partial wave amplitude analysis and requires analytical continuation off the real axis to the complex 
energy plane. This is an unconstrained problem that is approached by either using a microscopic model, {\it e.g.} the quark model or a hadronic 
EFT, or by exploring general analytical properties. The latter result from fundamental properties of the theory, such as unitarity or crossing 
relations. In principle, they can relate the physical behavior of the system, such as the existence of resonances, to amplitude singularities. 
Yet, another approach, which has been successfully used for other ill-defined inverse problems, is to use artificial neural networks in a supervised 
learning approach.

Light-front holography, based on superconformal symmetry, offers predictions for the masses and spin-parities of still unobserved tetraquark states 
from the known meson and baryon spectra. Searches for these exotic states will bridge efforts in the exploration of meson and baryon spectra.

\subsection{Elucidating $N^*$ Structure in Experiments with Electromagnetic Probes}

Analyses of CLAS results on the $N \to N^*$ electroexcitation amplitudes within continuum QCD approaches, quark models, and the coupled-channel 
approaches have revealed the structure of excited nucleon states as a complex interplay between the inner core of three dressed quarks and an external 
meson-baryon cloud~\cite{Burkert:2019bhp,Mokeev:2015lda}. The full range of length-scales through which the transition between the combined contribution 
from the meson-baryon cloud and quark core at intermediate $Q^2 \approx 2.0$~GeV$^2$ to quark core dominance at high $Q^2 > 5.0$~GeV$^2$, will be 
studied using CLAS12 data on exclusive $N\pi$, $N\eta$, $KY$, and $\pi^+\pi^-p$ electroproduction for 2.0~GeV$^2 < Q^2 < 6.0$~GeV$^2$. Precision 
data on the electroexcitation amplitudes of all prominent resonances at these photon virtualities will allow for exploration of the emergence of a 
deconfined cloud of mesons and baryons from the core of the three confined dressed quarks.

In order to achieve this objective, strong theory support is critical. Continuum QCD approaches have already demonstrated promising prospects for a 
description of the emergence of the $N\pi$ component of the meson-baryon cloud from the quark core. The meson-baryon cloud was incorporated or 
described effectively in quark models~\cite{Obukhovsky:2019xrs,Gutsche:2019yoo,Aznauryan:2016wwmO,Aznauryan:2015zta}. Accounting for the complexity 
of the subsequent meson-baryon final state interactions requires the full machinery developed in coupled-channel approaches for the global multi-channel 
analyses of hadro-, photo-, and electroproduction data~\cite{Kamano:2013iva}. A description of the $Q^2$-evolution of resonance electroexcitation 
amplitudes at $Q^2 < 3.0$~GeV$^2$ represents an important milestone for the LQCD efforts. 

Synergistic efforts between experimental studies of resonance electroexcitation amplitudes and theory support in describing emergence of the meson-baryon 
cloud are required to understand the complex interplay between the inner quark core and the external meson-baryon cloud in the structure of distinct 
excited nucleon states. This is especially true because such interactions are likely to depend sensitively on the quantum numbers of the states involved.

\subsection{Exposing the Emergence of Hadron Mass}

The Higgs mechanism accounts for only a small part of the proton's mass. The main part comes from the strongly interacting gluon field. In particular, 
the so called ``trace anomaly", whose strength is driven by gluon self-interactions,  is the most important part; yet, it is the least understood
\cite{XJi1995,Roberts:2016vyn,Lorce:2017xzd}. Owing to the computational difficulty, lattice QCD has no direct predictions thus far. It would be 
very valuable for the the lattice QCD community to deliver sound predictions in the near future. Experimentally, a potentially promising method is 
to measure heavy quarkonium production near threshold, which models have related to the trace anomaly~\cite{Kharzeev1996, Kharzeev1999}. Precision 
measurements of $J/\psi$ threshold production are planned with SoLID~\cite{Chen:2014psa, SoLID:Jpsi} and CLAS12 at JLab~\cite{jpsi_clas12}. Threshold 
$\Upsilon$ production measurements have been discussed for the EIC~\cite{Joosten2018}. These measurements may assist in revealing the trace anomaly's 
contribution to the proton mass.

Consistent results on the momentum dependence of the dressed quark mass from independent studies of nucleon elastic form factors and the 
electroexcitation amplitudes of the $\Delta(1232)3/2^+$ and $N(1440)1/2^+$ resonances have demonstrated the capability of gaining insight into the 
dynamics underlying the dominant part of hadron mass generation~\cite{Segovia:2014aza,Segovia:2015hra,Mokeev:2015lda}. In the near-term, experimental 
results on the resonance electroexcitation amplitudes will become available from CLAS exclusive meson electroproduction data for all prominent nucleon 
resonances in the mass range up to 2.0~GeV and at photon virtualities $Q^2$ up to 5.0~GeV$^2$~\cite{Mokeev:2018zxt,Mokeev:2019ron}. Future analyses 
of these experimental results within the continuum QCD approach will enable us to map out the dressed quark mass function from independent studies of 
the electroexcitation amplitudes of many different excited nucleon states. Mapping the momentum dependence of the dressed quark mass from the 
electroexcitation amplitudes of different resonances will either validate the universality of this function or demonstrate an environmental sensitivity 
of this key feature of strong QCD and the structure of hadrons.

The experimental results on $N^*$ electroexcitation amplitudes available and expected from the CLAS exclusive electroproduction data at 
$Q^2 < 5.0$~GeV$^2$ allow us to explore the dressed quark mass function at quark momenta $< 0.7$~GeV that correspond to distance scales with the 
fully dressed quarks bound within the quark core in the regime of quark-gluon confinement. The CLAS12 results on the electrocouplings of all 
prominent nucleon resonances in the mass range up to 3.0~GeV will become available at $Q^2 < 12$~GeV$^2$~\cite{clas12}. The first results on the 
resonance electroexcitation amplitudes within this $Q^2$ range will allow us to map out the momentum dependence of the dressed quark mass over the 
length scales whereupon the transition from quark-gluon confinement to pQCD regime is expected. The continuum QCD analyses of the CLAS12 results 
on nucleon resonance electroexcitation amplitudes at 5.0~GeV$^2 < Q^2 < 12$~GeV$^2$ will address a key open problem in the Standard Model on the 
emergence of the dominant part of hadron mass from QCD. Consistent results on the dressed quark mass function from the independent studies of 
electroexcitation amplitudes of many resonances with different structure will provide credible insight into the dynamics of hadron mass generation 
in a nearly model-independent way. Progress toward this challenging objective requires coordinated efforts between experimental research on 
resonance structure and QCD-connected theory analyses capable of describing nucleon resonance electroexcitation amplitudes.

The dressed quark mass function that unifies experimental results on nucleon elastic form factors and $N \to N^*$ transition electroexcitation 
amplitudes can also be used for the computation of pion electromagnetic form factors and PDFs within continuum QCD approaches. Hence, conclusions 
drawn about the dynamics of hadron mass generation through studies of the ground/excited-state nucleon structure can be validated via comparisons 
between experimental results on pion form factors/PDFs and the predictions from continuum QCD obtained with the same dressed quark mass function.

Studies of the pion elastic form factor and valence-quark PDFs using the Sullivan processes represent an important research thrust in Halls A/C 
during the JLab 12~GeV era~\cite{E12-07-105}. Independent results on the pion PDFs are expected from pion-induced Drell-Yan reactions in upcoming 
measurements with the new AMBER hadron facility at CERN~\cite{AMBER}. Future experiments with an EIC will further increase 
the kinematic coverage over $Q^2$ and $x$ in exploration of the structure of the pion and other mesons consisting of heavier quarks. Furthermore, 
studies with the EIC will reveal details of the interference between competing contributions from the dressed quark masses and the attractive 
interaction between quarks and anti-quarks that is constrained by DCSB to produce the very small physical 
pion mass. These results are crucial for understanding the dual nature of the pion as ({\it i}) a bound quark system and ({\it ii}) the Goldstone 
boson emerging with DCSB~\cite{Aguilar:2019teb}.

Reliable information on the dressed quark mass function for light $u$ and $d$ quarks will be crucial in the exploration of the momentum dependence 
of the $s$ quark mass function via experimental results on the kaon electromagnetic form factor and PDFs~\cite{Aguilar:2019teb}. Kaon structure will 
be explored in Sullivan processes in Halls A/C at JLab. Continuum QCD analyses of these results using the dressed $u$-quark mass function determined 
in studies of the structure of pions and ground/excited state nucleons will allow the $s$ quark mass function to be charted. Such exploration of the 
flavor dependence of hadron mass generation will deliver unique information on the evolution of hadron mass generating mechanisms from light $u$ and 
$d$ quarks, with the largest contributions from DCSB, to the heaviest $t$ quark, for which the Higgs mechanism is overwhelmingly dominant. With 
continuum QCD predicting an approximate balance between the DCSB and Higgs-mechanism contributions to the $s$-quark mass, studies of the structure of 
hadrons with $s$-quarks are particularly interesting.

The evolution of hadron mass generation with quark flavor represents a novel direction in hadron physics. The dressed quark mass function for $s$ and 
$c$ quarks can be mapped out in studies of the elastic and transition form factors in the time-like region for the mesons and baryons that carry these 
quark flavors and are produced in exclusive photo-/electroproduction channels. Such experiments require a further increase of the electron beam energy 
beyond the available 12~GeV at JLab, while achieving a luminosity not less than 10$^{35}$~cm$^{-2}$s$^{-1}$ presently achieved with the CLAS12 detector 
in Hall~B. Notably, this value of required luminosity is beyond that expected for the US Electron-Ion Collider. Further doubling of the JLab energy 
(up to 24~GeV) while maintaining the currently available luminosity in Hall~B will pave a new avenue in hadron physics, {\it viz.} charting the 
evolution of the dynamics of hadron mass generation with quark flavor.

Synergistic efforts in the exploration of meson and baryon structure in both the space- and time-like regions are very important for resolving key 
open problems in the Standard Model; namely, solutions to the puzzles of: the emergence of hadron mass; the evolution of hadron mass generation 
dynamics with quark flavor; and understanding quark gluon confinement and its probable intimate connection with DCSB.

\subsection{The Structure of Atomic Nuclei from Strong QCD}

Studies of monopole, quadrupole, and rotational collective states of atomic nuclei have demonstrated the relevance of approximate SL(3,R) symplectic 
symmetry as the leading organizing principle for the structure of the atomic nuclei over a broad range of nuclear mass numbers $A$ from light to 
intermediate ranges. The important expectation from this symmetry group supported by experimental studies in low-energy nuclear physics is that 
nuclear deformation and shape coexistence dominate the entire nuclear landscape~\cite{Dytrych_2008,Launey:2016fvy}. The above background on 
symmetry informed advances in {\it ab initio} approaches for studying the structure of atomic nuclei is coupled with progress in gaining a description 
of the ground state structure of nucleons from a strong QCD perspective within continuum QCD approaches. These studies address two important 
questions: ({\it i}) whether the ground state of a nucleon in its intrinsic frame is spherical or deformed, and ({\it ii}) how do the interactions 
between nucleons within nuclei driven by strong QCD generate the dynamic deformation found in atomic nuclei?

Empirical access to nucleon deformation is provided by the pretzelosity TMD-distribution, with a non-zero value of the zeroth-order pretzelosity moment 
indicating a nucleon deformed in its intrinsic frame.

Theoretically, continuum QCD approaches, using well-constrained dressed quark mass functions and diquark correlation amplitudes, mapped out in studies 
of experimental results on nucleon elastic form factors and $N \to N^*$ transition electroexcitation amplitudes, are capable of evaluating the nucleon's 
light-front wave function. This quantity can yield a complete theoretical description of the nucleon's shape in its intrinsic frame. Moreover, 
addressing the emergence of the dynamical deformation seen in the structure of atomic nuclei, continuum QCD approaches can also describe the pion-exchange 
part of the $NN$-interaction with a $NN\pi$ vertex inferred from strong QCD within the same framework used for the description of nucleon shape. Building 
with these results, theoretical approaches for the description of the structure of atomic nuclei as multi-nucleon bound systems can relate such 
expectations from continuum QCD to structural features of atomic nuclei, which are explored in low-energy nuclear physics experiments.

Synergistic efforts in the experimental studies of the ground state nucleon and $N^*$ structure, as well as TMD-pretzelosity, combined with developments 
in continuum QCD approaches and the theory of atomic nuclear structure, pave the way towards understanding how the structure of atomic nuclei emerges 
from strong QCD. The predictions from these studies on particular features in the structure of light and medium-heavy atomic nuclei will promote 
experimental efforts in low-energy nuclear physics, motivating experiments at FRIB and at other nuclear physics facilities around the world.

\subsection{The Longitudinal Structure of Nucleons}

There are a number of recent and upcoming experiments at JLab to study nucleon longitudinal structure in the high $x_B$ (valence quark) region, including
measurements of the $d/u$ ratio~\cite{Tkachenko14,Afnan03,SoLID-PVDIS} and the spin asymmetries $A_1^n$~\cite{Zheng:2003un,E12-06-110} and $A_1^p$ (that 
allow extractions of $\Delta u/u$ and $\Delta d/d$). High-$x_B$ provides a clean region to test theory and model predictions; in particular, to study 
the interplay between strong and perturbative QCD. These precision data will provide stringent tests of our understanding of strong QCD when comparing 
with theoretical calculations.

Moments of the spin structure functions can be directly compared to theoretical predictions through sum rules. Recent and forthcoming measurements
\cite{E97010,Chen2010} of the zeroth moments (spin sum rules) and second moments (spin and color polarizabilities) provide direct comparisons with 
theoretical calculations (LQCD and chiral effective theory), helping to gain insight on chiral symmetry and its breaking pattern in QCD.

The progress achieved in the exploration of nucleon resonance electrocouplings opens up new prospects to gain insight into nucleon PDFs at large values 
of Bjorken $x_B$ within the resonance excitation region. For the first time, the resonance contributions to the inclusive structure functions $F_1$ and 
$F_2$ have been evaluated from experimental results on the resonance electroexcitation amplitudes from CLAS~\cite{HillerBlin:2019hhz}. Knowledge of 
the resonant contributions will enable us to extend results on the unpolarized, spin-averaged nucleon PDFs, summed over all parton flavors, toward large 
$x_B$ in the resonance region. The inclusive $F_1$ and $F_2$ structure functions can be computed as the sum of the term evaluated from the parameterized
PDFs~\cite{Accardi:2012qut} plus the resonant contribution evaluated with resonance electrocouplings from the CLAS results, as described in
Ref.~\cite{HillerBlin:2019hhz}, and compared with experimental data. A combined fit of the PDF parameters to the data, both in the resonant and DIS 
regions, will make it possible to gain new information about the nucleon PDFs in the resonance region at the highest achievable $x_B$ values, limited 
by the applicability of the factorization theorem.

These studies are strongly motivated by theory advances in the description of the nucleon PDFs starting from the QCD Lagrangian within the novel quasi- 
and pseudo-PDF concepts~\cite{Radyushkin:2017cyf,Chambers:2017dov,Ma:2014jla}. This new avenue can forge synergistic efforts between experimentalists,
phenomenologists, and theorists toward understanding the emergence of the unpolarized, spin-averaged PDF from the QCD Lagrangian with coverage over 
$x_B$ values in the resonance region for the first time. These studies will also reveal the interplay between the resonant and non-resonant contributions 
in inclusive structure functions, allowing exploration of quark-hadron duality and its evolution at distance scales from the regime of quark-gluon 
confinement to pQCD. Data taken during the 12~GeV era at JLab will establish the photon virtuality range within which nucleon resonances remain a 
relevant contributor to inclusive electron scattering.

Light-front analyses have made interesting predictions on the presence of partons heavier than the $u$ and $d$ quarks in the ground state nucleon at 
large $x_B$. This prediction is based on the behavior of the light-front wave functions that support minimal off-shellness for these partons at large
$x_B$~\cite{Brodsky:hq,Brodsky:hq1}. This effect was not expected from the usual gluon splitting mechanism. Studies of $\phi$ and $J/\psi$ semi-inclusive 
meson electroproduction in experiments at the 12~GeV JLab and further extension of these efforts with the EIC offer prospects to check this important 
and novel prediction.

\subsection{3D Nucleon Structure and its Emergence from QCD}

Advances in the exploration of DVCS and DVMP have already provided initial insight into the structure of the nucleon in the 3D space defined by $x_B$ 
and the two spatial coordinates in the plane transverse to the virtual photon. This progress with DVCS~\cite{Burkert:2018bqq} promises to deliver a 
chart of the pressure distribution in nucleons, extending access to the ground state nucleon energy-momentum tensor. Progress in SIDIS studies offers 
a complementary picture of the nucleon structure in 3D momentum space. 3D nucleon femtography represents a central direction in the JLab 12~GeV 
physics program. Synergistic interactions between different areas in hadron physics that use electromagnetic probes are of particular importance as 
attempts are made to extend the range of insights into the 3D structure of the nucleon and, potentially, its excited states; especially in connecting 
observations with emergent phenomena in QCD.

So far, GPD structure functions have been related to DVCS/DVMP observables assuming the contribution from handbag diagrams and, in the case of DVCS, 
also from the Bethe-Heitler amplitude. However, the DVCS/DVMP exclusive channels at $W < 2.5$~GeV should also include contributions from 
well-established resonances excited in the virtual photon/proton $s$-channel. Use of the resonance electroexcitation amplitudes determined independently 
from the CLAS/CLAS12 exclusive meson electroproduction data (see Section~\ref{nstar_structure_intro}) offers a realistic evaluation of the resonant 
contributions to the DVCS/DVMP processes at the amplitude level. Accounting for nucleon resonance contributions to DVCS/DVMP will allow for extension 
of the scope of the 3D nucleon structure studies toward smaller invariant masses of the virtual photon/proton system. Adding the resonant contributions 
to the diagrams traditionally used in DVCS/DVMP analyses will enable exploration of how stable the information on ground state nucleon GPD structure 
functions is against the implementation of the other relevant amplitudes. Moreover, these studies will reveal whether there is a need for further 
extension of the analysis frameworks used for the extraction of GPDs.

Available information on the 3D structure of the ground state nucleon will be augmented by novel results on the 3D structure of nucleon excited states 
as soon as the transition $N \to N^*$ GPDs are determined from exclusive $\gamma_vp \to N\pi\gamma_r$ electroproduction data. Studies with CLAS have 
demonstrated pronounced resonance-like peaks in the $N\pi$ invariant mass for these final states at mass values corresponding to the $\Delta$ resonance, 
and to the second and third resonance regions seen in inclusive/exclusive electroproduction. These structures in the $N\pi$ invariant mass are suggestive 
of contributions from the processes $\gamma_vp \to \gamma_rN^* \to N\pi\gamma_r$. The $\gamma_vp \to \gamma_rN^*$ amplitudes contain the transition 
$N \to N^*$ GPDs, allowing them to be constrained by fitting to observables in the $\gamma_vp \to \gamma_rN\pi$ exclusive channel. This activity 
requires input from theory on modeling the off-diagonal $N \to N^*$ transition GPDs. Experimental information on the $N^*$ electroexcitation amplitudes 
from CLAS/CLAS12 will provide valuable constraints on the modeling of transition GPDs.

Continuum and lattice QCD approaches promise a connection between the 3D images of the nucleon, from the GPD and TMD distributions, and the strong QCD 
dynamics underlying formation of the ground state nucleon. Indeed, continuum QCD approaches are capable of computing the nucleon's light-front wave 
function by employing dressed quark mass functions and diquark correlation amplitudes evaluated with a traceable connection to the QCD Lagrangian. Such 
a step is valuable because all GPDs and TMDs can be evaluated from the nucleon's light-front wave function. 

Studies of GPDs and TMDs within continuum QCD approaches may not only elucidate the emergence of hadron mass, but also allow for investigation of the 
relevance and emergence of diquark correlations in the structure of ground state nucleons. As was discussed in Section~\ref{cqcd_prosp}, continuum 
QCD approaches predict particular features in the nucleon and Roper resonance PDAs related to the generation of two types of diquark correlation with 
spin-parities $0^+$ and $1^+$. In the computed PDAs, the momenta of two of the quarks are much closer in comparison with the momentum of the third 
quark, suggesting that two of the quarks are correlated and the third quark is uncorrelated. Furthermore, implementation of the two types of diquark 
correlations, $0^+$ and $1^+$, shifts the continuum QCD expectations for the PDA moments more into line with existing LQCD results (see details in 
Section~\ref{cqcd_prosp}). Studies of ground state nucleon structure in 3D offer experimental opportunities to check these expectations, obtained 
within two conceptually different approaches to solving the problem posed by the QCD Lagrangian.

Knowledge of the ground state nucleon light-front wave function will allow us to evaluate the GPD and TMD structure functions. They can be then be 
inserted into the reaction models that relate the GPDs/TMDs to the DVCS/SIDIS observables. The successful description of many different DVCS and TMD 
observables will validate the reaction models used in order to connect the GPDs and TMDs to the measured observables.

The first results on the pressure distribution in the nucleon and results on the moments from different combinations of GPDs will experimentally 
constrain many components of the energy-momentum tensor within the ground state nucleon. Sound theory predictions for these quantities are therefore
much in demand. Meeting this requirement is a challenging task; but it is essential if science is to fully capitalize on planned experiments and 
facilities. 

Completing a 3D femtography program relating to ground and excited state nucleons, including charting their mechanical properties, will considerably 
enhance understanding of the strong QCD dynamics underlying baryon generation. Success requires the combined efforts of experimentalists, 
phenomenologists, and theorists, both from $N^*$ and DIS physics.

\section*{Acknowledgments}

We are grateful to the following people for constructive comments during the preparation of this manuscript:
I.~Aznauryan
D.~Binosi,
L.~Chang,
X.~Chen,
F.~De~Soto,
A.~Deur,
M.~Ding,
R.~Ent,
F.-X.~Girod
B.-L.~Li,
T. Liu,
N.~Markov,
M. Nielsen,
J.~Papavassiliou,
E.~Pasyuk,
K.~Raya,
E.~Santopinto
S.M.~Schmidt,
C.~Shi,
R.S. Sufian,
S.-S.~Xu,
P.-L.~Yin,
S.~Zafeiropoulos,
J.-L.~Zhang, and
L. Zou.

We also express our gratitude to SURA/JSA and Deputy Director of Jefferson Lab R. McKeown for generous support of this workshop, as well as 
to the JLab Event Services group for their efficient help in the workshop organization. Work supported by: the United States Department of Energy 
under DOE Contracts DE-AC05-06OR23177 and DE-AC02-76SF00515; National Natural Science Foundation of China, under grant no.\ 11805097; Jiangsu 
Province Natural Science Foundation, under grant no.\ BK20180323; Jiangsu Province {\it Hundred Talents Plan for Professionals}; the Spanish 
Ministerio de Econom\'ia, Industria y Competitividad under contract No. FPA2017-86380-P and the Junta de Andaluc\'ia under contract No. UHU-1264517;
the UK Science and Technology Facilities Council (STFC); and the European Research Council (ERC) under the European Union’s Horizon 2020 Research 
and Innovation Program (grant no. 647981, 3DSPIN).

\end{document}